\newif\ifelsevier
\elseviertrue

\ifelsevier

\documentstyle [epsf,bezier]{elsart}
\else
\documentstyle [floats,epsf,preprint,aps,eqsecnum,bezier]{revtex}
\tightenlines
\fi

\let\scr=\cal

\def\dim{\nu}
\def\r{{\bf r}}
\def\k{{\bf k}}
\def\p{{\bf p}}
\def\q{{\bf q}}
\def\Li{{\rm Li}}

\def\Scl{S_{\rm cl}}
\def\Sone{\Delta S}
\def\Sct{S_{\rm ind}}
\def\Seff{S_{\rm eff}}
\def\Sint{S_{\rm int}}
\def\Sind{S_{\rm ind}}
\def\Wone{S_{\rm int}}
\def\n#1#2{n^{\ifx0#10\else(#1)\fi}_{#2}}
\def\half{\coeff12}
\def\coeff#1#2{{\textstyle{#1\over #2}}}
\def\expect#1{\left\langle #1 \right\rangle_\beta}
\def\avg#1{\left<\!\left<#1\right>\!\right>}
\def\vol{{\scr V}}
\def\Gqm{{\scr K}}
\def\gbar{\Gamma}
\def\Gam#1#2{\Gamma_{#1#2}}
\def\Gamprime#1#2{\Gamma'_{#1#2}}
\ifelsevier
\def\bxi{{\bf \xi}}		
\else
\def\bxi{\mbox{\boldmath$\xi$}}
\fi

\global\arraycolsep=2pt         

\ifelsevier
\makeatletter
\@addtoreset{equation}{section}%
\@namedef{theequation}{\thesection.\arabic{equation}}
\makeatother
\advance\hoffset -1.0cm
\else
\makeatletter	
\def\footnotesize{\@setsize\footnotesize{9.5pt}\xpt\@xpt
\abovedisplayskip 10pt plus2pt minus 5pt
\belowdisplayskip \abovedisplayskip
\abovedisplayshortskip \z@ plus 3pt
\belowdisplayshortskip 6pt plus 2pt minus 2pt
\def\@listi{\topsep 6pt plus 2pt minus 2pt
\parsep 3pt plus 2pt minus 1pt \itemsep \parsep}}
\makeatother
\fi

\begin{document}
\def\floatpagefraction{.1}

\ifelsevier
\begin{frontmatter}
\fi
\title {Effective Field Theory for Highly Ionized Plasmas}

\author {Lowell S. Brown and Laurence G. Yaffe}

\address {Department of Physics, University of Washington,
	Seattle, Washington 98195--1560}

\date{\today}

\ifelsevier\else
\maketitle
\thispagestyle {empty}
\fi

\begin {abstract}
\ifelsevier\else
\leftskip=0.05\textwidth\rightskip\leftskip
\elvrm
\fi
We examine the equilibrium properties of
hot, non-relativistic plasmas.
The partition function and density correlation functions of
a plasma with several species are expressed in terms
of a functional integral over electrostatic potential distributions.
This is a convenient formulation for performing a
perturbative expansion.  The theory is made well-defined at every
stage by employing dimensional regularization which, among other
virtues, automatically removes the unphysical (infinite) Coulomb
self-energy contributions.
The leading order, field-theoretic tree approximation
automatically includes the effects of Debye screening.
No further partial resummations are needed for this effect.
Subleading, one-loop corrections are easily evaluated.
The two-loop corrections, however, have ultraviolet divergences.
These correspond to the short-distance, logarithmic divergence which is
encountered in the spatial integral of the Boltzmann exponential when
it is expanded to third order in the Coulomb potential.
Such divergences do not appear in the underlying quantum theory ---
they are rendered finite by quantum fluctuations.
We show how such divergences may be removed and the
correct finite theory obtained by introducing additional local interactions
in the manner of modern effective quantum field theories.
We compute the two-loop induced coupling by exploiting a
non-compact $su(1,1)$ symmetry of the hydrogen atom.
This enables us to obtain explicit results for density-density
correlation functions through two-loop order and thermodynamic quantities
through three-loop order.
The induced
couplings are shown to obey renormalization group equations, and these
equations are used to characterize all leading logarithmic contributions
in the theory.
A linear combination of pressure plus energy and number densities is
shown to be described by a field-theoretic anomaly.
The effective Lagrangian method that we employ yields a simple
demonstration that, at long distance,
correlation functions have an algebraic fall off
(because of quantum effects) rather than
the exponential damping of classical Debye screening.
We use the effective theory to compute, easily and explicitly,
this leading long-distance behavior of density correlation functions.

The presentation is pedagogical and self-contained.
The results for thermodynamic quantities
at three-loop [or $O(n^{5/2})$] order,
and for the leading long-distance forms of correlation functions,
agree with previous results in the literature, but they are obtained
in a novel and simple fashion using the effective field theory.
In addition to the new construction of the effective field theory
for plasma physics,
we believe that the results we report
for the explicit form
of correlation functions at two-loop order, as well as the
determination of higher-order leading-logarithmic contributions,
are also original.
\end {abstract}
\ifelsevier
\end {frontmatter}
\fi

\newpage
\ifelsevier
\bgroup
\parskip 2.4pt
\tableofcontents
\egroup
\else
\tableofcontents
\fi
\newpage

\section{Introduction and Summary}

Our work applies contemporary methods of effective quantum field
theory to the traditional problem of a multicomponent, fully ionized
hot (but non-relativistic) plasma. In this regime, a classical
description might appear to suffice. But the short-distance ($ 1/ r$)
singularity of the Coulomb potential gives rise to divergences in
higher-order terms.
Taming these divergences requires the introduction of
quantum mechanics. Quantum fluctuations smooth out the
short-distance singularity of the Coulomb potential so that the quantum,
many particle Coulomb system is completely finite.
This necessity for including quantum effects, even in a dilute plasma,
is discussed later in this introduction
when the relevant parameters which characterize the various
physical processes in the plasma are examined.

As we shall see, contemporary effective quantum field theory methods
simplify high-order perturbative computations and
generally illuminate the structure of the theory.
Effective quantum
field theories do, however, utilize a somewhat complicated formal
apparatus involving regularization, counter terms, and renormalization.
In an effort to make our work available to a wider audience,
we shall develop the theory in several stages,
and attempt to give a largely self-contained presentation.%
\footnote
    {%
    Other discussions of effective field theory techniques,
    applied to quite different physical problems,
    may be found in Refs.~\cite {eff-thy1,eff-thy2} and references therein.
    }
A brief review of some of the basic quantum field theory techniques
used in our paper is presented in Appendix \ref{someqft}.
We begin, in Section \ref{sec:classical}, by
casting the purely classical theory in terms of a functional integral.
We show how dimensional continuation is convenient even at this purely
classical level because it automatically and without any effort
removes the infinite Coulomb self-energy contributions of the
particles.  The simple saddle-point
evaluation of the functional integral ---
known as a tree approximation in quantum field theory ---
immediately gives
Debye screening of the long-distance Coulomb potential
without any need for the resummations appearing in traditional approaches.
The first sub-leading, so-called ``one-loop,''
corrections to the plasma thermodynamics and correlation functions are
also evaluated in this section.%
\footnote
    {
    The saddle-point of the functional integral defines
    the ``mean-field'' solution.
    The order of the perturbative expansion about this
    saddle-point is commonly referred to as the ``loop'' order
    in the quantum field literature because contributions at the $k$-th
    order in perturbation theory are graphically represented by
    $k$-loop diagrams.
    As will become clear later,
    in thermodynamic quantities such as the pressure,
    contributions of $k$-loop order correspond
    to terms which formally depend on the mean particle density $n$ as
    $O(n^{1+{k\over2}})$,
    up to logarithms of the density.
    }
These lowest-order results are also used to illustrate general relations
among correlations functions which are described more formally,
and systematically, in Appendix \ref{funmeth}.

The divergence associated with the singular, short-range behavior of
the Coulomb potential first arises at the subsequent, ``two-loop'' level of
approximation as shown in Section \ref{sec:quasi}.
This section explains
how the previous purely classical theory is obtained
from a limit of the quantum theory, and how the quantum corrections
that tame the classical divergences appear in the form of induced couplings
that contain compensating divergences.
This discussion uses various results on functional determinants and Green's
functions contained in Appendix \ref{app:det}.
Although it is relatively easy to construct
the ``counter terms'' that render the classical theory finite, it is
considerably more difficult to obtain the finite pieces in the
induced couplings that ensure that a calculation in the effective theory
correctly reproduces the corresponding result in the full quantum theory.
In the latter part of this section the ``matching conditions'' for the
leading two-loop induced couplings are derived,
and the two-loop induced couplings are explicitly evaluated.
The key to this evaluation is the exploitation of
an $su(1,1)$ symmetry of the Coulomb problem, which permits one to
derive a simple and explicit representation for the two-particle
contribution to the density-density correlation function.
This $su(1,1)$ symmetry, and its consequences, are presented in
Appendix \ref{app:SU(1,1)}.
Section III concludes with an examination of the
necessary inclusion in the effective theory of interactions
involving non-zero frequency components of the electrostatic potential.

It is worth noting that our determination of the induced
couplings is based on examining Fourier transforms of number-density
correlation functions at small but non-vanishing wave number.
We use this method because these functions --- at non-vanishing wave number
--- may be computed in a strictly perturbative fashion
without taking into account the Debye screening that is
necessary to make the zero wave number limit of these correlation functions
finite.%
\footnote
    {
    The utility of matching at small but non-vanishing wave number,
    thereby enabling one to ignore the effects of Debye screening,
    has been emphasized by Braaten and Nieto \cite {braaten}.
    }
Matching in this fashion enables us to use the
simple pure Coulomb potential for which the exact group-theoretical
techniques apply.
Our procedure is roughly equivalent to computing the
second-order virial coefficient for a pure Coulomb potential, except
that this coefficient has a long-distance, infrared divergence.
This logarithmic divergence is removed by Debye screening, but there is
always the difficulty of determining the constant under the logarithm.
Our method avoids this difficulty.
Years ago,
W.~Ebeling [and later Ebeling working together with collaborators]
computed the second-order
virial coefficient for a pure Coulomb potential with a long-distance
cutoff, and then related this quantity to other ladder approximation
calculations so as to obtain results that are, except for one term,
equivalent to, and consistent
with, our results for the induced couplings. Their work is summarized
in Ref.~\cite{ebeling}.
This seminal work is certainly very impressive and significant,
but it is much more complex than our approach,
and (at least in our view) is far more difficult to understand in detail.

With the leading induced couplings in hand,
we turn in Section \ref{sec:twolrs}
to compute all the thermodynamic quantities and the density-density
correlators to two-loop order.
As far as we have been able to determine,
the two-loop results for the density-density correlation functions
obtained in Section \ref{sec:twolrs} are new.
Various integrals required for these computations are evaluated in
Appendix \ref{required}, and
an alternative derivation of the two-loop thermodynamic results
using compact functional methods appears in Appendix \ref{funcal}.

The thermodynamic results are extended to the next, three-loop,
order in Section \ref{sec:threeloop}.
We give complete, explicit results for the pressure (or equation of state),
Helmholtz free energy, and internal energy,
as well as the relations between particle densities and
chemical potentials in a general multi-component plasma.
We also display the specializations of the equation of state for the
cases of a binary electron-proton plasma, and a one component plasma
(in the presence of a constant neutralizing background charge density).
As discussed at the end of this section,
a genuine classical limit exists only for the
special case of a one-component plasma.
As a check on our results,
Appendix \ref{qfluck} presents an independent, self-contained calculation
of the leading $O(\hbar^2)$ corrections to the equation of state
of a one-component plasma in the semi-classical regime.%
\footnote {This is a classic result which may also be found in \cite {hansen}.}

Prior results in the literature, corresponding to our three-loop
[or $O(n^{5/2})$] level of accuracy,
for the free energy and/or the equation of state
go back more than 25 years.
The book by Kraeft, Kremp, Ebeling, and R\"opke \cite {book}
quotes a result for the Helmholtz free energy which
is nearly correct but omits one term.%
\footnote
    {
    See Eq's.~(2.50)--(2.55) of Ref.~\cite {book}.
    See footnote \protect\ref {fn:misprint} on page
    \protect\pageref {fn:misprint} for details.
    The missing ``quantum diffraction'' term noted in that footnote
    was obtained for the electron gas in a neutralizing
    background by DeWitt. See Ref.~\cite{dewitt} and references therein.
    }
A fairly recent publication by Alastuey and Perez \cite {AP}
contains the complete, correct expression for the Helmholtz free energy,
to three loop order,
with which we agree.
Recent papers by
DeWitt, Riemann, Schlanges, Sakakura, and Kraeft~\cite{dewitt2,dewitt1,dewitt4}
report results for some, but not all, of the terms contained in the
three-loop pressure.
These partial results are consistent with our three-loop pressure,
once an unpublished erratum of J.~Riemann is taken into account.

Just as in any effective field theory, the induced couplings that must
be introduced to remove the infinities of the classical plasma theory
obey renormalization group equations. In Section~\ref{sec:higher} we
show how these renormalization group equations may be employed to
compute leading logarithmic terms in the partition function ---
terms involving powers of logarithms whose argument is the
(assumed large) ratio of the Debye screening length to the
quantum thermal wave length of the plasma.
We show that, in general, log$^k$ terms first arise at $2k$-loop order.
We derive a simple recursion relation which determines their explicit form.
We explicitly evaluate the coefficients of the logarithm-squared terms
which arise at four and five loop order, and the log-cubed terms which
appear at six loop order.
The existence of these higher powers of logarithms
which first appear at four loop order
(where they correspond to $n^3 \ln^2 n$ contributions to the pressure)
has often not been recognized.%
\footnote{For example, Refs.~\cite{AP} and \cite{kahlbaum}
imply that the next correction is of the form $n^3 \ln n$ rather than the
correct $n^3 \ln^2 n$.}
Recently, Ortner \cite{ortner} has examined a classical plasma consisting
of several species of positive ions moving in a fixed neutralizing
background of negative charge.  He computed the free energy to
four-loop (or $n^3$) order and correctly obtained the $n^3 \ln^2 n$ term.

Since Planck's constant, which carries the dimensions of action, does
not appear in classical physics, fewer dimensionless ratios can be
formed in a classical theory than in its quantum counterpart.
In particular, the partition function of the classical theory depends
upon a restricted number of dimensionless parameters, from which a
linear relation between the pressure, internal energy, and average
number densities follows. This relation is altered by the necessary
quantum-mechanical corrections. Section \ref{sec:higher} also explains
how this alteration of the linear relationship is connected to
``anomalies'' brought about by the renormalization procedure that
makes the classical theory finite.

We conclude our work, in Section \ref{sec:longdist},
with an examination of the long-distance behavior of the density-density
correlation function.
Despite the presence of Debye screening, it is known that
quantum fluctuations cause correlations to fall only
algebraically with distance \cite {alg-decay1,alg-decay2,alg-decay3,cornu}.
Using the effective theory, we compute the coefficient of the
resulting leading power-law decline in particle- and charge-density
correlators in a very simple and efficient fashion, and obtain
results which agree with previously reported asymptotic forms
\cite {alg-decay3,cornu}.

It should be emphasized that the major purpose of this paper is to
introduce the methods of modern effective field theory into the
traditional field of plasma physics. Although many of the results that
we derive and describe have been obtained previously, the methods that we
employ to obtain these results are new, and they substantially reduce
the computational effort as well as illuminating the general structure
of the theory. Although our work may have the length, it is not a
review paper; its length results from our desire to make the presentation
self-contained so that it may be read by someone who is neither an
expert in plasma physics nor quantum field theory.
Since our work is not a survey of the field,
we have not endeavored to provide anything resembling a
comprehensive bibliography.
For a recent review of known results concerning Coulomb plasmas
at low density,
including rigorous theorems and detailed discussion of the long
distance form of correlators,
we refer the reader to Ref.~\cite {brydges&martin}
and references therein.%
\footnote
    {
    During the final preparation of this report, we became aware of
    a paper by Netz and Orland \cite {netz&orland} that
    employs a functional integral representation for a
    classical plasma which is similar in spirit to our formulation.
    That paper considers only the special cases of one-component,
    or charge symmetric two-component plasmas
    and, moreover, does not address the inclusion
    quantum effects which we deal with using
    effective field theory techniques.
    }

\subsection {Relevant Scales and Dimensionless Parameters}

Various dimensionless parameters characterize the relative importance
of different physical effects in the plasma.
Before plunging into the details of our work,
we first pause to introduce these parameters and
discuss their significance.

Let $e$ and $n$ denote the charge and number density of a typical
ionic species in the plasma.  For simplicity of presentation in
this qualitative discussion, we shall assume that the charges and
densities of all species in the multicomponent plasma are roughly
comparable, and shall ignore the sums over different species which
should really be present in formulas such as (\ref {eq:kappasq})
below.  The subsequent quantitative treatment will, of course, remedy
this sloppiness.
We shall be concerned with neutral plasmas which are sufficiently dilute
so that the average Coulomb energy of a particle is small compared to
its kinetic energy.  We use energy units to measure the temperature
$T$ and write $ \beta = 1 / T $.  In the ideal gas limit,
the average kinetic energy is equal to  $\coeff 32T$.
The Coulomb potential is $e^2 / (4\pi r)$
in the rationalized units which we shall use.
So the typical Coulomb energy is $e^2 / (4\pi d)$
where $d \equiv n^{-1/3}$ denotes the mean inter-particle separation.
Hence, the dimensionless parameter
\begin {equation}
    \gbar \equiv {e^2 \over 4\pi d \, T}
    =
    {\beta e^2 \over 4\pi} \, n^{1/3}
\end {equation}
is essentially the ratio of the potential to kinetic energy in the
plasma, and it is an often used measure of the relative strength of Coulomb
interactions in a plasma.
However, we shall see that $\gbar$ is not the proper dimensionless
parameter which governs the size of corrections in the classical
perturbation expansion.

A charge placed in the plasma is screened by induced charges.
The screening length equals the inverse of the Debye wave number
which we denote as $\kappa$.  It is given
(to lowest order in a dilute plasma) by
\begin{equation}
    \kappa^2 = \beta e^2 n \,.
\label {eq:kappasq}
\end{equation}
A different measure of the strength of Coulomb interactions in the
plasma is defined by
\begin {equation}
    g = {\beta e^2 \kappa \over 4\pi} \,.
\end {equation}
This is the ratio of the electrostatic energy of two particles
separated by a Debye screening length to the temperature (which is
roughly the same as the average kinetic energy in the plasma).
Equivalently, it is ratio of the ``Coulomb distance''
\begin {equation}
    d_C \equiv {\beta e^2 \over 4\pi}
\label{Cdist}
\end {equation}
to the screening length $\kappa^{-1}$.
The Coulomb distance $d_C$ is the separation at which the
electrostatic potential energy of a pair of charges equals the temperature.%
\footnote
    {
    Dynamically, the Coulomb distance $d_C$ is also
    the impact parameter necessary for an $O(1)$ change
    in direction to occur during the scattering
    of a typical pair of particles in the plasma.
    }

The number of particles $N_\kappa$ contained within a sphere
whose radius equals the screening length $\kappa^{-1}$
is inversely related to $g$,
\begin {equation}
    N_\kappa = {4\pi\over 3} \, \kappa^{-3} \, n
    =
    {1 \over 3g} \,.
\end {equation}
Hence the weak coupling condition $g \ll 1$ is equivalent to the
requirement that the number of charges within a ``screening volume''
be large, $N_\kappa \gg 1$. In this case, a mean-field treatment of
Debye screening holds to leading order, and perturbation theory is
a controlled approximation.

It is easy to check that the two measures of interaction strength, $g$
and $\gbar$, are related by $g = \sqrt{4 \pi \, \gbar^{\,3}} $.
However, we shall show in our subsequent development that $g$, not
$\gbar$, is the dimensionless parameter whose increasing integer
powers characterize the size of successive terms in the classical
perturbative expansion for thermodynamic properties of the plasma.  As
we shall discuss, the classical perturbation series has a convenient
graphical representation in which contributions at $n$-th order in
perturbation theory are represented by graphs (or Feynman diagrams)
with $n$ loops.  We shall see that $g$ is the ``loop expansion''
parameter, such that contributions represented by $n$-loop graphs are
of order $g^n$.  Although $g$ and $\gbar$ are directly related as
noted above, we emphasize again that it is $g$ which is the correct
classical expansion parameter.

To bring out this point even more strongly, we note that the
screened Debye potential between two charges $e_a$ and $e_b$ a distance
$r$ apart is given by $ e_a e_b \, e^{ - \kappa r } / (4\pi r) $. The
modification of the self energy of a particle of charge $e_a$ when it
is brought into the plasma is given by half the difference between
the Debye potential and its Coulomb limit for the case of zero charge
separation,
$\half \lim_{r \to 0} \, ( e_a^2 / 4\pi r) [ e^{ - \kappa r } - 1 ] $,
which is $ - e_a^2 \, \kappa /(8\pi) $. Each
particle in the plasma makes this correction to the thermodynamic
internal energy of the plasma, and so including this leading order
correction to the energy density gives
\begin{eqnarray}
u &=& {\sum}_a \left\{ {3 \over 2} \, T \, n_a -
{ e_a^2 \kappa \over 8 \pi } \, n_a \right\}
\nonumber\\
&=&  T \, {\sum}_a \, n_a \left\{
{3 \over 2}  - {\beta  e_a^2 \kappa \over 8 \pi }  \right\}
\,,
\end{eqnarray}
which shows the appearance of $g$,
or more explicitly $\beta e_a^2 \kappa / (8\pi) $,
as the correct Coulomb coupling constant in this case. This result for
the internal energy which we have heuristically obtained agrees
with the correct one-loop result (\ref{oneu}) that is derived below.

A classical treatment of a plasma with purely Coulombic interactions
is, however, never strictly valid.  The classical partition function
fails to exist due to the singular short-distance behavior of the
Coulomb interaction.  This can be seen in an elementary fashion
directly from the divergence, for opposite signed charges, of the
Boltzmann-weighted integral over the relative separation of two
charges, $ \int (d^3\r) \> \exp \{\beta e^2 / 4\pi r\} $.  In the
perturbative expansion of the classical theory, this problem first
manifests itself at two-loop order through the diagram
\begin {equation}
\raisebox {-12pt}{
\begin {picture}(40,40)(0,0)
    \thicklines
    \put(20,20){\circle{40}}
    \put(0,20){\line(1,0){40}}
    \put(0,20){\circle*{5}}
    \put(40,20){\circle*{5}}
\end {picture}
}
\end {equation}
The three lines in this graph correspond to the three factors of the
Coulomb interaction energy $(e^2 / 4 \pi r )^3$ that appear in the
expansion of the Boltzmann exponential to third order.  This graph
represents a relative correction to the partition function of\,%
\footnote
    {Note that this contribution is indeed of order $g^2$,
    in accordance with its origin as a two-loop graph.
    }
\begin{equation}
    { 1 \over 3!} \, n \, \beta^3 \int (d^3\r)
    \left( { e^2 \over 4 \pi r } \right)^3 =
    { 1 \over 3! } \left( { \beta e^2 \kappa \over 4 \pi } \right)^2
    \int { dr \over r} \,.
\label {eq:div}
\end{equation}
Once screening effects are properly included,
the large-distance logarithmic divergence of this integral will be
cut off at the classical Debye screening length $\kappa^{-1}$.
But no classical mechanism exists to remove the short-distance divergence
of the integral.
To tame this divergence, one must include quantum effects.

The non-relativistic quantum-mechanical description
of a charged plasma is completely finite;
quantum fluctuations cut-off the short-distance divergences of the
classical theory. The de Broglie wavelength for a particle of
mass $m$ and kinetic energy comparable to the temperature is of order
\begin {equation}
    \lambda \equiv \hbar \sqrt {{2\pi \beta \over m}} \,.
\label{lambda}
\end {equation}
This is in accord with the average (rms) momentum of $\sqrt{ 3m /\beta}$
for a particle in a free gas at temperature $T = 1/\beta$.
We will refer to $\lambda$ as the ``thermal wavelength''.
This length sets the scale of the limiting precision with which
a quantum particle in the plasma can be localized.
Using the thermal wavelength as the lower limit
in the integral (\ref {eq:div}),
and the Debye length as the upper limit,
one obtains a finite result,
\begin{equation}
    \int_\lambda^{~~~\kappa^{-1}} { dr \over r} = - \ln(\lambda \kappa) \,,
\label{render}
\end{equation}
which replaces the infinity that would otherwise arise
in a purely classical treatment.
This logarithm of the ratio of a quantum wavelength to the screening
length will necessarily appear in coefficients of two-loop
(and higher order) contributions to all thermodynamic
quantities.%
\footnote
    {%
    A one-component plasma (with an inert,
    uniform background neutralizing charge density) has only repulsive
    Coulomb interactions. In this special case, the Boltzmann factor
    $\exp\{ - \beta e^2 / 4 \pi r \} $ itself provides a short-distance
    cutoff at the Coulomb distance $ d_C = \beta e^2 / 4 \pi $, resulting in
    logarithmic terms of the form $\ln ( d_C \kappa ) = \ln g$.
    [In this regard, see Eq.~(\ref{repulse}) and its discussion.]
    However, if the quantum thermal wavelength is larger than the
    Coulomb distance, $ \lambda > d_C $, then this purely classical removal
    of the would-be short-distance divergence is physically incorrect,
    for the quantum effects already come into play at larger distances,
    and the correct logarithmic term has the form $\ln ( \lambda \kappa ) $.
    The neutrality of a binary or multicomponent plasma requires
    that they have attractive as well as repulsive Coulomb interactions.
    These plasmas thus always require quantum-mechanical fluctuations
    to remove their potential short-distance divergences.\label{foot}}

This quick discussion shows that quantum mechanics must enter
into the description of the thermodynamics of a plasma ---
at least if two-loop or better accuracy is desired.
In addition to regularizing the divergences of the classical theory,
quantum mechanics also provides ``kinematic'' corrections
via the influence of quantum statistics.
To estimate the size of these effects, we recall
that for a free Bose ($-$) or Fermi ($+$) gas, the partition function
is given by
\begin{equation}
  {\ln Z \over \vol} = \mp g_S \int { (d^3\p) \over (2 \pi \hbar )^3 } \,
    \ln \left[ 1 \mp
	\exp \biggl\{ -\beta\biggl( { p^2 \over 2m} - \mu\biggr) \biggr\}
    \right] .
\label{qstat}
\end{equation}
Here $\vol$ is the volume containing the system, $g_S = 2S + 1$ is the
spin degeneracy factor, $\mu$ is the chemical potential of the
particle, and
\begin{equation}
    z \equiv e^{\beta \mu}
\end{equation}
is the corresponding fugacity.  The limit of classical
Maxwell-Boltzmann statistics is obtained when $ - \beta \mu \gg 1$ so
that the fugacity $ z \ll 1$.  Near this regime, the logarithm in
Eq.~(\ref{qstat}) may be expanded in powers of the fugacity, and the
resulting Gaussian integrals then yield
\begin{equation}
  {\ln Z \over \vol} = g_S \, \lambda^{-3} \, z \,
    \left[
	1 \pm { z \over 2^{5/2} } + { z^2 \over 3^{5/2} } + \cdots
    \right] .
\label{fugex}
\end{equation}
The corresponding number density defined by
$
n \, \vol  =  \partial \ln Z / \partial (\beta \mu )
$
is given by
\begin{equation}
    n = g_S \, \lambda^{-3} \, z
    \left[ 1 \pm {z \over 2^{3/2} } + {z^2 \over 3^{3/2} } + \cdots \right] .
\label {fugex-n}
\end{equation}
We shall always assume that the plasma is dilute,
\begin{equation}
    {n \, \lambda^3 \over g_S} \ll 1 \,,
\end{equation}
so that a fugacity expansion is appropriate.
This condition that the plasma be dilute can be stated in another way.
If all single-particle states in momentum space were filled up to a
(Fermi) momentum $p_F$, the density would take on the value
$ n = g_S \, p_F^3 \, / ( 6 \pi^2 \hbar^3) $
corresponding to a non-interacting Fermi gas at zero temperature.
The diluteness condition is equivalent to the
requirement that the Fermi energy $E_F = p_F^2 / 2m $
corresponding to the given density be small in comparison with the temperature,
\begin{equation}
    { E_F \over T }
    = {\beta \hbar^2 \over 2m} \left( {6\pi^2 n \over g_S} \right)^{2/3}
    = \left( { 9\pi \over 16} \, z^2 \right)^{1/3} \ll 1 \,.
\end{equation}
However, it is the fugacity $z$, not this ratio, that is the appropriate
expansion parameter.

Once quantum mechanics enters the analysis,
another dimensionless parameter
involving the ratio of two energies appears.
This is the Coulomb potential energy for two particles separated
by one thermal wavelength, divided by the temperature,
\begin{equation}
    \eta = { \beta e^2 \over 4 \pi \lambda } \,.
\end{equation}
Recalling the definition (\ref{lambda}) of the thermal wave length and
noting that the average (rms) particle velocity in a free gas is given
by $ \overline v = \sqrt{ 3 / \beta m}$, this ratio may
equivalently be expressed as
\begin{equation}
    \eta = \sqrt{ 3 \over 2\pi} \, { e^2 \over 4\pi \hbar \overline v } \,.
\end{equation}
This parameter is also related to the ratio of temperature to
binding energy of two particles in the plasma with equal and opposite%
\footnote
    {%
    For the general case of opposite but unequal charges,
    $e^2$ is replaced by the product of charges $-e_a e_b$.
    }
charge $e$ and reduced mass $m$.
The hydrogenic ground state of two such particles has a binding energy of
\begin{equation}
    \epsilon = \left( { e^2 \over 4 \pi } \right)^2 { m \over 2 \hbar^2} \,.
\end{equation}
The ratio of this energy to the temperature is just $\eta^2$
(up to a factor of $\pi$),
\begin{equation}
    \eta^2 = {1 \over \pi } \, \beta \, \epsilon \,.
\end{equation}
Note that the quantum parameter $\eta$ becomes small at sufficiently
high temperature, but that it diverges at low temperatures or in the
formal $\hbar \to 0$ or $ m \to \infty$ limits. We should also remark
that the quantum effects measured by $\eta$ only appear in two-loop
and higher-order processes. Thus these effects are suppressed by a
factor of $g^2$.

The quantum parameter $\eta$, together with the particle densities,
also provides an estimate of how many bound atoms are present in a
dilute plasma.
The Saha equation, which is simply the condition
for chemical equilibrium between bound atoms and ionized particles,
states that the fraction of bound atoms in the plasma is%
\footnote
    {
    This is just the requirement that the chemical potential plus
    binding energy of the lowest bound state equal the sum of the
    chemical potentials of the bound state constituents.
    Since an atom in free space has an infinite number of
    bound levels, and the presence of the surrounding particles in
    the plasma produces screening effects, the Saha equation
    only provides a rough indication of the numbers of bound atoms
    present.
    Indeed, the fraction of bound atoms in a plasma is
    intrinsically only an approximately defined concept.
        }
\begin {equation}
    n \, \lambda^3 \, e^{\beta \epsilon}
    =
    n \, \lambda^3 \, e^{\pi \eta^2} \,.
\end {equation}
Here $\lambda$ refers to the thermal wavelength corresponding to the
reduced mass of the two charges.  Thus, for a dilute plasma to be
(nearly) fully ionized, the parameter $\pi\eta^2$,
for opposite signed charges, must
be small compared to $-\ln n\lambda^3$.
If the plasma is sufficiently dense that the Debye screening length
becomes comparable to the size of isolated atoms,
then the Saha equation --- which neglects interactions with
the plasma --- breaks down.
Such plasmas can remain essentially fully ionized, even when the
Saha equation predicts a substantial number of bound atoms,
because Debye screening shortens the range of attractive interactions
and effectively prevents the formation of bound states.
The perturbative treatment which we shall develop applies only to the
case of well ionized plasmas.

Underlying any effective field theory, such as the one that we
develop in this paper, is a separation between the length scales of
interest and the scales of the underlying dynamics.
Our length scales of interest will be of order of
the Debye screening length $\kappa^{-1}$ or longer.
The relevant microscopic scales are the Coulomb distance
$d_C = \beta e^2 / 4\pi$ and the thermal wavelength $\lambda$.
The condition that the screening length
$\kappa^{-1}$ be much larger that the Coulomb distance $d_C$ is just
the statement that the classical loop expansion parameter must be small,
\begin{equation}
    g = { \beta e^2 \kappa \over 4 \pi } = { d_C \over \kappa^{-1} } \ll 1 \,.
\end{equation}
As noted above, the thermal wavelength $\lambda$ will provide the
short-distance cutoff in expressions, such as Eq.~(\ref {render}),
which diverge in the purely classical theory.
We assume that
\begin {equation}
    \lambda \kappa \ll 1 \,,
\end {equation}
so that there is a large separation between the scales of interest
and this short distance cutoff.
The quantum theory will generate additional corrections suppressed
by powers of $(\lambda\kappa)$ which, since $\lambda$ is proportional
to Planck's constant $\hbar$, represent an ascending series in
powers of $\hbar$, in contrast to the $\ln\hbar$ effects arising from the
short-distance cutoff.

The diluteness parameter $n \lambda^3 $
is not independent of our other dimensionless parameters since
\begin{equation}
    n \lambda^3 = {  (\kappa \lambda )^3 \over ( \beta e^2 \kappa ) }
= { ( \kappa \lambda )^3 \over 4 \pi g } \,,
\end{equation}
or
\begin{equation}
    n \lambda^3 = { (\kappa \lambda )^2 \over ( \beta e^2 / \lambda ) }
    = { (\kappa \lambda)^2 \over 4 \pi \eta }
    = { g^2 \over 4\pi \eta^3 }
    \,.
\label {eq:dil}
\end{equation}
In order to have a systematic expansion in which the size of different
effects can be easily categorized, we will treat the Coulomb parameter
$\eta = \beta e^2 / 4 \pi \lambda$ as a number that is formally of order one.
Consequently, if we regard $\kappa \lambda$ as the basic small
parameter which justifies the use of an effective field theory, then
$g = \beta e^2 \kappa/ 4 \pi = \eta (\kappa \lambda) $ is $O(\kappa
\lambda)$, while the diluteness parameter $n \lambda^3$ is
$O\left[(\kappa \lambda)^2\right]$, thus formally justifying the
inequalities $g \ll 1$ and $ n \lambda^3 \ll 1$.

The highly ionized plasma at the core of the Sun provides an example
of astrophysical interest. This plasma is mostly composed of electrons
and protons. We take the nominal values for the central temperature
as $T = 1.5 \times 10^7 \>$K, and the electron and
proton densities as $n_e = n_p = 5.0 \times 10^{25} / {\rm cm}^3 $.
Since this temperature is to be compared to atomic energies, electron
volts are far more convenient units, with $ T = 1.3 $ KeV. It is also
convenient to think of distances and densities in terms of the atomic
length unit, the Bohr radius $a_0 = 5.3 \times 10^{-9}$ cm. Thus $n_e
= n_p = 7.4 /a_0^3 $.  Since $ e^2 / 4 \pi a_0 = 27$ eV, and $ a_0 = 4
\pi \hbar^2 / m_e e^2 $, it is easy to find that the Debye wave number
at the Sun's center is given by $ \kappa = 2.0 / a_0 $ and that the
electron's quantum thermal wave length is $ \lambda_e = 0.36 \> a_0 $,
with the proton wave length a factor of $\sqrt{1840}$ smaller, $
\lambda_p = 8.4 \times 10^{-3} a_0 $.  Hence, at the center of the
Sun, the classical loop expansion parameter is quite small, $ g = \beta
e^2 \kappa / 4 \pi = 0.042 $.  For the proton,
\begin {equation}
    \kappa \lambda_p = 0.017 \,, \qquad
    n_p \lambda_p^3 = 4.4 \times 10^{-6} \,, \qquad
    {\beta e^2 \over 4\pi\lambda_p} = 2.4 \,,
\end {equation}
so the inequalities $\kappa \lambda \ll 1$ and $n \lambda^3 \ll 1$
are also well satisfied.
For the electron,
\begin {equation}
    \kappa \lambda_e = 0.72 \,, \qquad\quad
    n_e \lambda_e^3 = 0.35 \,, \qquad\qquad
    {\beta e^2 \over 4\pi\lambda_e} = 0.058 \,.
\end {equation}
While the proton fugacity is tiny, $ z_p = 2.2 \times 10^{-6}
$, the electron fugacity $ z_e = \exp( \beta \mu_e ) = n_e
\lambda_e^3 / 2 = 0.17 $ is small but not insignificant, which means
that the Fermi-Dirac correction to Maxwell-Boltzmann statistics for
the electron are a few percent. Although the Saha equation predicts
that there are 20\% or so neutral hydrogen atoms in the core of the
sun, this is wrong since the Debye screening length is half the Bohr
radius. The core of the Sun is essentially completely ionized.
The fact that $\kappa\lambda_e$ is only slightly less than one
means that the utility of the effective theory
(for describing electron contributions to the thermodynamics
at the core of the Sun) cannot really be judged until one
knows whether $\kappa\lambda$ or, for example, $\kappa\lambda/2\pi$
appears as the natural expansion parameter.
And there is only one way to find out ---
one must compute multiple terms in the perturbative
expansion and examine the stability of the series for the actual
parameters of interest.

\subsection {Utility of the Effective Theory}

For a sufficiently dilute ionized plasma,
all corrections to ideal gas behavior are negligible.
As the plasma density increases, the leading corrections are very well known
and come from either
the inclusion of quantum statistics for the electrons or
the first order inclusion of Debye screening.
At this ``trivial'' level of effort, the resulting equation of state
is easy to write down:
\begin {equation}
    {\beta p \over n}
    =
    1 + {n_e \over n}
    \left[
	 2^{-5/2} \, z_e
	+
	2 \, (2^{-5} {-} 3^{-5/2}) \, z_e^2
	+
	\cdots
    \right]
    - {\kappa^3 \over 24\pi n}
    \,.
\label {eq:eos-trivial}
\end {equation}
Here $n$ is the total particle density (ions plus electrons),
and $z_e$ is the electron fugacity, which is related to the
electron number density as shown in Eq.~(\ref {fugex-n}).
The electron fugacity corrections just come from combining
Eq's.~(\ref {fugex}) and (\ref {fugex-n})
[and noting that in the thermodynamic limit $\beta p = (\ln Z) / \vol$],
while the Debye screening correction will be derived in section
\ref {sec:classical} [Eq.~(\ref {eq:eos1})]. Since the ions are so
much more massive than the electrons, their fugacity will be very
small, and their quantum statistics corrections may be neglected.

The effective theory we construct incorporates
systematically higher-order interaction effects not contained in the trivial
equation of state (\ref {eq:eos-trivial}).
In sections \ref{sec:twolrs} and \ref{sec:threeloop}
we will give explicit forms for the complete second and third order
corrections to the equation of state expanded in powers of
the loop expansion parameter $g = \beta e^2 \kappa/4\pi$.
These results are valid
provided the temperature and density are
{\em not\/} in a regime where:
\begin {enumerate}\advance\itemsep -0.6\itemsep
\item
    \label {fug-breakdown}
    The electron density is so large that an expansion in electron
    fugacity is useless.
    This occurs when the electrons are nearly degenerate
    and their quantum degeneracy pressure becomes a dominant effect.
\item
    \label {low-temp}
    The temperature is so low that the loop expansion of the effective
    theory is useless.
    This happens when the plasma ceases to be nearly fully ionized.
\item
    \label {relativity}
    The temperature is so high that a non-relativistic treatment
    is inadequate.
    This requires that the temperature be small in comparison with
    the electron rest energy of 511 KeV.
\end {enumerate}

As a concrete test of the utility of our effective theory,
one may insert the numerical values of the density and temperature
quoted above as characteristic of the solar interior
($T \equiv 1.3\>$KeV, $n \equiv 15\>a_0^{-3}$)
into the third order result (\ref {eq:2comp}) for the equation of state.%
\footnote
    {%
    For this comparison, we assume that the plasma contains only protons
    and electrons.
    This is not realistic very near the center of the sun,
    where a significant abundance of helium is also present.
    }
Displaying the first, second, and third order corrections separately,
one finds that
\begin {equation}
    {\beta p \over n}
    =
    1 - 0.00693  + 0.01429 + 0.00074 + \cdots \,.
\end {equation}
All corrections to the ideal gas limit are small,
but the second order correction is larger than the first.
However, it is important to understand that our expansion
of the effective theory is based on formally treating the Coulomb parameters
$\eta = \beta e^2/4\pi\lambda$ of all species as numbers of order one.
As indicated in Eq.~(\ref {eq:dil}), this means that quantum statistics
corrections proportional to the $k$-th power of fugacity (or $n \lambda^3$)
are automatically included at $2k$-loop order in the effective theory.
For the solar plasma, because the electron fugacity is small, but
larger than the plasma coupling $g$,
the dominant correction to ideal gas behavior
comes from quantum statistics,
not from Debye screening.
Consequently, a more instructive comparison is to examine
the size of corrections generated by the effective theory
after removing (or resuming) the non-interacting quantum statistics corrections.
This comparison gives
\begin {equation}
    {\beta p \over n}
    -
    \left. {\beta p \over n} \right|_{\rm free}
    = - 0.006930 - 0.001516 + 0.000736 + \cdots \,,
\label {eq:eos-nontriv}
\end {equation}
where $(\beta p / n)|_{\rm free}$ denotes the equation of state
for non-interacting particles, but with quantum statistics for the electrons.
Expanding in electron fugacity, as in (\ref {eq:eos-trivial}),
and inserting the same characteristic parameters gives
\begin {equation}
    \left. {\beta p \over n} \right|_{\rm free}
    =
    1 + 0.01581 + 0.00105 + 0.00010 + \cdots \,.
\label {eq:eos-qstat}
\end {equation}
Both the quantum statistics series (\ref {eq:eos-qstat}),
and the effective theory expansion (\ref {eq:eos-nontriv})
are now quite well behaved.
For these parameter values, it appears that the
three-loop effective theory result (\ref {eq:eos-nontriv}),
combined with the first three terms%
\footnote
    {
    Adding the quadratic electron fugacity correction
    [that is, the $O(z_e^2)$ term in (\ref {eq:eos-trivial}),
    or the $10^{-3}$ term in (\ref {eq:eos-qstat})]
    to the three-loop effective field theory result is entirely
    reasonable since this quantum statistics correction is in fact
    the dominant part of the complete four-loop contribution of the
    effective theory when the Coulomb parameter for the electron
    is small, $\beta e^2 / 4\pi\lambda_e \ll 1$, as it is in the Sun.
    The last term of Eq.~(\ref {eq:eos-qstat})
    is the free-particle limit of the six-loop contribution
    in our expansion of the effective theory.
    This cubic fugacity correction,
    for our characteristic solar parameters,
    makes only a $10^{-4}$ correction to the
    equation of state.
    }
in the fugacity expansion (\ref {eq:eos-qstat}),
will correctly predict the equation of
state to within an accuracy of a few parts%
\footnote
    {
    We remind the reader that portions of the solar neutrino spectrum
    are exceptionally sensitive to the central temperature.
    So a very small change in the equation of state can potentially
    produce a measurable change in the solar neutrino flux.
    }
in $10^4$.

Missing from the above quantitative results,
and from our analysis in subsequent sections,
are relativistic corrections.
The leading ``kinematic'' relativistic effects may be obtained
by inserting the relativistic kinetic energy
$ E(p) = \sqrt{(pc)^2 + (mc^2)^2 } - mc^2 $
into the ideal gas partition function (\ref {qstat}).
The dominant effects come from the electrons, due to their small mass.
Expanding $E(p)$ in powers of momentum, one finds that
\begin {equation}
    \beta p
    =
    n_i
    +
    {2 z_e \over \lambda_e^3}
    \left[
	1 + {15 \over 8} \, {T \over m_e c^2}
	-
	{z_e \over 2^{5/2}}
	\left(
	    1 + {15 \over 16} \, {T \over m_e c^2}
	\right)
	+ O(z_e^2)
	+ O\biggl({T \over m_e c^2}\biggr)^2
    \right],
\end {equation}
where $n_i$ denotes the total density of ions.
The electron density,
$n_e \equiv  z_e \, \partial (\beta p) / \partial z_e$,
receives exactly the same ${15\over 8} {T \over m_e c^2}$ correction.
Hence, this correction (plus all further corrections to $\beta p$
which are linear in $z_e$) cancels in the equation of state.
However, other thermodynamic quantities, such as the internal energy,
do receive relative $O(T/m_e c^2)$ relativistic corrections.
For the equation of state, the first relativistic correction
which does contribute comes from the $O(T/m_e c^2)$ perturbation
to the $O(z_e^2)$ quantum statistics term, and one finds that
\begin {equation}
    \Delta_{\rm rel.}
    \left(
	{ \beta p \over n }
    \right)
    =
    {15 \over 16} \,
    {n_e \over n} \,
    {z_e \over 2^{5/2}} \,
    {T \over m_e c^2} \,.
\end {equation}
For the characteristic solar parameters used above, this correction
is less than a part in $10^4$,
\begin {equation}
    \Delta_{\rm rel.}
    \left(
	{ \beta p \over n }
    \right)
    =
    0.000036 \,.
\end {equation}

A hot plasma also contains black body radiation.
The contribution to the pressure
arising from this photon gas is given by the familiar formula
\begin{eqnarray}
    \Delta_{\rm photon} \left( { \beta p \over n } \right)
    &=&
	{ \pi^2 \over 45 }
	\left( { T \over \hbar c } \right)^3 { 1 \over n}
    =
    { \pi^2 \over 45 }
    \left( {2 \pi T \over m_e c^2 } \right)^{3/2}
    {1 \over n \lambda_e^3 }
\nonumber\\ &=&
    \left( { T \over 6.1 \> {\rm KeV} } \right)^3
    { 1 \over n a_0^3 } \,.
\end{eqnarray}
For the solar parameters that we have adopted,
\begin{equation}
    \Delta_{\rm photon} \left( { \beta p \over n } \right) = 0.00063 \,,
\end{equation}
which is the size of our third order correction. The relative
importance of this photon gas correction increases rapidly as the
temperature is increased, and it must be included in  some of the
regions discussed at the end of this section.

The transverse photons also interact with the charged particles to
alter the thermodynamic relations. This effect is dominated by the
coupling with the light electrons. It may be easily obtained by using
the radiation gauge to compute the first-order perturbation arising
from the `seagull' interaction Hamiltonian density $(e^2 / 2m_e c^2)
\psi^\dagger \psi {\bf A}^2 $ and taking the ${\bf j} \cdot {\bf A}$
interaction to second order.
Since the current involves $ e {\bf v} / c $, one expects
that the second-order ${\bf j} \cdot {\bf A}$ contribution is suppressed
by $ (v_e / c)^2 \sim T / m_e c^2 $ relative to the `seagull' term.
This is confirmed by a detailed computation.
A simple calculation expresses the (leading order in $ T / m_e c^2 $)
`seagull' contribution as
\begin{equation}
    \Delta_{\rm rad.}  \left( { \beta p } \right)
    =
	- \beta
	{ e^2 \over 2 m_e c^2 }
	\left[
	    \langle {\bf A}(0)^2 \rangle_{T} -
	    \langle {\bf A}(0)^2 \rangle_{T=0}
	\right]
	n_e
    =
	- {\alpha \, \pi \over 3} \,
	{ T \over m_e c^2 } \, n_e \,,
    \end{equation}
where $\alpha = e^2 / (4 \pi \hbar c) = 1 / 137.\cdots $ is the fine
structure constant.
Note that the vacuum, or $ T \to 0 $, contributions
are subtracted as they are completely absorbed by
renormalization of the bare electron parameters.
Since $n_e = \partial (\beta p) / \partial (\beta \mu_e)$,
this correction is equivalent to a shift in the electron
chemical potential of $\delta \mu_e = -{\pi \over 3} (\alpha T^2 / m_e c^2)$.
It modifies the chemical potential --- electron density relation
and thus has no effect on the equation of state.
However, the correction does affect other thermodynamic quantities
such as the internal energy.%
\footnote
    {%
    A recent paper \cite {opher} has attempted to argue that
    radiative corrections are far larger than this relative
    $O(\alpha T/m_e c^2)$ effect.
    The conclusions of this paper are not correct.
    }
The leading corrections to the equation of state
involving the interactions of transverse photons are actually
of relative order
$\alpha z_e (T / m_e c^2)^2$.
One finds that
\begin {equation}
    \Delta_{\rm rad.} \left( { \beta p \over n} \right)
    =
    -
    {\alpha \pi \over 3} \,
    {n_e \over n} \,
    {z_e \over 2^{3/2}} \,
    \left( {T \over m_e c^2} \right)^2 \,.
\end {equation}
For the characteristic solar parameters used above,
this is utterly negligible even at the part in $10^4$ level,
\begin {equation}
    \Delta_{\rm rad.} \left( { \beta p \over n_e } \right)
    =
    1.5 \times 10^{-9} \,.
\end {equation}

\begin {figure}[tp]
   \begin {center}
      \leavevmode
      \def\epsfsize #1#2{0.85#1}
      \epsfbox {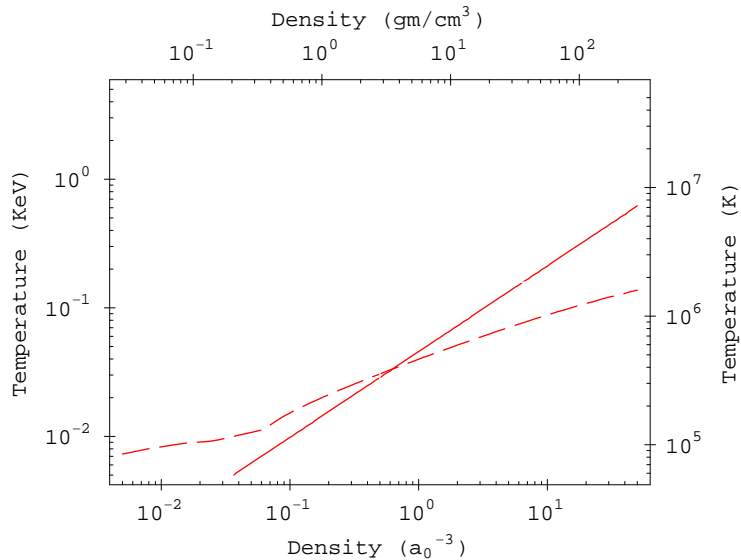}
         \end {center}
   \caption
	{%
	Region of validity
	of the effective theory for the case of a pure $Z=1$
	ionized hydrogen plasma.
	On the bottom axis, density denotes the total particle density
	(electrons plus protons)
	in units of the Bohr radius,
	while the top axis shows the corresponding mass density.
	The solid line shows where the fugacity expansion breaks down.
	The dashed line shows where the size of ``non-trivial''
	effective field theory
	corrections to the equation of state first exceed unity.
	(See the text for more precise descriptions.)
	The effective field theory is valid only in the region above
	both of these lines.
	}
\label {fig:validity1}
\end {figure}

\begin {figure}[pt]
   \begin {center}
      \leavevmode
      \def\epsfsize #1#2{0.85#1}
      \epsfbox {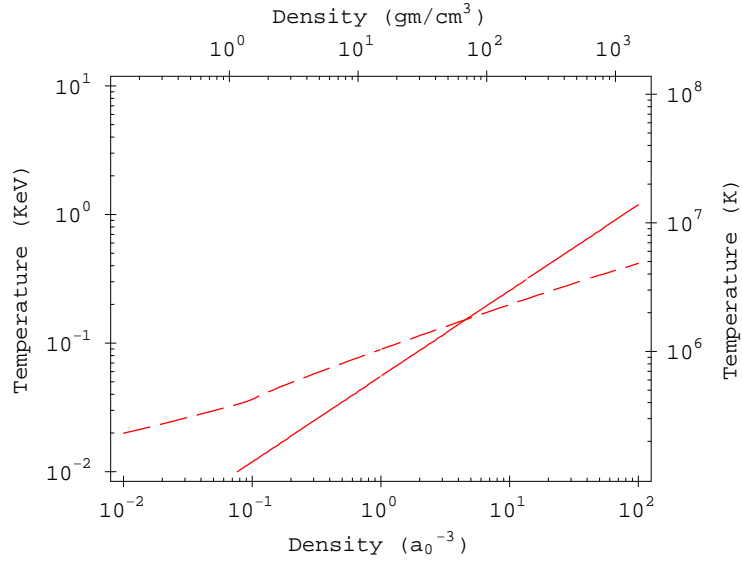}
   \end {center}
   \caption
	{%
	Region of validity for the effective theory for
	a pure $Z=2$ ionized helium plasma.
	The curves have the same meaning as in
	Fig.~\protect\ref{fig:validity1}.
	}
\label {fig:validity2}
\end {figure}

\begin {figure}[tp]
   \begin {center}
      \leavevmode
      \def\epsfsize #1#2{0.85#1}
      \epsfbox {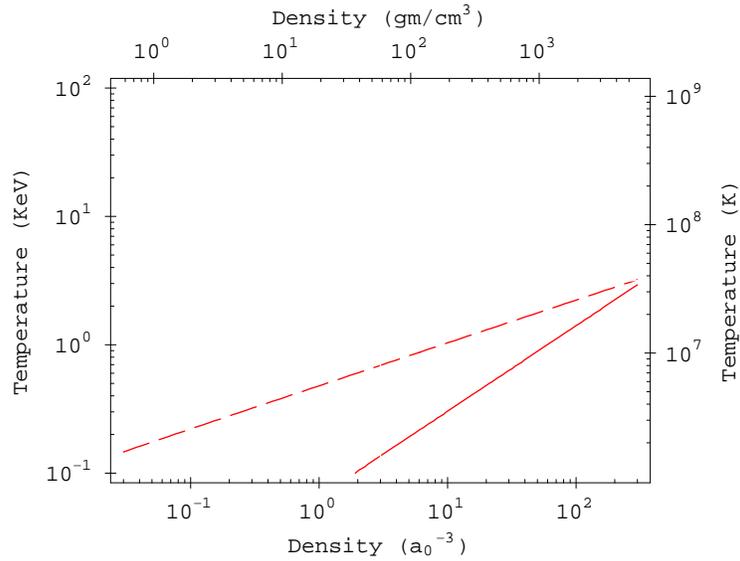}
   \end {center}
   \caption
	{%
	Region of validity for the effective theory for
	a pure $Z=6$ ionized carbon plasma.
	The curves have the same meaning as in
	Fig.~\protect\ref{fig:validity1}.
	}
\label {fig:validity6}
\end {figure}

\begin {figure}[tp]
   \begin {center}
      \leavevmode
      \def\epsfsize #1#2{0.85#1}
      \epsfbox {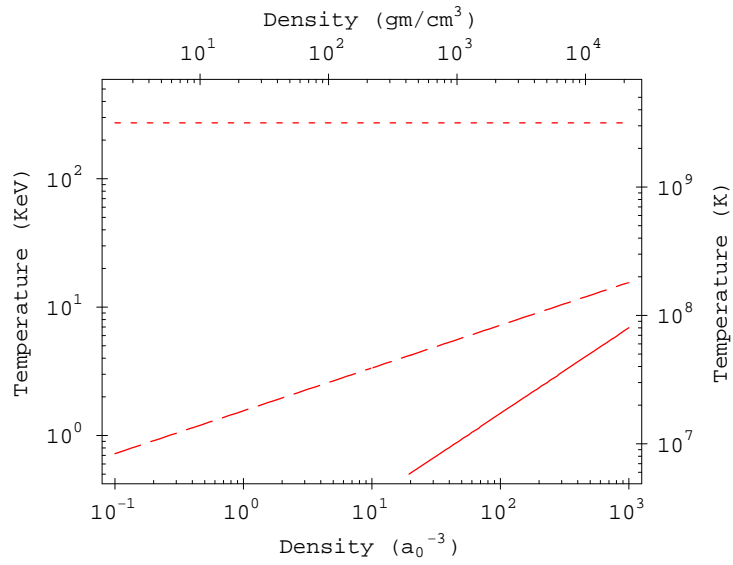}
   \end {center}
   \caption
	{%
	Region of validity for the effective theory for
	a pure $Z=13$ ionized aluminum plasma.
	The solid and dashed curves have the same meaning as in
	Fig.~\protect\ref{fig:validity1}.
	The dotted horizontal line shows where relativistic corrections
	to the electron pressure exceed unity;
	our non-relativistic treatment is valid only below this line.
	}
\label {fig:validity13}
\end {figure}

Depending on the mass and composition of a star,
the electron fugacity in stellar interiors may be relatively small
(as in the Sun), or may be large enough to completely invalidate a
quasi-classical treatment (as in white dwarfs or very massive stars).
Figures \ref {fig:validity1}--\ref {fig:validity13} represent an attempt
to delineate the region of validity of the effective theory
in the temperature-density plane
for the case of a pure $Z=1$ proton-electron plasma
[Fig.~\ref {fig:validity1}],
a pure $Z=2$ (ionized helium) plasma [Fig.~\ref {fig:validity2}],
a pure $Z=6$ (ionized carbon) plasma [Fig.~\ref {fig:validity6}],
and a pure $Z=13$ (ionized aluminum) plasma [Fig.~\ref {fig:validity13}].
The solid line shows where the second and third order corrections in the
fugacity expansion for electrons become equal in size.
This occurs before any of the individual first, second, or third order
fugacity corrections exceed unity, and provides a convenient signal
that the fugacity expansion is no longer well-behaved.
The dashed line shows where the size of effective field theory
corrections to the equation of state first exceed unity.%
\footnote
    {%
    More precisely, this line shows where any of
    the one-, two-, or three-loop corrections first exceed unity.
    To match the earlier discussion, the non-interacting quantum
    statistics portion of the two-loop correction is not included.
    }
This is taken as an indication that the perturbative expansion
of the effective field theory has broken down.
The effective field theory is valid only in the region above
(or to the left of) both of these lines.
In Fig.~\ref {fig:validity13}, the temperature range extends
into the relativistic domain.
The horizontal dotted line in this figure shows where the
${15 \over 8} {T \over m_e c^2}$ relative correction to the electron pressure
exceeds unity, and provides an indication of where relativistic
corrections invalidate our non-relativistic treatment.

\begin {figure}[tp]
   \begin {center}
   \vspace*{-15pt}
      \leavevmode
      \def\epsfsize #1#2{0.90#1}
      \epsfbox {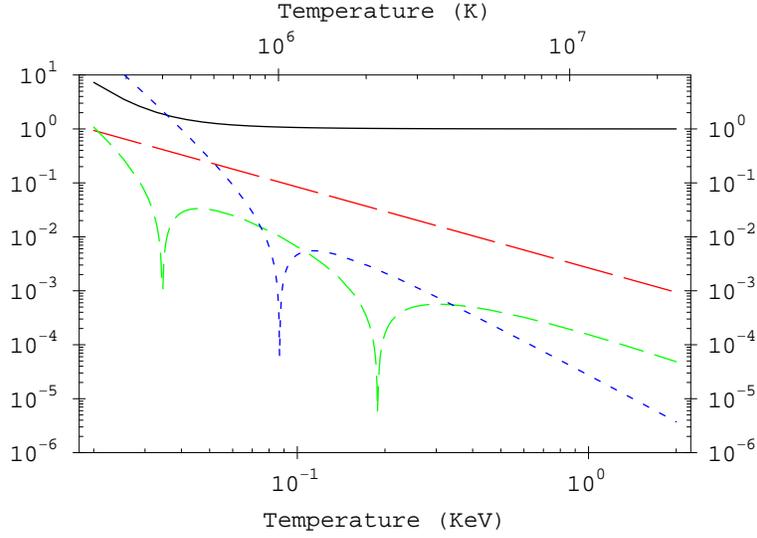}
   \vspace*{-10pt}
   \end {center}
   \caption
	{%
	Corrections to the equation of state, $\beta p/n$,
	as a function of temperature
	for a pure $Z=1$ plasma with a total density
	(electrons plus protons) of $1\> a_0^{-3}$.
	Here, and in the following related figures,
	the solid line shows the ideal gas result, including
	quantum statistics for the electrons but no interactions.
	The long dashed line shows the one-loop Debye screening correction,
	the medium dashed line shows the two-loop correction
	(minus its non-interacting quantum statistics piece),
	and the short dashed line shows the three-loop effective field
	theory correction.
	The absolute values of the various corrections are plotted.
	On the two- and three-loop curves,
	the ``cusps'' pointing downward
	show where these corrections cross zero and change sign.
	For this density, the effective field theory is only useful
	for temperatures above about 0.06 KeV.
	Below this temperature, the three-loop correction
	exceeds the size of the one-loop correction
	(and exceeds unity at temperatures below about 0.04 KeV),
	clearly showing that the perturbative expansion of
	the effective theory has ceased to be reliable.
	}
    \label {fig:slice1.1}
\end {figure}

\begin {figure}[tp]
   \begin {center}
   \vspace*{-20pt}
      \leavevmode
      \def\epsfsize #1#2{0.90#1}
      \epsfbox {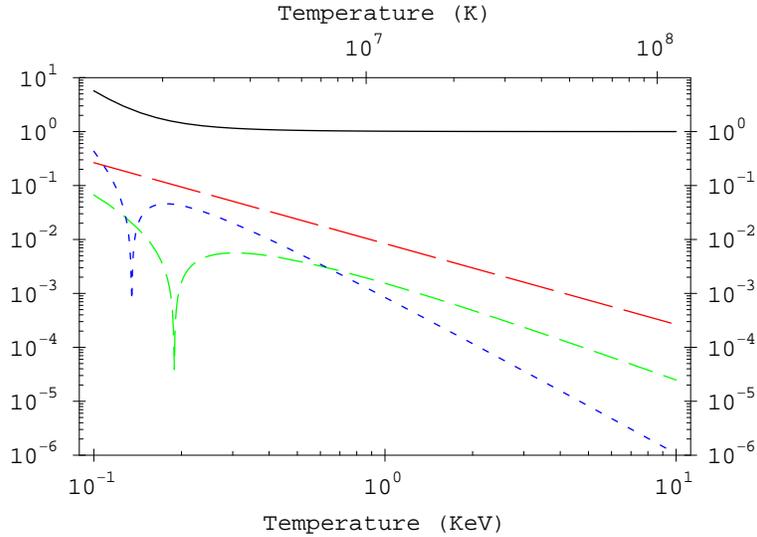}
   \vspace*{-10pt}
   \end {center}
   \caption
	{%
	Same as Fig.~\protect\ref {fig:slice1.1}, but at a total particle
	density of $10\> a_0^{-3}$.
	}
    \label {fig:slice1.10}
\end {figure}

\begin {figure}[thp]
   \begin {center}
      \leavevmode
      \def\epsfsize #1#2{0.90#1}
      \epsfbox {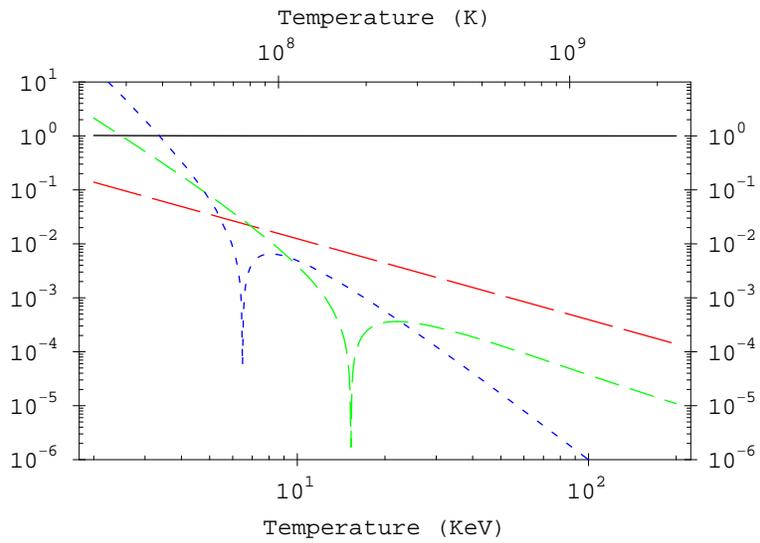}
   \end {center}
   \caption
	{%
	Same as Fig.~\protect\ref {fig:slice1.1}, but
	for a pure $Z=13$ plasma at a particle density of $10\> a_0^{-3}$.
	}
    \label {fig:slice13.10}
\end {figure}

For a given density (and composition),
if the effective field theory is to be useful,
then the temperature must be high enough so that the perturbative
expansion of the theory is valid, but not so high so that
all corrections to ideal gas behavior generated by the effective theory
are too small to be relevant.
In other words, the size of the effects produced by the effective
theory must be large enough to be interesting.
Figures~\ref {fig:slice1.1}--\ref {fig:slice13.10} show log plots
of the size of corrections to the equation of state for various
compositions and two different densities of the plasma.
In these plots, the
solid line shows the ideal gas result, including
quantum statistics for the electrons but no interactions.
The long dashed line shows the one-loop Debye screening correction,
the medium dashed line shows the two-loop correction
(minus its non-interacting quantum statistics piece),
and the short dashed line shows the three-loop effective field
theory correction.
Plotted are the absolute values of the various corrections.
The one-loop Debye screening correction is always negative.
The ``cusps'' pointing downward on the two- and three-loop
curves show where these corrections cross zero and change sign.
Asymptotically, for large temperature, the (non-trivial part of the)
two-loop correction is negative for $Z=1$ and positive for $Z \ge 2$,
while the three-loop correction is asymptotically positive in all
these plots.
Each plot begins at temperatures which are too low for
the effective theory to be valid, includes the region where
the effective theory can be useful, and ends at temperatures
sufficiently high that all corrections to ideal gas behavior
are tiny.
\goodbreak

\clearpage

\begin {figure}[thp]
   \begin {center}
      \leavevmode
      \def\epsfsize #1#2{0.90#1}
      \epsfbox {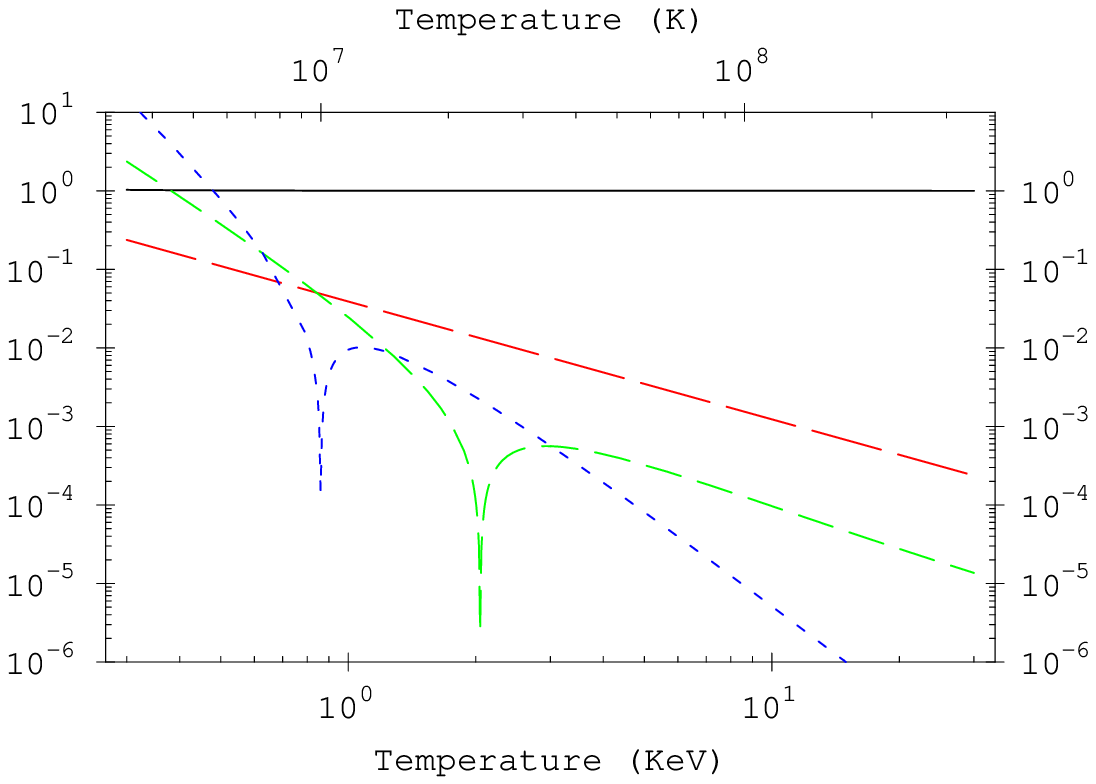}
   \end {center}
   \caption
	{%
	Same as Fig.~\protect\ref {fig:slice1.1}, but
	for a pure $Z=6$ plasma at a particle density of $1\> a_0^{-3}$.
	}
\end {figure}

\begin {figure}[thp]
   \begin {center}
      \leavevmode
      \def\epsfsize #1#2{0.90#1}
      \epsfbox {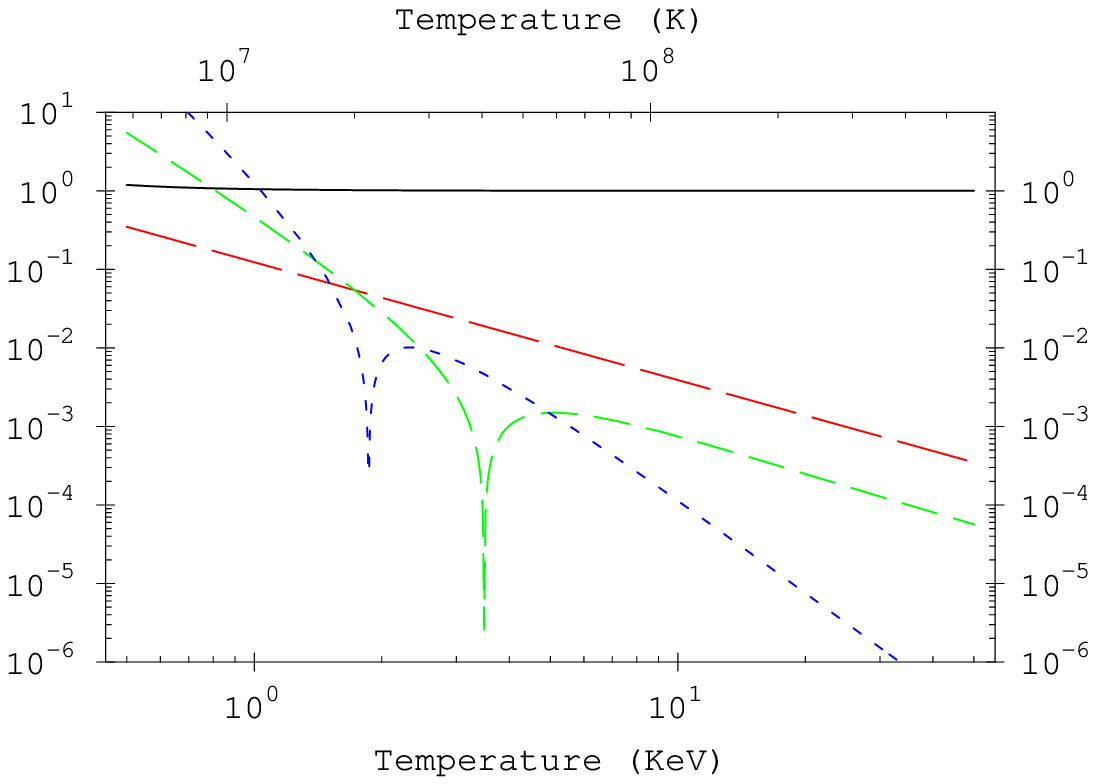}
   \end {center}
   \caption
	{%
	Same as Fig.~\protect\ref {fig:slice1.1}, but
	for a pure $Z=6$ plasma at a particle density of $10\> a_0^{-3}$.
	}
\end {figure}

\begin {figure}[thp]
   \begin {center}
      \leavevmode
      \def\epsfsize #1#2{0.90#1}
      \epsfbox {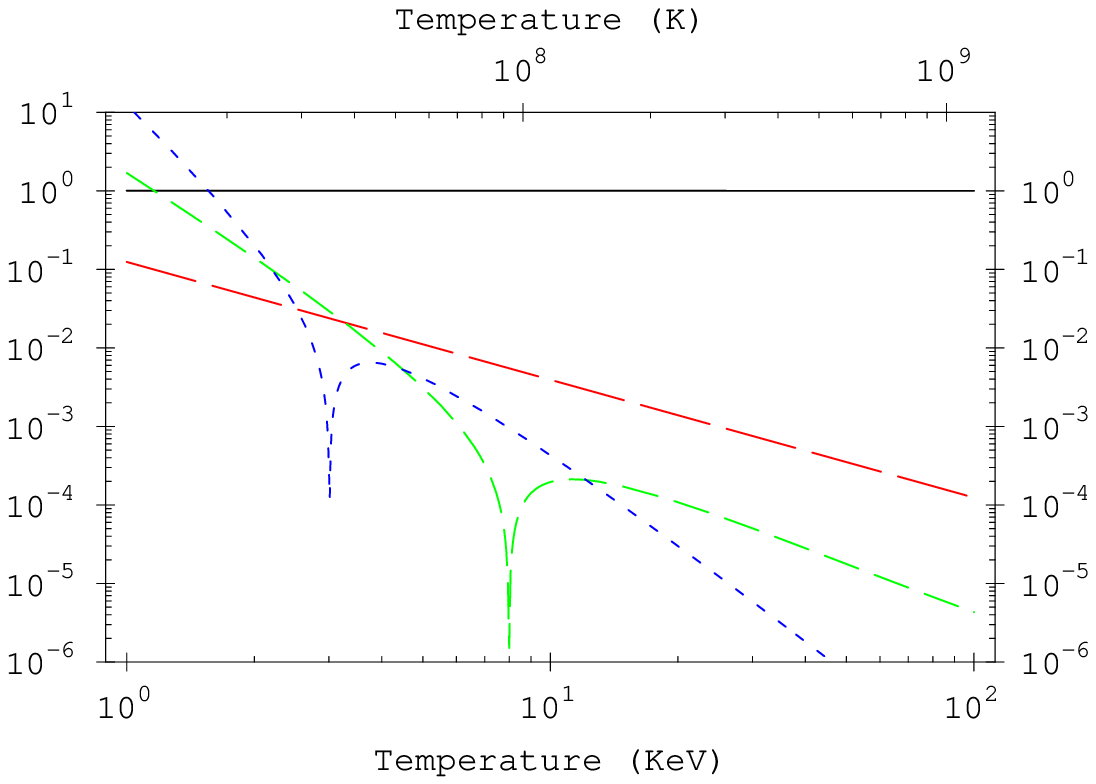}
   \end {center}
   \caption
	{%
	Same as Fig.~\protect\ref {fig:slice1.1}, but
	for a pure $Z=13$ plasma at a particle density of $1\> a_0^{-3}$.
	}
\end {figure}

\clearpage

\section {Classical Coulomb Plasmas}
\label{sec:classical}

We consider a plasma of $A$ different species of charged particles
(ions and electrons) and use the letters $a , b , \cdots = 1 , \cdots
, A$ to denote a specific species with charge $e_a$ and mass $m_a$.
In the classical limit, the particle mass only appears in the thermal
wavelength
\begin{equation}
\lambda_a^2 = { 2\pi \beta \hbar^2 \over m_a } \,,
\end{equation}
where $\beta$ is the inverse temperature measured in energy units, and
the thermal wavelength itself only serves to define the
free-particle density $\n0a$ in terms of the
chemical potential $\mu_a$ and spin degeneracy factor $g_a$
of the given species:
\begin{equation}
\n0a = g_a \, \lambda_a^{-3} \, e^{ \beta \mu_a} \,.
\label {eq:n0a}
\end{equation}
The grand canonical partition function for a free gas composed of
these species is given by
\begin{equation}
    Z_{\rm free} = \sum_{\{ N_a \} }
    \int d\sigma_1^{N_1} \cdots d\sigma_A^{N_A} \,,
\end{equation}
where $d\sigma_a^N$ is the $N$-particle measure for species $a$,
\begin {equation}
    d\sigma_a^N \equiv {1 \over N!} \>
    (d^3r_{a,1}) \> \n0a \cdots (d^3r_{a,N}) \> \n0a \,.
\end {equation}
The factors of $\lambda_a^{-3}$ hidden in the
$\n0a$ free-particle densities
in this measure come from performing the momentum integrals in the
equilibrium phase-space distribution,
\begin{equation}
\lambda_a^{-3} = \int { (d^3p) \over (2\pi\hbar)^3 } \>
\exp\{ - \beta p^2 /2 m_a \} \,,
\end{equation}
and the remaining parts of $\n0a$ arise from the degeneracy ($g_a$)
and fugacity ($e^{ \beta \mu_a}$) factors that enter into the
definition of the grand canonical ensemble. Introducing the total
volume of the system
\begin{equation}
\vol \equiv \int (d^3\r) \,,
\end{equation}
which we shall always assume is arbitrarily large, and carrying out
the summations, we get
\begin{equation}
Z_{\rm free} = \prod_{a=1}^A \>
    \sum_{N_a}
	{ (\vol \, \n0a)^{N_a} \over N_a ! }
    =
    \exp\biggl\{ \vol \sum_{a=1}^A \> \n0a \biggr\} \,.
\end{equation}

\subsection{Functional Integral for the Classical Partition Function}

The corresponding grand canonical partition function for a plasma
with Coulomb interactions between all the charged particles is
\begin{equation}
Z = \sum_{\{ N_a \} }
    \int d\sigma_1^{N_1} \cdots d\sigma_A^{N_A} \>
    \exp\biggl\{ - {\beta \over 2} \, \sum_{k \ne l} e_k e_l  \,
	V_C \left( \r_k {-} \r_l \right)
    \biggr\} \,.
\label{grandcoulb}
\end{equation}
Here the indices $k \,, l$ in the exponential run over all particles
of all the various types; $\r_k$ and $e_k$ denote the coordinates and
charge of any given particle, respectively.  We employ rational units,
so that the Coulomb potential for unit charges is given by
\begin{equation}
V_C(\r) = { 1 \over 4 \pi \, r } \,.
\end{equation}

We choose to work with the grand canonical ensemble because, as we
shall see, it has a simple functional integral representation which
leads to a very convenient diagrammatic form for perturbation theory
and allows easy use of effective field theory techniques.  However, we
are ultimately interested in calculating physical quantities as a
function of the particle densities, not chemical potentials, of the
various species.  Since the presence of interactions between particles
will modify the particle density --- chemical potential relation, we
will need to compute particle densities as a function of chemical
potential, and then invert this relation (order-by-order in
perturbation theory) to re-express results in terms of particle
densities.  The physical particle densities, which we will denote as
$\bar n_a$, satisfy charge neutrality,
\begin {equation}
   \langle Q \rangle_\beta = \vol \, \sum_a \> e_a \, \bar n_a = 0 \,,
\label {eq:charge neutrality}
\end {equation}
as required for a sensible thermodynamic limit.

It will be useful to regard the chemical potentials as temporarily
having arbitrary spatial variation, $\mu_a (\bf r)$. This extends the
partition function to be a functional of these generalized chemical
potentials, $Z \to Z[\mu]$, which is then the generating functional for
number density correlation functions.
The free-particle number density ---
chemical potential relation (\ref{eq:n0a}) is now generalized to
\begin{equation}
\n0a(\r) \equiv g_a \, \lambda_a^{-3} \, e^{\beta \mu_a(\r)} \,,
\end{equation}
with the variational derivative
\begin{equation}
{\delta \over \delta \beta \mu_b(\r')} \, \n0a(\r) =
\delta_{ab} \, \delta( \r {-} \r' ) \, \n0a(\r) \,.
\end{equation}
Here, and henceforth, variations in $\beta \mu_a$, and in $\beta$,
will be regarded as independent.
In other words, $\beta \mu_a$ is to be varied while holding $\beta$ fixed,
and vice-versa.
The density of particles of species $a$ is given by the variational
derivative of $\ln Z[\mu]$ with respect to the corresponding
generalized chemical potential,
\begin {equation}
    \expect { n_a(\r)}
    \equiv
   \expect { {\sum}_i \> \delta (\r {-} \r_{a,i})}
    =
    { \delta \over \delta \beta \mu_a(\r) }
    \ln Z[\mu] \,,
\label {eq:den}
\end {equation}
while two functional derivatives yield
the connected part of the density-density correlator,
\begin{eqnarray}
    K_{ab}(\r {-} \r')
&\equiv&
    \expect {n_a(\r) \> n_b(\r')}
    -
    \expect {n_a(\r)} \expect{n_b(\r')}
\nonumber
\\ &=&
    \expect { {\sum}_{i,j}
\delta (\r {-} \r_{a,i}) \> \delta (\r' {-} \r_{b,j})}
    -
    \expect { {\sum}_{i} \delta (\r {-} \r_{a,i})}
    \expect { {\sum}_{j}\delta (\r' {-} \r_{b,j})}
\nonumber
\\ &=&
    { \delta \over \delta \beta \mu_a(\r) } \,
    { \delta \over \delta \beta \mu_b(\r') } \,
    \ln Z[\mu] \,.
\label {eq:corr}
\label{corrl}
\end{eqnarray}
After the functional derivatives have been taken,%
\footnote{The derivation of the
  results (\ref{eq:den}) and (\ref{eq:corr})  from the spatially
  varying chemical potential extension of the standard form
  (\ref{grandcoulb}) of the partition function requires a little
  thought. These results are obvious however if one imagines the
  classical partition function to be given by the classical limit of
  the quantum form $ Z[\mu] = {\rm Tr} \> \exp\left\{ -\beta H + \int
    (d\r) \sum_a \beta \mu_a(\r) n_a(\r) \right\} $, with all operators
  commuting in this classical limit.}
it will be assumed
that the spatially-dependent, generalized chemical potentials
$\mu_c(\r)$ revert to the usual constant chemical potentials $\mu_c$.

The cumbersome form of the grand partition functional
(\ref{grandcoulb}) can be replaced by a much leaner functional
integral representation by using the Gaussian integral relation%
\footnote
    {
    The use of Gaussian integral relations such as this has
    a very long history in statistical physics,
    going back at least as far as
    Hubbard \cite {hubbard} and
    Stratonovitch \cite {stratonovitch}.
    }
\begin{eqnarray}
&&\int [d \phi] \exp\left\{ \beta \int (d^\dim\r) \left[ \half \, \phi(\r)
  \nabla^2 \phi(\r) + i \rho(\r) \phi(\r) \right]
\right\}
\nonumber\\
&& \qquad\qquad
= {\rm Det}^{- 1/2} \left[ \beta (- \nabla^2 ) \right]
\exp\left\{ - {\beta \over 2} \int (d^\dim\r) (d^\dim\r') \>
\rho(\r) \, V_\dim( \r {-} \r' ) \, \rho(\r') \right\} ,
\label{square}
\end{eqnarray}
which follows from completing the square in the functional integral
on the left.
The auxiliary field $\phi(\r)$ is nothing but the electrostatic
scalar potential.%
\footnote
    {%
    More precisely, $-i\phi$ is the normal electrostatic potential.
    Inserting an $i$ (or rotating the contour of
    the functional integral) is necessary to obtain
    an absolutely convergent functional integral.
    }
The relation above has been written in $\dim$ spatial dimensions with
\begin{equation}
V_\dim( \r {-} \r' ) \equiv \int { (d^\dim \k) \over
  (2\pi)^\dim } \>
  { e^{ i \k \cdot ( \r {-} \r') } \over \k^2 }
\end{equation}
the Coulomb potential in $\dim$ dimensions.
We choose to make a continuation in spatial dimensions at this
juncture because it automatically removes infinite
particle self-interactions.
Dimensional continuation is a regularization procedure
which introduces no external or extraneous dimensional
constants. Hence, since there is nothing available to make up the
correct dimensional quantity, in dimensional continuation%
\footnote
    {
    The dimensional regularization method is widely employed in relativistic
    quantum field theory calculations, and is discussed in many texts.
    For example, the book \cite{brown} contains a detailed treatment.
    The conclusion that $ V_\dim({\bf 0}) = 0 $ may be justified
    more explicitly by starting from the integral representation
    (\ref{gofr}) with $\kappa_0^2 = 0$, which shows that
    $V_\dim({\bf r}) \propto r^{2-\dim} $.
    Dimensional regularization is defined by the prescription
    that one first go to a region of spatial dimensions in which the
    quantity being examined is well defined, and thereafter analytically
    continue to the dimensionality of interest.
    In the present case, this requires going to $\dim < 2$ where
    $V_\dim({\bf 0}) = 0 $. Then
    one continues this result to arbitrary dimension, with zero of course
    remaining zero as $\dim$ varies, including $\dim = 3$.
    }
\begin{equation}
V_\dim({\bf 0}) = 0 \,,
\label{zero}
\end{equation}
and particle self-interactions vanish. We shall see how this works
out in practice as our development unfolds. We shall also need the
technique of dimensional continuation to deal
with the short-distance divergences of the classical Coulomb theory ---
the divergences that are removed by quantum fluctuations which we shall
later handle using effective field theory methods.
Hence one might as well get accustomed to dimensional
continuation at an early stage. At the end of our computations we
shall, of course, take $ \dim \to 3$. In view of the functional formula
(\ref{square}), it follows that the grand canonical partition function
may be written as
\begin{eqnarray}
Z[\mu] &=&
{\rm Det}^{1/2} \left[ \beta (- \nabla^2 ) \right]
\int [d \phi] \, \exp\left\{ - { \beta \over 2 } \int
  (d^\dim\r) \> \phi(\r) \left( - \nabla^2 \right) \phi(\r)
\right\}
\nonumber\\
&& \kern 1.45in {} \times
\exp\left\{ \sum_{a=1}^A \int (d^\dim\r) \>
    \n0a(\r) \> e^{i \beta e_a \phi(\r) } \right\} .
\label{fun}
\end{eqnarray}
Since $ - \nabla^2 $ is a positive operator, the first,
Gaussian, part of the integrand gives a well-defined and convergent
functional integral. Expanding the second exponential in a
power series in the free-particle densities $\n0a$,
and using the functional integration formula (\ref{square}),
it is easy to see that the result (\ref{fun})
does indeed reproduce the Coulomb plasma generating functional
(\ref{grandcoulb}). Note that this equivalence requires that the
self-interaction terms vanish, which is the case with our dimensional
regularization [Eq.~(\ref{zero})].
Combining the two exponentials of (\ref {fun}), one may write
the partition function in the concise form
\begin {equation}
    Z[\mu] =
    N_0
    \int [d \phi] \, e^{-\Scl[\phi;\mu]} \,,
\label {eq:fun}
\end {equation}
with an ``action'' functional defined by%
\footnote
    {
    In the special case of a binary plasma with equal fugacities,
    note that the action (\ref {eq:action}) reduces to the much-studied
    Sine-Gordon theory.
    See, for example, Ref.~\cite {samuel} and references therein.
    }
\begin {equation}
    \Scl[\phi;\mu] \equiv
    \int (d^\dim\r) \>
    \left\{
    {\beta \over 2} \left[ \nabla \phi(\r) \right]^2
    -\sum_{a=1}^A \n0a(\r) \> e^{i \beta e_a \phi(\r)}
    \right\} ,
\label {eq:action}
\end{equation}
and the overall normalization factor
\begin{equation}
    N_0 \equiv
    {\rm Det}^{1/2} \left[ \beta (- \nabla^2 ) \right] .
\end {equation}

    Varying the functional integral representation (\ref {eq:fun})
with respect to the chemical potential $\mu_a(\r)$
yields the representation
\begin {equation}
    \expect {n_a(\r)}
    =
    \left<\!\!\left< \n0a(\r) \, e^{i \beta e_a \phi(\r)} \right>\!\!\right>
\label {eq:den-fun}
\end {equation}
for the density of particles of type $a$,
where in general $\left<\!\left< \cdots \right>\!\right>$ denotes a
functional integral average,
\begin {equation}
    \left<\!\left<
	{\cal O}
    \right>\!\right>
    \equiv
    Z[\mu]^{-1} \> N_0 \int [d\phi] \> e^{-\Scl[\phi]} \>
    {\cal O} \,.
\end {equation}
With the generalized chemical potentials restricted to constant
values, Eq.~(\ref{eq:den-fun}) gives the functional integral
representation for the usual grand canonical average of the number
density of particles of species $a$.  A second variation with the
chemical potentials then restricted to constant values yields the
representation of the density-density correlation function (\ref
{eq:corr}),
\begin {eqnarray}
    K_{ab}(\r{-}\r')
    &=&
    \left<\!\!\left<
	\n0a \, e^{i \beta e_a \phi(\r)} \>
	\n0b \, e^{i \beta e_b \phi(\r')}
    \right>\!\!\right>
    - \left<\!\!\left< \n0a \, e^{i \beta e_a \phi(\r)} \right>\!\!\right>
      \left<\!\!\left< \n0b \,
       e^{i \beta e_b \phi(\r')} \right>\!\!\right>
\nonumber
\\
    && \quad {}
    + \delta_{ab} \, \delta(\r{-}\r')
    \left<\!\!\left<
	\n0a \, e^{i \beta e_a \phi(\r)} \>
    \right>\!\!\right> \,.
\label {eq:K}
\end {eqnarray}
The final contact term proportional to
$\delta(\r{-}\r')$ appears (when $a = b$)
because the functional integral naturally
generates correlators involving distinct particles,
\begin {equation}
    \left<\!\!\left<
	\n0a \, e^{i \beta e_a \phi(\r)} \>
	\n0a \, e^{i \beta e_a \phi(\r')}
    \right>\!\!\right>
    =
    \expect { {\sum}_{i \ne j} \>
     \delta (\r {-} \r_{a,i}) \, \delta (\r' {-} \r_{a,j}) } \,.
\end {equation}
This differs from the corresponding term in (\ref {eq:corr})
precisely by the single-particle contact term
$
 \expect { {\sum}_i
\delta (\r {-} \r_{a,i}) \, \delta (\r' {-} \r_{a,i}) }
    =
    \delta (\r{-}\r') \, \expect {n_a}
$.

    Since the functional integral of a total derivative vanishes,
\begin {equation}
0 = \int [d\phi] \> {\delta \over \delta \phi(\r)} \, e^{-\Scl[\phi;\mu]} \,,
\end {equation}
the field equation $\avg {\delta \Scl[\phi;\mu]/\delta \phi(\r) }=0$
is an exact identity.
For the action (\ref {eq:action}),
this is the Poisson equation
\begin {equation}
    \nabla^2 \avg {i\phi(\r)} = \avg {\rho(\r)}
\label {eq:EOM}
\end {equation}
with the charge density
\begin {equation}
    \rho(\r) \equiv
    \sum_{a=1}^A \> e_a \, \n0a(\r) \, e^{i \beta e_a \phi(\r)} \,.
\end {equation}
Integrating both sides of (\ref {eq:EOM}) over all space
yields the condition of total charge neutrality,
\begin {equation}
    0 = \expect {Q}
    = \sum_{a=1}^A e_a \int (d^\dim\r )\> \expect {n_a(\r)}\,.
\label {eq:Q=0}
\end {equation}
This identity holds for any choice of the generalized chemical
potentials $\mu_a(\r)$, in essence because the average value
of the electrostatic potential $\phi$ will always adjust itself
to produce a charge neutral equilibrium state.%
\footnote
    {%
    Assuming, of course, that the plasma contains both
    positively and negatively charged species.
    }

The fact that the chemical potentials enter the
action (\ref {eq:action}) only through the combination
$\n0a e^{i\beta e_a\phi}$
(with $\n0a \propto e^{\beta \mu_a}$) means that the theory
is completely unchanged if the electrostatic potential is
shifted by an arbitrary constant,
\begin {equation}
    i \phi \to i\phi + c \,,
\label {eq:shift-sym}%
\end {equation}
provided the chemical potentials are correspondingly adjusted,
\begin {equation}
    \mu_a \to \mu_a - e_a \, c \,.
\label {eq:shift-sym2}%
\end {equation}
Consequently, the values of the chemical potentials
are not uniquely determined by the physical particle densities.
This is also reflected in the fact that the conditions
\begin {equation}
    \bar n_a = \expect {n_a} \,, \qquad a = 1 , \ldots, A \,,
\end {equation}
only give $A-1$ linearly independent constraints on the chemical
potentials --- precisely because charge neutrality (\ref{eq:Q=0})
is an automatic identity.
To obtain uniquely defined chemical potentials (when they revert back
to their normal constant values),
one must remove the (physically irrelevant) freedom (\ref {eq:shift-sym})%
--(\ref {eq:shift-sym2})
to shift the mean value of the electrostatic potential.
We will make the obvious choice, and demand that the thermal average
of the electrostatic potential vanish,
\begin {equation}
     \avg {\phi} \equiv 0 \,,
\label {eq:mean phi = 0}
\end {equation}
to fix the chemical potentials uniquely.

\subsection {Mean Field Theory}
\label{mean}

Saddle-points of the functional integral (\ref {eq:fun})
correspond to solutions of the field equation
\begin{equation}
    { \delta \Scl[\phi;\mu] \over \delta\phi(\r) } = 0 \,,
\end{equation}
which, for the action (\ref {eq:action}), is just
the Debye-H\"uckel equation
\begin {equation}
    -\nabla^2 \phi(\r)
    =
    i \sum_{a=1}^A \> e_a \, \n0a(\r) \, e^{i\beta e_a \phi(\r) } \,.
\label {eq:debye-huckel}
\end {equation}
The leading saddle-point approximation corresponds to
neglecting all fluctuations in $\phi$ away from the saddle-point,
so that
\begin {equation}
    \ln Z_0[\mu] = - \Scl[\phi;\mu] \,,
\end {equation}
with $\phi$ solving the field equation (\ref {eq:debye-huckel}).  In
quantum field theory, this approximation is commonly called the tree
approximation because the classical action is the generating
functional of connected tree graphs.  In statistical mechanics it is
known as the mean field approximation.  In Appendix \ref{funmeth} we
shall describe the effective action functional $ \Gamma[\phi;\mu] $
which is the generalization of the classical action $ \Scl[\phi;\mu] $
that takes account of the thermal fluctuations about the mean field
which are described by the functional integral and thus provides an
exact description of the plasma. As will be shown in Appendix
\ref{funmeth}, the effective action method can be used to derive
general properties of the plasma physics. Our work now with the mean
field approximation will provide an introduction to the later use of
the more general effective action as well as illustrating basic plasma
properties.

For constant chemical potentials,
the field equation reduces to the
(lowest-order) charge neutrality condition,%
\footnote
    {%
    Note that this constraint does not
    have a perturbative solution
    that can be be expanded in powers of the electric charge.
    This lack of a perturbative solution
    occurs because $\phi$ appears only in the combination $e_a \phi$.
    Moreover, the lack of a perturbative solution and consequent
    condition of overall charge neutrality is related to the infinite
    range of the Coulomb potential. If, for example, the Coulomb
    potential were replaced by a Yukawa potential with range $1/m$, the
    classical field equation for constant fields would become
    $
	- i m^2 \phi = \sum_a e_a \, \n0a \, e^{ie_a \beta \phi } \,,
    $
    which imposes no constraint on the total charge and which does have a
    perturbative solution for $\phi$.
    }
\begin{equation}
    \sum_a  e_a \,\n0a \, e^{ie_a \beta \phi } = 0 \,,
\label{neutt}
\end{equation}
and
\begin{equation}
    \ln Z_0[\mu] = \vol \sum_a  \> \n0a \, e^{ie_a \beta \phi } \,.
\end{equation}
The mean-field number density --- chemical potential relation is given by
\begin{eqnarray}
    \bar n_a &=& \vol^{-1} \>
    \left. { \partial \ln Z_0 \over \partial \beta \mu_a} \right|_{\beta}
\nonumber\\
    &=&  \n0a \, e^{ie_a \beta \phi }
    + i \beta \left[ \sum_a  e_a \n0a e^{ie_a \beta \phi } \right]
    { \partial \phi \over \partial \beta \mu_a }
\nonumber\\
    &=&  \n0a \, e^{ie_a \beta \phi } \,,
\label{nochem}
\end{eqnarray}
with the last equality following from the charge neutrality
condition (\ref{neutt}).
If the free-particle densities satisfy ``bare'' charge neutrality,
\begin {equation}
    0 = \sum_a e_a \, \n0a \,,
\label{lowchn}
\end {equation}
then the saddle-point condition (\ref {neutt}) has
the trivial solution $\phi(\r) = 0$,
the physical densities $\bar n_a$,
within this mean field approximation,
will equal the free-particle densities $\n0a$,
and the mean-field partition function equals the usual ideal gas result,
\begin {equation}
    Z_0 = \exp \biggl\{ \vol \sum_{a=1}^A \n0a \biggr\} \,.
\label {eq:Z0}
\end {equation}

The average energy of our grand canonical ensemble is
the thermodynamic internal energy,
\begin{equation}
    U = \bar E = \langle E \rangle_\beta =
    - \left. { \partial \ln Z \over \partial \beta} \right|_{\beta\mu} \,.
\end{equation}
Since $\n0a \sim \beta^{-3/2}$,
varying the neutrality condition (\ref {neutt}) with respect to $\beta$
gives
\begin{equation}
-{3 \over 2} \, \beta^{-1} \sum_a  e_a \,\n0a \, e^{ie_a \beta \phi }
+ { i \over \beta } \, \kappa_0^2 \, { \partial (\beta\phi) \over \partial
  \beta } = 0 \,,
\label {eq:varbeta}
\end{equation}
where
\begin{equation}
    \kappa_0^2 \equiv \sum_a \beta \, e^2_a \, \n0a \, e^{ie_a \beta \phi }
\label{eq:kappa}
\end{equation}
will be seen to be the lowest-order (squared) Debye wave number.
The first term of (\ref {eq:varbeta}) again
vanishes by virtue of charge neutrality
(\ref{neutt}), and so
\begin{equation}
    { \partial (\beta\phi) \over \partial \beta } = 0 \,.
\end{equation}
Hence, to lowest order the average energy
\begin{equation}
    \bar E
    = { 3 \over 2} \, \beta^{-1} \vol \, \sum_a \bar n_a
    = { 3 \over 2} \, T \, \sum_a \bar N_a \,,
\end{equation}
which is just the familiar formula for an ideal gas.

Second derivatives of $\ln Z$ produce correlators. The second
derivative of $\ln Z_0$ with respect to the inverse temperature gives
the lowest-order result for the mean square fluctuation in energy,
\begin{equation}
\left\langle \left( E - \bar E \right)^2 \right\rangle_\beta = -
{\partial \bar E \over \partial \beta} = { 15 \over 4} \, T
\bar E \,.
\end{equation}
Mixed temperature --- chemical potential derivatives yield the
correlation between energy and particle number fluctuations,
\begin{equation}
    \left\langle \left( E - \bar E \right) \left( N_a - \bar N_a \right)
    \right\rangle_\beta = - { \partial \bar N_a \over \partial \beta}
    = { 3 \over 2 } \, T \, \bar N_a \,.
\end{equation}
These are again just the results for a free gas.
But for fluctuations in particle numbers,
given by second derivatives with respect to the chemical
potentials,
one must account for the fact that varying the
chemical potentials will cause the mean field to vary.
Since the charge neutrality constraint (\ref {neutt}) holds
for arbitrary chemical potentials, varying it with respect
to the chemical potentials yields
\begin{equation}
    e_a \,\bar n_a + i \,\kappa_0^2 \, { \partial \phi
    \over \partial \beta \mu_a } = 0 \,,
\label{phimuvar}
\end{equation}
Hence,
\begin{eqnarray}
    \left\langle
	\left( N_a - \bar N_a \right) \left( N_b - \bar N_b \right)
    \right\rangle_\beta
    =
    {\partial N_a \over \partial \beta \mu_b}
&=&
    \delta_{ab} \bar N_a
    +
    \bar N_a i e_a \, \beta \, { \partial \phi \over \partial \beta \mu_b}
\nonumber\\
&=&
    \delta_{ab} \bar N_a
    - e_a \bar N_a { \beta \over \kappa_0^2 } \, e_b \, \bar n_b \,.
\label {eq:nfluct}
\end{eqnarray}
The physical implications of this result,
which differs from the ideal gas result,
will be discussed below in subsection \ref {den-den}.

\subsection {Loop Expansion}

The saddle-point (or ``loop'') expansion of the functional integral
(\ref{eq:fun}),
incorporates corrections beyond mean field theory and
systematically generates the perturbative expansion for
physical quantities of interest.
In the development that follows, we shall assume that all of the
desired functional derivatives with respect to the generalized,
spatially varying chemical potentials which produce the insertions in
the functional integral, as shown in the previous number density
(\ref{eq:den-fun}) and density-density correlator (\ref{eq:K}), have
already been taken. Thus, we henceforth restrict our considerations to
constant chemical potentials.  In the lowest-order approximation, the
free-particle densities $\n0a$ will equal the physical densities $\bar
n_a$, which are charge neutral (\ref {eq:charge neutrality}).
However, perturbative corrections to the chemical
potential --- number density relation will shift the free-particle
densities away from the physical densities, and therefore displace the
true saddle point away from $\phi = 0$. Even though the bare
neutrality constraint (\ref{lowchn}) no longer holds in higher orders,
it will be most convenient to expand the functional integral about
$\phi = 0$ instead of the true saddle-point value. At each stage of
this (loop) expansion, further corrections to the bare (tree
approximation) charge neutrality constraint (\ref{lowchn}) appear
which alter the relation amongst the chemical potentials that arises
from charge neutrality. Expanding the action in powers of $\phi$ and
separating the quadratic and constant terms gives
\begin{equation}
    \Scl[\phi;\mu] = S_0[\phi;\mu] + \Sone[\phi;\mu] \,,
\end{equation}
where
\begin{equation}
    S_0[\phi;\mu] \equiv
	\int (d^\dim\r) \> \biggl\{
	    -\sum_{a=1}^A \n0a
	    + {\beta \over 2} \,
	    \phi(\r) \left[ -\nabla^2 + \kappa_0^2 \right] \phi(\r)
	\biggr\} \,,
\label {eq:S0}
\end{equation}
and
\begin{eqnarray}
    \Sone[\phi;\mu] &\equiv&
	-\int (d^\dim\r) \> \sum_{a=1}^A \n0a
	\left\{
	    e^{i \beta e_a \phi(\r)} - 1 + \half \beta^2 e_a^2 \phi(\r)^2
	\right\}
\nonumber
\\&=&
	-\int (d^\dim\r) \> \sum_{a=1}^A \n0a
	\left\{
	    [i \beta e_a \phi(\r )]
	    + \coeff 1{3!} [i \beta e_a \phi(\r )]^3
	    + \coeff 1{4!} [i \beta e_a \phi(\r )]^4
	    + \cdots
	\right\} .
\label{Sint}
\end {eqnarray}
In Eq.~(\ref {eq:S0}), $\kappa_0^2$ is
the lowest-order Debye wave number previously defined in Eq.~(\ref {eq:kappa}).
Since the bare neutrality condition is modified by loop corrections,
$\sum_a e_a \n0a$ will not vanish beyond the mean field approximation.
Consequently, $\Sone$ contains a piece linear in the field $\phi$
and $\phi = 0$ does not remain a saddle point in higher orders.

Evaluating the action at $\phi = 0$
gives the ideal gas partition function
The first (``one-loop'') correction is obtained by
neglecting%
\footnote
    {%
    As discussed in the next subsection,
    the term in $\Sone$ linear in the field
    may be counted as being of one-loop order.
    However, because it is odd in $\phi$, its first order contribution
    to the functional integral vanishes (just like the $\phi^3$ term)
    and so it does not contribute to one-loop result (\ref{eq:Z1}).
    }
$\Sone$
and integrating over fluctuations in $\phi$ with
just the quadratic action $S_0$.
This gives the Gaussian functional integral
\begin{eqnarray}
    Z_1 &=&
    Z_0 \>
     {\rm Det}^{1/2} \left[ \beta (- \nabla^2 ) \right]
    \int [d \phi] \, \exp\left\{ - { \beta \over 2 }
	\int (d^\dim\r) \>
	\phi(\r) \left( - \nabla^2 + \kappa_0^2 \right) \phi(\r)
    \right\}
\nonumber\\
    &=&
    Z_0 \>
    {\rm Det}^{-1/2} \left[ 1 + {1 \over - \nabla^2 } \, \kappa_0^2 \right]
    \,.
\label {eq:Z1}
\end{eqnarray}
The product of the determinant produced by the Gaussian
integration with the prefactor (which may be written as the inverse
determinant of the operator inverse) produces the determinant
shown on the second line. This functional determinant will
be evaluated shortly.
The correlation function of potential fluctuations
$\avg {\phi(\r)\phi(\r')}$,
to lowest order,
is given by the Green's function for the linear operator
$\left(-\nabla^2 + \kappa_0^2\right)$ appearing in $S_0$,
\begin {equation}
    \beta\avg{\phi(\r) \phi(\r')}^{(0)}
    =
    {N_0 \over Z_1} \int [d\phi] \> e^{-S_0} \> \beta \phi(\r) \phi(\r')
    =
    G_\dim(\r{-}\r') \,.
\label{phiphiexp}
\end {equation}
Here $G_\dim( \r {-} \r')$ denotes
the Debye Green's function (in $\dim$-dimensions), which satisfies
\begin{equation}
    \left[ -  \nabla^2 + \kappa^2_0 \, \right] G_\dim( \r {-} \r' )
    = \delta ( \r {-} \r' ) \,,
\end{equation}
and has the Fourier representation
\begin{equation}
    G_\dim( \r {-} \r' ) = \int { (d^\dim\k) \over (2\pi)^\dim } \>
    { e^{ i \k \cdot ( \r {-} \r' ) } \over \k^2 + \kappa_0^2 }
    \,.
\label{frep}
\end{equation}
Expanding the functional integral (\ref {eq:fun}) in powers of
$\Sone$ will lead to Feynman diagrams in which each line
represents a factor of this Debye Green's function times $1/\beta$,
with vertices joining $k$ lines representing factors of
$\sum_a \n0a \, (i \beta e_a)^k$.

A convenient integral representation for the Debye Green's function in
a space of arbitrary dimensions is obtained by writing the
denominator in (\ref{frep}) as a parameter integral of an
exponential, interchanging the parameter and wave number integrals,
and completing the square to perform the wave number integral:
\begin{equation}
G_\dim( {\bf r} ) = \int_0^\infty ds \> e^{ - \kappa_0^2 s}
 \int { (d^\dim\k) \over (2\pi)^\dim } \> e^{- \k^2 s} e^{ i \k \cdot \r}
  = \int_0^\infty ds \>
\left( 4 \pi s \right)^{- \dim /2} \, e^{- \kappa_0^2 s - (\r^2 / 4 s)} \,.
\label{gofr}
\end{equation}
The coincident limit of the Debye Green's function $G_\dim({\bf 0})$
will be needed in the following sections. In this limit, the
representation (\ref{gofr}) becomes the standard representation of the
Gamma function, and we have
\begin{equation}
G_\dim({\bf 0}) =
    { \kappa_0^{\dim-2} \over (4 \pi)^{\dim/2} } \,
    \Gamma \left( 1 {-} { \dim \over 2} \right) \,.
\label{gooo}
\end{equation}
Since $ \Gamma( - \half ) = - 2 \sqrt \pi $, the $ \dim \to 3 $ limit of
$G_\dim( {\bf 0} )$ is perfectly finite and yields
\begin{equation}
\lim_{\dim\to3} G_\dim( {\bf 0} ) = - { \kappa_0 \over 4 \pi } \,.
\label{perf}
\end{equation}
Comparing this with the Debye Green's function fixed
at three dimensions,
\begin{equation}
G_3( \r ) =  { e^{ - \kappa_0 r } \over 4 \pi r } \,,
\end{equation}
one sees that
\begin{equation}
\lim_{r \to 0} { 1 \over 4 \pi r } \left[ e^{ - \kappa_0 r } - 1
\right] = -  { \kappa_0 \over 4 \pi } \,.
\end{equation}
In other words, the dimensional regularization method automatically
deletes the vacuum self-energy contribution that comes from the pure
Coulomb potential.

\subsection {Particle Densities}

Although the densities of the various particle species may be obtained
simply by differentiating the partition function with respect to the
corresponding chemical potential --- which we shall do
subsequently --- one may directly evaluate these densities
using  diagrammatic perturbation theory. We shall do this through
one-loop order to illustrate the working of the perturbation theory
and charge neutrality. In perturbation theory,
the density of particles of a given species is evaluated by
expanding the exponential in (\ref {eq:den-fun}) in powers of $\phi$
yielding, to one loop order,
\begin {eqnarray}
    \expect {n_a}^{(1)} &=&
    \avg{ \n0a \, e^{i \beta e_a \phi(\r)} }
\nonumber
\\ &=&
    \n0a
    \left[
	1
	+
	i \beta e_a \avg{\phi(\r)}
	-
	\half \beta^2 e_a^2 \avg{\phi(\r)^2}
    \right]
\nonumber
\\ &=&
    \n0a
    \left[
	1 + i \beta e_a \avg{\phi}^{(1)}
	- \half \beta \, e_a^2 \, G_\dim({\bf 0})
    \right].
\label{denone0}
\end {eqnarray}

\begin {figure}[t]
   \begin {center}
      \leavevmode
      \def\epsfsize #1#2{0.60#1}
      \epsfbox {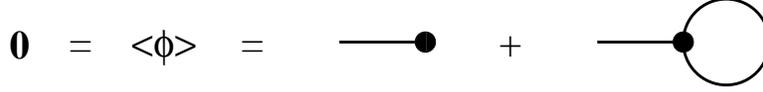}
   \end {center}
   \caption
	{%
	One-loop order contributions to $\expect {\phi}$.
	Unlabeled blobs (or vertices) represent insertions
	of $-\Sone$ taken to some order in $\phi$;
	a vertex joining $k$ lines stands for a factor
	of $\sum_a \n0a \, (i\beta e_a)^k$.
	Each line represents a factor of the Debye Green's
	function divided by $\beta$,
	and the contribution of each diagram is to be multiplied
	by the appropriate symmetry factor
	which, for the diagram above containing a loop, the ``tadpole
	graph,'' is $1/2$.
	[The basic rules for diagrammatic perturbation theory,
	corresponding to a saddle-point expansion of the functional integral,
	are discussed in virtually all textbooks on quantum field theory.
	See, for example, Refs.~\protect\cite {brown,peskin,negle}.]
	The condition $\avg{\phi}=0$ taken to one-loop order implies
	that the one-legged vertex (---\llap{$\bullet$}) must cancel
	the one-loop ``tadpole''.  Hence this one-legged vertex should
	be counted here as being a one-loop contribution.
	Two-loop diagrams (and beyond) generate further higher-order
	corrections to the one-legged vertex $ i \beta {\sum}_a e_a \, n_a^0$.
	}%
\label {fig:tadpole}
\end {figure}

In the tree approximation with $\phi = 0$, the charge neutrality
condition (\ref{neutt}) requires that the chemical potentials are
arranged such that $\sum e_a \, n^0_a = 0 $. Thus, this sum should be
considered to start out at one-loop order.  The one-legged vertex, the
coefficient of the term in the interaction part of the action
(\ref{Sint}) linear in $\phi$, is proportional to this sum, and hence
it also should be considered to start at one-loop order.  Thus
computing the expectation value of $\phi$ to one-loop order requires
expanding $e^{-\Sone}$ in powers of $\phi$ and keeping the linear and
cubic terms. This expansion, shown in the graphs of figure
\ref{fig:tadpole}, gives
\begin {eqnarray}
    \avg{\phi}^{(1)}
    &=&
    {N_0 \over Z_1} \int [d\phi] \> e^{-S_0} \>
    \phi({\bf 0})
    \int (d^\dim\r) \>
    \sum_a \n0a
    \left[
	(i\beta e_a \phi(\r))
	+
	\coeff 1{3!}
	(i\beta e_a \phi(\r))^3
    \right]
\nonumber\\
&=&
    {i \over \kappa_0^2} \> \sum_{a=1}^A \> e_a \, \n0a
    \left[
	1 - \half \beta \, e_a^2 \, G_\dim({\bf 0})
    \right].
\label{phione}
\end {eqnarray}%
This calculation is spelled out in greater detail in the derivation of
Eq.~(\ref{phionee}) in Appendix \ref{someqft}.  Note that the first
term in Eq.~(\ref{phione}), the tree approximation, is obtained by
expanding the tree level neutrality condition (\ref{neutt}) to zeroth
and first order in $\phi$.

\begin {figure}[t]
   \begin {center}
      \leavevmode
      \def\epsfsize #1#2{0.60#1}
      \epsfbox {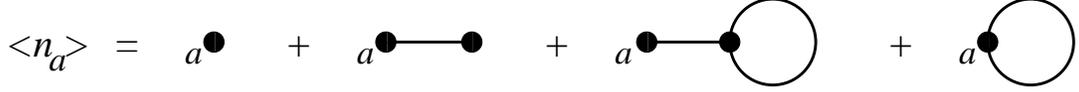}
   \end {center}
   \caption
	{%
	One-loop order contributions  to
	the mean particle density $\expect {n_a}$.
	Labeled blobs (${}_a\bullet$) refer to insertions
	of the number density $\n0a \, e^{i \beta e_a \phi}$
	for a given species;
	a labeled blob radiating $k$ lines stands
	for a factor of $\n0a \, (i \beta e_a)^k$.
	The condition $\avg{\phi}=0$ implies that the second
	and third diagrams cancel.
	More generally, the condition $\avg{\phi} = 0$
	implies that such ``tadpole'' diagrams
	cancel in the expansion of any quantity,
	and such diagrams may simply be neglected. This cancelation
	is described more fully in Appendix~\protect\ref{funmeth}.
	 }%
\label {fig:den1}
\end {figure}%

Imposing the condition (\ref {eq:mean phi = 0})
that the mean electrostatic potential vanish now requires,
to this order, that
\begin {equation}
     \sum_{a=1}^A \> e_a \,\n0a
    =
	\half \beta \, G_\dim({\bf 0}) \sum_{a=1}^A e_a^3 \, \n0a \,,
\label {eq:phi1=0}
\end {equation}
which alters the tree level neutrality constraint (\ref{lowchn}) on
the chemical potentials, making the sum on the left-hand side of
Eq~(\ref{eq:phi1=0})
equal to the one-loop contribution on the right-hand side.
This confirms the statement above that the sum on the left-hand should
be considered to start out at one-loop order.
With the imposition of the one-loop
constraint (\ref{eq:phi1=0}), the expression (\ref {denone0}) for the
one-loop densities simplifies to
\begin {equation}
    \expect {n_a}^{(1)} =
    \n0a
    \left[ 1 - \half \beta \, e_a^2 \, G_\dim({\bf 0}) \right].
\label {denone}
\end {equation}

The discussion of the density that we have just given is illustrated
in figure
\ref{fig:den1}. Inverting the one-loop density relation
(\ref{denone}) to express the bare density
$\n0a$ in terms of the physical density $\bar n_a$ gives
\begin {equation}
    \n0a
= \bar n_a \left[ 1 - \half \beta \, e_a^2 \, G_\dim({\bf 0})
\right]^{-1}
= \bar n_a \left[ 1 + \half \beta \, e_a^2 \, G_\dim({\bf 0}) \right] ,
\label {eq:n0-1loop}
\end {equation}
to one-loop order.
Note that $e_a^2 G_\dim( {\bf 0})/2$ is the self-energy of a charge
$e_a$ in the Debye screened plasma,
and so the right-hand side of Eq.~(\ref {denone})
may be recognized as the first order expansion of the Boltzmann factor
$\exp\{-\beta e_a^2 G_\dim({\bf 0})/2\}$.
Other effects besides this simple exponentiation of course appear in
higher orders.
Also note that the mean charge density (computed to one-loop order)
vanishes, as it must, even before the imposition of the constraint
(\ref {eq:phi1=0}), for it follows from Eq's.~(\ref{denone0})
and (\ref{phione}) and the definition (\ref{eq:kappa}) of the
lowest-order Debye wave number that
\begin {eqnarray}
    \expect {\rho}^{(1)}
    &=& \sum_{b=1}^A \> e_b \expect {n_b}^{(1)}
\nonumber
\\ &=&
    \sum_{b=1}^A \> e_b \, \n0b
    \left[ 1 - \half \beta \, e_b^2 \, G_\dim({\bf 0}) \right]
    +
    \sum_{b=1}^A \> \beta \, e_b^2 \, \n0b \> \avg{i\phi}^{(1)}
\nonumber
\\ &=&
    \sum_{a=1}^A \> e_a \, \n0a
    \left[ 1 - \half \beta \, e_a^2 \, G_\dim({\bf 0}) \right]
    \left( 1 -
    \sum_{b=1}^A \> {\beta \, e_b^2 \, \n0b \over \kappa_0^2}
    \right)
\nonumber
\\ &=& 0 \,.
\end {eqnarray}

\subsection{Loop Expansion Parameter}

We have just seen that the size of one-loop corrections is measured,
in $\dim$ dimensions, by the dimensionless parameter
$ \beta e^2 G_\dim({\bf 0}) \sim \beta e^2 \kappa_0^{\dim-2}$,
which reduces to $ \beta e^2 \kappa_0$ in three dimensions.
This parameter is the essentially the ratio of the Coulomb energy
for two particles separated by a Debye screening distance to
their typical kinetic energy in the plasma.
Since $\kappa_0^2 \sim \beta e^2 / d^3 $, where $d$ is the average
interparticle spacing, this expansion parameter is also
$[ \,\beta e^2 / d ]^{3/2}$ --- the $ 3/2 $ power of the ratio of the
average Coulomb energy in the plasma to the kinetic energy in the plasma.

At higher orders in the perturbative expansion, the relative
contribution of any Feynman diagram containing $\ell$ loops will be
suppressed by $[ \,\beta e^2 \kappa_0^{\dim-2} ]^\ell$, or in three
dimensions, by $[\,\beta e^2 \kappa_0]^\ell$.
A detailed proof of this appears in section 3 of Appendix \ref{someqft}.%
\footnote
    {%
    Here is a brief version.
    The rescaling $\phi = \tilde\phi / (\beta e )$, \r = $\tilde \r
    / \kappa_0 $ in the functional integral (\ref {eq:fun}) conveniently
    reveals the dimensionless loop expansion parameter $g = \beta e^2
    \kappa_0^{\dim-2}$:  the integrand acquires the canonical form
    $e^{-\tilde S[\tilde\phi]/g}$, with all dependence on the
    dimensionless parameter $g$ isolated in the explicit
    prefactor which controls the validity of a saddle-point
    expansion.
    }
In other words, the loop expansion parameter
is $[\,\beta e^2 \kappa_0]$ (up to some $O(1)$ numerical factor).
In fact, we shall find in our explicit calculations
that $[\beta e^2 \kappa / 4 \pi]$ appears as
the most natural loop expansion parameter.

\subsection {Thermodynamic Quantities}

All thermodynamic quantities may be derived from the grand canonical
partition function. In particular, the internal energy density $u$ is
given by
\begin{equation}
u \vol = - \left. { \partial \ln Z \over \partial \beta }
\right|_{\beta\mu} \,,
\label{enden}
\end{equation}
where, as indicated  the partial derivative is taken with the all the
$\beta \mu_a$ fixed, while the chemical potential --- number density
relation is given by
\begin{equation}
\bar n_a \vol = \left. { \partial \ln Z \over \partial \beta \mu_a }
\right|_\beta \,,
\label{barbahn}
\end{equation}
where now $\beta$ is held fixed in the partial differentiation.
The grand potential $\Omega(\vol,T,\{ \mu_a \})$ is related to
the partition function of the grand canonical ensemble by
\begin {equation}
    Z = e^{-\beta \Omega} \,.
\end {equation}
The grand potential is extensive for a macroscopic volume,
and it is simply related to the pressure, $\Omega = -p \vol$,  or
\begin {equation}
    p = {\ln Z \over \beta \vol} \,.
\end {equation}
The Legendre transform of the grand potential gives
the Helmholtz free energy,
$F(\vol,T,\{N_a\}) = \Omega(\vol,T,\{\mu_a\}) + \sum_a \mu_a N_a$.
Hence the free energy density is given by
\begin {equation}
    f = -p + \sum_a \mu_a \, \bar n_a \,.
\end {equation}

The previous zeroth order and one-loop results (\ref{eq:Z0}) and
(\ref{eq:Z1}) express the partition function through one-loop order as
\begin {equation}
    Z_1 = \exp \biggl\{ \vol \sum_{a=1}^A \n0a \biggr\} \,
    {\rm Det}^{-1/2} \left[ 1 + {1 \over - \nabla^2 } \, \kappa_0^2 \right]
    \,.
\end{equation}
To evaluate the determinant,
one may apply the general variational formula
\begin{equation}
\delta \ln {\rm Det} X = {\rm Tr} X^{-1} \delta X
\end{equation}
to a variation of $\kappa_0^2$, to show that
\begin{equation}
    \delta \ln {\rm Det}
    \left[ 1 + { 1 \over - \nabla^2 } \, \kappa_0^2 \right]
    = \int (d^\dim\r) \> G_\dim( {\bf 0} ) \, \delta \kappa_0^2 \,.
\end{equation}
Since this is homogeneous in $\kappa_0$ of degree $\dim{-}2$, it implies that
\begin{equation}
    \ln {\rm Det} \left[ 1 +  { 1 \over - \nabla^2 } \kappa_0^2 \right]
    =
    {2 \over \dim} \, G_\dim( {\bf 0} ) \, \kappa_0^2 \, \vol \,,
\label{dett}
\end{equation}
and thus%
\footnote
    {
    This result assumes that the chemical potentials
    (and temperature) are constrained so that $\avg {\phi} = 0$ (to one
    loop order). If this constraint is violated, as it apparently is in
    varying $\beta$ to obtain the internal energy by
    Eq.~(\ref{enden}) or varying $\beta\mu_a$ to obtain the density of
    particles of species $a$ by Eq.~(\ref{barbahn}), then additional
    terms are present in the complete one-loop result.
    These additional terms do not contribute to the first variations
    yielding the energy or number densities
    and hence may be neglected for these terms, but they do contribute
    to second or higher variations that define correlation functions.
    This is discussed more fully in Appendix \ref{funmeth};
    see in particular Sections 1 and~3.
    }
\begin{equation}
Z_1 = \exp\left\{ \left[ {\sum}_a n_a^0 -  { 1 \over \dim} \, \kappa_0^2 \,
     G_\dim( {\bf 0} ) \right] \vol \right\} ,
\label{zzone}
\end{equation}

Let us now go over to the physical limit $\dim \to 3$. Using
Eq.~(\ref{perf}) for $G_3({\bf 0})$, we have
\begin{equation}
Z_1 = \exp\left\{ \left[ {\sum}_a n_a^0 +  {  \kappa_0^3
     \over 12 \pi } \right] \vol \right\} ,
\label{zzzone}
\end{equation}
Since
\begin {eqnarray}
\left. {\partial \n0a \over \partial \beta \mu_a}\right|_\beta &=& \n0a \,,
\qquad
\left. {\partial \kappa_0^2 \over \partial \beta \mu_a} \right|_\beta
 = \beta\, e_a^2\, \n0a\,,
\end {eqnarray}
it follows from Eq.~(\ref{zzzone}) that the number density to one
loop order is given by
\begin {eqnarray}
 \bar n_a =   \expect { n_a }^{(1)} &=&
    \n0a \left[ 1 + {\beta \, e_a^2 \kappa_0 \over 8 \pi}  \right] ,
\label{barn}
\end {eqnarray}
in agreement with the physical $\dim \to 3$ limit of the previous
direct calculation (\ref {denone}).  To one-loop order, the pressure
is given by
\begin{equation}
p_1 = T \vol^{-1} \ln Z_1 =
    T \, {\sum}_a\>  \n0a
    \left[ 1  + { \beta \, e_a^2 \, \kappa_0 \over 12 \pi} \right]
    \,.
\label {pone}
\end {equation}
Re-expressing the one-loop pressure in terms of physical particle densities
using Eq.~(\ref{barn}) produces
\begin {eqnarray}
    p_1 &=&
    T \>
 {\sum}_a \bar n_a \biggl[1 -
     { \beta e_a^2 \, \kappa_0 \over 24 \pi}  \biggr] \,.
\label {eq:eos1}
\end{eqnarray}
This is the equation of state of the plasma to
one-loop order.

Using
\begin {eqnarray}
   - \left. {\partial \n0a \over \partial \beta }\right|_{\beta\mu}
 &=&   {3 \over 2} \, T \n0a \,,
\qquad
   - \left. {\partial \kappa_0^2 \over \partial \beta
       }\right|_{\beta\mu}
 =        { 1 \over 2}  T \kappa_0^2 \,,
\end {eqnarray}
it follows from Eq.~(\ref{zzzone})
that the internal energy to one-loop order is given by
\begin{equation}
u_1 = T \, {\sum}_a\>  \n0a
    \left[ { 3 \over 2}  + { \beta \, e_a^2 \, \kappa_0 \over 16 \pi }  \right] .
\end {equation}
or, in terms of the physical density $\bar n_a$,
\begin{equation}
u_1 = T \, {\sum}_a \> \bar n_a
\left[ {3 \over 2}   - {  \beta e_a^2 \kappa_0 \over 8 \pi }
 \right] .
\label{oneu}
\end {equation}
And finally, the Helmholtz free energy density, to one-loop order, is
\begin {equation}
    f_1 = T \, {\sum_a} \> \bar n_a
    \left[
	-1 + \ln (\bar n_a \lambda_a^3/g_a)
	- {\beta e_a^2 \kappa_0 \over 12\pi}
    \right] .
\end {equation}

\subsection{Density-Density Correlators}
\label {den-den}

We now compute the density-density correlator $K_{ab}(\r{-}\r')$
through one loop order.
Expanding about $\phi = 0$,
the first non-vanishing (``tree'' graph) contribution
appears when $\Sone$ is neglected and the explicit
exponentials in (\ref {eq:K}) are expanded to linear order, yielding
\begin {eqnarray}
    K^{\rm tree}_{ab}(\r{-}\r')
    &=&
    \delta_{ab} \delta(\r{-}\r') \, \n0a
    - \beta \, \n0a \n0b \, e_a e_b \> G_\dim(\r{-}\r') \,.
\label {eq:Ktree}
\end {eqnarray}
Fourier transformation produces the density-density correlation
as a function of wave number,
\begin {eqnarray}
    \tilde K^{\rm tree}_{ab}(\k)
    &=&
    \delta_{ab} \, \n0a
    - \beta \,{ e_a \n0a \, e_b \n0b  \over \k^2 + \kappa_0^2}
    \,.
\label {eq:KtreeFT}
\end {eqnarray}

Multiplying this result by $\vol$ and taking the limit $\k \to 0$
gives the tree or mean-field approximation to the total particle number
fluctuations for the various species:
\begin{equation}
\left\langle \left( N_a - \bar N_a \right) \left( N_b - \bar N_b \right)
\right\rangle_\beta^{\rm tree}
= \delta_{ab} \bar N_a - e_a \bar N_a { \beta \over \kappa_0^2 } \, e_b
\bar n_b \,,
\label{avgNN}
\end{equation}
in agreement with the previous result (\ref {eq:nfluct}).
The second term on the right-hand side of this equality is a
consequence of charge neutrality. It involves the ratio of charges,
and shows that one cannot naively expand in powers of charges.
It causes the number
fluctuations to depart from Poisson statistics even in this
lowest-order approximation.
Its presence ensures that
\begin{equation}
\left\langle \left( N_a - \bar N_a \right) Q \right\rangle_\beta^{\rm tree}
= {\sum}_b e_b
\left\langle \left( N_a - \bar N_a \right) \left( N_b - \bar N_b \right)
\right\rangle_\beta^{\rm tree} = 0 \,,
\label{flucneut}
\end{equation}
where in the first equality we made use of total
average charge neutrality,
\begin{equation}
\langle Q \rangle_\beta = {\sum}_a e_a \bar N_a = 0 \,.
\end{equation}
Multiplying Eq.~(\ref{flucneut}) by $e_a$ and summing over $a$ shows
that%
\footnote{In this regard, it is worth noting that
  $\tilde K_{ab}^{\rm tree}({\bf 0})$ is a symmetrical, real, positive,
  semi-definite matrix whose only vanishing eigenvalue appears for the
  eigenvector whose components are the electric charges $e_a$
  (provided all densities $\n0a$ are non-zero).
  These properties are easily demonstrated explicitly.
  First define the matrix $ {\cal N}_{ab} \equiv \delta_{ab}
  \sqrt{\n0a} $ and then the matrix
  $
      {\cal L} \equiv {\cal N}^{-1} \tilde K^{\rm tree}({\bf 0}) {\cal N}^{-1}
  $,
  so that $ {\cal L}_{ab} = \delta_{ab} - v_a \, v_b $ with
  $ v_a \equiv e_a \sqrt{ \beta \n0a} / \kappa_0 $.
  The claimed properties
  hold because $v$ is a unit vector.}
\begin{equation}
\expect { Q^2 }^{\rm tree} = 0 \,.
\label{qtwo}
\end{equation}
Thus, at least at tree level, there is no fluctuation in the total
charge of the ensemble described by our functional integral. The usual
grand canonical ensemble is modified by the long-range Coulomb
potential so that only subsectors of totally neutral particle
configurations appear in the sum over configurations. The general
structure of the number density correlation function described
below [in particular Eq.~(\ref{fluctneut})] shows that the vanishing
of charge fluctuations (\ref{qtwo}) holds to all orders, and thus, in
general, only neutral configurations contribute to the ensemble.
Finally, we note that, to lowest order, charge neutrality
also ensures that the fluctuation of the total
number of particles $ N = {\sum}_a N_a $ in the grand canonical
ensemble is Poissonian,
\begin{equation}
\left\langle \left( N - \bar N \right)^2 \right\rangle_\beta^{\rm tree} =
\sum_{a,b} \left\langle \left( N_a - \bar N_a \right)
\left( N_b - \bar N_b \right)
\right\rangle_\beta^{\rm tree} = \bar N \,.
\end{equation}
As shown in Eq.~(\ref{nonP}) below,
higher-order corrections alter this result.

\begin {figure}[tb]
   \begin {center}
      \leavevmode
      \def\epsfsize #1#2{0.55#1}
      \epsfbox {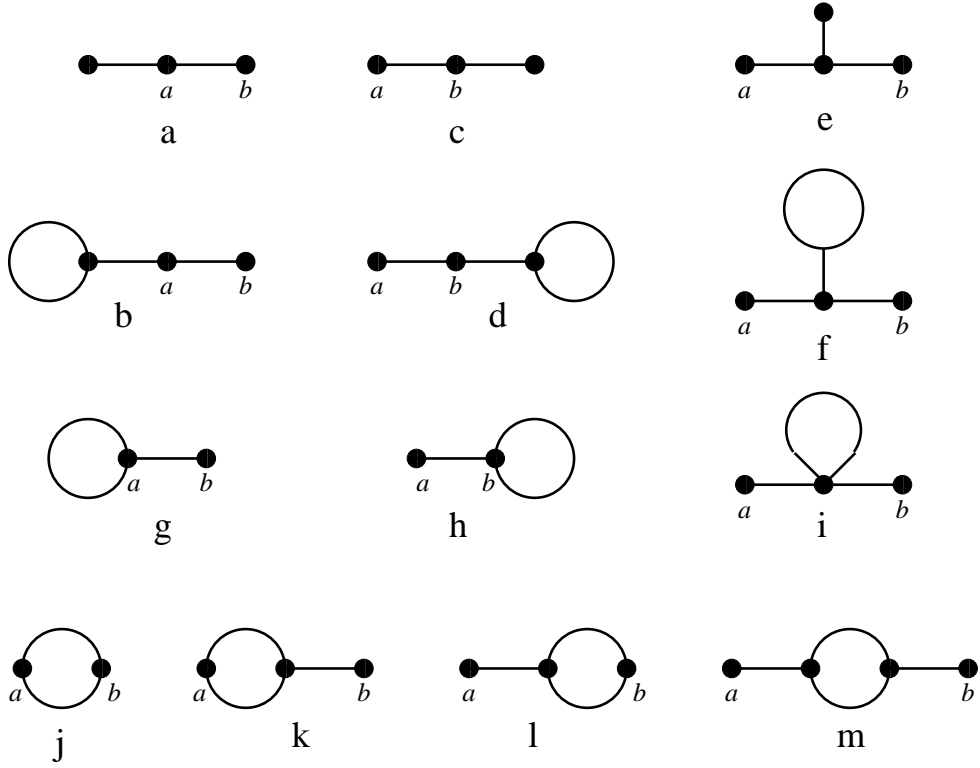}
   \end {center}
   \caption
       {%
       One loop diagrams contributing to the connected density-density
       correlation function
       $K_{ab}(\r{-}\r') = \expect {n_a(\r) n_b(\r')}^{\rm conn}$.
       Diagrams $a$--$f$ are all tadpole diagrams which
       cancel and hence can be neglected.
       Diagrams $g$--$i$ merely serve to correct the
       bare densities appearing in the lowest order result.
       Diagrams $j$--$m$ involve the essentially new contribution
       $C_{ab}^{(1)}$ discussed in the text.
       }%
\label {fig:corr1}
\end {figure}

One-loop corrections to the density correlator are obtained by
expanding both $e^{-\Sone}$ and the exponentials in the density
operator insertions of (\ref {eq:corr}) in powers of $\phi$, and
retaining all next-to-leading order corrections.  This leads to the
one-loop contributions shown graphically in Fig.~\ref{fig:corr1}.
There are three classes of diagrams:
those which cancel, those which simply serve to replace bare densities
by the physical densities (to one-loop order), and the rest.
Diagrams $a$ and $b$ cancel,
as do $c$ \& $d$, and $e$ \& $f$,
because their sum is proportional to $\expect {\phi} \equiv 0$.
Here, as well as in higher orders,
all such ``tadpole'' diagrams can simply be neglected. That these
single-particle reducible graphs%
\footnote
    {
    A graph is `single-particle
    reducible' if it can be separated into two disjoint pieces by
    cutting a single line.
    }
cancel to all orders is proven in Appendix
\ref{funmeth}.
Diagrams $g$ and $h$ correct
the explicit bare densities in (\ref {eq:Ktree})
by
\begin{equation}
\Delta n_a^{(1)} = \langle n_a \rangle_\beta^{(1)} - n_a^0 \,,
\end{equation}
giving the one-loop contribution
\begin {equation}
    -\beta
    \left[  \Delta n_a^{(1)}  \, \n0b +
	    \n0a \,  \Delta n_b^{(1)}
    \right]
    e_a e_b \, G_\dim(\r{-}\r') \,.
\end {equation}
Diagram $i$ corrects the Debye wave number
which appears in the Green's function $G_\dim(\r{-}\r')$;
explicitly it produces
\begin {equation}
    -\beta \, \n0a \n0b \, e_a e_b \>
    {\partial G_\dim(\r{-}\r') \over \partial \kappa_0^2} \>
    {\sum}_a \beta e_a^2 \Delta n_a^{(1)} \,,
\end{equation}
or in Fourier space,
\begin{equation}
   { \beta \, \n0a \n0b \, e_a e_b \over
    (\k^2 + \kappa_0^2)^2} \> {\sum}_a \beta e_a^2 \Delta n_a^{(1)} \,.
\end {equation}
The net effect of these two classes of diagrams
(plus the one loop correction to the
$\delta_{ab} \expect {n_a(\r)}$ contact term)
is to replace, through one loop order,
the particle densities and Debye wave number
appearing in (\ref {eq:KtreeFT}) with their physical values,
\begin {equation}
    \tilde K_{ab}^{\rm tree}(\k)
    \to
    \tilde K_{ab}^{\rm tree,(1)}(\k)
    \equiv
    \delta_{ab} \, \bar n_a
    - \beta \,{ e_a \bar n_a \, e_b \bar n_b  \over \k^2 + \bar\kappa^2} \,.
\label {eq:Ktree1}
\end {equation}
Here $\bar\kappa^2$ is the Debye wave number computed with
physical particle densities,
\begin {equation}
    \bar\kappa^2
    \equiv
    \sum_a \> \beta \, e_a^2 \, \bar n_a  \,.
\label {eq:kappa-phys}
\end {equation}
The second part of Eq.~(\ref{eq:Ktree1}) involves
\begin{equation}
\tilde {\cal G}^{\rm tree,(1)}(\k) = { \beta^{-1}
 \over \k^2 + \bar\kappa^2} \,,
\end{equation}
which is just the Fourier transform of the tree level
electrostatic potential correlator
$\avg {\phi(\r)\phi(\r')}$
as given in Eq.~(\ref{phiphiexp}), but with the physical Debye wave
number $\bar\kappa$.
Understanding the general structure of the number density
correlation function will be facilitated if (\ref{eq:Ktree1}) is
rewritten in the form
\begin {equation}
    \tilde K_{ab}^{\rm tree,(1)}(\k) =
    \delta_{ab} \, \bar n_a
    - (\beta \, e_a \bar n_a )\,\tilde {\cal G}^{\rm tree,(1)}(\k) \,
       (\beta \, e_b \bar n_b ) \,.
\label{smartform}
\end {equation}

The remaining graphs $j$--$m$ give non-trivial corrections.
Diagram $j$ may be viewed as generating a correction
to the first, `contact' term part of (\ref{smartform}),
\begin{equation}
    \delta_{ab} \, \bar n_a \to \tilde C_{ab}(\k)
\label{lowestc}
\end{equation}
where, to one-loop order,
\begin{equation}
    \tilde C^{(1)}_{ab}(\k) =
    \delta_{ab} \, \bar n_a +
    \half (\beta e_a^2 \bar n_a) \, D^{(2)}_\dim(\k) \,(\beta e_b^2 \bar n_b)
    \,,
\label{cone}
\end{equation}
with
\begin{equation}
    D^{(2)}_\dim(\k) \equiv
    \int (d^\dim\r) \> e^{-i \k \cdot \r } \, G_\dim (\r)^2 \,.
\end{equation}
This function represents the loop which is common to diagrams $j$--$m$.
Graphs $k$ and $l$ correspond to making the
corrections
\begin{equation}
    e_a \bar n_a \to \sum_c e_c \, \tilde C_{ca}^{(1)}(\k)\,,
    \qquad
    e_b \bar n_b  \to \sum_c e_c \, \tilde C_{cb}^{(1)}(\k)\,,
\end{equation}
in the factors flanking $\tilde{\cal G}^{\rm tree,(1)}(\k)$
in Eq.~(\ref{smartform}).
Physically,
these diagrams may be viewed as generating corrections
to the coupling between the particle density operators and fluctuations
in the electrostatic potential.
The final graph $m$ is a one-loop polarization (or `self-energy')
correction to the electrostatic potential correlator
\begin {equation}
    {\cal G}(\r{-}\r') = \avg {\phi(\r)\phi(\r')} \,.
\end {equation}
This graph, together with higher order graphs in which the
same ``bubble'' is inserted two or more times,
produce a change in the (Fourier transformed) potential correlator given by
\begin{equation}
    {\tilde{\cal G}^{\rm tree,(1)}(\k)}^{-1} \to
    {\tilde{\cal G}(\k)}^{-1}
    \equiv \beta \biggl[ k^2 + \beta \sum_{a,b} e_a e_b
    \tilde C_{ab}(\k) \biggr] \,,
\label{gginver}
\end{equation}
with the same one-loop result (\ref{cone}) for $\tilde C_{ab}(\k)$.
Note that, according to Eq.~(\ref{cone}),
\begin{equation}
\beta \sum_{a,b} e_a e_b \, \tilde C^{(1)}_{ab}(\k) = \bar\kappa^2 +
    {\beta \over 2} \sum_{a,b}
    (\beta e_a^3 \bar n_a) \, D^{(2)}_\dim(\k) \, (\beta e_b^3 \bar n_b) \,,
\end{equation}
showing that this `self-energy' contribution includes the previous
squared Debye wave number $\bar\kappa^2$ as well as the loop
contribution described by graph $m$.
Putting the pieces together,
we find that the one-loop
corrections conform to the general structure
\begin{equation}
\tilde K_{ab}(\k) = \tilde C_{ab}(\k)
- \left[ \beta {\sum}_c e_c \tilde C_{ca}(\k) \right]
 \,\tilde {\cal G}(\k) \,
\left[ \beta {\sum}_c e_c \tilde C_{cb}(\k) \right].
\label{kstruct}
\end{equation}
That this form holds to all orders is proven in Appendix
\ref{funmeth}, with this result given in Eq.~(\ref{exactk}). This
Appendix shows that $\tilde C_{ab}(\k)$ is a single-particle
irreducible function, symmetric in $a$ and $b$,
and provides its definition in terms of an
effective action functional.
Section \ref{ffunone} of that appendix
also demonstrates how the complete one-loop calculation
may be easily performed using somewhat more sophisticated
functional techniques.

The explicit form of the one-loop function
$ D^{(2)}_\dim(\k) $ is easily evaluated in
three dimensions since
\begin{equation}
G_3^2(\r) = { e^{-2 \kappa_0 r } \over ( 4 \pi r )^2 } =
{ 1 \over 4 \pi} \int_{2 \kappa_0}^\infty d\mu \> { e^{- \mu r} \over 4
  \pi r} \,.
\end{equation}
Thus taking the Fourier transform and interchanging integrals yields
the dispersion relation representation
\begin{equation}
D^{(2)}_3(\k) = { 1 \over 4 \pi} \int_{2 \kappa_0}^\infty
{ d\mu \over k^2 + \mu^2 } \,,
\end{equation}
which is readily evaluated to give
\begin{equation}
    D^{(2)}_3(\k) = (4\pi k)^{-1} \arctan {k \over 2 \kappa_0} \,.
\label {D2}
\end{equation}

The $\k = 0$ limit of $\tilde K_{ab}(\k)$ characterizes
the fluctuations in particle numbers,
\begin {equation}
    \left<
	\left( N_a - \bar N_a \right)
	\left( N_b - \bar N_b \right)
    \right>_\beta
    =
    \vol \, \tilde K_{ab}({\bf 0}) \,.
\end {equation}
The one-loop result for $\tilde K_{ab}({\bf 0})$ is easily generated by
inserting (\ref {D2}) into (\ref {cone}) and thence into (\ref {kstruct}).
In particular, for the total particle number $N = \sum_a N_a$,
one finds to one-loop order
\begin {equation}
    \left<
	\left( N - \bar N \right)^2
    \right>_\beta
    =
    \bar N + {\vol \, \bar \kappa^3 \over 16\pi} \,,
\label{nonP}
\end {equation}
which explicitly shows that the Coulomb interactions generate
non-Poissonian statistics for fluctuations in total particle number.

\subsection{Charge Correlators and Charge Neutrality}

As noted earlier,
the charge neutrality condition (\ref {eq:Q=0})
holds in the presence of arbitrary chemical potentials $\mu_a(\r)$.
Consequently, a corollary of (\ref{eq:Q=0}) is an identity for the
correlator of the number density of some species $a$ with the total
charge:
\begin {eqnarray}
    0
    =
    {\delta \over \delta \mu_a(\r)} \, \expect {Q}
    &=&
    \sum_{b=1}^A  \, \int (d^\dim\r')\> K_{ab}(\r {-}  \r') \, e_b
\nonumber\\
&=&
    \sum_{b=1}^A  \, \tilde K_{ab}({\bf 0}) \, e_b \,.
\label{qvan}
\end {eqnarray}
It follows from the general structure (\ref{kstruct}) of the density
correlator and the form (\ref{gginver}) of the inverse Green's
function that
\begin{eqnarray}
{\sum}_b \tilde K_{ab}(\k) \, e_b &=& {\sum}_b \tilde C_{ab}(\k) \, e_b
\Biggl\{ 1 -  \beta \, \tilde {\cal G}(\k) \,
\Biggl[ \beta \sum_{c,d} e_c e_d \, \tilde C_{cd}(\k) \Biggr] \Biggr\}
\nonumber\\
&=& {\sum}_b \tilde C_{ab}(\k) \, e_b \, \beta \, k^2 \,\tilde{\cal G}(\k) \,,
\label{fluctneut}
\end{eqnarray}
which does indeed vanish in the limit $ k^2 \to 0$ in accordance
with Eq.~(\ref{qvan}).

The charge density --- charge density correlation function is given
by
\begin{eqnarray}
\tilde K(\k) &=& \sum_{a,b} e_a \, \tilde K_{ab}(\k) \, e_b
\nonumber\\
&=& { \, k^2 \left[ \sum_{a,b} e_a \,\tilde C_{ab}(\k) \, e_b \right]
\over k^2 + \beta \left[ \sum_{a,b} e_a \,\tilde C_{ab}(\k) \, e_b
\right] } \,,
\label{chcorr}
\end{eqnarray}
or equivalently
\begin {equation}
    \tilde K(\k) = k^2 \, T - k^4 \, \tilde {\cal G}(\k) \,,
\end {equation}
where $T = 1 / \beta$ is the temperature in energy units.
It has the small wave number limit
\begin{equation}
    \tilde K(\k)     = k^2 \, T + O(k^4) \,.
\label{lowk}
\end{equation}
This relation is known as the Stillinger-Lovett sum rule
\cite {stillinger,martin}.
This limit,
which follows directly from the structure (\ref{kstruct}) that is
established in Appendix \ref{funmeth}, also follows from examining the
coupling of the plasma to a static external electric potential.
The static
dielectric function of the plasma $\epsilon(\k)$ is related to the
charge density correlation function by
\begin{equation}
\tilde K(\k) = { k^2 \over \beta } \left[ 1 - { 1 \over \epsilon(\k) }
\right] .
\label{ekrel}
\end{equation}
This will be derived in the following section [{\em c.f.} Eq.~(\ref{epsk})].
Thus, the small wave number limit (\ref{lowk}) implies that
$\epsilon(\k) \to \infty$ as $ k \to 0$. But this is just the
statement that the plasma is a conductor --- when an external uniform
electric field is applied to
the plasma, charges move and the plasma becomes polarized in such a way
as to completely screen the constant external field.
The small wave number
behavior of the static dielectric function is made explicit by inserting
Eq.~(\ref{chcorr}) in Eq.~(\ref{ekrel}) to obtain
\begin{equation}
\epsilon(\k) = 1 + {\beta \over k^2}
  \Biggl[ \, \sum_{a,b} e_a \,\tilde C_{ab}(\k) \, e_b \Biggr] \,.
\label {eq:eCrel}
\end{equation}

\section {Effective Field Theory}
\label {sec:quasi}

We have just worked out the statistical mechanics of
a classical, multicomponent plasma through one-loop order.
One cannot go to higher order in this purely
classical theory.
Ultraviolet divergences appear at two-loop order and beyond.
For example, the pressure in two-loop order
receives a contribution from the diagram
\begin {equation}
\begin {picture}(40,40)(0,5)
    \thicklines
    \put(20,20){\circle{40}}
    \put(0,20){\line(1,0){40}}
    \put(0,20){\circle*{5}}
    \put(40,20){\circle*{5}}
\end {picture}
\label {fig:2-loop}
\end {equation}
which is proportional to the integral of the cube of the Debye Green's
function,
$
    \int (d^\dim\r) \> G_\dim(\r)^3
$.
In three-dimensions, the short-distance part of
this integral behaves as $\int (d^3\r) \, / r^3$,
which is logarithmically divergent.
This divergence can be seen in an elementary fashion
directly from the divergence (for opposite signed charges)
of the Boltzmann-weighted integral over the relative separation
of two charges,
$
    \int (d^3\r) \> \exp\{-\beta e_a e_b V_C(\r)\}
$.
Diagram (\ref {fig:2-loop}) is just the third-order term in the
expansion of this integral in powers of the charges.
These ultraviolet divergences of the classical theory are
tamed by quantum-mechanics ---
quantum fluctuations smear out the short distance singularities.
To reproduce the effects of this quantum mechanical smearing,
we must augment our previous dimensionally regulated classical
theory with additional local interactions which
both serve to cancel the divergences present
in diagrams such as (\ref{fig:2-loop}),
and reproduce quantum corrections which are suppressed
by powers of $\hbar$ (or equivalently $\kappa\lambda$).
The coefficients of some of these induced interactions will
diverge in the $\dim \to 3$ limit.
The finite parts of these coefficients (or ``induced couplings'')
will then be determined by matching predictions of this
effective quasi-classical theory
with those of the underlying quantum mechanical theory.

\subsection {Quantum Theory}

The full (non-relativistic) many-body quantum theory
generates the grand canonical partition function
--- extended to be a number density generating functional
$Z_{\rm QM}[\mu]$ by the introduction of the generalized, spatially
varying chemical potentials $\mu_a(\r)$ ---
as a trace over all states,
\begin{equation}
Z_{\rm QM}[\mu] = {\rm Tr} \> \exp\biggl\{ - \beta \biggl[ H - \sum_{a=1}^A
   \int(d^3\r) \> \mu_a(\r) \, n_a(\r) \biggr] \biggr\} \,,
\end{equation}
where $n_a(\r)$ is the number density operator for
particles of species $a$.
The multi-particle Hamiltonian of the complete system has the structure
\begin{equation}
H = \sum_{a=1}^A K_a + \sum_{a,b=1}^A \, H^{\rm Coul}_{ab}
\,,
\end{equation}
where $K_a$ represents the kinetic energy of all particles of
species $a$
and $H^{\rm Coul}_{ab}$ is the Coulomb energy between particles
of types $a$ and $b$.
In second-quantized notation,
\begin {equation}
    K_a =  {1 \over 2m_a} \int (d^3\r) \>
    \nabla \hat\psi_a(\r)^\dagger \cdot \nabla \hat\psi_a(\r) \,,
\end{equation}
and
\begin{equation}
    H^{\rm Coul}_{ab} =
    {e_a e_b \over 2} \int (d^3\r) (d^3\r') \>
    \hat\psi_a(\r)^\dagger \hat\psi_b(\r')^\dagger \>
    V_C(\r{-}\r') \>
    \hat\psi_b(\r') \hat\psi_a(\r) \,.
\end {equation}

The quantum-mechanical partition function $Z_{\rm QM}[\mu]$ may be
expressed as a functional integral
involving $A$ pairs of fields
$\psi_a^* (\r,\tau)\,,\, \psi_a(\r,\tau)$
defined on the imaginary time interval $[0,\beta]$.%
\footnote
    {%
    These fields may be either complex fields satisfying periodic
    boundary conditions, $\psi_a(\r,\tau{+}\beta) = \psi_a(\r,\tau)$,
    or anti-commuting Grassmann algebra valued fields satisfying antiperiodic
    boundary conditions, $\psi_a(\r,\tau{+}\beta) = -\psi_a(\r,\tau)$.
    The first case describes the quantum mechanics of Bosons,
    while the second describes Fermions.
    The following discussion is applicable to either case.
    }
Just as in the previous section,
the Coulomb interaction between charges
can be written in terms of a Gaussian functional integral over an
auxiliary electrostatic potential.
Therefore,%
\footnote
    {%
    It should be emphasized that our functional integral representation
    (\ref {eq:Zquant})
    involves an integral over the Gaussian distributed
    electrostatic potential $\phi$
    together with functional integrals over matter fields
    that generate the ideal gas partition function of a second-quantized theory
    in the presence of the background field $\phi$.
    This is in marked contrast to approaches
    using first-quantized representations
    in which one writes a path integral ({\em i.e.}, Feynman-Kac formula)
    over the trajectories of individual particles.
    For a short discussion of this Feynman-Kac approach
    see, for example, section 3.3 of Ref.~\cite {alastuey},
    and references therein.
    }
\begin{eqnarray}
Z_{\rm QM}[\mu] &=& {\rm Det}^{1/2} \left[ - \nabla^2 \right] \int [ d \phi ]
  \> \exp\left\{ - {1 \over 2} \int_0^\beta d\tau \int (d^\dim\r) \>
  \, \Bigl( \nabla \phi(\r, \tau) \Bigr)^2 \right\}
\nonumber\\
&& \qquad\qquad \times
    \prod_{a=1}^A \int [d\psi_a^* d\psi_a] \>
    \exp\left\{ - \int_0^\beta d\tau \int (d^\dim\r) \> {\cal L}_a \right\} ,
\label {eq:Zquant}
\end{eqnarray}
where
\begin{equation}
{\cal L}_a = \psi_a^*(\r, \tau)
\left\{ { \partial \over \partial \tau} - { \nabla^2 \over 2 m_a} -
  \mu_a(\r) - i e_a \phi(\r , \tau) \right\} \psi_a(\r, \tau) \,.
\label {eq:Lqm}
\end{equation}
The integrations are now over $\dim<3$ spatial dimensions,
since we work with the dimensionally regulated theory.
As explained earlier, the dimensionally continued Coulomb potential
vanishes at vanishing spatial separation
[Eq.~(\ref{zero})], and so there are no infinite particle
self-energies with this regularization scheme.

If the generalized chemical potentials have arbitrary variation in
both space and imaginary time, then $\ln Z_{\rm QM}[\mu]$ is the
generating functional for connected time-ordered correlation functions
of the density operators
\begin{equation}
n_a(\r,\tau) = \psi_a^*(\r,\tau) \, \psi_a(\r,\tau) \,.
\end{equation}
These correlation functions are periodic in the imaginary time $\tau$
with period $\beta$. Thus they have a Fourier series representation
with frequencies $\omega_n \equiv 2\pi n / \beta = 2\pi n T / \hbar$, where
in the last equality we have restored Planck's constant $\hbar$.
In the $\hbar \to 0$
classical limit, all these frequencies run off to infinity save for
the static $n=0$ mode. Thus the classical limit involves zero-frequency
correlators and, correspondingly, generalized chemical potentials that
are independent of the imaginary time%
\footnote
    {
    Generalized chemical
    potentials that depend upon both space and {\em real\/} time do,
    however, have a role to play in the classical theory since they may
    be used to probe the response to time-dependent disturbances.
    }
$\tau$.
If one is only interested in thermodynamic quantities, or zero frequency
correlators, then one may restrict the generalized
chemical potentials to be time-independent.%
\footnote
    {%
    We will only need to use time-dependent chemical potentials
    when discussing the difference between zero-frequency and equal
    time correlators in section \ref {sec:longdist}.
    }
Since the extended
Hamiltonian of the system including the chemical potential terms
is then time independent, the ensemble averages remain time-translationally
invariant. Thus $ \expect { n_a (\r ,\tau) } $ is independent of $\tau$,
and it may be replaced by the $\tau = 0$ form $ \expect { n_a (\r) } $.
Accordingly,
\begin{equation}
    {\delta \over \delta \beta\mu_a(\r)} \> \ln Z_{\rm QM}[\mu]
= { 1 \over \beta } \int_0^\beta d\tau \, \expect { n_a (\r ,\tau) }
= \expect { n_a (\r) } \,.
\end{equation}
The variational derivative of this result now yields%
\footnote
    {%
    The final form shown for the second variation (\ref {eq:Kqm})
    involves an integral over
    imaginary time of the time-ordered correlation function
    $
	K_{ab}(\r,\tau;\r',\tau') =
	{\delta \over \delta \mu_a(\r,\tau)}
	{\delta \over \delta \mu_b(\r',\tau')}
	\ln Z[\mu]
    $
    which is symmetric,
    $
	K_{ab}(\r,\tau;\r',\tau')=K_{ba}(\r',\tau';\r,\tau)
    $,
    periodic in imaginary time,
    $
	K_{ab}(\r,\tau;\r',\tau')=
	K_{ab}(\r,\tau{-}\beta;\r',\tau')=
	K_{ab}(\r,\tau;\r',\tau'{-}\beta)
    $,
    and (when evaluated at constant chemical potentials),
    time-translation invariant,
    $
	K_{ab}(\r,\tau;\r',\tau')=
	K_{ab}(\r,\tau{-}\tau';\r',0)
    $.
    Since the integral in (\ref {eq:Kqm}) has $\tau > 0$,
    the product of density operators appearing in the integrand is trivially
    time-ordered.
    }
\begin{eqnarray}
    K_{ab}(\r,\r') &=&
    {\delta \over \delta \beta\mu_a(\r)} \,
    {\delta \over \delta \beta\mu_b(\r')} \> \ln Z_{\rm QM}[\mu]
\nonumber\\
&=&    \beta^{-1} \int_0^\beta d\tau \>
    \left[
    \expect { n_a (\r,\tau) \, n_b(\r',0) }
    -
    \expect { n_a (\r) }
    \expect { n_b(\r') }
    \right] .
\label {eq:Kqm}
\end {eqnarray}

If every chemical potential is shifted by an amount proportional to
the corresponding charge, $\mu_a(\r)  \to \mu_a(\r) + e_a \lambda(\r)$,
then derivatives of the partition function with respect to $\lambda(\r)$
generate correlation functions of the charge density
$\rho(\r,\tau) \equiv \sum_a e_a \, n_a(\r,\tau)$,
\begin{equation}
    \expect { \rho (\r) }
    =
    \sum_a e_a \, \expect {n_a(\r)}
    =
    \left.
    {\delta \over \delta \beta\lambda(\r)} \> \ln Z_{\rm QM}[\mu+ e\lambda]
    \right|_{\lambda=0}
    \,,
\end{equation}
and
\begin{eqnarray}
    K(\r,\r')
    &=&
    \sum_{a,b} \, e_a \, e_b \, K_{ab}(\r,\r')
    =
    \left.
    {\delta \over \delta \beta\lambda(\r)} \,
    {\delta \over \delta \beta\lambda(\r')} \> \ln Z_{\rm QM}[\mu+e\lambda]
    \right|_{\lambda=0}
\nonumber\\
&=&
    \beta^{-1} \int_0^\beta d\tau \>
    \left[
    \expect { \rho (\r,\tau) \rho(\r',0) }
    -
    \expect { \rho (\r) }
    \expect { \rho(\r') }
    \right].
\end {eqnarray}
Alternatively, if one makes a compensating change of variables
$\phi \to \phi + i \lambda$ in the functional integral (\ref {eq:Zquant}),
all dependence on $\lambda$ disappears from the charged field
Lagrangian ${\cal L}_a$, and the net effect is merely to shift the
Gaussian measure for the electrostatic potential,
\begin {eqnarray}
    \exp\left\{-{1\over2} \int d\tau \, (d^\dim\r) \>
      [ \nabla \phi ]^2 \right\}
    &\to&
    \exp\left\{-{1\over2} \int d\tau \, (d^\dim\r) \>
	[ \nabla (\phi + i \lambda) ]^2 \right\}
    \nonumber
    \\ &=&
    \exp\left\{-{1\over2} \int d\tau \, (d^\dim\r) \>
	\left[
	( \nabla \phi) ^2
	- 2 i \phi \nabla^2 \lambda
	- (\nabla \lambda)^2 \right]
	\right\}.
\end {eqnarray}
Hence,
\begin {eqnarray}
    Z_{\rm QM}[\mu + e \lambda]
    &=&
    Z_{\rm QM}[\mu] \>
    e^{ \beta \int (d^\dim\r) \> {1\over2}(\nabla \lambda)^2} \,
    \avg{ e^{i \int d\tau \, (d^\dim\r) \> \phi\nabla^2 \lambda } },
\end {eqnarray}
and consequently
\begin {eqnarray}
    \expect {\rho(\r)}
    &=&
\left.
    {\delta \over \delta \beta\lambda(\r)} \> \ln Z_{\rm QM}[\mu+ e\lambda]
\right|_{\lambda=0}
=
    \nabla^2 i \avg { \phi(\r) } ,
\label {eq:PoissonQM}
\end {eqnarray}
and
\begin {eqnarray}
    K(\r,\r')
    &=&
    \left.
    {\delta \over \delta \beta\lambda(\r)} \,
    {\delta \over \delta \beta\lambda(\r')} \> \ln Z_{\rm QM}[\mu+e\lambda]
    \right|_{\lambda=0}
    \nonumber
    \\&=&
    - \beta^{-1} \nabla_\r^2 \, \delta(\r{-}\r')
    - \nabla_\r^2 \, \nabla_{\r'}^2 \, {\cal G}(\r,\r') \,,
\label {eq:KQM}
\end {eqnarray}
where $\cal G$ is the zero-frequency correlator of
fluctuations in the electrostatic potential,
\begin {equation}
    {\cal G}(\r,\r')
    \equiv
    \beta^{-1}
    \int_0^\beta d\tau \>
    \Bigl[ \avg {\phi(\r,\tau) \phi(\r',0) }
    - \avg {\phi(\r)} \avg {\phi(\r') }\Bigr].
\end {equation}
The relation (\ref {eq:PoissonQM}) is just the Poisson equation (now
derived in the full quantum theory).  When the chemical potentials
have no spatial variation, $\avg {\phi(\r)}$ is constant, the charge
density $\expect {\rho(\r)}$ vanishes, and the correlation
functions $K(\r,\r')$ and ${\cal G}(\r,\r')$ depend only on
$\r{-}\r'$.  In this case, Eq.~(\ref {eq:KQM}) becomes a simple
relation between the Fourier transformed correlators,
\begin {equation}
    \tilde K(\k)
    =
    \beta^{-1} k^2 - (k^2)^2 \, \tilde {\cal G}(\k) \, .
\label {eq:KQM2}
\end {equation}
Because of screening, $\tilde {\cal G}(\k)$ is bounded as $\k \to 0$.
Hence,
\begin {equation}
    \tilde K(\k) = T k^2 + O(k^4) \,,
\label {eq:Klim}
\end {equation}
and we have an alternative derivation of the Stillinger-Lovett relation
(\ref{lowk}) discussed in the previous section.

The charge density correlator $\tilde K(\k)$
is directly related to the static dielectric function of the plasma.
To see this,
note that $Z[\mu + e \lambda]$ is precisely the partition function
in the presence of an applied electrostatic potential $-\lambda(\r)$.
The variation of charge density with respect to $\lambda$ is
just the charge density correlator times $\beta$,
$\delta \!\expect{\rho(\r')} / \delta \lambda(\r) = \beta K(\r,\r')$.
Hence, the Fourier transform of the
charge density induced by this applied potential,
to first order in the applied field, is
$
    \tilde \rho_{\rm ind}(\k)
    =
    \beta \tilde K(\k) \, \tilde \lambda(\k)
$,
or equivalently the induced electric field is
\begin {equation}
    \tilde {\bf E}_{\rm ind}(\k)
    =
    - {i \k \over k^2} \> \tilde \rho_{\rm ind}(\k)
    =
    -{\beta \tilde K(\k) \over k^2} \> \tilde {\bf D}(\k) \,,
\end {equation}
where ${\bf D}(\r) = {\bf \nabla} \lambda(\r)$ is the applied field.
The ratio of the applied field to the total field (at a given wave number)
defines the static dielectric function $ \epsilon(\k) $,
\begin {equation}
    \tilde {\bf D}(\k) = \epsilon(\k) \left\{
    \tilde {\bf D}(\k) + \tilde {\bf E}_{\rm ind}(\k) \right\} .
\end{equation}
Thus
\begin{equation}
\epsilon(\k) =
    \left[ 1 - {\beta \tilde K(\k) \over k^2} \right]^{-1}
    =
    {1 \over k^2 \, \tilde {\cal G}(\k)} \,.
\label{epsk}
\end {equation}
The first equality is equivalent to Eq.~(\ref{ekrel}) asserted previously.
The condition (\ref {eq:Klim}) implies
that $\epsilon(\k)$ diverges as $\k \to 0$.
This, of course, reflects the fact that the plasma
is a conducting medium which exactly screens uniform
applied electric fields.
Finally, expressing the correlator $\tilde{\cal G}(\k)$
in terms of the self-energy (or polarization tensor),
$\tilde{\cal G}(\k)^{-1} = \beta [k^2 + \Pi(\k)]$,
shows that $\Pi(\k)/k^2$ and $\epsilon(\k)$ are related by
\begin {equation}
    \epsilon(\k) -1
    =
    {\Pi(\k) \over k^2} \,.
\end {equation}
Appendix \ref {funmeth} (as quoted in Eq.~(\ref{gginver}))
shows that the self-energy
$\Pi(\k) = \beta \sum_{a,b} e_a e_b \tilde C_{ab}(\k)$.
Inserting this form yields the previously quoted
relation (\ref{eq:eCrel})
between the dielectric function and $\tilde C_{ab}(\k)$.

\subsection {Classical Limit}

In the limit in which the
thermal wavelength $\lambda_a$ is much smaller
that the scale of spatial variation in the electrostatic potential,
$\lambda_a |\nabla \ln\phi(\r)| \ll 1$,
the functional integral over the charged fields $\psi_a^*$ and $\psi_a$
may be performed explicitly.
Appendix \ref {app:det} presents this calculation in detail.
Neglecting corrections suppressed by powers of $\lambda_a$,
one finds that
\begin{equation}
    \int [d\psi_a^* d\psi_a] \exp\biggl\{
    - \int_0^\beta d\tau \int (d^\dim\r) \> {\cal L}_a \biggr\} =
    \exp\left\{ \int (d^\dim\r) \> \n0a(\r) \> e^{ i e_a  \int_0^\beta d\tau \,
    \phi(\r, \tau) } \right\} .
\label{funint}
\end{equation}
This is just the classical limit of the quantum partition function
for particles moving in a background potential $-i\phi(\r,\tau)$.
Here
$\n0a(\r)$ is the free-particle density of species $a$ (in $\dim$ dimensions),
\begin{equation}
    \n0a(\r) = g_a e^{\beta \mu_a(\r)}
    \left( { m_a \over 2 \pi \beta } \right)^{\dim/2}
    =  g_a \, \lambda^{-\dim}_a \, e^{\beta \mu(\r)} \,,
\end{equation}
(which reduces to (\ref {eq:n0a}) when $\dim\to3$).
Notice that the result (\ref{funint}) only depends on the
time-integral of the electrostatic potential.%
\footnote
    {%
    This will not be true when sub-leading terms are included,
    as discussed later in this section.
    }
Consequently, it is useful to make a Fourier series expansion of the
electrostatic potential on the imaginary time interval $ 0 < \tau <
\beta $. We separate out the zero frequency mode by writing
\begin{equation}
\phi(\r , \tau) = \phi(\r) + \sum_{n \neq 0} \phi^n(\r) \,
e^{-i \omega_n \tau } \,,
\label {eq:fourier-phi}
\end{equation}
where
\begin{equation}
    \omega_n \equiv { 2 \pi n \over \beta } \,.
\end{equation}
Since $\phi(\r,\tau)$ is real, the zero mode part is real, $\phi(\r)^*
= \phi(\r)$, while $\phi^{-n}(\r) = \phi^n(\r)^*$.  The non-zero
frequency modes do not contribute to the functional integral result
(\ref{funint}). Hence, in this classical limit, the non-zero frequency
modes only appear in the initial Gaussian functional integral in
Eq.~(\ref{eq:Zquant}), and they may be trivially integrated out.
Their only effect is to change the determinantal prefactor in
Eq.~(\ref{eq:Zquant}) from its implicit $\dim+1$ dimensional form to
an $\dim$-dimensional form which just normalizes the Gaussian
functional integral of the zero modes to unity if there were no other
factors.  Hence, in the classical limit one finds that
\begin {equation}
    Z[\mu] = N_0 \int [ d \phi ]
    \> e^{-S_{\rm cl}[\phi;\mu]} \,,
\end {equation}
where
\begin{eqnarray}
    S_{\rm cl}[\phi;\mu] &=&
      \int (d^\dim\r) \>
      \left\{
      {\beta \over 2} \, \phi(\r) \left[ - \nabla^2 \right] \phi(\r)
	-
	\sum_{a=1}^A \> \n0a(\r) \, e^{ i \beta e_a \phi(\r) }
    \right\}.
\label {eq:Zclass}
\end{eqnarray}
This is precisely the representation (\ref{fun}) for the classical
partition function derived in the preceding section.
We have just seen that this form emerges naturally as the limit
of the quantum partition function.

\subsection {Induced Couplings}
\label {sec:induced}

But this ``derivation'' of (\ref {eq:Zclass}) as the classical limit
of the quantum partition function (\ref {eq:Zquant}) is {\em wrong\/}!
As emphasized earlier,
the classical partition function (\ref {eq:Zclass}) is singular
when $\dim \to 3$, while the quantum partition function
(\ref {eq:Zquant}) is completely regular in three dimensions.
It is impossible for the classical partition function (\ref{eq:Zclass})
to equal the quantum partition function up to negligible corrections.
What went wrong was the use of Eq.~(\ref {funint}),
which is valid for a sufficiently slowly varying background
$\phi(\r,\tau)$, inside a functional integral over fluctuations
in $\phi$ --- which includes fluctuations on scales
comparable to the typical de Broglie wavelengths of the charged particles.
In other words,
the contributions of short distance fluctuations in $\phi$
were mangled when going from the quantum partition function
(\ref {eq:Zquant}) to the classical partition function (\ref {eq:Zclass}).
To fix this error one may, in principle, integrate exactly
over the charged fields $\psi(\r,\tau)$ together with the
non-zero frequency modes $\phi^n(\r)$ of the electrostatic potential,
to produce a non-local, effective action $S_{\rm QM}[\phi;\mu]$
for the remaining zero-frequency mode $\phi(\r)$ such that
\begin{equation}
Z[\mu] = N_0 \int [d\phi] \> e^{-S_{\rm QM}[\phi;\mu]} \,.
\label{qmfi}
\end{equation}
However, explicitly constructing or dealing with this non-local action
is impossible.
Our aim is to construct a local approximation to $S_{\rm QM}$
which retains those parts of the complete non-local action
which must be added to the classical theory to obtain finite,
correct results to a given order in powers of the ratio
of scales $\kappa \lambda$.
To do this, the first step is to regulate the theory
by working in $\dim < 3$ dimensions
and then add to the classical action (\ref {eq:action})
additional local terms, referred to as induced interactions,
which both serve to fix the incorrect short-distance behavior
of the classical theory,
and incorporate quantum effects suppressed%
\footnote
    {%
    It will also be necessary to include non-linear
    interactions involving the non-zero frequency
    modes $\phi^n(\r)$.
    This will be discussed at the end of this section.
    }
by powers of $\hbar$,
\begin{equation}
    \Scl[\mu] \to
    \Seff[\phi;\mu] \equiv \Scl[\phi;\mu] + \Sct[\phi;\mu] \,.
\label {eq:Seff}
\end{equation}
The induced interactions $\Sct$ may, in general, include arbitrary
combinations of the field $\phi(\r)$ and its derivatives at a point $\r$,
integrated over all space.  However, only terms which are consistent
with the symmetries of the original underlying theory can appear.  Of
particular importance is the invariance $\phi \to \phi -i c$ and
$\mu_a \to \mu_a - e_a c$. The discussion of
Eq.~(\ref{eq:shift-sym}) shows that this is an invariance of the
classical theory. In view of the structure (\ref {eq:Lqm}) of the
quantum Lagrangian, this shift is also an invariance of the full
quantum theory.
In addition, a constant shift in any chemical
potential by an integer multiple of $2\pi i/\beta$ causes no
change in the grand canonical partition function,
and hence must also not change the effective theory.%
\footnote
    {
    Because the total particle number operators $\hat N_a$ 
    have only integer eigenvalues,
    a shift in $\beta\mu_a$ by an integer multiple of $2\pi i$
    leaves unchanged the exponential of $\beta \mu_a \hat N_a$
    appearing in the grand canonical partition function.
    }
Consequently, only the combination $\n0a \, e^{i\beta e_a \phi}$,
which is invariant under the combined shift of $\phi$
and $\mu_a$, plus spatial derivatives of $ \mu_a(\r) {+}
i e_a \phi(\r) $,
will appear in $\Sct$. As a result, the induced interactions have the
general structure
\begin{eqnarray}
    && \Sct[\phi;\mu]
=
    \sum_{p=2}
    \sum_{a_1 \cdots a_p} \int (d^\dim\r) \>
    g^0_{a_1 \cdots a_p} \,
    \beta^{3(p-1)} \,
    n_{a_1}^0(\r) \, e^{i\beta e_{a_1} \phi(\r) }
    \cdots
    n_{a_p}^0(\r) \, e^{i\beta e_{a_p} \phi(\r) }
\nonumber\\
&& \quad {} +
    \sum_{p=1}
    \sum_{a_1 \cdots a_p} \! \int (d^\dim\r) \>
    h^0_{a_1 \cdots a_p} \,
    \beta^{3p+1}
    \,\Bigl[ \nabla
	\Bigl( \mu_{a_1}(\r) {+}i e_{a_1} \phi(\r) \Bigr)
    \Bigr]^2
    n_{a_1}^0(\r) \, e^{i\beta e_{a_1} \phi(\r) }
    \cdots
    n_{a_p}^0(\r) \, e^{i\beta e_{a_p} \phi(\r) }
\nonumber\\
&& \quad {} +
    \sum_{p=2}
    \sum_{a_1 \cdots a_p} \int (d^\dim\r) \>
    k^0_{a_1 \cdots a_p} \,
    \beta^{3p+1}
    \,\Bigl[
	\nabla \Bigl( \mu_{a_1}(\r) {+}i e_{a_1} \phi(\r) \Bigr)
	\cdot
	\nabla \Bigl( \mu_{a_2}(\r) {+}i e_{a_2} \phi(\r) \Bigr)
    \Bigr]
\nonumber\\ && \kern 2.5in {} \times
    n_{a_1}^0(\r) \, e^{i\beta e_{a_1} \phi(\r) }
    \cdots
    n_{a_p}^0(\r) \, e^{i\beta e_{a_p} \phi(\r) }
\nonumber\\
&& \quad {} +
    \cdots \,,
\label{Sind}
\end{eqnarray}
where the final ellipsis $\cdots$ stands for similar
terms with four or more derivatives.

For calculations to a given loop order,
only a finite number of the induced interactions are needed.
The classification of the various terms according to the order
in which they first contribute will be spelled out below.

Interactions involving only a single density
(that is, the classical $-\n0a e^{i \beta e_a \phi}$ interaction,
the two-derivative term proportional to $h^0_a$,
and corresponding higher derivative terms)
have coefficients which are finite in three dimensions,
and are simply determined by expanding the charged field
functional integral (\ref {funint}) as described in Appendix \ref {app:det}.
The result (\ref {eq:S2}) of this appendix,
also shown in Eq.~(\ref{eq:lnzFT}), gives
\begin {equation}
    h^0_a = {\lambda_a^2 \over 48\pi \, \beta^2} =
{ \hbar^2 \over 24 \beta \, m_a} \,.
\label {eq:h0a}
\end {equation}

The induced couplings $g^0_{a_1 \cdots a_p}$
(as well as $h^0_{a_1 \cdots a_p}$, {\em etc}.)
multiplying two or more densities
will contain poles in $\dim{-}3$ which serve to cancel poles
at $\dim{=}3$ generated by two-loop and higher order graphs
generated by the classical interaction
(or the single-density induced interactions).
The ``infinite'' parts of $\Sct[\phi;\mu]$,
(that is, the residues of these pole terms) are relatively
easy to calculate --- they are precisely the terms needed
to make the complete theory finite (as it must be).
This will be illustrated explicitly in the following subsection.
The remaining ``finite'' parts of these induced couplings,
the non-pole terms, can only be obtained by matching results
for some physical quantity computed in this effective theory with
corresponding results for the same quantity computed in the original
(full quantum) theory.
The first such matching for an induced coupling
will be performed at the end of this section.
Once the required matching is done, to a given loop order,
the effective theory may then be used to
calculate any other physical quantity.

To ascertain the loop order of the
various induced interactions, we note that since $\int (d^\dim\r) \> \n0a$
is dimensionless, $g^0_{a_1 \cdots a_p}$ times the remaining
$p{-}1$ factors of $\beta^3 \n0a$ must be dimensionless.
In the physical $\dim \to 3$ limit, the particle density $\n0a \sim 1/ d^3$,
where $ d $ is the interparticle spacing,
$ e^2 / d $ has the dimensions of energy,
and $ \beta e^2 / d $ is dimensionless.
Hence $g^0_{a_1 \cdots a_p}$ must be a pure number%
\footnote
    {%
    More precisely, a dimensionless function of the various
    quantum parameters $\beta e_a^2 / \lambda_a$.
    }
times $e^{6(p-1)}$
(where, by $e^6$ we mean six factors of the various charges $e_a$),
so that
each of the $p{-}1$ densities is accompanied by a factor of $\beta^3 e^6$.
Equivalently, each of the $p{-}1$ densities appears in the form
$\beta^2 e^4 (\beta e^2 \n0a) \sim [\beta e^2 \kappa_0]^2$.
Recalling that $\beta e^2 \kappa_0$ is just the loop-counting
parameter, we see that the $g^0_{a_1\cdots a_p}$
interaction with no derivatives
and $p$ densities will first contribute at $2(p{-}1)$ loop order.
Similarly, for the interactions with two derivatives,
$h^0_{a_1 \cdots a_p}$ and $k^0_{a_1 \cdots a_p}$
must both be dimensionless functions of the quantum parameters
times $e^{6p-2}$ in $\dim{=}3$ dimensions.
This is because
each particle density is again accompanied by a factor of $\beta^3 e^6$,
so that the $p$-density two-derivative interactions involve the
dimensionless quantity
$(\beta e^2 \kappa_0)^{2p} \int (d\r) \> \beta (\nabla \phi)^2$.
Consequently, the induced couplings
$h^0_{a_1 \cdots a_p}$ and $k^0_{a_1 \cdots a_p}$
first contribute to correlation functions at $2p$ loop order.
Induced interactions with four or more derivatives,
which were not displayed explicitly in (\ref {Sind}),
are only needed for calculations at four loop order or beyond.
Note that there are no induced couplings which first
contribute at any odd loop order.

The multiple-density induced couplings have poles at $\dim{=}3$, and so the
dimensionality $\dim$ must be kept away from three until all terms of
a given order have been combined.  The extra dimensional factors
needed away from $\dim{=}3$ have the form of factors of $\lambda^{3 -
  \dim}$, where $\lambda$ stands for a characteristic thermal
wavelength of particles in the plasma. Since the Coulomb
potential in $\dim$ dimensions has the coordinate dependence $ r^{2
  -\dim}$, an extension of the analysis in the previous paragraph
shows that the induced coupling
$g^0_{a_1 \cdots a_p} \propto \lambda^{-2(p-1)(\dim-3)}$
while both $h^0_{a_1 \cdots a_p}$
and $k^0_{a_1 \cdots a_p}$ are proportional to $\lambda^{-2p(\dim-3)}$.

    Because the interactions depend on the chemical potentials,
physical particle densities in the effective theory (\ref {eq:Seff}) are
not equal to the functional integral average of $\n0a \, e^{i \beta e_a \phi}$,
as in the original classical theory.
Rather,
\begin {equation}
    \expect { n_a }
    \equiv
    {\delta \ln Z \over \delta \beta\mu_a }
    =
    -\avg { {\delta \Seff \over \delta \beta\mu_a} }
    =
    \avg { \n0a \> e^{i \beta e_a \phi}
    - {\delta \Sct \over \delta \beta\mu_a} } \,,
\end {equation}
and similarly for the density-density correlator,
\begin {eqnarray}
    K_{ab}(\r,\r')
    &\equiv&
    {\delta^2 \ln Z \over \delta \beta \mu_a(\r) \, \delta \beta \mu_b(\r') }
    \nonumber
    \\&=&
    \avg {
	{\delta \Seff \over \delta \beta \mu_a(\r)}
	{\delta \Seff \over \delta \beta \mu_b(\r')}
	}
    -
    \expect {n_a} \expect {n_b}
    -
    \avg {
    {\delta^2 \Seff \over \delta \beta \mu_a(\r) \, \delta \beta \mu_b(\r') }
    } \,.
\label {eq:Keff}
\end {eqnarray}

\subsection {Renormalization}
\label {sec:renorm}

The residue of a pole in an induced coupling
may be determined by calculating a suitable $n$-point density correlator
to a given loop order, and requiring that the result be finite
as $\dim\to3$. Once this has been done for all the couplings that appear
in a given order, then any other process will be finite to this order.
In addition to the pole terms in the induced couplings, there are,
of course, finite remainders. These finite terms are determined by
matching a result computed in our effective theory to the same result
computed in the full quantum theory. We shall take up the matching
problem later. Here we shall exhibit the nature of the (infinite) pole
terms by examining several examples.

At two-loop order, the induced coupling $g^0_{ab}$ contributes
through the last term in (\ref {eq:Keff})
to the irreducible part $\tilde C_{ab}(\k)$ of the density-density correlator.
The only other contributions at this order
which are singular as $\dim\to3$ are the diagrams:
\begin {equation}
\raisebox{-10pt} {
\begin {picture}(120,40)(-40,5)
    \thicklines
    \put(20,20){\circle{40}}
    \put(0,20){\line(1,0){40}}
    \put(0,20){\circle*{5}}
    \put(40,20){\circle*{5}}
    \put(-5,20){\vector(-1,0){25}}
    \put(70,20){\vector(-1,0){25}}
    \put(-39,18){$\k$}
    \put(73,18){$\k$}
    \put(-9,8){$a$}
    \put(44,8){$b$}
\end {picture}
}
\qquad
\raisebox{-15pt} {
\begin {picture}(85,50)(-25,-5)
    \thicklines
    \put(20,20){\circle{40}}
    \put(20,0){\line(0,1){40}}
    \put(20,0){\circle*{5}}
    \put(20,40){\circle*{5}}
    \put(9,0){\vector(-1,0){25}}
    \put(55,0){\vector(-1,0){25}}
    \put(-25,-2){$\k$}
    \put(58,-2){$\k$}
    \put(18,-10){$a$}
\end {picture}
}
\label {fig:K2}
\end {equation}
[The full set of two-loop diagrams contributing to $\tilde C_{ab}(\k)$
is shown in figure \ref{fig:self2} of the following section.]
The second diagram of (\ref {fig:K2}) generates a contact term
proportional to $\delta_{ab}$ and independent of the external
momentum $\k$.
The contribution to $\tilde C_{ab}(\k)$
of these diagrams, plus the $g^0_{ab}$ interaction, is
\begin {eqnarray}
    \tilde C_{ab}^{(2,\rm sing)}(\k) =
    \beta^3 \, \n0a \, \n0b
    \left[
	-2 g^0_{ab}
	- { e_a^3 e_b^3 \over 3! } \, D^{(3)}_\dim (\k)
    \right]
    +
    \beta^3 \,
    \delta_{ab}
    \sum_{c=1}^A \n0a \, \n0c
    \left[
	-2 g^0_{ac}
	- { e_a^3 e_c^3 \over 3! } \, D^{(3)}_\dim ({\bf 0})
    \right],
\label{clcorr}
\nonumber\\
\end{eqnarray}
where $D^{(3)}_\dim(\k)$ denotes the Fourier transform of the cube
of the Debye potential,
\begin {equation}
    D^{(3)}_\dim(\k) \equiv
    \int (d^\dim \r) \, e^{-i \k \cdot \r } \>
    G_\dim(\r)^3 \,.
\end {equation}
The function $D^{(3)}_\dim(\k)$ has a simple pole in $\dim{-}3$,
which arises from the existence, in 3 dimensions,
of a non-integrable $1/r^3$ short-distance singularity in the integrand.
The long-distance behavior of the integral is effectively cut-off
by the larger of the Debye wave number $\kappa$ or the external
wave-vector $\k$.
This function is evaluated explicitly in section 2 of
Appendix \ref {required} [{\em c.f.} Eq.~(\ref {eq:D3})]
but for our present purposes all we need is the residue of
the pole in $\dim{-}3$.
Since this pole arises solely from the short-distance behavior,
its residue does not depend on whether $\k$ or $\kappa$
controls the long distance behavior.
Using the result (\ref{eq:D3})
and neglecting pieces that are non-singular when $\dim = 3$ gives
\begin {equation}
    D^{(3)}_\dim(\k) =
    {1 \over 2 \, (4\pi)^2} \,
    {1 \over 3-\dim } \,
    (\kappa^2)^{\dim-3}
    \left[ 1 + O(\dim{-}3) \right] .
\label {eq:D3sing}
\end {equation}
Note that it makes no difference whether
the factor which provides the correct dimensions
is written as $(\kappa^2)^{\dim-3}$, as $(k^2)^{\dim-3}$,
or as a power of some arbitrary wave vector $\mu$,
since different choices merely correspond to a change
in the non-singular part of $D^{(3)}_\dim(\k)$.
For example,
\begin {equation}
    {\kappa^{2(\dim-3)} \over 3-\dim}
    =
    {\mu^{2(\dim-3)} \over 3-\dim}
    +
    \ln \left( {\mu^2 \over \kappa^2} \right)
    +
    O(\dim{-}3) \,.
\end {equation}
As will be seen explicitly later on, it is generally very convenient to
make use of this arbitrary scale in the pole residue and
write all such divergent quantities in terms of a single, standard, but
arbitrary parameter $\mu$ with the dimensions of wave number or
inverse length, a parameter that is also used to exhibit the
extra dimensions that arise when the parameters are extended
beyond $\dim = 3$ dimensions. Thus we write the induced coupling
$g^0_{ab}$ as
\begin{equation}
    g^0_{ab} = \mu^{2(\dim-3)}
    \left[
	- { 1 \over 4!} \,
	{\left(e_a e_b \right)^3 \over (4\pi)^2} \, { 1 \over
	  3-\dim} \,
	+ g_{ab}(\mu)
    \right] .
\label {eq:Zab}
\label {eq:g0ab}
\end{equation}
In view of the result (\ref {eq:D3sing}),
the first pole term in this expression cancels the singular
contributions arising from $D^{(3)}_\dim(\k)$ in Eq.~(\ref {clcorr}).
The prefactor
$ \mu^{2(\dim-3)} $ absorbs the variation in dimension when
$\dim$ departs from $\dim = 3$, and so the remaining finite coupling
$g_{ab}(\mu)$ always retains its $\dim = 3$ dimensions. This finite
(or ``renormalized'') coupling must depend upon $\mu$ in such a way as
to ensure that the bare coupling $g^0_{ab}$ is independent of the
arbitrary value of $\mu$. Thus we have defined the finite coupling $g_{ab}$
to be a scale-dependent ``floating'' coupling, and we shall later exploit
the renormalization group results that follow from the arbitrary character
of $\mu$. For now, we simply note that $g_{ab}(\mu)$ will soon be
determined by matching the effective theory to the underlying
microscopic theory.

Similar considerations apply to the induced couplings of higher loop order.
At four-loop order,
the irreducible correlator $\tilde C_{ab}(\k)$ receives contributions
from the $h^0_{ab}$ and $k^0_{ab}$ derivative interactions
which are proportional to $k^2$.
Therefore, to determine the pole parts of these couplings,
it is sufficient to focus just on those contributions
to $\tilde C_{ab}(\k)$ (at four-loop order)
which are also proportional to $k^2$ and singular as $\dim\to3$.
[There are additional singular contributions to $\tilde C_{ab}(\k)$
at four-loop order which are proportional to $\kappa^2$.
The renormalization of these terms requires the four-loop
coupling $g^0_{abc}$ in addition to $h^0_{ab}$ and $k^0_{ab}$.
The determination of $g^0_{abc}$ is discussed below.]

There is only one four loop diagram constructed from the classical
interaction which contributes to $\tilde C_{ab}(\k)$ and contains a
term singular as $\dim\to3$ that is proportional to $k^2$:
\begin {equation}
\raisebox{-13pt}{
\begin {picture}(200,40)(-50,0)
    \thicklines
    \put(20,20){\circle{40}}
    \put(0,20){\line(1,0){40}}
    \bezier{80}(0,20)(20,40)(40,20)
    \bezier{80}(0,20)(20,0)(40,20)
    \put(0,20){\circle*{5}}
    \put(40,20){\circle*{5}}
    \put(-5,20){\vector(-1,0){25}}
    \put(70,20){\vector(-1,0){25}}
    \put(-39,18){$\k$}
    \put(73,18){$\k$}
    \put(-9,8){$a$}
    \put(44,8){$b$}
\end {picture}
}
\label {fig:K4}
\end {equation}
In addition, the following four-loop order diagrams involving the
classical interaction plus the finite $h^0_a$ induced interaction
contain terms proportional to $k^2$ which are singular as
$\dim \to 3$:
\begin {equation}
\raisebox{-32pt}{%
  \leavevmode
  \def\epsfsize #1#2{0.35#1}
  \epsfbox {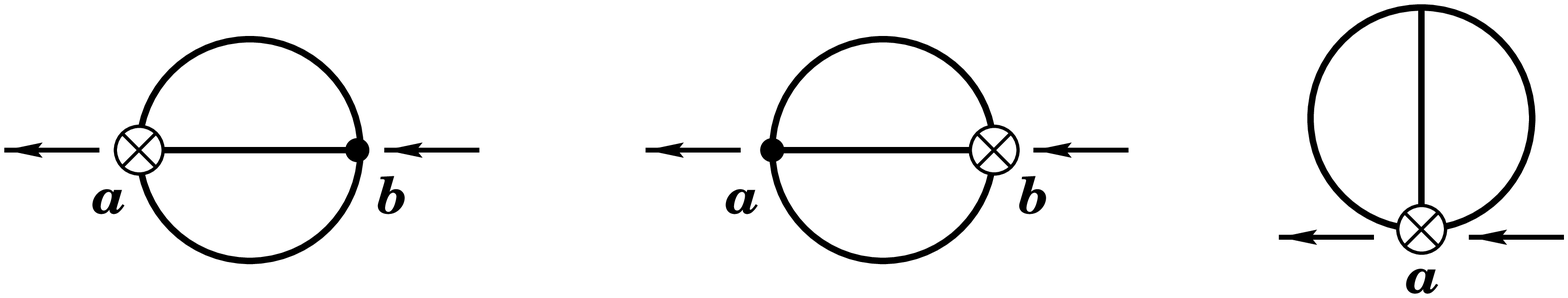}
}
\label {fig:K4bc}
\end {equation}
Here, the circled `X' denotes the vertex generated by the
$h^0_a$ induced interaction.
Since $h^0_a$ itself counts as a two-loop factor, these diagrams
contribute to the correlator at four-loop order.
The contributions of the $h^0_{ab}$ and $k^0_{ab}$ interactions,
plus the above graphs, give
\begin {eqnarray}
    \tilde C^{(4,\rm sing)}_{ab}(\k)
    &=&
    \beta^5 \n0a \, \n0b
    \left[
    2 \, k^2 \, k^0_{ab}
    - k^2 \, {e_a^3 e_b^3 \over 3!} \,
	\left( h^0_a + h^0_b \right) D^{(3)}_\dim(\k)
    - { e_a^5 e_b^5 \over 5! } \, D^{(5)}_\dim(\k)
    \right]
\nonumber
\\ &+&
    k^2 \, \delta_{ab} \, \beta^5
    \sum_{c=1}^A \n0a \, \n0c
    \left(
    2 h^0_{ac} - 2 h^0_a \, {e_a^3 e_c^3 \over 3!} \, D^{(3)}_\dim({\bf 0})
    \right)
    + \cdots \,,
\label{5clcorr}
\end {eqnarray}
where only the pieces proportional to $k^2$ have been displayed, and
where
\begin {equation}
    D^{(5)}_\dim(\k) \equiv
    \int (d^\dim \r) \, e^{-i \k \cdot \r } \> G_\dim(\r)^5 \,.
\end {equation}
This integral may be evaluated explicitly using
the methods of Appendix \ref {required}.
However, the part of the integral which is proportional to $k^2$
and singular as $\dim\to3$ arises solely from
the short-distance singularity in the integrand.
To extract just this portion of the integral,
it is sufficient to use unscreened Coulomb potentials instead
of the Debye potential.
The resulting Fourier transform of $V_\dim(\r)^5$
is evaluated in Appendix \ref {required}
[{\em c.f.} Eq.~(\ref {ftvfive})]
where it is shown that
\begin{eqnarray}
    \int (d^\dim \r) \, e^{-i \k \cdot \r } \> V_\dim(\r)^5
    &=&
    - {k^2 \over 4! \, (4\pi)^4}
    \left( {k^2 \over 4 \pi} \right)^{2(\dim-3)}
    {1 \over 3-\dim} + {\rm finite} \,.
\end{eqnarray}
Using this result plus (\ref {eq:D3sing}), it is easy to see that
the four-loop $O(k^2)$ part of $\tilde C_{ab}(\k)$ will be finite
as $\dim \to 3$ provided
\begin {equation}
    h^0_{ab}
    = \mu^{2(\dim -3)} \Bigg[
    -
    h^0_a \,
    {\pi \over 3}
    \left( {e_a e_b \over 4\pi} \right)^3 \,
    {1 \over 3-\dim} + h_{ab}(\mu) \Bigg] \,,\kern 0.6in
\end{equation}
and
\begin{eqnarray}
    k^0_{ab}
    &=& \mu^{2(\dim -3)} \Bigg[
    - {2\pi \over 4! \cdot 5!}
    \left( {e_a e_b \over 4\pi} \right)^5 \,
    {1 \over 3-\dim}
\nonumber \\ && \kern 0.55in{}
    +
    {\pi \over 3!}
    \left( {e_a e_b \over 4\pi} \right)^3 \,
    \left( h^0_a + h^0_b \right)
    {1 \over 3-\dim} + k_{ab}(\mu) \Bigg] \,.
\end {eqnarray}
Just as before with $g_{ab}(\mu)$, the finite renormalized couplings
$h_{ab}(\mu)$ and $k_{ab}(\mu)$ can only be determined by matching
with the full quantum theory.

The final four loop induced coupling $g^0_{abc}$ multiplies
three factors of bare particle densities.
The most convenient way to determine the poles in this coupling is to
consider the (irreducible part of the) triple density correlator,%
\begin {equation}
    \tilde K_{abc}(\k,\q)
    \equiv
    \int (d^\dim\r)(d^\dim\r')(d^\dim\r'') \>
    e^{i (\k \cdot \r + \q \cdot \r' - (\k+\q) \cdot \r'')} \>
    {\delta^3 \ln Z \over
    \delta \beta\mu_a(\r) \,
    \delta \beta\mu_b(\r') \,
    \delta \beta\mu_c(\r'')} \,.
\end {equation}
This correlator receives a contribution of
$-3! \, g^0_{abc} \, \beta^6 \, \n0a \, \n0b \, \n0c$
from the $g^0_{abc}$ induced coupling.
It also receives contributions,
which are singular as $\dim\to3$,
from the four-loop diagrams:
\begin {equation}
\raisebox{-42pt}{%
\begin {picture}(100,100)(-50,-25)
    \thicklines
    \put(20,20){\circle{40}}
    \bezier{80}(0,20)(15,28.66)(30,37.3)
    \bezier{80}(30,37.3)(30,20)(30,2.7)
    \bezier{80}(0,20)(15,11.34)(30,2.7)
    \put(0,20){\circle*{5}}
    \put(30,37.3){\circle*{5}}
    \put(30,2.7){\circle*{5}}
    \put(-5,20){\vector(-1,0){25}}
    \put(33,41){\vector(1,2){10}}
    \put(43,-21){\vector(-1,2){10}}
    \put(-39,18){$\k$}
    \put(43,64){$\q$}
    \put(45,-26){$\k{+}\q$}
    \put(-9,11){$a$}
    \put(36,35){$b$}
    \put(36,-1){$c$}
\end {picture}
}
\qquad
\raisebox{-42pt}{%
\begin {picture}(100,100)(-50,-25)
    \thicklines
    \put(20,20){\circle{40}}
    \bezier{80}(0,20)(11,32.66)(30,37.3)
    \bezier{80}(0,20)(20,20)(30,37.3)
    \bezier{80}(0,20)(15,11.34)(30,2.7)
    \put(0,20){\circle*{5}}
    \put(30,37.3){\circle*{5}}
    \put(30,2.7){\circle*{5}}
    \put(-5,20){\vector(-1,0){25}}
    \put(33,41){\vector(1,2){10}}
    \put(43,-21){\vector(-1,2){10}}
    \put(-39,18){$\k$}
    \put(43,64){$\q$}
    \put(45,-26){$\k{+}\q$}
    \put(-9,11){$a$}
    \put(36,35){$b$}
    \put(36,-1){$c$}
\end {picture}
}
\qquad
\raisebox{-42pt}{%
\begin {picture}(100,100)(-50,-25)
    \thicklines
    \bezier{80}(0,20)(11,32.66)(30,37.3)
    \bezier{80}(0,20)(20,20)(30,37.3)
    \bezier{80}(0,20)(5,45)(30,37.3)
    \bezier{80}(0,20)(11,7.33)(30,2.7)
    \bezier{80}(0,20)(20,20)(30,2.7)
    \bezier{80}(0,20)(5,-5)(30,2.7)
    \put(0,20){\circle*{5}}
    \put(30,37.3){\circle*{5}}
    \put(30,2.7){\circle*{5}}
    \put(-5,20){\vector(-1,0){25}}
    \put(33,41){\vector(1,2){10}}
    \put(43,-21){\vector(-1,2){10}}
    \put(-39,18){$\k$}
    \put(43,64){$\q$}
    \put(45,-26){$\k{+}\q$}
    \put(-9,11){$a$}
    \put(36,35){$b$}
    \put(36,-1){$c$}
\end {picture}
}
\label {fig:K3}
\end {equation}
(plus 5 other versions of the second diagram, and 2 other versions
of the third diagram, in which the labels are permuted in various ways).
In addition, there is a singular four-loop contribution
involving the two-loop coupling $g^0_{ab}$.%
\footnote
    {%
    There are also ``contact'' terms proportional to
    $\delta_{ab}$, $\delta_{bc}$ or $\delta_{ac}$, analogous to
    the second term appearing in (\ref {clcorr}).
    However, the required pole terms in $g^0_{abc}$ may be entirely
    inferred from the non-contact terms in $\tilde K_{abc}(\k,\q)$
    which are proportional to $\n0a \, \n0b \, \n0c$.
    The resulting value of $g^0_{abc}$ necessarily also renders
    the contact terms finite, just as seen explicitly at two-loop
    order in (\ref {clcorr}).%
    }
Provided the external momenta $\k$ and $\q$ are non-zero,
one may replace the Debye potentials in all these diagrams
by unscreened Coulomb potentials without changing the residue
of the $1/(\dim{-}3)$ poles.
In order for the sum of these contributions to be finite,
the four-loop coupling $g^0_{abc}$ must have both single
and double poles in $\dim{-}3$.
The resulting structure for $g^0_{abc}$,
and yet higher-order couplings,
will be discussed further in section \ref {sec:higher}.

\subsection {Matching}
\label {sec:matching}

    The most direct approach to determine the finite part of the two-loop
coupling $g^0_{ab}$ given in Eq.~(\ref {eq:Zab}) is to compare
the density-density correlator $\tilde K_{ab}(\k)$ in the effective
theory and the original quantum theory.
Because the induced coupling $g^0_{ab}$ makes a contribution to
$\tilde C_{ab}(\k)$ (and hence to the full correlator $\tilde K_{ab}(\k)$)
proportional to $\n0a \, \n0b$,
it is sufficient to retain in both the effective and fundamental theories
only those contributions with the same $\n0a \, \n0b$ dependence
on the bare densities.
Since it is the
short-distance contributions which must be correctly matched, Debye
screening may be completely ignored \cite {braaten} if one compares the correlator
evaluated at a non-zero wave number $\k$.
Consequently, to determine the two-loop coupling $g^0_{ab}$
it is sufficient to work just to second order in the fugacity expansion.
And because the induced coupling $g^0_{ab}$ makes a momentum-independent
contribution to the correlator (\ref {clcorr}),
it is also sufficient to work in the limit of small momentum
$\k \ll \lambda^{-1}$
and neglect all contributions which vanish as $\k \to 0$.%
\footnote
    {%
    To carry out the matching for the four-loop derivative coupling $k^0_{ab}$,
    one would need to evaluate and compare the $O(k^2)$ terms
    in the density-density correlator.
    }

The tree and one-loop contributions to the correlator are given
by Eq's.~(\ref {cone}) and (\ref {gginver})--(\ref {kstruct}).
Two-loop contributions to $\tilde K_{ab}(\k)$ which are proportional
to $\n0a \, \n0b$ arise in two ways.
The one-particle irreducible part $\tilde C_{ab}(\k)$
receives such a contribution from the first term in (\ref {clcorr}).
In addition, there is a one-particle reducible contribution arising
from the two-loop $h^0_a$ interaction appearing in Eq.~(\ref {Sind}).
This may be seen as follows.
The $h^0_a$ interaction generates, through the last term of (\ref{eq:Keff}),
a two-loop contribution of
$
    -2 \, \beta^2 \, \delta_{ab} \, h^0_a \, \n0a \, \k^2
$
to the irreducible correlator $\tilde C_{ab}(\k)$.
A two-loop
reducible contribution to the full correlator $\tilde K_{ab}(\k)$ of\,%
\footnote
    {%
    This term may equivalently be described as arising from the
    first term of (\ref{eq:Keff}) when one variation acts on
    the $\nabla \phi \cdot \nabla \mu_c$ part of the $h^0_c$
    interaction in Eq.~(\ref{Sind})
    and the other variation acts on the classical interaction.
    }
\begin {equation}
    2\beta^4 \, \n0a \, \n0b \, e_a \, e_b \, (h^0_a + h^0_b) \,
    \k^2 \, \tilde {\cal G}(\k) \,.
\end {equation}
is then generated by the two-loop cross term in Eq.~(\ref {kstruct})
which results from this irreducible contribution together with
the lowest-order piece contained in Eq.~(\ref {cone}).
As noted above, for this matching calculation (only), we may
neglect Debye screening by sending $\kappa\to0$.
In this limit, the electrostatic potential correlator ${\cal G}(\k)$ is,
to lowest order, just $1/(\beta \k^2)$.
In other words, the $1/\k^2$ of the (Fourier transformed) Coulomb potential
cancels the $\k^2$ appearing from the two derivatives in the $h^0_a$
interaction, leading to result which (with the neglect of Debye screening)
is non-vanishing as $\k\to0$.

Consequently, the relevant portion of the complete correlator
$\tilde K_{ab}(\k)$ in the effective theory is
\begin {eqnarray}
    \tilde K_{ab}^{(2)}(\k)
    &=&
    \n0a \, \n0b \>
    \Biggl[
	- {\beta e_a e_b \over k^2}
	+ {(\beta e_a e_b)^2 \over 2} \, D^{(2)}_\dim(\k)
	- {(\beta e_a e_b)^3 \over 3!} \, D^{(3)}_\dim(\k)
\nonumber\\&&\kern 0.5in {}
	- 2 \beta^3 \, g^0_{ab}
	+ 2 \beta^3 \, e_a \, e_b \, (h^0_a + h^0_b) \,
    \Biggr] \>
    + \cdots
\label{redeff}
\end {eqnarray}
where $\cdots$ denotes irrelevant
terms with different dependence on the bare densities.%
\footnote
    {
    An independent way to derive the $h^0_a$
    terms in the result (\ref {redeff}), which illuminates the character
    of the theory, is as follows.
    In our construction of the interaction terms in the effective theory
    (\ref {Sind}), we fixed the meaning of the functional integration
    field $\phi$ by requiring that the invariance $\phi \to \phi - ic $,
    $\mu_a \to \mu_a - e_a c$ be maintained, implying that this field
    and the chemical potentials always appear in the combination
    $\mu_a + i e_a \phi$.
    This requirement casts the theory in its most useful form.
    However, since $\phi$ is simply a dummy integration variable,
    one is free to make field redefinitions that violate this restriction,
    and it is sometimes convenient to do so temporarily.
    Since $n^0_a(\r) \propto \exp\{\beta \mu_a(\r) \}$,
    the cross term in the $h^0_a$ interaction involving
    $\nabla \phi \cdot \nabla \mu_a$
    may equivalently be written as
    $
	2 \beta^3 \, {\sum}_a i e_a h^0_a \, e^{i\beta e_a \phi }
	    \nabla\phi \cdot \nabla n^0_a \,.
    $
    To first order in $h^0_a$, which is all that concerns us,
    the field redefinition
    $$
	\phi \to \phi - 2 \beta^2 \, {\sum}_b i e_b h^0_b \, n^0_b \,
	    e^{i \beta e_b \phi}
    $$
    in the kinetic term $ \coeff \beta 2 (\nabla \phi)^2$ removes the cross
    term.
    The effect of the same field redefinition on
    the classical interaction term
    $
	- {\sum}_a n^0_a \, e^{i\beta e_a \phi}
    $,
    again to leading order in $h^0_a$,
    is a change in the action that is equivalent
    to the induced coupling alteration
    $$
	g^0_{ab} \to g^0_{ab} - e_a e_b \, ( h^0_a + h^0_a ) \,.
    $$
    This combination is precisely what appears in Eq.~(\ref{redeff}),
    and serves as an independent check on the validity of that result.
    \label{happy}
    }
With the neglect of Debye screening, the integrals $D^{(m)}_\dim(\k)$
reduce to Fourier transforms of powers of the original Coulomb potential,
\begin {equation}
    C^{(m)}_\dim(\k)
    =
    \lim_{\kappa \to 0} D^{(m)}_\dim(\k)
    =
    \int (d^\dim\r) \> e^{-i \k\cdot\r} \, V_\dim(\r)^m \,.
\end {equation}
The $\kappa\to0$ limit of (\ref {D2}) immediately gives
\begin {equation}
    C^{(2)}_3(\k) = {1 \over 8k} \,,
\end {equation}
and in the first part of Appendix \ref {required}
it is shown [{\em c.f.} \ref {ftvcube}] that
\begin{eqnarray}
    C^{(3)}_\dim(\k)
    &=&
    {1 \over 2\,(4\pi)^2}
    \left( {k^2 \over 4 \pi} \right)^{\dim-3}
    \left\{ { 1 \over 3-\dim} + 3 - \gamma + O(\dim{-}3) \right\} ,
\end {eqnarray}
where $\gamma = 0.57721\cdots$ is Euler's constant.
Inserting $g^0_{ab}$ from (\ref {eq:Zab}),
and using
\begin {equation}
    \lim_{\dim\to3} \>
    { 1 \over 3-\dim}
    \left[
	\left(\mu^2\right)^{\dim-3}
	-
	\left( {k^2 \over 4 \pi} \right)^{\dim-3}
    \right]
    =
    \ln \!\left({k^2 \over 4\pi\mu^2}\right)
\end {equation}
to take the physical $\dim\to3$ limit yields
\begin{eqnarray}
    \tilde K^{(2)}_{ab}(\k) &=&
    \n0a \, \n0b \>
    \Biggl\{
	- {\beta e_a e_b \over k^2}
	+ {(\beta e_a e_b)^2 \over 16 \, k}
	+ { \pi \over 3}
	\left( {\beta e_a e_b \over 4\pi } \right)^3
	\left[
	    \ln \!\left({k^2 \over 4\pi\mu^2}\right) -3 + \gamma
	\right]
\nonumber\\ && \kern 1.86in {}
	- 2 \beta^3 g_{ab}(\mu)
	+ {\beta \, e_a e_b \, \lambda_{ab}^2 \over 24\, \pi}
    \Biggr\} \>
    + \cdots
\label {eq:Kclass}
\end{eqnarray}
for the $\n0a \, \n0b$ piece
of the density-density correlator, neglecting Debye screening,
to two-loop order.
In writing the last term of (\ref {eq:Kclass}) we have made use
of the definition (\ref {eq:h0a}) of $h^0_a$ to express
\begin {equation}
    h^0_a + h^0_b
    =
    {\lambda_a^2 + \lambda_b^2 \over 48\pi \, \beta^2}
    =
    {\lambda_{ab}^2 \over 48\pi \, \beta^2} \,,
\end {equation}
in which $\lambda_{ab}$ is the thermal wavelength for the reduced mass
$m_{ab} = [1/m_a + 1/m_b]^{-1}$.

We write the corresponding result in the underlying quantum theory as
the Fourier transform of the density-density correlator%
\footnote
  {We are glossing over a subtlety here, for
  Eq.~(\ref {eq:Kqmdef}) involves the equal time
  expectation value $ \expect { n_a(\r_1,0) \, n_b(\r_2,0) } $,
  whereas our desired correlator is the zero-frequency correlation
  function (\ref {eq:Kqm}).
  The difference between these two is just the sum of correlations at all
  non-zero Matsubara frequencies $\omega_n$.
  However, as discussed at the end of this section,
  non-zero frequency correlators are proportional to $\k^2$
  (due to current-conservation),
  and hence do not affect the matching for the $g^0_{ab}$
  interaction in the effective classical theory,
  which may be extracted from the $\k\to0$ behavior
  of the density-density correlator.
  On the other hand, the difference between the equal-time and zero-frequency
  correlators would be essential for the matching of the four-loop
  couplings $k^0_{ab}$ since these depend on the $O(\k^2)$ terms
  of the zero-frequency correlator.
  }
\begin {eqnarray}
    \Gqm_{ab}(\r_1,\r_2) &\equiv&
    \expect { n_a(\r_1) \, n_b(\r_2) } -
    \expect { n_a(\r_1) } \expect { n_b(\r_2) }
    \nonumber
    \\ &=&
    Z^{-1} \> {\rm Tr}
	\left[
	    e^{-\beta H + \sum_a \beta \mu_a N_a} \> n_a(\r_1) \, n_b(\r_2)
	\right]
    - \expect { n_a(\r_1) } \expect { n_b(\r_2) } \,.
\label {eq:Kqmdef}
\end {eqnarray}
The subtraction of $\expect {n_a} \expect {n_b}$ removes what would
otherwise be a delta function contribution to the Fourier transform
at $\k = 0$, and is completely ignorable when working at $\k \ne 0$.
We specifically want the second-order contribution in the fugacity
expansion of $\Gqm_{ab}(\r_1,\r_2)$.
For our purposes, this is most conveniently obtained by using an
(old-fashioned) expansion of the trace in terms of ordinary
quantum-mechanical multi-particle states rather than using
many-body quantum field theory.
The desired second-order terms in the fugacity expansion come from the
two-particle subspace of the thermodynamic trace over all particle states,
so that
\begin{eqnarray}
    \Gqm_{ab}^{(2)}(\r_1,\r_2)
    &=&
    {1 \over 2} \sum_{cd} \,
    e^{\beta \mu_c} \,
    e^{\beta \mu_d}
    \!\int (d^3\r) (d^3\r') \,
    \langle \r c , \r' d | e^{-\beta H_{cd} } \,
    n_a(\r_1) \, n_b(\r_2)
\nonumber\\ && \kern 3.0in
    \left[ \, | \r c, \r' d \rangle \pm | \r' d , \r c \rangle \right].
\label {eq:Kqm2}
\end{eqnarray}
Here, $|\r c, \r' d\rangle$ denotes the (un-symmetrized) two-particle
basis ket with one particle of species $c$ at $\r$
and one of species $d$ at $\r'$,
$H_{cd}$ is the (first-quantized) two particle Hamiltonian,
\begin {equation}
    H_{cd} =
    {\p_1^2 \over 2m_c} +
    {\p_2^2 \over 2m_d} +
    {e_c \, e_d \over 4 \pi |\r_1 {-} \r_2|} \,,
\end {equation}
and the $\pm$ sign in the final combination of ket vectors accounts
for Bose $(+)$ or Fermi $(-)$ statistics.  To avoid a clutter of
notation, we temporarily use the indices $a,b$ to denote spin
components as well as species labels.
Now
\begin{eqnarray}
    && n_a(\r_1) \, n_b(\r_2) \>
    \half \left[ \, | \r c, \r' d \rangle \pm | \r' d , \r c \rangle \, \right]
    \nonumber\\
    && \quad {} =
    \left\{
	\delta_{ac} \, \delta(\r_1 {-} \r) +
	\delta_{ad} \, \delta(\r_1 {-} \r')
    \right\}
    \left\{
	\delta_{bc} \, \delta(\r_2 {-} \r) +
	\delta_{bd} \, \delta(\r_2 {-} \r')
    \right\}
    \half \left[ \, | \r c, \r' d \rangle \pm | \r' d , \r c \rangle \right]
    \nonumber\\
    && \quad {} \to
    \left\{
	\delta_{ab} \, \delta(\r_1 {-} \r_2) \,
	\delta_{ac} \, \delta(\r_1 {-} \r) +
	\delta_{ac} \, \delta(\r_1 {-} \r) \,
	\delta_{bd} \, \delta(\r_2 {-} \r')
    \right\}
    \left[ \, | \r c, \r' d \rangle \pm | \r' d , \r c \rangle \right] ,
\end{eqnarray}
where in the last line terms which become equivalent when inserted
into (\ref {eq:Kqm2}) have been combined.
Since we are only interested in terms proportional to $\n0a \, \n0b$
(or equivalently $e^{\beta \mu_a} e^{\beta \mu_b}$)
the contact term involving $\delta_{ab} \, \delta(\r_1 {-} \r_2)$
may be neglected.%
\footnote
    {%
    As it stands, this contact term is infrared divergent since Debye
    screening, which involves an arbitrary number of particles, is
    needed to provide the long-distance cut off which makes the contact
    term infrared finite.  It is precisely because the required value
    of the induced coupling $g_{ab}$ can be deduced solely from the
    non-contact part of the correlator that it is permissible to ignore
    Debye screening in this matching calculation and just use a
    fugacity expansion.
    }
The density operators $n_a$ (as well as the Hamiltonian $H_{cd}$) are
spin independent, so that the sum over particle spins just produces
the spin degeneracies $g_a$ and $g_b$.  Hence, reverting to the
previous notation in which the indices $a,b$ label only different
species without regard to spin, the required piece of the quantum
mechanical density-density correlator is given by
\begin {equation}
    \Gqm_{ab}^{(2)}(\r_1,\r_2)
    =
    g_a \, e^{\beta\mu_a} \, g_b \, e^{\beta\mu_b} \,
    \left[
    \langle \r_1,\r_2 | e^{-\beta H_{ab}} | \r_1,\r_2 \rangle
    \pm
    \left( \delta_{ab} / g_a \right)
    \langle \r_1,\r_2 | e^{-\beta H_{ab}} | \r_2,\r_1 \rangle
    \right] .
\label {eq:Gab}
\end {equation}
At this point,
it is convenient to write the two-particle Hamiltonian in terms of
center-of-mass ${\bf R}$ and relative ${\bf r}$ coordinates,
\begin{equation}
H_{ab} = H_{ab}^{\rm cm} + H_{ab}^{\rm rel} \,,
\end{equation}
with
\begin{equation}
    H_{ab}^{\rm cm} = {{\bf P}^2 \over 2M_{ab}} \,,
\end{equation}
the Hamiltonian for center-of-mass motion with total mass
$M_{ab} \equiv m_a + m_b$, and
\begin{equation}
    H_{ab}^{\rm rel} = {\p^2\over 2m_{ab}} + {e_a e_b \over 4\pi |\r|}
\end{equation}
the Hamiltonian for relative motion
with reduced mass $m_{ab}^{-1} \equiv  m_a^{-1} + m_b^{-1}$.
The Fourier transform now reads
\begin {eqnarray}
    \tilde \Gqm_{ab}^{(2)}(\k)
    =
    \!\int (d^3\r) \, e^{-i \k \cdot \r} \>
    g_a e^{\beta\mu_a} \, g_b e^{\beta\mu_b} \,
    \langle {\bf R} | e^{-\beta H_{ab}^{\rm cm}} | {\bf R} \rangle\!
    \left[
    \langle \r | e^{-\beta H_{ab}^{\rm rel}} | \r \rangle
    \pm \left( \delta_{ab} / g_a \right)
    \langle \r | e^{-\beta H_{ab}^{\rm rel}} | {-}\r \rangle
    \right] \!.
\label {eq:qmcorr}
\nonumber\\
\end {eqnarray}
Our goal is to compare $\tilde \Gqm_{ab}^{(2)}(\k)$ with
the effective theory result (\ref {eq:Kclass}), and to
adjust the finite coupling $g_{ab}(\mu)$ in the effective theory
so that both results coincide up to corrections that vanish
as $\k \to 0$.

The center-of-mass matrix element is just
\begin{equation}
    \langle {\bf R} | e^{ - \beta H_{ab}^{\rm cm} } | {\bf R} \rangle =
    \int { (d^3 {\bf P}) \over (2\pi)^3 } \>
    \exp\left\{ - \beta {P^2 \over 2 M_{ab} } \right\}
    = \Lambda^{-3}_{ab} \,,
\end{equation}
where
\begin{equation}
    \Lambda_{ab} = \hbar \left({  2 \pi \beta \over M_{ab}} \right)^{1/2}
\end{equation}
is the thermal wavelength of the center-of-mass motion.
We shall also make use of the thermal wavelength of the relative
motion,
\begin {equation}
 \lambda_{ab} = \hbar \left({ 2\pi \beta \over m_{ab} }\right)^{1/2} \,.
\end {equation}
Note that since the product of the reduced mass $m_{ab}$ and the total
mass $M_{ab}$ is just the product of the separate masses, $m_{ab} \,
M_{ab} = m_a \, m_b $, the corresponding relation also holds for the
thermal wavelengths, $\Lambda_{ab} \, \lambda_{ab} = \lambda_a \,
\lambda_b$. Hence
\begin{equation}
g_a e^{\beta \mu_a} \, g_b e^{\beta \mu_b} \, \Lambda_{ab}^{-3} \,
 = \n0a \, \n0b \, \lambda_{ab}^3 \,,
\end{equation}
and we may write
\begin{equation}
    \tilde \Gqm_{ab}^{(2)}(\k)
     = \n0a \, \n0b \, \lambda_{ab}^3 \>
     \left[ F_+(\k) \pm \left(\delta_{ab} / g_a \right)
 F_-(\k) \right] \,,
\label{qmcorr}
\end{equation}
with
\begin {equation}
    F_\pm(\k) =
    \int (d^3\r) \> e^{-i \k \cdot \r} \>
    \langle \r | e^{-\beta H_{ab}^{\rm rel}} | {\pm}\r \rangle \,.
\end {equation}

As shown in Appendix \ref {app:SU(1,1)},
an explicit representation for
the matrix elements $F_\pm(\k)$ may be found
by expressing the relative Hamiltonian in terms
of the generators of an $su(1,1)$ algebra.
The result for the direct term,
given in Eq.~(\ref{suresult}),
is
\begin {eqnarray}
    F_+(\k)
    &=&
    \lambda_{ab}^{-3}
    \Bigg\{
    -
    {\beta e_a e_b \over k^2}
    +
    {(\beta e_a e_b)^2 \over 16\,k}
\nonumber\\
&& \qquad{}
  +
    {\pi \over 3}
    \biggl({\beta e_a e_b \over 4\pi}\biggr)^3
    \left[
    \ln \biggl({\lambda_{ab}^2 k^2 \over 4\pi}\biggr) -3 + \gamma
    + {8\pi \lambda_{ab}^2 \over (\beta e_a e_b)^2}
       +
      f \biggl({\beta e_a e_b \over 4\pi\lambda_{ab}}\biggr)
    \right]
    + O(k)
    \Bigg\} ,
\label {neato}
\end {eqnarray}
where the function $f(y)$ has the (convergent) power series expansion
\begin {eqnarray}
    f(y)
    &=&
    -
    {3 \over 4y}
    -
    {3\sqrt \pi \over 2}
    \sum_{n=1}^\infty
    \left(-\sqrt \pi \, y\right)^n
    {\zeta(n{+}1) \over \Gamma((n{+}5)/2)} \,.
\label {eq:f}
\end {eqnarray}
The asymptotic behavior of this function as $y\to\pm\infty$ is spelled
out in detail in Eq's.~(\ref {eq:ffpos}) and (\ref {eq:fneg}).
Here we note that in the case of strong repulsive interactions
corresponding to $y \to + \infty$, $f(y)$ increases only as $\ln y$,
with
\begin{equation}
    f(y)
    \sim
	\ln (4 \pi y^2 ) + 3 \gamma - \frac {8}3
    - {1 \over 4 \pi  y^2}
    + O(y^{-4}) \,.
\label{strep}
\end{equation}
For the case of strong attractive interactions with the resulting
deeply bound Coulombic states when $y \to - \infty$, $f(y)$ grows
very rapidly,%
\footnote
    {
    This asymptotic result corresponds exactly to the contribution of the
    hydrogen-like intermediate ground state term in Eq.~(\ref {eq:qmcorr}).
    }
\begin {equation}
    f(y)
    \sim
    {3 \over \pi \, y^3} \, \exp \left\{ \pi \, y^2 \right\} \,.
\label{statt}
\end{equation}
The corresponding result for the exchange piece,
given in Eq.~(\ref {eq:Fminus}), is
\begin{equation}
    F_-(\k)
    =
    {\pi \over 3}  \,
    \lambda_{aa}^{-3} \,
    \biggl({\beta e_a^2 \over 4\pi}\biggr)^3 \,
    \tilde f \biggl({\beta e_a^2 \over 4\pi \lambda_{aa}}\biggr) + O(k) \,,
\label {neato2}
\end{equation}
where the function $\tilde f(y)$ has the (convergent) expansion
\begin{eqnarray}
    \tilde f(y)
    =
   { 3 \over 8 \pi y^3 } - { 3 \over 2 \pi y^2 } + { 3 \ln 2 \over 2 y }
    -{ 3\sqrt\pi \over 2 } \sum_{n=0}^\infty \,
    \left(-\sqrt\pi \, y\right)^n \left[ 1 - { 1 \over 2^{1+n}} \right] \,
    { \zeta(n{+}2) \over \Gamma((n{+}5)/2)} \,.
\label {eq:tildef}
\end{eqnarray}
The $y\to\infty$ asymptotic behavior is given in Eq.~(\ref {eq:tilde-f-asym}),
and yields the strong decrease
\begin{equation}
   \tilde f(y)
    \sim  { 2 \sqrt 3  \over y^2 \, \pi }
      \> \exp\left\{ - {3 \pi \over 2} \left( 2 y^2 \right)^{1/3}
      \right\} \,.
\label{stdec}
\end {equation}
Note that since the $\lambda_{ab}$ were defined in terms of the
reduced mass of the $a$---$b$ system, $ \lambda_{aa} = {\sqrt 2}
\lambda_a$.

Inserting the results (\ref{neato}) and (\ref {neato2})
into Eq.~(\ref{qmcorr}), and comparing to the result (\ref {eq:Kclass})
computed in the effective quasi-classical theory,
we see that the two results coincide provided that
\begin {equation}
    g_{ab}(\mu)
    = -
    {\pi \over 6}
    \left( { e_a e_b \over 4 \pi } \right)^3
    \left[
	\ln \left(\mu^2 \lambda_{ab}^2 \right)
	+ \Gam ab
    \right] ,
\label {eq:gab}
\end {equation}
where%
\footnote
    {
    Previous work \cite{ebeling,book,AP}
    makes use of dimensionless parameters $ \xi_{ab} $
    (also called $x_{ab}$) related to our notation by
    $\xi_{ab}  = - \sqrt{4\pi} \, \eta_{ab}$,
    and functions $Q(\xi_{ab})$, $E(\xi_{ab})$ of these parameters.
    To establish contact with this prior work (which also does not use
    our rationalized Gaussian electrostatic units), we note that
    $$
    \Gamma(\eta_{ab}) + \gamma + \ln9 -1
    =
    { 6 \over \sqrt2}
    \left( { 4\pi \over \beta e_a e_b } \right)^3
    \left( { \lambda^2_{ab} \over 2\pi } \right)^{3/2}
    \left\{
	{1 \over 6} \, \xi_{ab}
	+ Q(\xi_{ab})
	\pm {\delta_{ab} \over g_a} \, E(\xi_{ab})
    \right\} \,.
    $$
    Here, $\lambda_{ab} = \hbar [2\pi \beta/m_{ab}]^{1/2}$ is our
    definition of the thermal wavelength;
    various previous work uses the same symbol to denote either
    $\hbar \sqrt {\beta/m_{ab}}$ or
    $\hbar \sqrt {\beta/2m_{ab}}$.
    Note that our $e_a e_b / (4\pi)$ becomes just $e_a e_b$
    when converting to unrationalized electrostatic units.
    }
\begin {equation}
    \Gam ab \equiv
    f \left(\eta_{ab}\right) \pm \left(\delta_{ab} / g_a \right)
 \tilde f (\eta_{aa}) \,,
\label {eq:Gam}
\end {equation}
with $\eta_{ab}$ denoting the quantum parameter for species $a$ and $b$,
\begin {equation}
    \eta_{ab}
    \equiv
    {\beta e_a e_b \over 4\pi\lambda_{ab}} \,,
\label {eq:etaab}
\end {equation}
and where,
as usual, the exchange term in (\ref{eq:Gam}) comes in
with a plus (minus) sign if species $a$ is a Boson (Fermion).

In the limit of strong repulsion, $\eta_{ab} \to + \infty$,
Eq's.~(\ref{strep}) and (\ref{stdec}) inform us that
\begin {equation}
    g_{ab}(\mu)
    \sim -
    {\pi \over 6}
    \left( { e_a e_b \over 4 \pi } \right)^3
    \left[
	\ln \left( {\beta^2 e_a^2 e_b^2 \, \mu^2 \over 4 \pi }  \right)
	 + 3 \gamma - \frac {8}3
    \right] \,.
\label{repulse}
\end {equation}
Note that this limit does not involve Planck's constant $\hbar$: The
argument of the logarithm entails the classical ratio of the Coulomb
energy of two charges a distance $\mu^{-1}$ apart to the temperature.
(When this coupling is inserted in physical quantities, it will appear
with a $\ln(\kappa/\mu)$ term which turns the arbitrary distance
$\mu^{-1}$ into the Debye length $\kappa^{-1}$.)%
\footnote
    {
    Writing the
    result in terms of the Coulomb distance (\ref{Cdist}), but for the
    specific charges $e_a$, $e_b$, $d_{ab} = \beta e_a e_b / 4 \pi $,
    gives
    $$
	g_{ab}(\mu) \sim - { 1 \over 2}
	{4\pi \over 3!}
	\left( { e_a e_b \over 4 \pi } \right)^3
	\left[ \ln \left( d_{ab} \mu  \right) + \cdots \right] \,.
    $$
    This form is in precise accord with the remarks made in
    footnote \ref{foot}. Namely, the coefficient of
    $\ln(d_{ab}\mu)$ exactly corresponds to the two-particle part of the
    partition function, with the exponential of the Coulomb interaction
    expanded to third order and the integration over the relative
    coordinate cut off at the short distance $d_{ab}$ and at the long
    distance $\mu^{-1}$.\label{ffoot}
    }
In view of
Eq.~(\ref{strep}), the first correction to this classical limit is of
order $\hbar^2$.
On the other hand, in the limit of strong attraction,
$\eta_{ab} \to - \infty$, the exponential blow-up exhibited in
Eq.~(\ref{statt}) shows that our perturbative development breaks down,
as it must, since in this limit the ionized plasma must condense into
neutral atoms. This is, of course, a highly quantum-mechanical regime.
Finally, for small $ \eta_{ab} $, the exchange term (\ref{eq:tildef})
dominates and, with $ \lambda_{aa} = {\sqrt 2} \lambda _a$, one has
\begin{equation}
\Gam ab \sim \pm { \delta_{ab} \over g_a }
\left( { 4 \pi \over \beta e_a^2 } \right)^3 \,
{ 3 \lambda_a^3 \over 2 \sqrt{2} \pi } \,.
\label{shortly}
\end{equation}
Noting that this multiplies $ (n_a^0)^2 e_a^6 $, the result appears as an
exchange term independent of the particle's charge. Indeed, we shall
shortly see in the following Sec. IV that this is just the usual free
particle exchange correction that is quadratic in the fugacity. The
next term of order $\eta_{aa}^{-2} $ in the exchange contribution
$\tilde f(\eta_{aa})$ gives the familiar order $e^2$ exchange
correction to the plasma.

With the single two-loop coupling $g^0_{ab}$ completely determined
by Eq's.~(\ref {eq:g0ab}) and (\ref {eq:gab}),
one may now use the quasi-classical effective theory to compute
thermodynamics, or other quantities of interest, to two or three loop order.
Before four-loop calculations of physical quantities can be performed,
the finite renormalized parts of the four loop couplings $g^0_{abc}$,
$h^0_{ab}$, and $k^0_{ab}$
would need to be determined by an analogous higher order matching calculation.
This we have not attempted to do.
However, it should be noted that
determining the finite part of the $g^0_{abc}$
couplings requires a fully quantum-mechanical three-body calculation
which (almost certainly) is not possible to do analytically.
Only when this three-body matching is accomplished
will it be possible to extend the perturbative analysis
of Coulomb plasmas to four (or higher) loop order.%
\footnote
    {%
    This same issue arises regardless of whether one is using
    our effective field theory approach, or more traditional methods.
    Results for some contributions to the pressure at
    four-loop order [or $O(n^3)$] have recently been reported
    \cite {kahlbaum}.
    However, these partial results do not include the most difficult
    contributions which are sensitive to the short-distance behavior
    of three-body Coulomb systems.
    }

\subsection {Non-zero Frequency Modes}
\label {sec:freq}

    Up to this point, the effects of the non-zero frequency
components of the potential $\phi(\r,\tau)$,
defined by the Fourier series (\ref{eq:fourier-phi})
and repeated here for convenience,
\begin{equation}
    \phi(\r , \tau) = \phi(\r) + \sum_{m \neq 0} \phi^m(\r) \,
    e^{-i \omega_m \tau } \,, \qquad\quad
    \omega_m =  {2 \pi m \over \beta} \,,
\end{equation}
have been ignored.
These components, which obey the reality constraint
\begin{equation}
\phi^m(\r)^* = \phi^{-m}(\r) \,,
\end{equation}
characterize quantum fluctuations in the electrostatic potential.
They decouple from the zero-frequency degrees of freedom
and could be trivially integrated out
in the leading-order classical limit.
But in higher orders, this is no longer true.
To examine the effects which result from non-zero frequency fluctuations,
we return to the full quantum theory whose
functional integral representation (\ref {eq:Zquant})
may be rewritten as
\begin{eqnarray}
    Z_{\rm QM}[\mu]
    &=&
	N_0'
	\int [ d \phi ]
	\exp\left\{ - {1 \over 2} \int_0^\beta d\tau \int (d^\dim\r) \>
	\Bigl( \nabla \phi(\r, \tau) \Bigr)^2 \right\}
	\exp\{ -\Wone[\phi;\mu] \}
\nonumber\\
	&=&
	N_0'
	\int [d\phi] \! \prod_{m\ne0} [ d\phi^m ] \,
	\exp \Biggl\{\!
	-{\beta \over 2} \int (d^\dim\r) \,
	\biggl[
	|\nabla\phi(\r)|^2 + \!\sum_{m\ne 0} |\nabla \phi^m(\r)|^2
	\biggr]\!
	\Biggr\}
	\exp\{ -\Wone[\phi;\mu] \} .
\nonumber\\
&&
\label{fullquant}
\end{eqnarray}
In the first line, the integration measure $[d\phi]$
represents functional integration over the space-time dependent
field $\phi(\r,\tau)$. In the second line,
$[d\phi]$ now stands for functional integration over just
the time-independent (static mode)
$\phi(\bf r)$, while in the following product
$[d\phi^m]$ denotes functional integration
over the remaining non-zero frequency modes. The prefactor $N_0'$
involves the square root of the functional determinant of the
Laplacian operator for all of the modes, $N_0' = {\rm Det}^{1/2}
\left[ - \beta \nabla^2 \right]$. The final factor of $e^{-\Wone}$
denotes the product of Gaussian functional integrals for
each charged species,
\begin {eqnarray}
    \exp\{ -\Wone[\phi;\mu] \}
    &\equiv&  \prod_{a=1}^A
    \int [d\psi_a^* d\psi_a] \>
    \exp\biggl\{ -\int_0^\beta d\tau \int (d^\dim\r) \> {\cal L}_a \biggr\} \,,
\label {eq:za}
\end{eqnarray}
with
\begin{eqnarray}
    {\cal L}_a
    =
    \psi_a^*(\r, \tau)
    \Biggl[ { \partial \over \partial \tau} - { \nabla^2 \over 2 m_a} -
    \mu_a(\r,\tau) - i e_a \phi(\r , \tau) \Biggr] \psi_a(\r, \tau) \,.
\label {eq:La}
\end{eqnarray}
We are now allowing arbitrary temporal, as well as spatial, variation
in the chemical potentials so that the resulting partition function
may be used to generate time-dependent number density correlators.
This will be of use in section \ref {sec:longdist}.
In Appendix \ref{app:det}, we derive the large mass
asymptotic expansion of $\Wone$, and give the explicit form for
both first and second order corrections in $\lambda^2$ (or inverse mass).
For our present purposes,
only the classical term and the first order corrections
contained in Eq.~(\ref{eq:S2}) are required.
They are
\begin {eqnarray}
    \Wone[\phi;\mu]
    &=& \sum_{a=1}^A
    \int (d^\dim {\bf r}) \>
    n_a^0(\r) \, e^{ i e_a \beta \, \phi ({\bf r}) } \>
    \Bigg\{ -1
    +
    {\beta^2 \lambda_a^2 \over 48 \pi} \,
    \Bigl[ \nabla\mu_a(\r)+i e_a \nabla\phi(\r)\Bigr]^2
\nonumber\\
&& \qquad\qquad {}
    -
    {\beta^2\lambda_a^2 \over 16\pi^3}
    \sum_{m\ne 0} {1 \over m^2}\,
    \left[ \nabla\mu_a^{\vphantom-m}(\r) + i e_a \nabla \phi^m(\r) \right]
    \cdot
    \left[ \nabla\mu_a^{-m}(\r) + i e_a \nabla \phi^{-m}(\r) \right]
\nonumber\\
&& \qquad\qquad {}
    + \cdots \,
    \Bigg\} \,.
\label {eq:lnzFT}
\end {eqnarray}
The zero-frequency parts which appear in (\ref {eq:lnzFT})
have already been included in the effective theory.
These are precisely the classical $e^{i e_a \beta\phi}$ interaction,
plus the first derivative interaction in (\ref {Sind})
involving
$
    h^0_a
    \left[ \nabla\mu_a(\r){+}i e_a \nabla\phi(\r)\right]^2
$.
Because of the presence of the exponential factor $e^{i e_a \beta \phi(\r)}$,
note that the third term in (\ref {eq:lnzFT})
generates a coupling between the non-zero frequency modes
of $\phi$ and the static mode.
The final ellipsis denotes higher order terms containing
at least four spatial derivatives acting on various
factors of $\beta (\mu + i e \phi)$
[in such a way that every field $\phi$ is differentiated at least once],
with each derivative accompanied by a factor of some thermal wavelength
$\lambda$.
The four-derivative $O(\lambda^4)$ terms, which are formally of four loop
order, are exhibited explicitly in Eq.~(\ref {eq:S2}).

The expansion (\ref {eq:lnzFT}) is valid if the
potential $\phi(\r,\tau)$ varies slowly on the scale of
the thermal wavelength $\lambda_a$.
As discussed earlier,
inserting this expansion (truncated at some order)
into the functional integral (\ref {eq:za}) completely mangles
the effects of short-distance fluctuations in~$\phi$.
However, as with any effective field theory,
the resulting errors are compensated,
to any given order in $\kappa\lambda$,
by including the requisite induced interactions
and suitably adjusting their coefficients.

At this point, one may contemplate completely integrating
out the non-zero frequency modes of $\phi$ in order to generate
an effective theory containing only the static potential $\phi(\r)$.
But doing so would be a mistake.
Integrating out the non-zero frequency
modes is no longer trivial because of the coupling
between the static and non-zero frequency modes.
More importantly, the resulting functional of $\phi(\r)$
could not be adequately approximated by any set of local interactions.
Correlations of the non-zero frequency components of $\phi$
only decrease like $1/r$
(due to the long-range nature of Coulomb interactions),
and are not Debye screened.
This will be demonstrated below.
In physical terms,
the absence of Debye screening in the potential correlations
at non-zero (Matsubara) frequencies reflects the effect of
inertia on the response of charges in the plasma.
Consequently,
if one completely integrates out the non-zero frequency components
of $\phi$, then the resulting
theory will contain complicated non-local interactions
which decrease only algebraically with distance.
To produce a useful effective theory,
that can be approximated by local interactions,
one must explicitly retain in the effective theory
all degrees of freedom with long distance correlations---%
including the non-zero frequency components of $\phi$.
In other words,
the complete effective theory must have the form
\begin {equation}
    Z[\mu] = N_0' \int [d\phi] \> \prod_{m\ne0} \left[ d\phi^m \right] \,
    \exp \Bigl\{
	-\Scl[\phi;\mu]
	-S_{\rm ind}[\phi;\mu]
	-S_{\rm non-static}[\phi,\phi^m;\mu]
    \Bigr\} \,,
\label {eq:neweff}
\end {equation}
where $\Scl$ and $S_{\rm ind}$ are given in
Eq's.~(\ref{eq:Zclass}) and (\ref {Sind}), respectively, and
\begin{eqnarray}
    S_{\rm non-static}[\phi,\phi^m;\mu]
    &=&
    \int (d^\dim\r)
    \sum_{m\ne 0}
    {\beta \over 2} \>
    \Biggl\{
    |\nabla \phi^m|^2
\nonumber\\ && \qquad {}
    -
    {\beta\over 8\pi^3 m^2}
    \sum_a
    \lambda_a^2 \, n_a^0 \, e^{ i e_a \beta \, \phi } \>
    \!\left[ \nabla\mu_a^{\vphantom-m} + i e_a \nabla \phi^m \right]
    {\cdot}
    \left[ \nabla\mu_a^{-m} + i e_a \nabla \phi^{-m} \right]
\nonumber\\ && \qquad {}
    + \cdots
    \Biggr\} \> .
\label {eq:non-static}
\label {eq:Shhb}
\end{eqnarray}
Here $\mu_a^m(\r)$ denotes the non-zero frequency components of the
chemical potential,
\begin{equation}
    \mu_a(\r , \tau) \equiv \mu_a(\r) + \sum_{m \neq 0} \mu_a^m(\r) \,
    e^{-i \omega_m \tau } \,.
\end{equation}
The final ellipsis in (\ref {eq:non-static}) denotes yet higher-order terms
involving four or more derivatives, as well as non-zero frequency
induced interactions involving $|\nabla\phi^m|^2 $
multiplying products of two or more densities.

At leading order
(when all interaction terms in $S_{\rm non-static}$ are neglected),
the correlator of the non-zero frequency
components of $\phi$ (times $\beta$)
is given by an unscreened Coulomb potential,
\begin {equation}
    \beta \, {\cal G}_m^{(0)}(\r,\r')
    \equiv
    \beta \left\langle \phi^m(\r) \phi^m(\r')^* \right\rangle_0
    =
    {1 \over 4 \pi |\r{-}\r'|} \,,
\label {eq:Gm}
\end {equation}
since this is the Green's function of $-\nabla^2$.
In other words, the Fourier transformed correlator is given by
\begin{equation}
   \beta  \, \tilde {\cal G}_m^{(0)}(\k)
    =
    {1 /  \k^2}  \,.
\end{equation}
Because the sub-leading interaction in (\ref {eq:non-static}) involves
the gradient $\nabla\phi^m(\r)$, and not $\phi^m(\r)$ itself,
this interaction does not cause non-zero frequency correlations to
develop a finite correlation length.  Rather, it merely produces an
$O\left[(\kappa\lambda)^2\right]$ change in the residue of the
$1/\k^2$ pole of $\tilde {\cal G}_m(\k)$.  Recalling that $\omega_m =
2 \pi m / \beta $ (which has units of energy in our notation) and noting
that the (lowest-order) plasma frequency $\omega_P$ is defined by
\begin{equation}
\omega_P^2 = \sum_a { e_a^2 \, n_a^0 \over m_a} =
\sum_a { e_a^2 \, \lambda_a^2 \, n_a^0 \over 2\pi \, \beta \, \hbar^2 } \,,
\end{equation}
we find that with this correction,
\begin{equation}
    \left[ \beta \, \tilde{\cal G}_m(\k) \right]^{-1} =
    \k^2 \, \left[ 1 + { \hbar^2 \, \omega_P^2 \over \omega_m^2 } \right] \,.
\label{nodebye}
\end{equation}
This same result is obtained from the $\k \to 0$ limit of the sum of
ring diagrams contributing to this correlator, which generates a
denominator involving the one-loop polarization function $\Pi(\k,\omega)$.
This well-known function is presented in Eq.~(\ref{PI}) of Appendix
\ref{app:det},
as well as in many textbooks (such as Ref.~\cite{fetter}).
The $\k \to 0$ limit corresponds to the classical
limit, and the continuation $ \omega_m / \hbar \to i ( \omega - i
\epsilon) $, $ \epsilon \to 0^+$, further produces the classical
retarded response function. The resulting pole at $\omega = \omega_P$
corresponds to the propagation of classical longitudinal plasma waves,
waves whose resonant frequency is independent of their wave number.

The lack of Debye screening of the non-zero frequency fluctuations
in the electrostatic potential is an exact result.
It is a consequence of electromagnetic current conservation,
${d \rho / dt} + \nabla \cdot {\bf j} = 0 $,
or equivalently gauge invariance.
The fundamental quantum theory (\ref {eq:Zquant}) is,
in particular, invariant under time-dependent, but space-independent,
gauge transformations,
\begin {equation}
    \phi(\r,\tau) \to \phi(\r,\tau) + {d \chi(\tau) \over d\tau} \,, \qquad
    \psi_a(\r,\tau) \to e^{i e_a \chi(\tau)} \> \psi_a(\r,\tau) \,.
\label {eq:gauge-inv}
\end {equation}
The effective theory must necessarily share this invariance.
But in the effective theory,
where the charged fields have been integrated out,
these gauge transformations reduce to arbitrary constant shifts in the
non-zero frequency components of $\phi$,
\begin {equation}
    \phi^m(\r) \to \phi^m(\r) + i \omega_m \, \chi^m \,,
\label {eq:eff-gauge-inv}
\end {equation}
where $\chi^m$ are the Fourier components of $\chi$.  This means that
the effective theory (\ref {eq:non-static}) can never depend on the
non-zero frequency fields $\phi^m(\r)$ other than through their
gradients.  And this implies that arbitrarily long wavelength
fluctuations in the non-zero frequency components of $\phi$ must have
arbitrarily low action, which in turn implies that the Fourier
transform of the correlation function $\langle \phi^m(\r)\,
\phi^m(\r')^* \rangle$ will diverge as $\k \to 0$.  In other words,
the interactions of the effective theory cannot cause the pole in the
non-zero frequency correlator to shift away from%
\footnote
    {%
    This is completely analogous to Goldstone's theorem proving the
    presence of long range fluctuations in any theory with a
    spontaneously broken continuous symmetry.
    Since the symmetry (\protect\ref {eq:eff-gauge-inv})
    shifts $\phi^m$, it is impossible for
    the expectation values $\langle \phi^m \rangle$ to be invariant
    under this symmetry.
    Consequently, $\phi^m$ must have long range correlations.
    }
$\k = 0$.  A detailed explanation of these points is given in
Appendix \ref{timeapp}.

\begin {figure}[t!]
    \begin {center}
    \leavevmode
    \def\epsfsize #1#2{0.4#1}
    \epsfbox {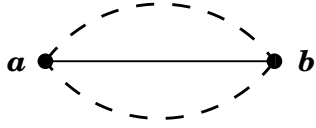}
    \end {center}
\caption
    {%
    First non-vanishing correction to $\ln Z$ involving the non-zero
    frequency modes of $\phi(\r,\tau)$.
    The dashed lines represent the long-range, unscreened Coulomb
    Green's functions of the non-zero frequency modes.
    Each vertex represents one insertion of $S_{\rm non-static}$
    (specifically that part which is linear in the static mode $\phi(\r)$).
    The relative size of the resulting contribution is
    $O[(\lambda\kappa)^4 (\beta e^2 \kappa)^2]$,
    since each vertex contains an overall factor of $\lambda^2$,
    and the two loops of the diagram generate two powers of the
    loop expansion parameter $\beta e^2 \kappa$.
    Because we are treating the quantum parameters $\beta e_a^2/\lambda_a$
    as fixed numbers of order one, the net result is a
    contribution of six-loop order.
    }
\label {fig:nonz}
\end {figure}

    The first interaction term in $S_{\rm non-static}$ is formally
$O\!\left[(\kappa \lambda)^2\right]$ smaller than the leading
$|\nabla\phi^m|^2$  term and thus is of two-loop order.  As
noted above, this term produces a relative change of this size in non-zero
frequency correlators.  However, it does not affect thermodynamic
quantities, or static correlators, at two-loop order because
\begin {equation}
    \avg{ \nabla \phi^m(\r) \cdot \nabla \phi^{-m}(\r)}
    =
    \beta^{-1} \int {(d^\dim \k) \over (2\pi)^\dim}
    =
    0
\end {equation}
in our dimensional continuation regularization.%
\footnote
    {%
    This identity would not hold
    if we had chosen to employ a different regularization scheme,
    such as a wave-number cutoff, or a hard core interaction,
    in defining the effective theory.
    Had we done so, it would be necessary to adjust the coefficient
    of the classical $\n0a \, e^{i \beta e_a \phi}$ interaction
    in order to compensate for cutoff-dependent effects
    resulting from fluctuations of the non-zero frequency modes.
    }
In fact, the non-zero frequency interaction term first affects the
thermodynamic quantities at six-loop order, through the diagram
illustrated in Fig.~\ref{fig:nonz}.
However,
even though the non-zero frequency interactions are suppressed by
numerous powers of $\kappa\lambda$, they fundamentally alter the
long-distance behavior of the static density-density correlator.
Instead of exhibiting classical Debye-screened exponential decay,
the correlator acquires a long-distance tail which decreases only
algebraically with distance.  This happens at five-loop order as
shown in section \ref {sec:longdist}, where a simple but explicit
evaluation of the resulting long-distance limit is given.

\newpage

\section {Two-Loop Results}
\label{sec:twolrs}

\begin {figure}[ht]
   \begin {center}
      \leavevmode
      \def\epsfsize #1#2{0.45#1}
      \epsfbox {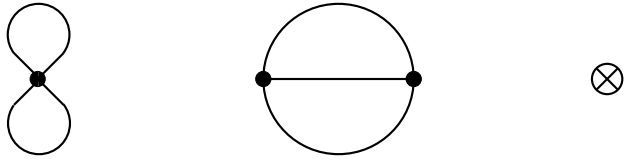}
   \end {center}
   \caption
       {%
       Two-loop diagrams contributing to $\ln Z$.
       The circled `X' denotes the $\phi$-independent part
       of the two-loop $g^0_{ab}$ induced interaction.
       }%
\label {fig:P2}
\end {figure}

The two loop contributions to $\ln Z$ are given by the diagrams shown
in figure \ref {fig:P2}. They correspond to the analytic expression
\begin{eqnarray}
    {\ln Z_2 \over \vol}
    &=&
    {\ln Z_1 \over \vol}
    + {1 \over 2^3} \sum_a \n0a \left[\beta e_a^2 \, G_\dim({\bf 0})\right]^2
    -
    \sum_{a,b} \beta^3 \, \n0a \, \n0b \,
    \left[
	{e_a^3 e_b^3 \over 2 \cdot 3!} \int (d^\dim\r) \> G_\dim(\r)^3
	+ g^0_{ab}
    \right] .
\label{twolps}
\end{eqnarray}
As noted earlier in Eq.~(\ref {perf}), the $\dim\to3$ limit of $
G_\dim({\bf 0}) $ is finite and equals $-\kappa_0 / (4 \pi)$. The
integral of the cube of the Debye Green's function,
\begin{equation}
D^{(3)}_\dim({\bf 0}) = \int (d^\dim\r) \> G_\dim(\r)^3 \,,
\end{equation}
is the vanishing wave number limit of the corresponding Fourier
transform (which is the reason for the notation used here). It is computed
in Appendix \ref {debye ints} [{\em c.f.} Eq.~(\ref{d30})] and
shown to be
\begin{equation}
    D^{(3)}_\dim({\bf 0}) =
    { 1 \over (4\pi)^2}
    \left( {9 \kappa^2 \over 4 \pi } \right)^{\dim-3}
    {1 \over 2}
    \left\{
    {1 \over 3-\dim} + 1 - \gamma + O(\dim{-}3)
    \right\} .
\label {d30a}
\end {equation}
The two-loop induced coupling $g^0_{ab}$, given in (\ref{eq:Zab}),
cancels the $\dim\to3$ pole of $D^{(3)}_\dim({\bf 0})$.
It is convenient to write the final bracket in (\ref {twolps})
as
\begin{equation}
    \left[
	{e_a^3 e_b^3 \over 2 \cdot 3!} \int (d^\dim\r) \> G_\dim(\r)^3
	+ g^0_{ab}
    \right] =
    {1\over 12} \,
    e_a^3 e_b^3 \, D^{(3)}_{\rm R}({\bf 0};\mu) + g_{ab}(\mu)\,,
\label {eq:D3+g}
\end{equation}
where, in view of Eq's.~(\ref{eq:Zab}) and (\ref{d30a}),
\begin {equation}
D^{(3)}_{\rm R}({\bf 0};\mu) =
    { 1 \over 32\pi^2} \left\{ \left[ \left( { 9 \kappa_0^2 \over
      4 \pi } \right)^{\dim-3} - \left( \mu^2 \right)^{\dim - 3} \right]
      { 1 \over 3 - \dim} + 1 - \gamma + O(\dim{-}3)
    \right\} .
\label{d3r0}
\end {equation}
The physical $\dim \to 3$ limit is finite, as it must be, and gives
\begin{equation}
D^{(3)}_{\rm R}({\bf 0};\mu) =
    { -1 \over 32\pi^2 } \left\{
    \ln \left( { 9 \kappa_0^2 \over 4\pi \mu^2} \right)
    -1 + \gamma \right\} .
\label{D3R}
\end{equation}
Note that the coefficient of the induced interaction
that produces this finite result was determined from a
different physical quantity, the density-density correlator.
Nevertheless the structure of the effective theory
[in particular, the shift symmetry (\ref{eq:shift-sym})]
guarantees that the single two-loop $g^0_{ab}$ interaction
removes the cutoff-dependence in any physical quantity
computed to either two or three loop order.

Putting the pieces together, including the previous
one-loop result (\ref{pone})
and the value (\ref {eq:gab}) for the renormalized coupling $g_{ab}(\mu)$,
produces $\ln Z$ to two-loop order (as a function of bare particle densities),
\begin{eqnarray}
    { \ln Z_2 \over \vol }
    &=&
    \sum_{a=1}^A \n0a
    \left[ 1
    + {1 \over 3} \left({ \beta e_a^2 \kappa_0 \over 4 \pi } \right)
    + {1 \over 8} \left({ \beta e_a^2 \kappa_0 \over 4 \pi } \right)^2
    \right]
\nonumber\\
&& {}
    + { \pi \over 6} \sum_{a,b=1}^A \n0a \, \n0b
    \left({ \beta e_a e_b  \over 4 \pi } \right)^3
    \left[ \ln \left( {9 \lambda_{ab}^2 \kappa_0^2 \over 4\pi} \right)
    -1 + \gamma + \Gam ab \right] .
\label{ztwolps}
\end{eqnarray}

As was remarked above, the leading term in $\Gam ab$ when $\eta_{ab}$
becomes small comes from the exchange contribution.  Using the limiting
form (\ref{shortly}) of $\Gam ab$ gives the exchange correction
\begin{equation}
    { \ln Z_2^{\rm exch} \over \vol }
    = \pm { 1 \over 4 \sqrt 2 } \sum_{a=1}^A \, n_a^0
    \left( \lambda_a^3 \, n_a^0 / g_a \right)
    =
    \pm
    \sum_{a=1}^A \, n_a^0 \>
    {e^{\beta \mu_a} \over 2^{5/2}} \,,
\end{equation}
which is just the first quantum statistics correction shown in
Eq.~(\ref{fugex}).

\subsection {Number Density}

The particle number density of species $a$ is given by
\begin{equation}
    \langle n_a \rangle_\beta = { \partial \over \partial \beta \mu_a}
    {\ln Z \over \vol} \,,
\label{nmu}
\end{equation}
where the partial derivative is computed at fixed $\beta$.
Inserting the result (\ref {ztwolps}) and using
$$
    {\partial \n0b \over \partial \beta \mu_a} = \delta_{ab} \, \n0a \,,
    \qquad\qquad
    {\partial \kappa_0 \over \partial \beta \mu_a} =
    { \beta e_a^2 \n0a \over 2 \kappa_0 } \,,
$$
and $\partial \eta_{bc}/\partial (\beta\mu_a) = 0$,
yields
\begin{eqnarray}
    \langle n_a \rangle_\beta
    &=&
    \n0a \left[ 1 +
    \left({ \beta e_a^2 \kappa_0 \over 8 \pi } \right) + {1 \over 2}
    \left({ \beta e_a^2 \kappa_0 \over 8 \pi } \right)^2 \right]
    + {1 \over 8} \left(\n0a \, \beta e_a^2\right) \sum_{b=1}^A \n0b
    \left( {\beta e_b^2 \over 4 \pi } \right)^2
\nonumber\\
    && {}
    +
    {\pi \over 6} \, { \n0a \beta e_a^2 \over \kappa_0^2 }
    \sum_{b,c=1}^A \n0b \n0c \left( {\beta e_b e_c \over 4\pi}\right)^3
\nonumber\\
    && {}
    +
    { \pi \over 3} \, \n0a \sum_{b=1}^A \n0b
    \left({ \beta e_a e_b  \over 4 \pi } \right)^3
    \left[
    \ln\!\left( {9 \lambda_{ab}^2 \kappa_0^2 \over 4\pi} \right)
    -1 + \gamma + \Gam ab
    \right] .
\label{ntwolps}
\end{eqnarray}
The first bracket contains the first three terms in the expansion
of the exponential
\begin{equation}
 \exp\left( { \beta e_a^2 \kappa_0 \over 8\pi} \right)
  = \exp\left[ - \half { \beta e_a^2 } \, G_3( {\bf 0} ) \right] ,
\label{expp}
\end{equation}
which is just the Boltzmann factor for the polarization correction to
the self-energy of a species $a$ particle when it is placed in the plasma.
The next term is just
\begin{equation}
    \half \n0a \left( { \beta e_a^2 \delta \kappa \over 4 \pi } \right) ,
\end{equation}
where
\begin{equation}
    {\delta \kappa \over 4\pi } = { 1 \over 4} \sum_{b=1}^A \n0b
    \left( { \beta e_b^2 \over 4 \pi } \right)^2
\end{equation}
is the change in the lowest-order Debye wave number induced by
the first-order density correction.
Thus, through the order we have computed, our result is equivalent to
\begin{eqnarray}
    \bar n_a = \expect {n_a}
    &=&
    \n0a \Bigg\{
    \exp\left( { \beta e_a^2 \over 2 } {\bar\kappa \over 4\pi} \right)
    +
    { \beta^2 e_a^2 \over 4! \, \kappa_0^2 }
    \left[ \sum_{b=1}^A \n0b \left({\beta e_b^3 \over 4\pi}\right)
    \right]^2
    \nonumber\\ && \quad {}
    +
    {\pi \over 3} \,  \sum_{b=1}^A \n0b
    \left({ \beta e_a e_b  \over 4 \pi } \right)^3
    \left[ \ln \!\left( {9 \lambda_{ab}^2 \kappa_0^2 \over 4\pi} \right)
	-1 + \gamma + \Gam ab
    \right]
    \Bigg\} \,,
\label{number}
\end{eqnarray}
where $\bar\kappa = \left[ \sum_a \beta e_a^2 \bar n_a \right]^{1/2}$
is the Debye wave number computed with the physical particle densities.
Inverting this result to express the bare densities in terms
of the physical densities is now easy since, to this order,
the bare quantities in the remaining two-loop terms
may simply be replaced by physical quantities,
\begin{eqnarray}
    \n0a &=&
    \bar n_a \Bigg\{
    \exp\left( -{ \beta e_a^2 \over 2 } {\bar\kappa \over 4\pi} \right)
    -
    { \beta^2 \, e_a^2 \over 4! \, \bar\kappa^2 }
    \left[ \sum_{b=1}^A \bar n_b \left({\beta e_b^3 \over 4\pi}\right)
      \right]^2
    \nonumber\\
    && \quad {}
    -{\pi \over 3} \, \sum_{b=1}^A \bar n_b
    \left({ \beta e_a e_b  \over 4 \pi } \right)^3
    \left[ \ln \!\left( {9 \lambda_{ab}^2 \bar\kappa^2 \over 4\pi} \right)
	-1 + \gamma + \Gam ab
    \right]
    \Bigg\} \,.
\end{eqnarray}

\subsection {Energy Density}

The internal energy density in the plasma is given by
\begin{equation}
    u = \expect {H\over\vol}
    = - { \partial \over \partial \beta } {\ln Z \over \vol} \,,
\end{equation}
where now $\beta\mu_a$ is kept fixed for all $a$.
Noting that $ \n0a \propto \lambda_a^{-3} \propto \beta^{-3/2} $,
$\kappa_0^2 \propto \beta^{-1/2} $, and $\eta_{ab} \propto \beta^{1/2}$,
one finds at two loop order
\begin{equation}
    \beta u =
    \sum_{a=1}^A \n0a
    \left[
	{3 \over 2} +
	{1 \over 4} \left({ \beta e_a^2 \kappa_0 \over 4 \pi } \right)
    \right]
    - { \pi \over 12}
    \sum_{a,b=1}^A \n0a \n0b
    \left({ \beta e_a e_b  \over 4 \pi } \right)^3
    \left[ 1 + \eta_{ab} \, \Gamprime ab \right],
\end {equation}
or in terms of the physical densities,
\begin {eqnarray}
    \beta u &=&
    {3\over 2} \sum_{a=1}^A \bar n_a
    - {\bar\kappa^3 \over 8\pi}
    - {\pi \over 2} \sum_{a,b=1}^A \bar n_a \bar n_b \!
    \left({ \beta e_a e_b  \over 4 \pi } \right)^3
    \!\Biggl\{
	\ln \!\left( {9 \lambda_{ab}^2 \bar\kappa^2 \over 4\pi} \right)
	-\coeff 13 + \gamma + \Gam ab
	+ \coeff 16 \, \eta_{ab} \, \Gamprime ab
    \Biggr\} .
\end{eqnarray}
Here (and henceforth),
\begin {equation}
    \Gamprime ab \equiv f'(\eta_{ab}) \pm \delta_{ab} \tilde f'(\eta_{ab}) \,,
\label {eq:Gamprime}
\end {equation}
with the functions $f$ and $\tilde f$ given in
Eq's.~(\ref {eq:f}) and (\ref {eq:tildef}), respectively.

\subsection {Pressure and Free Energy Density}

The pressure, re-expressed in terms of physical densities, is the
equation of state. To two loop order
\begin{eqnarray}
    \beta p = { \ln Z \over \vol}
    &=&
    \sum_{a=1}^A \bar n_a
    -{\bar\kappa^3 \over 24\pi}
    - {\pi \over 6} \, \sum_{a,b=1}^A \bar n_a \bar n_b
    \left({ \beta e_a e_b  \over 4 \pi } \right)^3
    \left[ \ln \!\left( {9 \lambda_{ab}^2 \bar\kappa^2 \over 4\pi} \right)
	+ \gamma + \Gam ab
    \right].
\end{eqnarray}
And the two-loop Helmholtz free energy density is
\begin{eqnarray}
    \beta f
    &=&
    \sum_{a=1}^A \bar n_a \left[ -1 + \ln (\bar n_a \lambda_a^3/g_a)\right]
    -{\bar\kappa^3 \over 12\pi}
\nonumber\\ && {}
    - {\pi \over 6} \, \sum_{a,b=1}^A \bar n_a \bar n_b
    \left({ \beta e_a e_b  \over 4 \pi } \right)^3
    \left[ \ln \!\left( {9 \lambda_{ab}^2 \bar\kappa^2 \over 4\pi} \right)
	+ \gamma + \Gam ab - 1
    \right].
\end{eqnarray}

\subsection {Number Density Correlators}

\begin {figure}[t]
   \begin {center}
      \leavevmode
      \def\epsfsize #1#2{0.35#1}
      \epsfbox {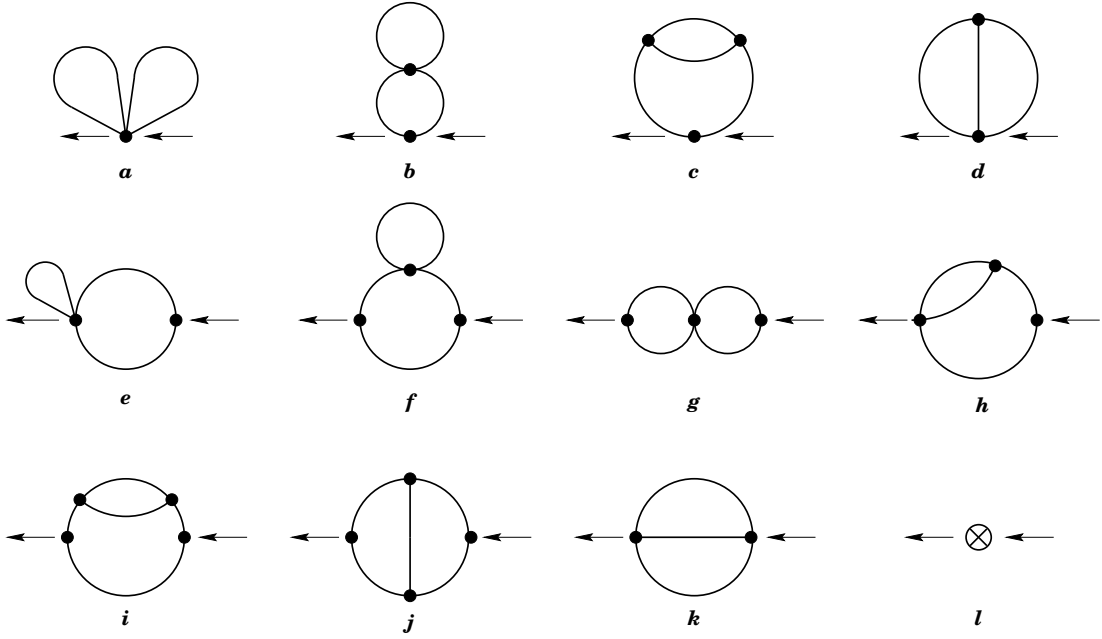}
   \end {center}
   \caption
	{%
	Two-loop diagrams contributing to the irreducible part
	$\tilde C_{ab}(\k)$ of the density-density correlator.
	The arrows merely serve to indicate where external momentum
	flows in and out of each diagram;
	the vertices where momentum flows in and out should
	be regarded as having attached species labels $a$ and $b$,
	respectively.
	The circled `X' in diagram $l$ denotes the contribution
	from the two-loop interactions proportional to either
	$g^0_{ab}$ or $h^0_a$.
	Diagrams $a$--$d$ are two loop contributions
	to the contact term $\delta_{ab} \expect {n_a}$. These terms,
	together with part of diagram $l$,
	simply provide the two-loop correction to the number density
	$\expect {n_a}$.
	Not shown are reflected versions of diagrams $e$ and $h$ which
	differ only by an interchange of incoming and outgoing
	vertices. Diagram $e$ and its reflection are one-loop
	corrections to the densities that appear in the vertex factors in
	the one-loop result for $\tilde C_{ab}(\k)$.
	Diagram $f$ is a one-loop
	density correction to the Debye wave number in the Debye
	Green's function appearing in the one-loop $\tilde C_{ab}(\k)$.
	Thus, the effect of diagrams $a$ through $f$ (plus a part of $l$)
	is merely to put the correct physical densities, to two-loop order,
	in the previous one-loop $\tilde C_{ab}(\k)$. Only
	diagrams $g$ through $k$ (plus the remaining part of $l$)
	give non-trivial, two-loop corrections to $\tilde C_{ab}(\k)$.
	}%
\label {fig:self2}
\end {figure}

The two-loop diagrams contributing to the irreducible part
$\tilde C_{ab}(\k)$ of the number density correlation function
are shown in figure \ref {fig:self2}.
Diagrams $ a $--$ d $ represent (momentum independent) contributions
to the contact term $\delta_{ab} \expect{n_a}$.
As noted earlier, only diagrams $ d $ and $ k $
(plus the induced coupling contribution $ l $)
are singular as $\dim\to3$; all the other diagrams may be evaluated
directly in three dimensions.
The explicit contributions of these diagrams to $\tilde C_{ab}(\k)$
are given by%
\footnote
    {%
    There are two versions each of diagrams $ e $ and $ h $,
    differing only by the interchange $a\leftrightarrow b$
    and $\k\to-\k$.
    The contributions of both diagrams of each pair are
    included in $\Sigma_e$ and $\Sigma_h$.
    }
\begin {eqnarray}
    \Sigma_a &=&
	\delta_{ab} \,
	\beta^2 \, \n0a \, \,
	 e_a^4 \,\,
	{ 1 \over 2^3} \, G_3({\bf 0})^2 \,,
\\
    \Sigma_b &=&
	-\delta_{ab}
	\sum_{c}
	\beta^3 \, \n0a \, \n0c \,\,
	e_a^2 \, e_c^4 \,\,
	{ 1 \over 2^2} \, D^{(2)}_3({\bf 0}) \, G_3({\bf 0}) \,,
\\
    \Sigma_c &=&
	\delta_{ab}
	\sum_{c,d}
	\beta^4 \, \n0a \, \n0c \, \n0d \,\,
	e_a^2 \, e_c^3 \, e_d^3 \,\,
	{1 \over 2^2} \, D^{(121)}_3({\bf 0}) \,,
\\
    \Sigma_d &=&
	-\delta_{ab}
	\sum_{c}
	\beta^3 \, \n0a \, \n0c \,\,
	 e_a^3 \, e_c^3 \,\,
	{1 \over 3!} \, D^{(3)}_\dim({\bf 0}) \,,
\\
    \Sigma_e &=&
	-\beta^3 \, \n0a \, \n0b \,
	\left( e_a^4 \, e_b^2 + e_a^2 \, e_b^4 \right)
	{1 \over 2^2} \, D^{(2)}_3(\k) \, G_3({\bf 0}) \,,
\\
    \Sigma_f &=&
	\sum_{c}
	\beta^4 \, \n0a \, \n0b \, \n0c \,\,
	 e_a^2 \, e_b^2 \, e_c^4 \,\,
	{ 1 \over 2} \, D^{(111)}_3(\k) \, G_3({\bf 0}) \,,
\\
    \Sigma_g &=&
	\sum_{c}
	\beta^4 \, \n0a \, \n0b \, \n0c \,\,
	 e_a^2 \, e_b^2 \, e_c^4 \,\,
	 {1 \over 2^2} \, D^{(2)}_3(\k)^2 \,,
\\
    \Sigma_h &=&
	\sum_{c}
	\beta^4 \, \n0a \, \n0b \, \n0c \,
	\left( e_a^3 \, e_b^2 + e_a^2 \, e_b^3 \right) e_c^3 \,\,
	{1 \over 2} \, D^{(211)}_3(\k) \,,
\\
    \Sigma_i &=&
	-\sum_{c,d}
	\beta^5 \, \n0a \, \n0b \, \n0c \, \n0d \,\,
	 e_a^2 \, e_b^2 \, e_c^3 \, e_d^3 \,\,
	{1 \over 2} D^{(1211)}_3(\k) \,,
\\
    \Sigma_j &=&
	-\sum_{c,d}
	\beta^5 \, \n0a \, \n0b \, \n0c \, \n0d \,\,
	 e_a^2 \, e_b^2 \, e_c^3 \, e_d^3 \,\,
	{1 \over 2} D_J(\k) \,,
\\
    \Sigma_k &=&
	-\beta^3 \, \n0a \, \n0b \,\,
	 e_a^3 \, e_b^3 \,\,
	{1 \over 3!} \, D^{(3)}_\dim(\k) \,,
\\
    \Sigma_l &=&
	-2 \beta^3 \, g^0_{ab}\, \n0a \, \n0b \,
	-2\delta_{ab}
	\sum_{c}
	\beta^3 \, g^0_{ac} \, \n0a \, \n0c \vphantom {{1\over 1}}
	- 2 \, \delta_{ab} \, \beta^2 \, h^0_a \, \n0a \, \k^2 \,,
\end {eqnarray}%
where the required integrals are
\begin {eqnarray}
    D^{(n)}_\dim(\k) &\equiv&
	\int (d^\dim\r) \> e^{-i \k \cdot \r} \, G_\dim(\r)^n \,,
\label {Dn}
    \\
    D^{(lmn)}_\dim(\k) &\equiv&
	\int (d^\dim\r)(d^\dim\r_1) \>
	e^{-i \k \cdot \r} \,
	G_\dim(\r{-}\r_1)^l G_\dim(\r_1)^m G_\dim(\r)^n \,,
\label {Dnlm}
    \\
    D^{(klmn)}_\dim(\k) &\equiv&
	\int (d^\dim\r)(d^\dim\r_1)(d^\dim\r_2) \>
	e^{-i \k \cdot \r} \,
	G_\dim(\r{-}\r_1)^k G_\dim(\r_1{-}\r_2)^l G_\dim(\r_2)^m
	G_\dim(\r)^n \,,
\label {Dnlmq}
    \\
    D_J(\k) &\equiv&
	\int (d^3\r)(d^3\r_1)(d^3\r_2) \>
	e^{-i \k \cdot \r} \,
	G_3(\r{-}\r_1) G_3(\r{-}\r_2) G_3(\r_1{-}\r_2)
	G_3(\r_1) G_3(\r_2) \,.
\end {eqnarray}%
These integrals are evaluated in Appendix \ref {debye ints}
(with help from Ref.~\cite {rajantie}).

By examining the graphical structure, it is easy to see that the
irreducible density correlator has the form
\begin{equation}
  \tilde C_{ab}(\k) = \delta_{ab} \, \bar n_a \, \tilde F_a(\k)
  + \half \left( \beta e_a^2 \bar n_a \right) \tilde F_{ab}(\k)
    \left( \beta e_b^2 \bar n_b \right)
\label{genform}
\,,
\end{equation}
which generalizes the one-loop result (\ref{cone}).
The derivative interaction involving the induced coupling $h^0_a$
is responsible for generating the $\k$-dependence in the $\delta_{ab}$
contact term,
\begin {equation}
    \tilde F_a^{(2)}(\k)
    =
    1 - 2 \beta^2 \, h^0_a \, \k^2 \,.
\end {equation}
The terms $\Sigma_a$ through $\Sigma_d$, together with the second part of the
renormalization term $\Sigma_l$, just give the two-loop corrections to
the number density $\langle n_a \rangle_\beta$ that appears in
$\delta_{ab} \, \bar n_a$ in the general form above.
It is straightforward to show that these terms are just the
two-loop parts in the previous result (\ref{ntwolps}) for the number density.
The one-loop correction, given in Eq.~(\ref{cone}), involves
\begin{equation}
\tilde F_{ab}^{(1)}(\k) = D_3^{(2)}(\k) \,.
\label {eq:F1}
\end{equation}
Recalling that the one-loop density correction reads
\begin{equation}
\delta \langle n_a \rangle_\beta = - \half \beta \, e_a^2 \, G_3({\bf
  0}) \, n^0_a \,,
\end{equation}
we see that the two-loop term $\Sigma_e$ gives the one-loop correction
for each of the two explicit density factors appearing in the second term of
(\ref{genform}), with $\tilde F_{ab}(\k)$ taking on its one-loop
value $ D_3^{(2)}(\k) $. Writing out $ D^{(111)}_3(\k)$ in terms of
Fourier integrals [as is done explicitly in Eq.~(\ref{d111k})], it is
easy to see that
\begin{equation}
     D^{(111)}_3(\k)
    =
  - { 1\over 2} {d D^{(2)}_3(\k)\over d\kappa_0^2} \,.
\end{equation}
Therefore $\Sigma_f$ accounts for the correction to the one-loop $
\tilde F_{ab}^{(1)}(\k) $ brought about by replacing the bare
Debye wave number $\kappa_0$ by its one-loop corrected value.
In summary, the two-loop terms $\Sigma_a$ through
$\Sigma_f$, plus the second piece of $ \Sigma_l$, just provide simple
density corrections to the one-loop $\tilde C_{ab}^{(1)}(\k)$, and all
of these terms may be omitted if the correct physical densities $\bar
n_a$ are used in the construction of $\tilde C_{ab}^{(1)}(\k)$.

To assemble the remaining terms in the two-loop, irreducible
correlator, we first use the explicit form (\ref{eq:Zab})
of the induced coupling $g^0_{ab}$
to write the sum of $\Sigma_k$ and the first piece of
$\Sigma_l$ as
\begin{equation}
    \Sigma_k(\k) -2 \beta^3 \, g^0_{ab} \, \n0a \, \n0b
    =
    -\beta^3 \, \n0a \, \n0b
    \left[
	{1 \over 3!} \, \left( e_a^3 e_b^3 \right)
	D^{(3)}_{\rm R} (\k;\mu) + 2 g_{ab}(\mu)
    \right]
\end{equation}
where
\begin{equation}
    D^{(3)}_{\rm R} (\k;\mu)
    \equiv
    D^{(3)}_\dim (\k)
    -
    { 1 \over 2 (4\pi)^2} \, { \mu^{2(\dim-3)} \over 3 - \dim} \,.
\end{equation}
This is the generalization to non-vanishing wave number of
$D_{\rm R}^{(3)}({\bf 0};\mu) $
previously introduced in Eq.~(\ref{d3r0}). Using
the result (\ref{d3kd}) for $D_\dim^{(3)}(\k)$
in dispersion relation form and taking the
physical $\dim \to 3$ limit yields
\begin{equation}
    D_{\rm R}^{(3)}(\k;\mu)
    =
    - { 1 \over (4\pi)^2 } \, { 1 \over 2}
    \Bigg\{
	\ln\left( { 9 \kappa^2 \over 4 \pi \mu^2 } \right)
	- 3 + \gamma
	+
	2 \int_{3\kappa}^\infty d s
	\left( { k^2 \over s} + 3 \kappa \right) { 1 \over k^2 + s^2 }
    \Bigg\} \,.
\end{equation}
Alternatively, using the result (\ref{eq:D3}) for $D_\dim^{(3)}(\k)$
evaluated in terms of elementary functions gives the $\dim = 3$ limit
\begin{equation}
    D^{(3)}_{\rm R}(\k;\mu)
    =
    - { 1 \over (4\pi)^2} \,
    {1 \over 2}
    \left\{ \ln\left( { 9 \kappa^2 \over 4 \pi \mu^2 } \right)
     - 3 + \gamma
    + {6\kappa \over k} \arctan{k \over 3\kappa}
    + \ln \biggl[1 + {k^2 \over 9 \kappa^2}\biggr]  \right\} .
\end{equation}
Combining the one-loop result (\ref {eq:F1}) with
this renormalized contribution of $\Sigma_k$
plus the other non-trivial two-loop
terms $\Sigma_g$ through $\Sigma_j$,
and recalling the definition (\ref{gab}) of the coupling $g_{ab}(\mu)$,
leads to
\begin{eqnarray}
    \tilde F_{ab}^{(2)}(\k)
    &=&
    D_3^{(2)}(\k)
    \Bigl[
	1 + \half \sum_c \beta^2 \, \bar n_c \, e^4_c \, D_3^{(2)}(\k)
    \Bigr]
    +
    \sum_c \beta^2 \, \bar n_c \, \left( e_a + e_b \right) \, e_c^3 \,
	D_3^{(211)}(\k)
\nonumber\\ && \quad
    -
    \sum_{c,d} \beta^3 \, \bar n_c \, \bar n_d \, e^3_c \, e^3_d \,
    \left[ D_3^{(1211)}(\k) + D_J(\k) \right]
\nonumber\\ && \quad
    -
    \coeff 13 \, \beta \, e_a \, e_b
    \left[
	D_{\rm R}^{(3)}(\k)
	-
	\half \, (4\pi)^{-2}
	\left( \ln \mu^2 \lambda_{ab}^2 + \Gam ab \right)
    \right] .
\end{eqnarray}
The functions $D_3^{(2)}(\k)$, $D_3^{(211)}(\k)$, and $D_3^{(1211)}(\k)$
are given in dispersion relation
form in Eq's. (\ref{eq:D2}), (\ref{eq:d211}), and (\ref{eq:d1211}) of
Appendix \ref{required}. The same functions are also expressed in terms
of elementary functions and the Euler dilogarithm (or Spence function)
in Eq's.~(\ref{eq:D2}), (\ref{eq:D211}), and
(\ref{eq:D1211}). The function $D_J(\k)$ is not so tractable.
However, it has
been expressed in terms of a one-dimensional integral by Rajantie
\cite {rajantie}. His result is quoted in
Eq.~(\ref{eq:J}) of Appendix~\ref{required}.

The $\k\to0$ limit of the irreducible correlator is related to the
particle number fluctuations. Using the results (\ref{d20}),
(\ref{d2110}), (\ref{d12110}), and (\ref{dj0}) for
$D_3^{(2)}({\bf 0})$,
$D_3^{(211)}({\bf 0})$,
$D_3^{(1211)}({\bf 0})$
and $D_J({\bf 0})$, plus
Eq.~(\ref{D3R}) for $D_{\rm R}^{(3)}({\bf 0})$, we have
\begin {eqnarray}
    \tilde F^{(2)}_{ab}({\bf 0})
    &=&
    {1 \over 8\pi\kappa} \left[ 1 + {1 \over 2} \sum_c { \beta^2 \,
	\bar n_c \, e_c^4 \over 8\pi\kappa } \right]
    +
    {1 \over 6}
    \sum_c
    \bar n_c \,
    {\beta^2 \, \left( e_a {+} e_b \right)
    e_c^3 \over (4\pi)^2 \, \kappa^2}
\nonumber\\
&& \qquad {}
    -
    {1 \over 12}
    \sum_{c,d}
    \bar n_c \, \bar n_d \,
    {\beta^3 e_c^3 e_d^3 \over (4\pi)^2 \, \kappa^4 }
+
    {1 \over 6} \,
    {\beta e_a e_b  \over (4\pi)^2}
	\left[
       \ln \!\left({ 9 \lambda_{ab}^2 \kappa^2 \over 4 \pi } \right)
      -1  + \gamma + \Gam ab
	\right] .
\label{fab0}
\end {eqnarray}
To check this result, we note that, as is shown in Eq.~(\ref{Cab0}) of
Appendix \ref{funmeth}, there is a simpler way to obtain the same
result, namely:
\begin{equation}
\tilde C_{ab}({\bf 0}) = - { \partial \langle n_a \rangle_\beta
     \over \partial \beta \mu_b }
= - { \partial \langle n_b \rangle_\beta
     \over \partial \beta \mu_a } \,,
\end{equation}
where the partial derivatives are to be computed at fixed $\beta$. It
is a straight forward matter to take the $\beta \mu_b$ derivative of
the two-loop result (\ref{ntwolps}) for the density $\langle n_a
\rangle_\beta $ and confirm [via Eq.~(\ref{genform})] that the result
(\ref{fab0}) is indeed correct.

\goodbreak
\ifelsevier\newpage\fi

\section {Three-Loop Thermodynamics}
\label{sec:threeloop}

\begin {figure}[ht]
   \begin {center}
      \leavevmode
      \def\epsfsize #1#2{0.4#1}
      \epsfbox {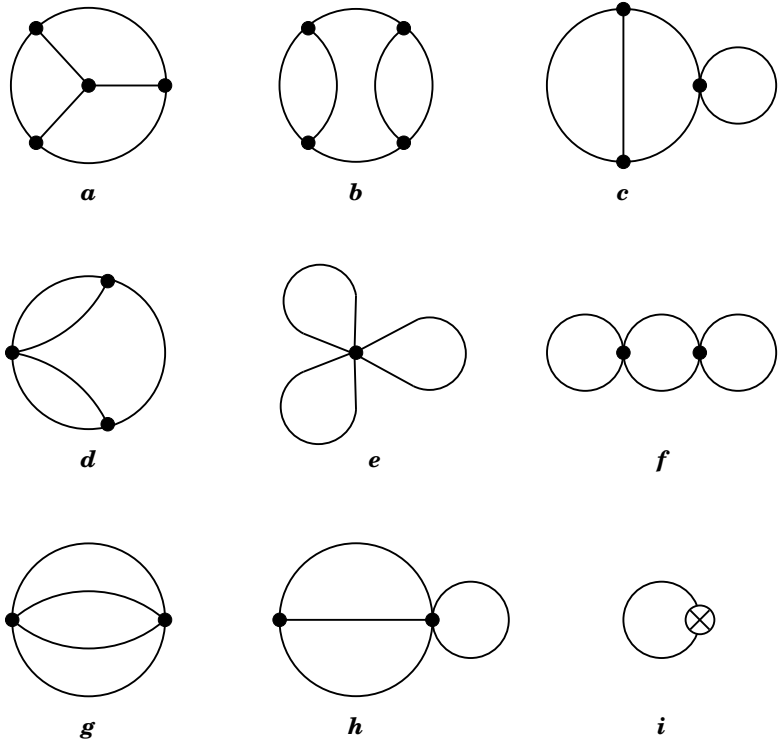}
   \end {center}
   \caption
       {%
       Three-loop diagrams contributing to the $\ln Z$.
       The circled `X' denotes the insertion of the part of
       the two-loop $g^0_{ab}$ and $h^0_a$ interactions
       which are quadratic in $\phi$.
       }%
\label {fig:P3}
\end {figure}

There are nine diagrams, shown in figure \ref {fig:P3},
which contribute to $\ln Z$ (or equivalently the pressure)
at three-loop order.
Diagrams $a$--$f$ are finite as $\nu\to3$ and may be computed
directly in 3 dimensions.
The explicit contributions of these diagrams to $(\ln Z)/\vol$
are:
\begin {eqnarray}
    \Gamma_a + \Gamma_b &=&
	\sum_{a,b,c,d}
	\beta^6 \, \n0a \, \n0b \, \n0c \, \n0d \,\,
	e_a^3 \,e_b^3 \, e_c^3 \, e_d^3 \,\,
	\left[ {1 \over 4!} \, D_M +
	  {1 \over 2^4} \, D^{(2121)}_3({\bf 0}) \right] ,
\\
    \Gamma_c + \Gamma_d &=&
	-\sum_{a,b,c}
	\beta^5 \, \n0a \, \n0b \, \n0c \,\,
	 e_a^4 \, e_b^3 \, e_c^3 \,\,
	\left[ { 1 \over 2^3 } \,D^{(211)}_3({\bf 0}) \, G_3({\bf 0})
	    + { 1 \over 2^3 } \,D^{(221)}_3({\bf 0}) \right] ,
\\
    \Gamma_e &=&
	-\sum_{a}
	\beta^3 \, \n0a \,\,
	 e_a^6 \,\, { 1 \over 2^3 \cdot 3!} \,
	G_3({\bf 0})^3
\\
    \Gamma_f + \Gamma_g &=&
	\sum_{a,b}
	\beta^4 \, \n0a \, \n0b \,\,
	 e_a^4 \, e_b^4 \,
	\left[ { 1 \over 2^4} \, D^{(2)}_3({\bf 0}) \, G_3({\bf 0})^2
       + {1 \over 2 \cdot 4!} \, D^{(4)}_\dim({\bf 0}) \right] ,
\\
    \Gamma_h &=&
	\sum_{a,b}
	\beta^4 \, \n0a \, \n0b \,\,
	e_a^5 \, e_b^3 \,\, { 1 \over 2 \cdot 3! } \,
	D^{(3)}_\dim({\bf 0})\,G_\dim({\bf 0}) \,,
\\
    \Gamma_i &=&
	\sum_{a,b}
	\beta^4
	\, g^0_{ab} \, \n0a \, \n0b
	\left( e_a {+} e_b \right)^2 \, { 1 \over 2} \,
	 G_\dim({\bf 0})
	 -
	 \sum_a
	 \beta^4 \, h^0_a \, \n0a \, e_a^2 \, \kappa_0^2 \, G_\dim({\bf 0})
	 \,.
\label {eq:Gami}
\end {eqnarray}
The last term of Eq.~(\ref {eq:Gami})
comes from
\begin {equation}
    \beta \avg {\nabla \phi(\r)^2}^{(0)}
    =
    -\kappa_0^2 \, G_\dim({\bf 0})
    \mathop{\longrightarrow}_{\dim\to3}
    {\kappa_0^3 \over 4\pi} \,,
\label {eq:gulp}
\end {equation}
which easily follows using the same method
that lead to Eq.~(\ref {perf}).
The integral
\begin {eqnarray}
    D_M &\equiv&
	\int (d^3\r)(d^3\r')(d^3\r'') \>
	G_3(\r) \,
	G_3(\r') \,
	G_3(\r'') \,
	G_3(\r{-}\r') \,
	G_3(\r'{-}\r'') \,
	G_3(\r''{-}\r)
\end {eqnarray}%
corresponds to the ``Mercedes'' graph $a$ of
Fig.~\ref{fig:P3}, and the other
required integrals are defined in Eq's.~(\ref {Dn})--(\ref{Dnlmq}).
These integrals are evaluated in Appendix \ref {debye ints} (with help
from Refs.~\cite {rajantie,broadhurst}).  Once again, the
final contribution $\Gamma_i$ involving the single two-loop coupling
$g^0_{ab}$ removes the short-distance singularities in both the
three-loop graphs $\Gamma_g$ and $\Gamma_h$ that diverge in three
dimensions. Note that this cancelation of divergences involves the
detailed structure of the induced coupling interaction with its
exponential dependence upon the potential $\phi$.
As discussed earlier, the
basic (``primitive'') divergences which the induced couplings must cancel
appear only in even loop order.
The subsidiary divergences at this three-loop order are canceled by
the non-trivial potential dependence of the two-loop induced coupling.
Inserting the explicit results for these integrals produces
the physical $\dim\to3$ limit of
$\ln Z$ to three loop order:
\begin {eqnarray}
    {\ln Z_3 \over \vol}
    =
    {\ln Z_2 \over \vol}
    &+&
    {1 \over 4! \, (4\pi)^3} \,
    \Biggl\{ \,
	\sum_{a,b,c,d} \beta^6 \, \n0a \, \n0b \, \n0c \, \n0d \,
	\,  e_a^3 \, e_b^3 \, e_c^3 \, e_d^3 \,\, {C_1 \over \kappa_0^3}
    \nonumber\\ &&\quad\qquad {}
	+
	\sum_{a,b,c} \beta^5 \, \n0a \, \n0b \, \n0c \,
	\, e_a^4 \, e_b^3 \, e_c^3 \,\, {C_2 \over \kappa_0}
    \nonumber\\ &&\quad\qquad {}
	+
	\sum_{a} \beta^3 \, \n0a \,\,  e_a^6 \,\, \kappa_0^3 \,
	\left[
	    C_3
	    +
	    { 8\pi \lambda_{a}^2 \over \beta^2 \, e_a^4}
	\right]
    \nonumber\\ &&\quad\qquad {}
	+
	\sum_{a,b} \beta^4 \, \n0a \, \n0b \,\, e_a^4 \, e_b^4 \,\,
	\kappa_0
	\left[ C_4 +
	    \ln \!\left({ 9 \lambda_{ab}^2 \bar\kappa^2 \over 4 \pi} \right)
	    + \Gam ab
	\right]
    \nonumber\\ &&\quad\qquad {}
	+
	\sum_{a,b} \beta^4 \, \n0a \, \n0b \,\, e_a^5 \, e_b^3 \,\,
	\kappa_0
	\left[ C_5 +
	     \ln \!\left( {9\lambda_{ab}^2 \bar\kappa^2 \over 4\pi} \right)
	    + \Gam ab
	\right]
    \Biggr\},
\label {eq:lnZ3}
\end {eqnarray}
with
\begin {eqnarray}
	C_1 &=&
	    C_M
	    + \coeff 34 \, \Li_2(-\coeff 13)
	    + \ln \coeff 34
	    + \coeff{\pi^2}{16}
	=
	    0.119131 \cdots
	\,,
\\
	C_2 &=&
	    -3 \, \Li_2(-\coeff 13)
	    + \half
	    - \coeff{\pi^2}{4}
	=
	    -1.04030 \cdots
	\,,
\label{eq:C2}
\\
	C_3 &=& \half
	\,,
\label{half}
\\
	C_4 &=& \gamma - \coeff 94 -2 \ln \coeff 34
	=
	    -1.09742 \cdots
	\,,
\\
	C_5 &=& \gamma -1
	=
	    -0.422784 \cdots
	\,,
\end {eqnarray}%
where
\begin {equation}
    \Li_2(-z) \equiv -\int_0^z {dt\over t} \> \ln(1+t) \,.
\end {equation}
is Euler's dilogarithm, $C_M$ is given in Eq.~(\ref {eq:C_M}),
and $\Gamma_{ab}$ was defined in Eq.~(\ref{eq:Gam}).%
\footnote{The closed form expression (\ref{eq:C2}) for $C_2$
involving the Euler dilogarithm was obtained previously by
Kahlbaum \cite{kahlbaum}.}

Differentiating Eq.~(\ref {eq:lnZ3}) with respect to $\beta$ yields
the internal energy, in terms of bare parameters, to three loop order:
\begin {eqnarray}
    \beta u &=&
    \sum_{a=1}^A \n0a \>
    \Biggl\{
	{3\over2}
	+
	{1\over4} \left( {\beta e_a^2 \kappa_0 \over 4\pi}\right)
	-
	{C_3 \over 32} \left( {\beta e_a^2 \kappa_0 \over 4\pi}\right)^3
	+
	{1 \over 48}
	\left( {\kappa_0^2 \lambda_a^2 \over 4\pi} \right)
	\left( {\beta e_a^2 \kappa_0 \over 4\pi}\right)
    \Biggr\}
    \nonumber\\ &-&
    \sum_{a,b=1}^A \n0a \, \n0b
    \left\{
	{1 \over 48}
	\left({\beta^3 e_a^3 e_b^3 \over (4\pi)^2}\right)
	\left[ 1 + \eta_{ab} \Gamprime ab \right]
    \right.
    \nonumber\\ && \kern 0.7in
    \left. {}
	+
	{1 \over 32}
	\left({\beta^4 e_a^5 e_b^3 \kappa_0 \over (4\pi)^3}\right)
	\left[
	    C_5 +\coeff 23
	    +\ln \!\left( {9 \lambda_{ab}^2 \bar\kappa^2 \over 4\pi} \right)
	    + \Gam ab + \coeff 23 \eta_{ab} \Gamprime ab
	\right]
    \right.
    \nonumber\\ && \kern 0.7in
    \left. {}
	+
	{1 \over 32}
	\left({\beta^4 e_a^4 e_b^4 \kappa_0 \over (4\pi)^3}\right)
	\left[
	    C_4 +\coeff 23
	    +\ln \!\left( {9 \lambda_{ab}^2 \bar\kappa^2 \over 4 \pi }\right)
	    + \Gam ab + \coeff 23 \eta_{ab} \Gamprime ab
	\right]
    \right\}
    \nonumber\\ &-&
    \sum_{a,b,c=1}^A \n0a \, \n0b \, \n0c
    \left\{
	{C_2 \over 32}
	\left({\beta^5 e_a^4 e_b^3 e_c^3 \over \kappa_0 \, (4\pi)^3}\right)
    \right\}
    \nonumber\\ &-&
    \sum_{a,b,c,d=1}^A \n0a \, \n0b \, \n0c \, \n0d
    \left\{
	{C_1 \over 32}
	\left(
	    {\beta^6 e_a^3 e_b^3 e_c^3 e_d^3 \over \kappa_0^3 \, (4\pi)^3}
	\right)
    \right\},
\end {eqnarray}
while differentiating with respect to the chemical potentials yields
the particle densities (in terms of bare parameters) to three loop
order:%
\footnote
    {%
    Graphs with $n$ ``clover leafs'' produce a factor of
    $[G_3({\bf 0})]^n$. The first graph shown in Fig.~\ref{fig:P2} is a
    two-loop clover leaf graph; graph $e$ of Fig.~\ref{fig:P3} is a
    three-loop clover leaf graph, and it yields the value $C_3 = \half$
    given in Eq.~(\ref{half}). Formula (\ref{twelve}) shows that these
    graphs form part of the generic density correction factor $\exp \{
    \beta e^2 \kappa_0 / 8 \pi \}$, extending the result quoted in
    Eq.~(\ref{expp}).
    }
\begin {eqnarray}
\bar n_s =
    \expect {n_s} &=&
    \n0s
    \left[
	1
	+  \left({ \beta e_s^2 \kappa_0 \over 8 \pi } \right)
	+ {1 \over 2} \left({ \beta e_s^2 \kappa_0 \over 8 \pi } \right)^2
	+ {2 C_3 \over 3!} \left({ \beta e_s^2 \kappa_0 \over 8 \pi } \right)^3
	+ {1 \over 12} \left( {\beta e_s^2 \kappa_0 \over 4\pi}\right)
			\left( {\kappa_0^2 \lambda_s^2 \over 4\pi} \right)
    \right.
\nonumber\\ && {}
    +
    \sum_{a=1}^A \n0a
    \left\{
	{1 \over 8}
	\left( {\beta^3 e_s^2 e_a^4 \over (4 \pi)^2 } \right)
	+
	{1 \over 12}
	\left({ \beta^3 e_s^3 e_a^3  \over (4 \pi)^2 } \right)
	\left[
	     \gamma -1
	     + \ln\!\left( \coeff 94 \lambda_{as}^2 \kappa_0^2 / \pi \right)
	     + \Gam as
	\right]
    \right.
\nonumber\\ && \kern 1.5in {}
	+
	{C_3 \over 16}
	\left( {\beta^4 e_s^2 e_a^6 \kappa_0 \over (4 \pi)^3 } \right)
	+
	{\kappa_0 \over 8}
	\left( {\beta^2 e_s^2 e_a^2 \lambda_a^2 \over (4\pi)^2}\right)
\nonumber\\ && \kern 1.5in {}
	+
	{\kappa_0\over 12}
	\left( {\beta^4 e_s^4 e_a^4 \over (4 \pi)^3 } \right)
	\left[
	    C_4
	    +\ln \!\left( \coeff 94 \lambda_{as}^2 \bar\kappa^2 / \pi \right)
	    + \Gam as
	\right]
\nonumber\\ && \kern 1.5in
    \left. {}
	+
	{\kappa_0\over 24}
	\left( {\beta^4 [e_s^3 e_a^5 {+} e_s^5 e_a^3] \over (4 \pi)^3 } \right)
	\left[
	    C_5
	    + \ln \!\left( \coeff 94 \lambda_{as}^2 \bar\kappa^2 / \pi \right)
	    + \Gam as
	\right]
    \right\}
\nonumber\\ && {}
    +
    \sum_{a,b=1}^A \n0a \n0b
    \left\{
	{1 \over 24}
	\left( {\beta^4 e_s^2 e_a^3 e_b^3 \over \kappa_0^2\, (4\pi)^2}\right)
	+
	{C_2 \over 24}
	\left( {\beta^5 e_s^4 e_a^3 e_b^3 \over \kappa_0 \, (4\pi)^3}\right)
	+ {C_2 \over 12}
	\left( {\beta^5 e_s^3 e_a^4 e_b^3 \over \kappa_0 \, (4\pi)^3}\right)
    \right.
\nonumber\\ && \kern1.5in {}
	+ {1 \over 48}
	\left( {\beta^5 e_s^2 e_a^5 e_b^3 \over \kappa_0 \, (4\pi)^3}\right)
	\left[
	    C_5 + 2
	    + \ln \!\left( \coeff 94 \lambda_{ab}^2 \bar\kappa^2 / \pi \right)
	    + \Gam ab
	\right]
\nonumber\\ && \kern1.5in
    \left. {}
	+ {1 \over 48}
	\left( {\beta^5 e_s^2 e_a^4 e_b^4 \over \kappa_0 \, (4\pi)^3}\right)
	\left[
	    C_4 + 2
	    + \ln \!\left( \coeff 94 \lambda_{ab}^2 \bar\kappa^2 / \pi \right)
	    + \Gam ab
	\right]
    \right\}
\nonumber\\ && {}
    -
    \sum_{a,b,c=1}^A \n0a \, \n0b \, \n0c
    \left\{
    {C_2\over 48}
    \left( {\beta^6 e_s^2 e_a^4 e_b^3 e_c^3 \over \kappa_0^3\,(4\pi)^3} \right)
    - {C_1\over 6}
    \left( {\beta^6 e_s^3 e_a^3 e_b^3 e_c^3 \over \kappa_0^3\,(4\pi)^3} \right)
    \right\}
\nonumber\\ &&
    \left. {}
    -
    \sum_{a,b,c,d=1}^A \n0a \, \n0b \, \n0c \, \n0d
    \left\{
    {C_1\over16}
    \left(
	{\beta^7 e_s^2 e_a^3 e_b^3 e_c^3 e_d^3 \over \kappa_0^5 \,(4\pi)^3}
    \right)
    \right\}
    \right] .
\label{twelve}
\end {eqnarray}

Inverting the relation between physical and bare densities,
and inserting the explicit values for $C_3$, $C_4$ and $C_5$
(because this simplifies the subsequent results), yields
\begin {eqnarray}
   \n0s &=&
    \bar n_s
    \left[
	1
	-  \left({ \beta e_s^2 \bar\kappa \over 8 \pi } \right)
	+ {1 \over 2} \left({ \beta e_s^2 \bar\kappa \over 8 \pi } \right)^2
	- {1 \over 3!} \left({ \beta e_s^2 \bar\kappa \over 8 \pi } \right)^3
	- {1 \over 6} \left({ \beta e_s^2 \bar\kappa \over 8\pi} \right)
		    \left({\bar\kappa^2 \lambda_s^2 \over 4\pi}\right)
    \right.
\nonumber\\ && {}
    -
    \sum_{a=1}^A \bar n_a
    \left\{
	{1 \over 12}
	\left({ \beta^3 e_s^3 e_a^3  \over (4 \pi)^2 } \right)
	\left[
	     \gamma -1 + \ln( {\coeff 94 \lambda_{as}^2 \bar\kappa^2 / \pi})
	     + \Gam as
	\right]
    \right.
\nonumber\\ && \;\quad\qquad \left.{}
	+
	{\bar\kappa\over 12}
	\left( {\beta^4 e_s^4 e_a^4 \over (4 \pi)^3 } \right)
	\left[
	    \gamma -3
	    + \ln \!\left( 4 \lambda_{as}^2 \bar\kappa^2 / \pi \right)
	    + \Gam as
	\right]
	\right.
\nonumber\\ && \;\quad\qquad \left.{}
	-
	{\bar\kappa\over 24}
	\left( {\beta^4 e_s^5 e_a^3 \over (4 \pi)^3 } \right)
	\left[ \gamma -1
		+ \ln\!\left(\coeff 94 \lambda_{as}^2\bar\kappa^2/\pi \right)
		+ \Gam as
	\right]
	+ {\bar\kappa \over 8}
	\left( {\beta^2 e_s^2 e_a^2 \lambda_a^2 \over (4\pi)^2} \right)
    \right\}
\nonumber\\ && {}
    -
    \sum_{a,b=1}^A \bar n_a \bar n_b
    \left\{
	{1 \over 24}
	\left( {\beta^4 e_s^2 e_a^3 e_b^3 \over \bar\kappa^2\, (4\pi)^2}\right)
	+
	{C_2{-}1 \over 24}
	\left( {\beta^5 e_s^4 e_a^3 e_b^3 \over \bar\kappa \, (4\pi)^3}\right)
	+ {C_2{-}\half \over 12}
	\left( {\beta^5 e_s^3 e_a^4 e_b^3 \over \bar\kappa \, (4\pi)^3}\right)
    \right.
\nonumber\\ && \;\;\qquad\qquad
    \left. {}
	+ {1 \over 48}
	\left( {\beta^5 e_s^2 e_a^4 e_b^4 \over \bar\kappa \, (4\pi)^3}\right)
	\left[ \gamma -1
	    + \ln \!\left( 4 \lambda_{ab}^2 \bar\kappa^2 / \pi \right)
	    + \Gam ab
	\right]
    \right\}
\nonumber\\ && {}
    +
    \sum_{a,b,c=1}^A \bar n_a \, \bar n_b \, \bar n_c
    \left\{
    {C_2{-}\half\over 48}
    \left({\beta^6 e_s^2 e_a^4 e_b^3 e_c^3 \over \bar\kappa^3\,(4\pi)^3} \right)
    - {C_1\over 6}
    \left({\beta^6 e_s^3 e_a^3 e_b^3 e_c^3 \over \bar\kappa^3\,(4\pi)^3} \right)
    \right\}
\nonumber\\ &&
    \left. {}
    +
    \sum_{a,b,c,d=1}^A \bar n_a \, \bar n_b \, \bar n_c \, \bar n_d
    \left\{
    {C_1\over16}
    \left(
	{\beta^7 e_s^2 e_a^3 e_b^3 e_c^3 e_d^3 \over \bar\kappa^5 \,(4\pi)^3}
    \right)
    \right\}
    \right] .
\end {eqnarray}
Using this result to express the pressure in terms of
physical densities gives the equation of state%
\footnote
    {
    In footnote \ref{happy} it
    was remarked that a field redefinition could be performed that removes
    the $\nabla\mu_a \cdot \nabla\phi $ cross term in the $h^0_c$
    interaction which contributes to the reducible part of the
    density-density correlator.
    For thermodynamic quantities, it is the $ (\nabla\phi)^2 $ term in the
    $ h_a^0 $ coupling of the induced interactions (\ref{Sind}) that
    contributes, since the chemical potentials are now constants.
    To independently check the $h^0_a$ contributions to the pressure
    or free energy, one may
    follow the logic of footnote \ref{happy} in a slightly different way.
    To leading order in $h^0_a$, the $(\nabla\phi)^2$ part of this
    interaction is removed by the field redefinition
    $
	\phi \to \phi - \beta^2 \, {\sum}_b i \, e_b \, h^0_b \, n^0_b
	e^{i\beta e_b\phi} \,.
    $
    The effect of this redefinition on the classical interaction,
    again to leading order in $ h^0_a$,
    is equivalent to the alteration of the induced coupling
    $
	g^0_{ab} \to g^0_{ab} - {1\over2} \, e_a e_b \, ( h^0_a + h^0_b ) \,.
    $
    In view of the evaluation (\ref{eq:h0a}) of $h^0_a$, and the relation
    (\ref{eq:gab}) between $g_{ab}$ and $\Gamma_{ab}$, this substitution is
    equivalent to the change
    $$
	{ \pi \over 6} \left( { e_a e_b \over 4\pi } \right)^3 \Gamma_{ab}
	\longrightarrow
	{ \pi \over 6} \left( { e_a e_b \over 4\pi } \right)^3 \Gamma_{ab}
	+
	{1 \over 2}
	\left({ e_a e_b \over 48 \pi \beta^2 }\right)
	\left( \lambda^2_a + \lambda^2_b \right) \,.
    $$
    Due to the charge neutrality condition $\sum e_a \bar n_a = 0$,
    this change has no effect at two loop order.
    It is easy to check that the
    three loop terms in the equation of state (\ref{eq:p3}) and the
    Helmholtz free energy (\ref{eq:h3}) are in agreement with the
    corrections produced by this redefinition.
    It is worth noting that the effect of the $h^0_a$ interaction for
    the original partition function written in terms of the bare densities
    $n^0_a$ is {\em not} produced by the change given above.
    The field redefinition changes the dependence of the partition
    function on the bare densities, and also changes the relation between
    physical and bare densities.
    However, when these modified results are re-expressed in terms of
    the physical densities, the same physical equation of state emerges,
    as it must.
    }
\begin {eqnarray}
    \beta p =
    {\ln Z \over \vol}
    &=&
    \sum_{a=1}^A \bar n_a
    \left\{
	1
	- {1 \over 6} \left({\beta e_a^2 \bar\kappa \over 4\pi}\right)
	- {1 \over 8}
	\left( {\beta e_a^2 \bar\kappa \over 4\pi}\right)
	\left( {\bar\kappa^2 \lambda_a^2 \over 4\pi} \right)
    \right\}
    \nonumber\\ &-&
    \sum_{a,b=1}^A \bar n_a \, \bar n_b
    \left\{
	{1 \over 24}
	\left({\beta^3 e_a^3 e_b^3 \over (4\pi)^2}\right)
	\left[
	    \gamma
	    + \ln\!\left(\coeff 94 \lambda_{ab}^2\bar\kappa^2/\pi \right)
	    + \Gam ab
	\right]
    \right.
    \nonumber\\ && \;\;\quad\qquad
    \left.{}
	+
	{1 \over 16}
	\left({\beta^4 e_a^4 e_b^4 \bar\kappa \over (4\pi)^3}\right)
	\left[
	    \gamma -\coeff 73
	    + \ln \!\left( 4 \lambda_{ab}^2 \bar\kappa^2 / \pi \right)
	    + \Gam ab
	\right]
    \right\}
    \nonumber\\ &-&
    \sum_{a,b,c=1}^A \bar n_a \, \bar n_b \, \bar n_c
    \left\{
	{1 \over 16}
	\left({\beta^5 e_a^4 e_b^3 e_c^3 \over \bar\kappa \, (4\pi)^3}\right)
	\left[ C_2 - \half \right]
    \right\}
    \nonumber\\ &-&
    \sum_{a,b,c,d=1}^A \bar n_a \, \bar n_b \, \bar n_c \, \bar n_d
    \left\{
	{1 \over 16}
	\left(
	    {\beta^6 e_a^3 e_b^3 e_c^3 e_d^3 \over \bar\kappa^3 \, (4\pi)^3}
	\right)
	\left[ \, C_1 \right]
    \right\} .
\label {eq:p3}
\end {eqnarray}
The internal energy expressed in terms of the physical
densities is given by
\begin {eqnarray}
    \beta u &=&
    \sum_{a=1}^A \bar n_a
    \left\{
	{3\over 2}
	- {1 \over 2} \left({\beta e_a^2 \bar\kappa \over 4\pi}\right)
	- {7 \over 24}
	\left( {\beta e_a^2 \bar\kappa \over 4\pi}\right)
	\left( {\bar\kappa^2 \lambda_a^2 \over 4\pi} \right)
    \right\}
    \nonumber\\ &-&
    \sum_{a,b=1}^A \bar n_a \, \bar n_b
    \left\{
	{1 \over 8}
	\left({\beta^3 e_a^3 e_b^3 \over (4\pi)^2}\right)
	\left[
	    \gamma - \coeff 13
	    + \ln(\coeff 94 \lambda_{ab}^2\bar\kappa^2/\pi)
	    + \Gam ab + \coeff 16 \eta_{ab} \Gamprime ab
	\right]
    \right.
    \nonumber\\ && \;\quad\qquad
    \left.{}
	+
	{3 \over 16}
	\left({\beta^4 e_a^4 e_b^4 \, \bar\kappa \over (4\pi)^3}\right)
	\left[
	    \gamma -\coeff {23}{9}
	    + \ln \!\left( 4 \lambda_{ab}^2 \bar\kappa^2 / \pi \right)
	    + \Gam ab + \coeff 19 \eta_{ab} \Gamprime ab
	\right]
    \right\}
    \nonumber\\ &-&
    \sum_{a,b,c=1}^A \bar n_a \, \bar n_b \, \bar n_c
    \left\{
	{3 \over 16}
	\left({\beta^5 e_a^4 e_b^3 e_c^3 \over \bar\kappa \, (4\pi)^3}\right)
	\left[ C_2 - \half \right]
    \right\}
    \nonumber\\ &-&
    \sum_{a,b,c,d=1}^A \bar n_a \, \bar n_b \, \bar n_c \, \bar n_d
    \left\{
	{3 \over 16}
	\left(
	    {\beta^6 e_a^3 e_b^3 e_c^3 e_d^3 \over \bar\kappa^3 \, (4\pi)^3}
	\right)
	\left[ \, C_1 \right]
    \right\},
\end {eqnarray}
while the Helmholtz free energy is
\begin {eqnarray}
    \beta f &=&
    \sum_{a=1}^A \bar n_a
    \left\{
	-1 + \ln (\bar n_a \lambda_a^3/g_a)
	- {1 \over 3} \left({\beta e_a^2 \bar\kappa \over 4\pi}\right)
	- {1 \over 12}
	\left( {\beta e_a^2 \bar\kappa \over 4\pi}\right)
	\left( {\bar\kappa^2 \lambda_a^2 \over 4\pi} \right)
    \right\}
    \nonumber\\ &-&
    \sum_{a,b=1}^A \bar n_a \, \bar n_b
    \left\{
	{1 \over 24}
	\left({\beta^3 e_a^3 e_b^3 \over (4\pi)^2}\right)
	\left[
	    \gamma - 1
	    + \ln(\coeff 94 \lambda_{ab}^2\bar\kappa^2/\pi)
	    + \Gam ab
	\right]
    \right.
    \nonumber\\ && \;\;\quad\qquad
    \left.{}
	+
	{1 \over 24}
	\left({\beta^4 e_a^4 e_b^4 \, \bar\kappa \over (4\pi)^3}\right)
	\left[
	    \gamma -3
	    + \ln \!\left( 4 \lambda_{ab}^2 \bar\kappa^2 / \pi \right)
	    + \Gam ab
	\right]
    \right\}
    \nonumber\\ &-&
    \sum_{a,b,c=1}^A \bar n_a \, \bar n_b \, \bar n_c
    \left\{
	{1 \over 24}
	\left({\beta^5 e_a^4 e_b^3 e_c^3 \over \bar\kappa \, (4\pi)^3}\right)
	\left[ C_2 - \half \right]
    \right\}
    \nonumber\\ &-&
    \sum_{a,b,c,d=1}^A \bar n_a \, \bar n_b \, \bar n_c \, \bar n_d
    \left\{
	{1 \over 24}
	\left(
	    {\beta^6 e_a^3 e_b^3 e_c^3 e_d^3 \over \bar\kappa^3 \, (4\pi)^3}
	\right)
	\left[ \, C_1 \right]
    \right\}.
\label{eq:h3}
\end {eqnarray}
This result agrees with the corresponding result in
Alastuey and Perez \cite{AP}.%
\footnote
    {
    In particular, we agree with the result for the Helmholtz free
    energy given in Eq.~(7.3) of Ref.~\cite{AP}.
    Their result involves two constants, also called $C_1$ and $C_2$.
    The relations between their and our parameters reads
    $ C_{1,\rm AP} = - \pi^2 [ C_2 - \half] $, and
    $ C_{2,\rm AP} = - 4\pi^3 C_1$.
    Using our numbers gives $C_{1,\rm AP} = 15.2021$,
    to be compared with their value $ 15.201 \pm 0.001$,
    and $C_{2,\rm AP} = -14.7752 $, to be compared to their
    $ -14.734 \pm 0.001 $.
    Also, as will be shown explicitly in the next section,
    the error estimate given at the end of their equation,
    which reads $O(n^3\ln n)$ in our notation, should actually be
    $O(n^3 \ln^2 n) $.
    }
It also agrees,
except for some small
misprints and the omission of one quantum-mechanical term,
with the result to be found in Ref.~\cite {book}.%
\footnote
    {
    The last term in the first line of our Eq.~(\ref{eq:h3})
    is often referred to as a ``quantum diffraction term''.
    It is missing from Eq.~(2.52) of Ref.~\cite {book}.
    In addition,
    the coefficient of the term involving $(1 - \ln \coeff 43)$
    should be $1/6$, not $\pi/3$.
    [Finally, there is a typographical error in the free energy
    for a non-interacting gas, which lacks the spin degeneracy factor
    inside the logarithm appearing in Eq.~(2.50).]
    \label {fn:misprint}
    }
Note that the pressure
is related to the Helmholtz free energy
by the thermodynamic identity
$p = -\left. \partial F / \partial \vol \right|_{T,N}$.
It is easy to apply this identity and verify that Eq.~(\ref {eq:p3})
follows from (\ref {eq:h3}).

As explained in the next section, the combination
\begin {equation}
A =  \beta (3 p - u)
- \coeff 32 \sum_{a=1}^A  \bar n_a
\end{equation}
would vanish identically if the plasma could be treated entirely in a classical
manner. However, the results above give
\begin{eqnarray}
A = -
    { 1 \over (4\pi)^2}
    \left\{
    {1 \over 24}
    \sum_{a,b=1}^A \bar n_a \, \bar n_b \,
    \beta^3 e_a^3 e_b^3
    \left[ 1 - \half \, \eta_{ab} \, \Gamprime ab \right]
    \left[ 1 + { \beta \, e_a e_b \, \bar\kappa \over 4 \pi} \right]
    +
    {1 \over 12}
    \sum_{a=1}^A \bar n_a \, \beta \, e_a^2 \, \bar\kappa^3 \, \lambda_a^2
    \right\}
    .
\nonumber\\
\label{eq:3lanom}
\end {eqnarray}
The next section shows how $A$ may be independently computed from quantum
corrections to the virial theorem for a classical Coulomb plasma,
and discusses how $A$ plays a role analogous to the anomalies which appear in
relativistic quantum field theory.

\subsection {Binary Plasma}

These expressions simplify considerably for a two-component plasma
such as an electron-proton plasma where $-e_e = e_p \equiv e$, and charge
neutrality requires that $ \bar n_e = \bar n_p \equiv n/2$. For example,
the three-loop equation of state becomes
\begin{eqnarray}
    { \beta \, p \over n}
    =
    1 &-& { 1 \over 3}
    \left( { \beta e^2 \bar\kappa \over 8 \, \pi } \right)
    \left\{
	1
	+
	{3 \over 8} \,
	{\bar\kappa^2 (\lambda_e^2 {+} \lambda_p^2)\over 4\pi}
    \right\}
\nonumber\\
    &-&  { 1 \over 24}
    \left( { \beta e^2 \bar\kappa \over 8 \, \pi } \right)^2
    \left\{ \ln \Biggl[ { 4 \, m_p \, m_e \over ( m_p {+} m_e)^2 } \Biggr]
    + \Gam ee + \Gam pp - 2 \, \Gam ep \right\}
\nonumber\\
    &-&  { 1 \over 8}
    \left( { \beta e^2 \bar\kappa \over 8 \, \pi } \right)^3
    \Biggl\{ 4 \gamma - \coeff {28}3 +
    2 \ln \left[ \lambda_{ee} \lambda_{pp}
    \lambda_{ep}^2 \left( 4 \bar\kappa^2 / \pi \right)^2 \right]
    + \Gam ee + \Gam pp + 2 \, \Gam ep \Biggr\} \,.
\label{eq:2comp}
\end{eqnarray}
This expression simplifies a bit more if the very small
electron/proton mass ratio is neglected, which is to say that the
formal $m_p \to \infty$ limit is taken.  In this limit,
Eq's.~(\ref{strep}), (\ref{stdec}), (\ref{eq:Gam}), and
(\ref{eq:etaab}) yield:
\begin{eqnarray}
 \ln\left[ { 4 m_p m_e \over (m_p + m_e )^2 } \right] + \Gamma_{pp}
	&\mathop{\longrightarrow}\limits_{m_p\to\infty}&
	\ln\left[ 8\pi \left( { \beta e^2 \over 4\pi \lambda_e}
	\right)^2 \right] + 3 \gamma - { 8 \over 3} \,,
\\
\noalign{\hbox {and}}
\nonumber\\
	 2 \ln \lambda_{pp} \bar\kappa + \Gamma_{pp}
	&\mathop{\longrightarrow}\limits_{m_p\to\infty}&
	\ln\left[ 4\pi \left( { \beta e^2 \bar\kappa \over 4 \pi}
	\right)^2 \right] + 3 \gamma - { 8 \over 3} \,,
\end{eqnarray}
so that
\begin{eqnarray}
    \lim_{m_p\to\infty} { \beta \, p \over n}
    =
    1 - { g \over 6}
    &-&
    { g^2 \over 96}
    \left\{
    \ln \!\left[ 8\pi \!\left({\beta e^2 \over 4\pi \lambda_e }\right)^2 \right]
	+ 3 \gamma - {8 \over 3} + \Gam ee - 2 \, \Gam ep
    \right\}
\nonumber\\
    &-&  { g^3 \over 64} \,
    \Biggl\{
    \ln \!\left[
	{g^2 \over 2 \pi^3}
	\left({4 \bar\kappa \lambda_e}\right)^6
     \right]
    + 7 \gamma - 12
    + \Gam ee + 2 \, \Gam ep
    + {16 \pi \lambda_e^2 \over (\beta e^2)^2}
    \Biggr\} \,,
\end{eqnarray}
where $g = \beta e^2 \bar\kappa / (4\pi)$ is the dimensionless
Coulomb coupling parameter.
We note that the proton mass $m_p$ disappears and this limit is
well-behaved.

\subsection {One-Component Plasma}

\def\B{{\hbox {\tiny B}}}
Another special case is the ``jellium'' model,
in which a single
charged particle species moves in the presence of a neutralizing,
uniform background charge density.
This is the one-component plasma (OCP) which is much discussed in
the literature.  It may be obtained by
taking a limit of a plasma containing two species: one of charge $e$,
number density $n$, and mass $m$;
the other `spectator' species of charge $e_\B\equiv -z e$,
number density $n_\B\equiv n/z$, and mass $m_\B$, with $z \to 0$.
The charge of each spectator particle becomes vanishingly small,
but their density diverges, so as to preserve total charge neutrality.
The net result (for static equilibrium properties)
is that the spectator particles act like an smooth inert background charge
density.
The ideal gas pressure of the spectator particles
diverges as $z\to0$,
and must be subtracted from the total pressure before sending $z$ to zero.
If the
background, spectator particles are not taken to have a very large
mass, $m_\B \to \infty$, then they will also make quantum, exchange
contributions to the pressure.
To the three-loop order to which we compute,
these unwanted exchange contributions,
in the $z\to0$ limit, are given by
\begin{eqnarray}
    p^{\rm ex}_\B
    &=&
    \mp e_\B^2 \, n_\B^2
    \left[
	{1 \over 24} \left( { \beta {e_\B}^2 \over 4 \pi } \right)^2
	+
	{\bar\kappa \over 16} \left( { \beta {e_\B}^2 \over 4 \pi } \right)^3
    \right]
    { 1 \over g_\B} \, \tilde f(\eta_\B)
\nonumber\\
    &\to&
    \mp {n^2 \over 16\, g_\B}
    \left[
	{1 \over \beta z^2}
	\left(
	    {4 \pi \beta \hbar^2 \over m_\B}
	\right)^{3/2}
	-
	{4 e^2}
	\left( { \beta \hbar^2 \over m_\B }\right)
	+
	{3 \sqrt\pi} \, e^2 \bar\kappa
	\left({ \beta \hbar^2 \over m_\B }\right)^{3/2}
    \right] ,
\end{eqnarray}
where $g_\B$ is the spin degeneracy of the spectator particles.
These terms are also to be subtracted from the total pressure.
The resulting one-component equation of state,
to three-loop order, is given by
\begin{eqnarray}
   \left. { \beta \, p \over n} \right|_{\rm OCP}
    =
    1 - { g \over 6}
    &-&
    { g^2 \over 24}
    \left\{ \gamma + \ln \Biggl[ { 9 \kappa^2 \lambda_e^2 \over 2\pi } \Biggr]
    + \Gam ee(\eta_{ee})
    \right\}
\nonumber\\
    &-&  { g^3 \over 16} \,
    \Biggl\{ \gamma +
    \ln \left[ {8 \kappa^2 \lambda_{e}^2 \over \pi} \right]
    + \Gam ee(\eta_{ee})
    + {1 \over 4\pi \eta_{ee}^2}
    - \coeff {17}6 + C_1 + C_2
    \Biggr\} \,.
\label{eq:jelly}
\end{eqnarray}
Here $\kappa = \left( n \beta e^2 \right)^{1/2}$ is the Debye wavelength
due to the single charge species.
Once again,
we have written
the result in terms of ascending powers of
the {\em dimensionless} parameter $g = \beta e^2 \kappa / (4\pi)$
which characterizes the strength of Coulomb interactions in the plasma.
On the other hand, our result entails
no restriction on the size of the quantum parameter
\begin{equation}
\eta_{ee} = { \beta e^2 \over 4 \pi \lambda_{ee} } \,,
\end{equation}
which, together with $\lambda^2_{ee} = 2 \lambda_e$,
has been used to re-express
the order $g^3 \lambda_e^2$ term in terms  of $ 1 / \eta_{ee}^2$.

An often treated special case of the one-component plasma is its
classical limit. As already alluded to in footnotes \ref{foot} and
\ref{ffoot}, in this limit the Boltzmann factor with the repulsive
potential provides damping at the Coulomb distance $d_C = \beta e^2 /
4 \pi$, and the quantum-mechanical fluctuations are not required to
obtain a finite theory. (And, moreover, $d_C$ is the correct, physical
cutoff if $d_C > \lambda_e$.) The $\hbar \to 0$ limit takes $\eta_{ee}
\to \infty$, and Eq's.~(\ref{strep}), (\ref{stdec}), and
(\ref{eq:Gam}) give
\begin{equation}
\Gamma_{ee}(\eta_{ee}) = \ln\left( 4\pi \eta_{ee}^2 \right)
	+ 3 \gamma - { 8 \over 3} - { 1 \over 4\pi \eta_{ee}^2 }
			+ O(1 / \eta_{ee}^4) \,.
\end{equation}
Thus, we see explicitly that the short-distance cut off in the
logarithm now involves
\begin{equation}
\lambda_e^2 \eta_{ee}^2 = {1\over2}\left( { \beta e^2 \over 4\pi} \right)^2
 = {1\over2} \, d_C^2 \,,
\label{classic}
\end{equation}
and so, including the $O(\hbar^2)$ corrections which come from the
$ 1 / \eta_{ee}^2$ terms, we find that
\begin{eqnarray}
   \left. { \beta \, p \over n} \right|_{{\rm OCP} \,\, {\rm class}}
    =
    1 - { g \over 6}
    &-&
    { g^2 \over 24}
    \left\{
      2 \ln \left(  3 \, \kappa \, d_C \right)
    + 4 \gamma - { 8 \over 3}
    - { 1 \over 4\pi \eta_{ee}^2 }
    \right\}
\nonumber\\
    &-&  { g^3 \over 16} \,
    \Biggl\{
    2 \ln \left( 4 \, \kappa \, d_C \right)
    +  4 \gamma - { 11 \over 2} + C_1 + C_2
    \Biggr\} \,.
\label{good}
\end{eqnarray}
To obtain an independent check on this result,
the order $\hbar^2$ quantum correction
for the canonical partition function of the classical one-component
plasma is independently derived in Appendix \ref{qfluck}.  There it
is shown that there are no $\hbar^2$ corrections in three and higher
loop orders --- in agreement with the lack of an $\hbar^2$ correction
to the order $g^3$ term here --- while the two-loop $\hbar^2$
correction given in Eq.~(\ref{pfluck}) of that Appendix agrees exactly
with that in the $g^2$ term in Eq.~(\ref{good}), the term involving
$1 / \eta_{ee}^2$.

Riemann, Schlanges, DeWitt, and Kraeft \cite{dewitt2}
report an equation of state for a one-component plasma.
The terms in their formula which we classify as being of tree,
one-, and two-loop order --- the terms of order $g^0$, $g^1$,
and $g^2$ which appear in the first line of Eq.~(\ref{eq:jelly}) ---
agree precisely with our result.
They do not, however, present all the terms of three-loop, $g^3$ order,
but rather only include
terms ``up to the order $(ne^2)^{5/2}$''.
We note that such a statement has only a
formal significance since $ne^2$ bears dimensions, and hence there is
no physical significance in assuming that it is small.
The terms retained by Riemann {\em et al.} are only those parts
of the three-loop results which involve leading inverse powers
of the quantum parameter $\eta_{ee}$, namely
\begin{equation}
   \left.
   { \beta \, p \over n}
   \right|^{\hbox{\tiny $O(g^3),O(e^5)$}}_{\rm OCP} \!
    =
    -
    {g^3 \over 32 \pi}
    \left\{
	{1 \over 2 \eta_{ee}^2}
	\pm
	{3 \over g_e}
	\left[
	    {1 \over 4 \eta_{ee}^3} - {1 \over \eta_{ee}^2}
	\right]
    \right\} .
\end{equation}
Here, $g_e = 2$ is the spin degeneracy of the electron, and we have
chosen to separate the exchange contributions so as to facilitate
comparison with Ref.~\cite{dewitt2}.  Our result does not altogether
agree with formula (23) given by Riemann {\em et al.} \cite{dewitt2}
in that our exchange term of order $g^3 / \eta_{ee}^3 = O(e^3)$ is a
factor $1/2$ than theirs.  The earlier paper by DeWitt, Schlanges,
Sakakura, and Kraeft \cite{dewitt1} contains, in its Eq.~(15), the
same three-loop contributions, with the same discrepancies.  Their
two-loop terms are correct as far as they go, but in this paper the
two-loop terms also stop at the formal order of $(ne^2)^{5/2}$ rather
than containing the full dependence on $\beta e^2/\lambda$ as in the
later paper.
Recently,  we received an unpublished erratum from J.~Riemann
    in which the coefficient of the $O(g^3/\eta_{ee}^3)$ exchange term
    is corrected by a factor of two, and now all results are in
agreement.

\section {Higher Orders and the Renormalization Group}
\label {sec:higher}

\subsection {Renormalization Group Equations and Leading Logs}

In section \ref {sec:renorm} we introduced an arbitrary
scale $\mu$ in order to separate induced couplings
into pole terms and ``renormalized'' finite contributions.
For the first induced coupling this amounted to writing
\begin{equation}
    g^0_{ab} = \mu^{2(\dim-3)}
    \left[
	- { 1 \over 4!} \,
	{\left(e_a e_b \right)^3 \over (4\pi)^2} \, { 1 \over
	  3-\dim} \,
	+ g_{ab}(\mu)
    \right] .
\label {tworen}
\end {equation}
However, since the theory in general, and the bare coupling $g^0_{ab}$
in particular, knows nothing about the arbitrary value of the scale
$\mu$, we must have ${d g^0_{ab} / d\mu} = 0$.
This requires that
\begin{equation}
    \mu^2 { d \over d \mu^2} \> g_{ab}(\mu)
    =
    ( 3 - \dim ) \, g_{ab}(\mu)
    - {\left( e_a e_b \right)^3 \over 4! \, (4\pi)^2} \,,
\label{renormeq}
\end{equation}
which is the renormalization group equation for the renormalized
coupling $g_{ab}(\mu)$.
In the physical limit $\dim = 3$, the solution of the
renormalization group equation may be expressed as
\begin{equation}
    g_{ab}(\mu)
    =
    { 1 \over 4!} {\left(e_a e_b \right)^3 \over (4\pi)^2} \>
    \ln \!\left( { \mu^2_{ab} \over \mu^2 } \right) .
\label{gab}
\end{equation}
In other words, the form of the renormalized (or ``running'')
coupling $g_{ab}(\mu)$ is completely dictated by the pole terms,
which in turn depend only on the form of the effective theory.
It is only the integration constants,
which we have expressed as $ \mu^2_{ab} $,
that must be determined by matching the effective theory to the underlying
quantum theory.
The wave numbers $\mu_{ab}$ provide the quantum damping or
cutoff to the classical theory and hence must be proportional to
inverse thermal wave lengths.
The result of the matching (\ref{gab}) shows that
\begin {equation}
    \mu_{ab}^{-2} = \lambda_{ab}^2 \, e^{\Gamma_{ab}} \,,
\label {eq:muab}
\end {equation}
where $\Gamma_{ab}$, defined in Eq.~(\ref{eq:Gam}),
depends only the quantum parameter
$\eta_{ab} \equiv \beta e_a e_b / (4\pi\lambda_{ab})$.

In this section we are interested not in the precise results but rather
in exhibiting the leading logarithmic parts,
that is, those contributions that acquire arbitrarily large
logarithms in the limit of small thermal wave lengths.
Thus we introduce $\lambda$ to denote a
characteristic thermal wave length in the plasma,
and write $\mu^2_{ab} = c_{ab} / \lambda^2$
where $c_{ab}$ are dimensionless numbers that depend on the species
and on the quantum parameters $\eta_{ab}$, but formally are $O(1)$
and fixed.
Thus
$
    \ln(\mu_{ab}^2 / \mu^2) =
    - \ln ( \lambda^2 \mu^2 ) + \ln c_{ab}^2
$,
and the extra logarithm involving
$c_{ab}$ is negligible in the formal $\lambda \to 0$ limit.

The first contribution to $\ln Z$ involving a potentially large
logarithm arises at two-loop order.
 From Eq.~(\ref {ztwolps}),
the relevant two-loop part of $\ln Z$,
\begin {equation}
    { \ln Z^{(2)} \over \vol }
    =
    { \pi \over 6} \sum_{a,b=1}^A \n0a \, \n0b
    \left({ \beta e_a e_b  \over 4 \pi } \right)^3
    \left[ \ln \left( {9 \lambda_{ab}^2 \kappa_0^2 \over 4\pi} \right)
    -1 + \gamma + \Gam ab \right]
    + \cdots ,
\label {eq:z2ll}
\end {equation}
exhibits a term which depends logarithmically on the ratio of
scales $(\lambda\kappa)$.  If the plasma is sufficiently dilute, then
$\ln (\lambda\kappa)$ will be large compared to one, and the logarithmic
term will provide the dominant part of the entire two-loop correction.
The terms shown in (\ref {eq:z2ll}) come from the sum of the induced
coupling $g^0_{ab}$ contribution and the two-loop graph
\begin {equation}
\raisebox {-15pt}{%
\begin {picture}(40,33)(0,5)
    \thicklines
    \put(20,20){\circle{40}}
    \put(0,20){\line(1,0){40}}
    \put(0,20){\circle*{5}}
    \put(40,20){\circle*{5}}
    \put(-9,8){$a$}
    \put(44,8){$b$}
\end {picture}}%
\label {twoloopy}
\end {equation}
which together contribute
\begin {equation}
    -\sum_{a,b} \beta^3 \, \n0a \, \n0b
    \left[
	{e_a^3 \, e_b^3 \over 12} \, D_{\rm R}^{(3)}({\bf 0};\mu)
	+ g_{ab}(\mu)
    \right]
\label {eq:DR+g}
\end {equation}
to $(\ln Z)/\vol$,
with
\begin {equation}
D^{(3)}_{\rm R}({\bf 0};\mu) =
    { -1 \over 32\pi^2 } \left\{
    \ln \left( { 9 \kappa^2 \over 4\pi \mu^2} \right) -1 + \gamma \right\} ,
\end {equation}
as shown in Eq's.~(\ref {twolps}), (\ref {eq:D3+g}) and (\ref {D3R}).
The renormalization group equation (\ref{renormeq}) ensures that the
sum (\ref {eq:DR+g}) does not depend upon the arbitrary scale $\mu$.
It is, however, convenient to choose $\mu^2 = \kappa^2 / 4\pi$, for then the
entire logarithmic term in $\ln Z^{(2)}$ comes from the induced coupling
$g_{ab}(\mu) \sim - \ln ( \lambda^2 \kappa^2 / 4\pi) $,
rather than from the two-loop graph (\ref {twoloopy}).%
\footnote
    {%
    We could equally well have chosen $\mu^2$ to equal
    $9\kappa^2/4\pi$ or just $\kappa^2$, instead of
    $\kappa^2/4\pi$.
    Such $O(1)$ changes in the scale $\mu$ have no effect
    on the following discussion of higher-order leading-log results.
    }
Thus, the leading logarithmic piece of the two-loop
partition function may be expressed as
\begin{equation}
    {\ln Z^{(2,\rm ll)} \over \vol}
    =
    - \sum_{a,b} n_a^0 \, n_b^0 \, \beta^3 \, g_{ab}( \kappa^2 /4 \pi)
    =
    { 1 \over 4! } \sum_{a,b} n_a^0 \, n_b^0 \,
    {\left(\beta e_a e_b \right)^3 \over (4\pi)^2} \>
    \ln \!\left( { \lambda^2 \kappa^2 \over 4 \pi } \right) .
\end{equation}

The virtue of this simple observation is that it easily
generalizes to higher orders,
and allows one to determine the leading logarithmic contributions
to the pressure at any order with very little work.
To be concrete, we first consider four-loop contributions to $\ln Z$.
[Logarithmic contributions at odd-loop orders are discussed below.]
Pole terms in $\dim{-}3$ arise from (a) divergent four-loop graphs
(shown below), (b) the induced coupling $g^0_{abc}$ which first
contributes at four-loop order,
and (c) the two-loop induced coupling $g^0_{ab}$ inserted into
the two-loop graph
\begin {equation}
      \def\epsfsize #1#2{0.36#1}
      \vspace {10pt}
      \raisebox{-15pt}{\epsfbox {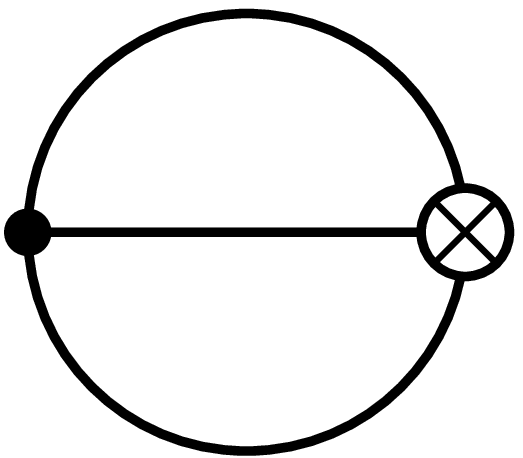}}
      \vspace {10pt}
\label {eq:I4gab}
\end {equation}
in which the left vertex represents the usual
classical interaction while the cross on the
right vertex denotes the insertion of the induced interaction with
coupling $g^0_{ab}$.
The contribution of the four-loop induced coupling $g^0_{abc}$ to
$(\ln Z)/\vol$ is just
\begin {equation}
    I^{(4)}_{g_{abc}}
    \equiv - \beta^6 \, \sum_{a,b,c} \n0a \, \n0b \, \n0c \> g^0_{abc} \,.
\label {eq:I4gabc}
\end {equation}
Using Eq.~(\ref{d30a}),
the leading divergence in the contribution of diagram (\ref {eq:I4gab})
to $(\ln Z)/ \vol$ is easily seen to be
\begin{equation}
    I^{(4)}_{g_{ab}}
    \equiv
    {1 \over 12} \,
    {1 \over 3-\dim}
    \left( {\kappa^2 \over 4 \pi} \right)^{\dim-3}
    \sum_{a,b,c}
    \n0a \, \n0b \, \n0c
    \left({\beta^3 \, ( e_a {+} e_b )^3 e_c^3 \over (4\pi)^2} \right)
    \beta^3 g^0_{ab} \>
    [ 1 + O(3{-}\dim) ] \,.
\label{double}
\end{equation}
Since the bare coupling $g^0_{ab}$ itself contains a single pole in
$\dim {-} 3$, this contribution has a double pole.
Various four-loop graphs also yield double poles in $\dim {-} 3$.
There is, however, no need to compute the double pole terms of
these four loop graphs because they are completely determined by
the renormalization group.
To prove this, we first note that $g^0_{abc}$ is completely symmetrical
in the indices $abc$, has the dimensions of
$\mu^{4(\dim-3)}$ times 12 powers of charges, and must cancel
the double poles (as well as lower order single poles)
in both diagram (\ref {eq:I4gab}) and the divergent four-loop graphs.
Consequently, $g^0_{abc}$ must have the form
\begin{eqnarray}
    g^0_{abc} &=&
    \mu^{2(\dim-3)}
    {1 \over (3-\dim)} \, {1 \over (24\pi)^2 }
	\left\{
	g^0_{ab} \, \left( e_a {+} e_b \right)^3 e_c^3 +
	g^0_{bc} \, \left( e_b {+} e_c \right)^3 e_a^3 +
	g^0_{ca} \, \left( e_c {+} e_a \right)^3 e_b^3
    \right\}
\nonumber\\
&+&
    \mu^{4(\dim -3)}
    \left\{
	{R^{(2)}_{abc} \over ( 3 - \dim)^2 } +
	{R^{(1)}_{abc} \over ( 3 - \dim) } +
	g_{abc} (\mu)
    \right\} \,.
\label{fourren}
\end{eqnarray}
The first set of terms removes the divergence in Eq.~(\ref{double}),
while the $R^{(2)}_{abc}$ and $R^{(1)}_{abc}$ terms cancel the double
and single poles generated by four-loop graphs, respectively.%
\footnote
    {%
Strictly speaking, the $R^{(2)}_{abc}$ and $R^{(1)}_{abc}$ terms
cancel primitively divergent four-loop graphs. Four loop graphs
containing divergent two-loop sub-graphs are rendered finite by
insertions of the $g^0_{ab}$ interaction in finite two-loop graphs.
    }
The remaining finite `renormalized' coupling is $g_{abc}(\mu)$.
Now $ d \, g^0_{abc} / d\mu = 0$
and $ d \, g^0_{ab} / d\mu =0 $,
while $ d \, g^{abc}(\mu) / d \mu $ must be
finite. Therefore, the single pole terms that
result when $\mu$ in Eq.~(\ref{fourren}) is varied must cancel,
\begin{equation}
    0 =
    \left. {1 \over (24\pi)^2}
    \left\{
	g^0_{ab} \left( e_a {+} e_b \right)^3 e_c^3 +
	g^0_{bc} \left( e_b {+} e_c \right)^3 e_a^3 +
	g^0_{ca} \left( e_c {+} e_a \right)^3 e_b^3
    \right\}
    \right|_{\rm pole} +
    \mu^{2(\dim -3)} \, { 2 R^{(2)}_{abc} \over 3 - \dim} \,,
\end{equation}
or, inserting the explicit form (\ref{tworen}) for $g^0_{ab}$,
\begin{equation}
    R^{(2)}_{abc} = { 1 \over 72} \,
    {e_a^3 \, e_b^3 \, e_c^3 \over 4! \, (4\pi)^4}  \>
    \left\{
	\left( e_a {+} e_b \right)^3 +
	\left( e_b {+} e_c \right)^3 +
	\left( e_c {+} e_a \right)^3
    \right\} .
\label{rtwo}
\end{equation}
This result is easily confirmed by direct computation. The graph
\begin {equation}
\begin {picture}(40,40)(10,5)
    \thicklines
    \put(20,20){\circle{40}}
    \put(0,20){\line(1,0){80}}
    \put(0,20){\circle*{5}}
    \put(40,20){\circle*{5}}
    \put(60,20){\circle{40}}
    \put(80,20){\circle*{5}}
\end {picture}
\label{sadly}
\end {equation}
produces a double pole contribution to $(\ln Z)/\vol$
involving the square of Eq.~(\ref{d30a}):
\begin{equation}
    \left[
	{1 \over 32\pi^2} \, { 1 \over 3-\dim} \,
	\left( {\kappa^2 \over 4 \pi} \right)^{\dim-3}
    \right]^2 \,
    { \beta^6 \over 2 \, (3!)^2 } \>
    \sum_{a,b,c}
    n_a^0 \, n_b^0 \, n_c^0
    \left( e_a^3 \, e_b^6 \, e_c^3 \right).
\label{33}
\end{equation}
The graph
\begin{equation}
\begin {picture}(40,40)(0,5)
    \thicklines
    \put(20,20){\circle{40}}
    \bezier{80}(0,20)(11,32.66)(30,37.3)
    \bezier{80}(0,20)(20,20)(30,37.3)
    \bezier{80}(0,20)(15,11.34)(30,2.7)
    \put(0,20){\circle*{5}}
    \put(30,37.3){\circle*{5}}
    \put(30,2.7){\circle*{5}}
\end {picture}
\label{smiley}
\end {equation}
also produces a double pole contribution to the partition function.
It is not difficult to show that the double pole in this graph,
without vertex and symmetry factors, is just $1/2$ times the square of
the single pole contribution (\ref{d30a}) of the two-loop graph.%
\footnote{An outline of the proof is as follows.
    Choose the left-hand 5-point vertex in (\ref{smiley}) to be the
    origin.
    Assign the upper 4-point vertex the coordinate ${\bf r}_1$,
  and the lower 3-point vertex the coordinate ${\bf r}_2$. The
  double pole contribution comes from the most singular integration region
  where
  $|{\bf r}_1| \ll |{\bf r}_2| \ll \kappa^{-1}$.
  In computing the leading contribution
  from this region, the right-hand line running
  between ${\bf r}_2$ and ${\bf r}_1$ can be replaced by a line
  that runs between ${\bf r}_2$ and the origin.
  Thus, as far as the
  leading singularity is concerned, the graph reduces to the graph
  (\ref{sadly}) except that the condition $|{\bf r}_1| < |{\bf r}_2|$
  must be imposed. Since the graph (\ref{sadly}) is symmetrical under
  the interchange of these two coordinates, imposing this condition
  merely multiplies the result by $1/2$.} A easy exercise
now shows that the double
pole contribution of this graph is given by
\begin{equation}
   {1 \over 2} \left[
	{1 \over 32\pi^2} \, { 1 \over 3-\dim} \,
	\left( {\kappa^2 \over 4 \pi} \right)^{\dim-3}
    \right]^2 \,
    { \beta^6 \over 2 \cdot 3!} \>
    \sum_{a,b,c}
    n_a^0 \, n_b^0 \, n_c^0
    \left( e_a^5 \, e_b^4 \, e_c^3 \right).
\label{321}
\end {equation}
It is a simple matter to verify that the double pole divergences in
Eq's.~(\ref{33}) and (\ref{321}) are indeed canceled by the contribution
(\ref{rtwo}) to the $g^0_{abc}$ coupling term.

The renormalization group equation for the finite coupling $g^{abc}(\mu)$
may now be obtained
by returning to the condition that $ d \, g^0_{abc} / d \mu = 0$.
Since the single pole terms in $d \, g^0_{abc} / d\mu$
have been shown to cancel,
this condition reduces to
\begin{eqnarray}
    0 &=&  { 1 \over 36 \, (4\pi)^2}
    \left\{
	g_{ab} \left( e_a {+} e_b \right)^3 e_c^3 +
	g_{bc} \left( e_b {+} e_c \right)^3 e_a^3 +
	g_{ca} \left( e_c {+} e_a \right)^3 e_b^3
    \right\}
\nonumber\\
&& \qquad\qquad
    + 2 R^{(1)}_{abc}
    + 2 \, (3 {-} \dim) \, g_{abc} - \mu^2 {d \over d \mu^2} \, g_{abc} \,.
\end{eqnarray}
Using Eq.~(\ref{gab}) for $g_{ab}$ and taking the physical limit $
\dim =3$, gives
\begin{eqnarray}
    \mu^2 {d \over d \mu^2} \, g^{abc}
    =
    2 R^{(1)}_{abc}
    +
    { 8 \over 27}
    { e_a^3 \, e_b^3 \, e_c^3 \over (16\pi)^4} \,
     \Bigg[
     \left( e_a {+} e_b \right)^3 \, \ln{ \mu^2_{ab} \over \mu^2 }
    +
    \left( e_b {+} e_c \right)^3 \, \ln{ \mu^2_{bc} \over \mu^2 }
    +
    \left( e_c {+} e_a \right)^3 \, \ln{ \mu^2_{ca} \over \mu^2 }
     \Bigg] .
\nonumber\\
\label{abcrun}
\end{eqnarray}
The integration of this renormalization group equation yields
\begin{eqnarray}
    g^{abc}(\mu)
    &=&
    -{ 4 \over 27}
    { e_a^3 \, e_b^3 \, e_c^3 \over (16\pi)^4}
    \left[
    \left( e_a {+} e_b \right)^3 \ln^2\biggl( {\mu^2_{ab} \over \mu^2} \biggr)
    +
    \left( e_b {+} e_c \right)^3 \ln^2\biggl( {\mu^2_{bc} \over \mu^2} \biggr)
    +
    \left( e_c {+} e_a \right)^3 \ln^2\biggl( {\mu^2_{ca} \over \mu^2} \biggr)
    \right]
\nonumber\\ && {}
    -
    2 R^{(1)}_{abc} \, \ln\biggl( {\mu^2_{abc} \over \mu^2} \biggr),
\label {eq:g3run}
\end {eqnarray}
where the integration constant has been written as a scale $\mu_{abc}$
which, once again, will be of order of (the inverse of) a typical
thermal wavelength $\lambda^{-1}$, but whose precise value can only be
determined by matching to the underlying quantum theory.
Note that the single pole residue $R^{(1)}_{abc}$ in the
renormalization group equation (\ref{abcrun}) gives
rise to single log terms in the running coupling (\ref {eq:g3run}),
which are subleading compared to the double log terms when
$\mu$ is much much less than $\lambda^{-1}$.
The residue $R^{(1)}_{abc}$ is determined by the less singular single-pole
terms of the previous double pole contributions, plus the single pole
produced by the graph
\begin {equation}
\begin {picture}(40,35)(0,5)
    \thicklines
    \put(20,20){\circle{40}}
    \bezier{80}(0,20)(15,28.66)(30,37.3)
    \bezier{80}(30,37.3)(30,20)(30,2.7)
    \bezier{80}(0,20)(15,11.34)(30,2.7)
    \put(0,20){\circle*{5}}
    \put(30,37.3){\circle*{5}}
    \put(30,2.7){\circle*{5}}
\end {picture}
\end{equation}
which has no double pole contribution.
Since our purpose here
is just to illustrate the character of the theory, we shall not bother
to compute $R^{(1)}_{abc}$ explicitly.

Recalling that the divergent terms in the classical theory have all
their non-integral dimensional dependence appearing as integer powers
of $\kappa^{\dim -3}$, we see that, just as in the previous two-loop
discussion, choosing $\mu^2 = \kappa^2 / 4\pi$ in the induced couplings
not only removes the poles in these classical loop graphs, it also prevents
the appearance of any additional large logarithms in the resulting finite
contributions of four-loop graphs.
Thus taking $\mu^2 = \kappa^2 / 4\pi$ and inserting (\ref
{eq:g3run}) into (\ref {eq:I4gabc}) immediately yields the leading
logarithmic contribution to the partition function at four-loop order:
\begin{eqnarray}
    {\ln Z^{(4,\rm ll)} \over \vol}
    =
    { 1 \over (16\pi)^4}
    \sum_{a,b,c} \beta^6 \, n_a^0 \, n_b^0 \, n_c^0
    \left\{
	{8 \over 9}
	\left[ e_a^3 \, e_b^6 \, e_c^3 + 3 \, e_a^5 \, e_b^4 \, e_c^3  \right]
	\ln^2 \!\left( {\lambda^2 \kappa^2 \over 4\pi} \right)
	+
	O \!\left[e^{12} \ln (\lambda\kappa)\right]
    \right\} .
\nonumber\\
\end{eqnarray}

\subsection {Leading Logs to All Orders}

    Exactly the same approach may be used to determine
the leading log contributions at higher orders.
Consider first the situation at an even loop order.
The induced coupling $g^0_{a_1 \cdots a_{p+1}}$ makes its
first contribution to $(\ln Z)/\vol$ at $2p$-loop order.
This contribution is
\begin {equation}
    I^{(2p)}_A
    \equiv
    -\beta^{3p} \sum_{a_1 \cdots a_{p+1}}
    \n0{a_1} \cdots \n0{a_{p{+}1}} \> g^0_{a_1\cdots a_{p{+}1}} \,.
\label {eq:I2p}
\end {equation}
For later convenience, we will refer to $g^0_{a_1 \cdots a_{p+1}}$
as the rank-$p$ coupling.
Poles in $\dim{-}3$ up to order $p$ are generated at $2p$-loop order and
must be canceled by the rank-$p$ induced coupling.
In particular, order $p$ poles are generated by diagrams of the form
\begin {equation}
      \def\epsfsize #1#2{0.36#1}
      \vspace {10pt}
      \raisebox{-15pt}{\epsfbox {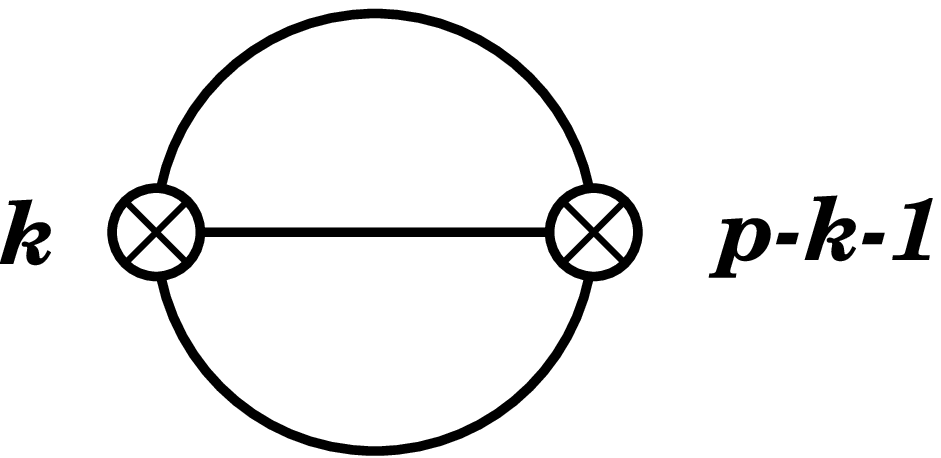}}
      \vspace {10pt}
\label {eq:Ieven}
\end {equation}
in which the left and right vertices represent insertions of the rank $k$
and rank $p{-}k{-}1$ couplings
$g^0_{a_1\cdots a_{k+1}}$ and $g^0_{a_{k+2}\cdots a_{p+1}}$, respectively.
The contribution of these diagrams to $(\ln Z)/\vol$ is
\begin {eqnarray}
    I^{(2p)}_B &=&
    -
    {\beta^{3p} \over 4! \, (4\pi)^2}
    \left({\kappa^2 \over 4\pi}\right)^{\dim-3}
    {1 \over 3-\dim}
    \sum_{k=0}^{p-1}
    \sum_{a_1 \cdots a_{p+1}}
    \n0{a_1} \cdots \n0{a_{p+1}} \>
    g^0_{a_1 \cdots a_{k+1}} \,
    g^0_{a_{k+2} \cdots a_{p+1}}
\nonumber
\\ && \kern 2.5in {} \times
    \left( e_{a_1 \cdots a_{k+1}} \right)^3
    \left( e_{a_{k+2} \cdots a_{p+1}} \right)^3 ,
\end {eqnarray}
where we have introduced the shorthand abbreviation
\begin {equation}
    e_{a_1 \cdots a_k} \equiv
    e_{a_1} + \cdots + e_{a_k} \,.
\end {equation}
By defining $g^0_{a} \equiv -1$, and including the terms
where $k=0$ and $k=p{-}1$, this expression also includes the
case where either vertex in the diagram (\ref {eq:Ieven})
represents the original classical interaction.
Since the rank-$k$ coupling $g^0_{a_1 \cdots a_{k+1}}$ contains
poles in $\dim{-}3$ up to order $k$
(and $g^0_{a_{k+2} \cdots a_{p+1}}$ has poles of order $p{-}k{-}1$),
this contribution does generate order-$p$ poles.
To cancel these poles,
the rank-$p$ coupling must have the form
\begin {eqnarray}
    g^0_{a_1 \cdots a_{p+1}}
    &=&
    -
    {1 \over 4!\, (4\pi)^2}
   \mu^{2(\dim-3)}
    {1 \over 3-\dim}
    \sum_{k=0}^{p-1} \>
    {\cal S}
    \left\{
	g^0_{a_1 \cdots a_{k+1}} \,
	g^0_{a_{k+2} \cdots a_{p+1}}
	\left( e_{a_1 \cdots a_{k+1}} \right)^3
	\left( e_{a_{k+2} \cdots a_{p+1}} \right)^3
    \right\}
\nonumber
\\&& {}
    + \cdots +
    \mu^{2p(\dim-3)}
    \left[
	\sum_{k=1}^p \>
	{R^{(k)}_{a_1 \cdots a_{p+1}} \over (3-\dim)^k}
	+
	g_{a_1 \cdots a_{p+1}} (\mu)
    \right] .
\label {eq:rank-p}
\end {eqnarray}
Here, $\cal S$ denotes a symmetrization operator which averages
over all permutations of the indices $a_1 \cdots a_{p+1}$.
The $R^{(k)}$ terms cancel the poles of order $k$ generated
by (primitively divergent) $2p$-loop graphs, and the unwritten ``$\cdots$''
pieces denote terms proportional to $\mu^{4(\dim-3)}$
which cancel the poles generated by induced couplings totaling rank
$p{-}2$ inserted into divergent four loop graphs,
terms proportional to $\mu^{6(\dim-3)}$
which cancel the poles generated by induced couplings of rank
$p{-}3$ inserted into divergent six loop graphs, {\em etc}.

The renormalization group condition $d \, g^0_{a_1 \cdots a_{p+1}} /
d\mu^2 = 0$ must hold as an exact identity.  The variation of (\ref
{eq:rank-p}) with respect to $\mu$ has poles in $\dim{-}3$ up to order
$p{-}1$.  The coefficients of each order pole must cancel
independently. As discussed before, this means that the residues
$R^{(k)}$, for $k = 2,\cdots,p$ are completely fixed by the structure
of the lower order diagrams.  The renormalization group equation for
the remaining finite terms, evaluated at $\dim=3$, becomes
\begin {eqnarray}
    \mu^2 {d \over d\mu^2} \, g_{a_1 \cdots a_{p+1}}
    =
    -{1 \over 4!\, (4\pi)^2}
    \sum_{k=0}^{p-1} \>
    {\cal S}
    \left\{
	g_{a_1 \cdots a_{k+1}} \,
	g_{a_{k+2} \cdots a_{p+1}}
	\left( e_{a_1 \cdots a_{k+1}} \right)^3
	\left( e_{a_{k+2} \cdots a_{p+1}} \right)^3
    \right\}
    +
    \cdots \,.
\nonumber\\
\label {eq:RGp}
\end {eqnarray}
The key point is that when $\mu^2$ is chosen to be of order $\kappa^2$,
the only source of large logarithms are the induced couplings themselves;
the rank-$k$ renormalized coupling $g_{a_1 \cdots a_k}$ is of order%
\footnote
    {%
    This has been shown explicitly for $k=2$ and 3.
    The current section may be regarded as an inductive proof
    of this assertion to all orders.
    }
$\left[ \ln (\lambda \mu) \right]^{k-1}$.
Consequently, the terms shown explicitly on the right hand side
of (\ref {eq:RGp}) are proportional to $\ln^{p-1} (\lambda\mu)$, while
all the unwritten ``$\cdots$'' terms have at most $p{-}2$
powers of $\ln (\lambda\mu)$.
Integrating (\ref {eq:RGp}), neglecting the sub-leading pieces, gives
\begin {equation}
    g_{a_1 \cdots a_{p+1}}(\mu)
    =
    -{\ln \lambda^2\mu^2 \over 4!\, (4\pi)^2 \, p} \,
    \sum_{k=0}^{p-1} \>
    {\cal S}
    \left\{
	g_{a_1 \cdots a_{k+1}} \,
	g_{a_{k+2} \cdots a_{p+1}}
	\left( e_{a_1 \cdots a_{k+1}} \right)^3
	\left( e_{a_{k+2} \cdots a_{p+1}} \right)^3
    \right\} .
\label {eq:gp}
\end {equation}
Equation (\ref {eq:gp}), together with the starting condition
$g_a \equiv -1$, provides a simple recursive recipe for determining
the leading-log contribution to the rank-$p$ induced coupling
$g_{a_1 \cdots a_{p+1}}$.
The leading-log contribution to $\ln Z$ at $2p$-loop order is then just
\begin {equation}
    {\ln Z^{(2p, \rm ll)} \over \vol}
    =
    - \beta^{3p} \sum_{a_1 \cdots a_{p+1}}
    \n0{a_1} \cdots \n0{a_{p+1}} \>
    g_{a_1 \cdots a_{p+1}} \,,
\label {eq:Z2pll}
\end {equation}
where the renormalized induced couplings are to be evaluated
at $\mu = O(\kappa)$.
The resulting contribution
is $O[(\beta e^2 \kappa)^{2p} \ln^p (\lambda\kappa)]$ in magnitude.%
\footnote{%
    The alert reader will have noticed that we have ignored
    the induced couplings for derivative interactions,
    $h^0_{a_1 \cdots a_p}$,
    $k^0_{a_1 \cdots a_p}$,
    {\em etc}., in this discussion.
    The four loop couplings $h^0_{ab}$ and $k^0_{ab}$ in the induced action
    (\ref{Sind}) give rise to only a single log at
    four loop order, not a double log,
    and so it does not contribute to the leading log result.
    Moreover, just as in the previous case of the $g_{a_1\cdots a_p}$
    couplings, these two-derivative couplings generate
    a sequence of higher powers of logs, but each
    member in this sequence of contributions is suppressed by one power
    of $\ln \kappa\lambda$ compared to the corresponding leading-log
    contribution.
    In the same manner, the four-derivative or higher terms schematically
    denoted by the ellipsis $\cdots$ in Eq.~(\ref{Sind}) give rise
    to still further subdominant logs.
    }

\begin {figure}[t]
   \begin {center}
      \leavevmode
      \def\epsfsize #1#2{0.4#1}
      \epsfbox {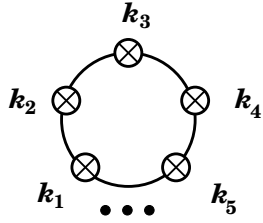}
   \end {center}
   \caption
       {%
       Leading-log contributions to $\ln Z$ at odd-loop order.
       Circled vertices labeled $k$ denote the insertion of the
       rank-$k$ induced interaction proportional to $g^0_{a_1 \cdots a_{k+1}}$.
       For order $2p+1$ contributions, the ranks of all the
       insertions around the loop must sum to $p$.
       }%
\label {fig:leading-log-odd}
\end {figure}

The leading-log contributions at odd-loop orders are also easily
determined.  When the scale $\mu$ is $O(\kappa)$, so that the only
source of large logarithms are the induced couplings themselves, the
largest number of logarithms at a given odd loop order will result
from diagrams where the maximal number of induced couplings are
inserted into a graph with only one explicit loop.  At loop order of
$2p+1$, this means a single insertion of the rank $p$ induced
coupling, or two insertions of rank $k$ and rank $p{-}k$ couplings, or
more generally, the insertion of any collection of induced couplings
whose ranks total $p$, as illustrated in figure \ref
{fig:leading-log-odd}.
Rather than following the cookbook method and
struggling to get the proper combinatorial factors to evaluate these
diagrams, it is much easier to simply return to the original
functional integral representation (\ref{qmfi}).  The sum of the
graphs in question just corresponds to the contribution of the order
$\phi(\r)^2$ terms in the first line of the induced interaction
(\ref{Sind}) to the action
\begin{equation}
    S^{(2)}_{\rm ind}
    =
    \Delta \kappa^2 \, {\beta \over 2} \int (d^\dim \r) \> \phi(\r)^2 \,,
\end{equation}
where
\begin {equation}
 \Delta \kappa^2
		=    -
    \sum_{p=2}^\infty
    \sum_{a_1 \cdots a_p}
    \beta^{3p-2} \>
    \n0{a_1} \cdots \n0{a_{p}} \>
    g_{a_1 \cdots a_{p}}
    \left( e_{a_1 \cdots a_{p}} \right)^2 \,.
\label {eq:kll}
\end {equation}
Thus, the total effect of these terms is to simply shift the
unperturbed (squared) Debye wave number,
\begin{equation}
\kappa_0^2 \to \kappa^2_{\rm ll} \equiv \kappa_0^2 + \Delta \kappa^2 \,.
\end{equation}
Referring back to the one-loop correction (\ref{zzzone}), we see the
sum of these odd-loop order leading logarithms plus the original
one-loop contribution is given by
\begin {equation}
    \sum_{p=0}^\infty
     \ln Z^{(2p+1,\rm ll)}
    =
    - {\kappa_{\rm ll}^3 \over 12\pi}  \, \vol \,.
\label {eq:odd-ll}
\end {equation}

A straightforward exercise
expanding (\ref {eq:odd-ll}) and (\ref {eq:kll}),
and iterating (\ref {eq:Z2pll}),
will yield the explicit leading-log contributions at any given order.
The results up to order 6 are:
\begin {eqnarray}
    {\ln Z^{(2,\rm ll)} \over \vol}
    &=&
    {\cal L}
    \sum_{a,b}
    \beta^3 \,
    \n0a \, \n0b \,
    e_a^3 \, e_b^3 \,,
\\
    {\ln Z^{(3,\rm ll)} \over \vol}
    &=&
    {\kappa_0 \, {\cal L} \over 8\pi}
    \sum_{a,b}
    \beta^4 \,
    \n0a \, \n0b \,
    e_a^3 \, e_b^3 \, (e_a {+} e_b)^2 \,,
\\
    {\ln Z^{(4,\rm ll)} \over \vol}
    &=&
    {\cal L}^2
    \sum_{a,b,c}
    \beta^6 \,
    \n0a \, \n0b \, \n0c \,
    e_a^3 \, e_b^3 \, e_c^3 \, (e_b {+} e_c)^3 \>,
\\
    {\ln Z^{(5,\rm ll)} \over \vol}
    &=&
    {{\cal L}^2 \over 32\pi \kappa_0}
    \sum_{a,b,c,d}
    \beta^8 \,
    \n0a \, \n0b \, \n0c \, \n0d \,
    e_a^3 \, e_b^3 \, e_c^3 \, e_d^2
\nonumber\\&& \qquad\qquad\qquad {} \times
    \left[
    4 (e_b {+} e_c)^3 \, (e_a{+}e_b{+}e_c)^2 +
    e_d \, (e_a {+} e_b)^2 \, (e_c{+}e_d)^2
    \right] ,
\\
    {\ln Z^{(6,\rm ll)} \over \vol}
    &=&
    {{\cal L}^3 \over 3}
    \sum_{a,b,c,d}
    \beta^9 \, \n0a \, \n0b \, \n0c \, \n0d \,
    e_a^3 \, e_b^3 \, e_c^3 \, e_d^3 \,
    \Bigl[ (e_a {+} e_b)^3 + 2 (e_b{+}e_c{+}e_d)^3 \Bigr] (e_c{+}e_d)^3 \,,
\end {eqnarray}
where
\begin {equation}
    {\cal L} \equiv
    {1 \over 4! \, (4\pi)^2} \,
    \ln \!\left({ \lambda^2 \kappa^2 \over 4\pi} \right) .
\label{logg}
\end {equation}

The expression for the partition function in leading logarithmic orders
reads
\begin{equation}
\ln Z^{(\rm ll)} = \vol \, \sum_{a=1}^A \, n^0_a
    + \vol \, {\kappa_0^3 \over 12\pi}
    + \sum_{k=2} \, \ln Z^{(k, \rm ll)} \,.
\end{equation}
The derivatives of this form with respect to $\beta \mu_a$ give the
leading logarithmic relations between the physical densities $\bar
n_a$ and the `bare' densities $n^0_b$. Solving for the bare densities
in terms of the physical quantities expresses
the leading-log partition function in the form
\begin{equation}
\ln Z^{(\rm ll)} = \vol \, \sum_{a=1}^A \, \bar n_a
    - \vol \, {\bar \kappa^3 \over 24\pi}
    + \sum_{k=2} \, \ln \bar Z^{(k, \rm ll)} \,,
\end{equation}
in which
\begin {eqnarray}
    {\ln \bar Z^{(2,\rm ll)} \over \vol}
    &=&
    -{\cal L}
    \sum_{a,b}
    \beta^3 \,
    \bar n_a \, \bar n_b \,
    e_a^3 \, e_b^3 \,,
\label{two}
\\
    {\ln \bar Z^{(3,\rm ll)} \over \vol}
    &=&
    -{3 \, \kappa \, {\cal L} \over 8\pi}
    \sum_{a,b}
    \beta^4 \,
    \bar n_a \, \bar n_b \,
    e_a^4 \, e_b^4 \,,
\label{three}
\\
    {\ln \bar Z^{(4,\rm ll)} \over \vol}
    &=&
    -
    12 {\cal L}^2
    \sum_{a,b,c}
    \beta^6 \,
    \bar n_a \, \bar n_b \, \bar n_c \,
    e_a^3 \, e_b^4 \, e_c^5 \>,
\label{four}
\\
    {\ln \bar Z^{(5,\rm ll)} \over \vol}
    &=&
    -
    {5{\cal L}^2 \over 16\pi\kappa}
    \sum_{a,b,c,d}
    \beta^8 \,
    \bar n_a \, \bar n_b \, \bar n_c \, \bar n_d \,
    e_a^2 \, e_b^3 \, e_c^4 \, e_d^4
    \left[
	e_a^2 \, e_b +
	12 e_b \, e_c \, e_d +
	16 e_b \, e_d^2 +
	12 e_c \, e_d^2
    \right]\!,\,
\\
    {\ln \bar Z^{(6,\rm ll)} \over \vol}
    &=&
    - 12 {\cal L}^3
    \sum_{a,b,c,d}
    \beta^9 \, \bar n_a \, \bar n_b \, \bar n_c \, \bar n_d \,
    e_a^3 \, e_b^3 \, e_c^4 \, e_d^5 \,
\nonumber
\\ && \kern 1.1in {} \times
    \Bigl[
	  3 e_a e_b e_c
	+ 3 e_b^2 e_c
	+ 3 e_c^2 e_d
	+ 3 e_c e_d^2
	+ 5 e_b e_d^2
	+ 14 e_b e_c e_d
    \Bigr] \,.
\end {eqnarray}
Note that for even-loop orders these leading-logarithmic contributions
always include a sum of particle densities weighted by an odd power
of the charge. Consequently,
the leading-logarithmic contributions at even-loop order
vanish in the special case of a neutral symmetric binary plasma,
such as a pure electron-proton plasma,
where the charges of the two species are equal and opposite and the
the physical densities are necessarily equal due to charge neutrality.
This is a general result,
which follows from the recursive structure of (\ref {eq:gp})
and the vanishing of its initial term.
The corresponding leading logarithmic expansion of the free energy $F$
is easily obtained from the thermodynamic relation
$
\beta \, p = \ln Z / \vol = - \partial (\beta F) / \partial \vol
$.
Using this relation, it is easy to confirm that the overall
coefficients of the first three logarithmic terms,
the two, three, and four loop terms,
(\ref{two}), (\ref{three}), and (\ref{four}), agree with the
corresponding free energy terms computed by Ortner \cite{ortner} and
displayed in his Eq.~(93). As we have noted before, Ortner works with
a plasma of various species of positive ions moving
in a fixed background of neutralizing negative charge. This model allows
him to use a purely classical description,
in which the effective short-distance cutoff is provided by
the essential singularities of the Boltzmann factors
with purely repulsive Coulomb interactions.
Thus, as was explained before in the discussion of
Eq.~(\ref{classic}), the quantum length $\lambda$ in the argument of
the logarithm in Eq.~(\ref{logg}) is replaced by the Coulomb length
$
d_C = \beta \, e^2 / (4\pi)
$
in Ortner's results.

\subsection {``Anomalous'' Virial Relation}

The grand canonical partition function may be regarded as a function of
the temperature and the (bare) density $\n0a$, charge $e_a$,
and thermal wavelength $\lambda_a$ of each species.
If one defines an average density $n$
which is the geometric mean of the bare densities $\n0a$
and thus entails $\beta^{-3/2}$ times the exponential of the average
of the chemical potentials,
then one may alternatively express a specific density in
terms of the average density and a relative density ratio $x_a$,
\begin {equation}
    \n0a \equiv n \, x_a \,.
\end {equation}
The charges of each species may similarly be written in terms
of some mean charge $e$ and a relative charge ratio $y_a$,
\begin {equation}
    e_a \equiv e \, y_a \,.
\end {equation}
Any dependence on the thermal wavelength $\lambda_a$ may be re-expressed
as dependence on the dimensionless quantum parameter
$\eta_a = \beta e_a^2 / 4\pi \lambda_a$.
Consequently,
any $n$-loop contribution will equal $(\beta e^2 \kappa_0)^n$ times
some function of the dimensionless variables $\{ x_a \}$, $\{ y_a \}$
and $\{ \eta_a \}$.
This is a precise version of the statement that the loop expansion
parameter (in the physical limit of three dimensions) is $\beta e^2 \kappa_0$.
The point to be emphasized is that the parameter $\beta e^2 \kappa_0$
captures the overall powers of the inverse temperature,
charge, and densities that appear in a given loop order.
Therefore,
the grand canonical partition function has the functional form
\begin{equation}
    \ln Z = {\cal F}(\beta e^2 \kappa_0,x,y,\eta)
    \left( \vol \, {\sum}_b n_b^0 \right) .
\end{equation}

Let us pretend, for the moment, that ${\cal F}$ does not depend on the
quantum parameters $\eta_a$ --- or that the purely classical theory exists.
We note that the differential operator
$
    \left[
    \beta {\partial \over \partial \beta}
    -
    \coeff 32 \sum_a {\partial\over \partial \beta\mu_a}
    \right]
$
annihilates the density ratios $x_a$, the charge ratios $y_a$,
and also $\beta e^2 \kappa_0$ because
\begin{equation}
\left[ \beta { \partial \over \partial \beta } - {  3 \over 2}
  {\sum}_a { \partial \over \partial \beta \mu_a} \right]
\beta^2 e^4 \kappa_0^2 \sim
\left[ \beta { \partial \over \partial \beta } - {  3 \over 2}
  {\sum}_a { \partial \over \partial \beta \mu_a} \right]
\beta^{3/2} {\sum}_b e^{\beta \mu_b} = 0 \,.
\end{equation}
Since
\begin{equation}
\left[ \beta { \partial \over \partial \beta } - {  3 \over 2}
  {\sum}_a { \partial \over \partial \beta \mu_a} \right]
n_b^0 = - 3 \, n_b^0 \,,
\label {eq:zapn}
\end{equation}
this shows that
\begin{equation}
\left[ 3 + \beta { \partial \over \partial \beta } - { 3 \over 2}
  {\sum}_a { \partial \over \partial \beta \mu_a} \right]
{\cal F}(\beta e^2 \kappa_0,x,y) \left(\vol {\sum}_b n_b^0 \right) = 0 \,.
\end{equation}
In other words, a purely classical partition
function must satisfy
\begin{equation}
\left[ 3 + \beta { \partial \over \partial \beta } - { 3 \over 2}
  {\sum}_a { \partial \over \partial \beta \mu_a} \right]
\ln Z = 0 \,.
\label {eq:zapcl}
\end{equation}
Recalling that
the pressure $p$ appears as
\begin{equation}
\ln Z[\mu] = \beta \, p \,  \vol \,,
\end{equation}
the thermodynamic, internal energy density $u$ is given by
\begin{equation}
- { \partial \ln Z[\mu] \over \partial \beta } = u \vol \,,
\label{ppdeff}
\end{equation}
and the number density $\bar n_a$ of species $a$ by
\begin{equation}
{ \partial \ln Z[\mu] \over \partial \beta \mu_a} = \bar n_a \vol \,,
\label{nndeff}
\end{equation}
the identity (\ref {eq:zapcl}) for a purely classical plasma
is equivalent to the relation
\begin{equation}
 3 \beta p - \beta u - {3 \over 2} \, {\sum}_a \bar n_a = 0 \,.
\end{equation}
Of course,
the purely classical plasma (with oppositely charged particles) does not exist.
The induced couplings necessary to render the theory finite give rise
to additional dependence on the quantum parameters $\eta_a$.
Hence, in fact,
\begin{equation}
    A \equiv 3 \beta p - \beta u - {3 \over 2} \, {\sum}_a \bar n_a \neq 0 \,.
\label{anomal}
\end{equation}
The non-vanishing of $A$ arises from the `anomalous' dependence
on the underlying quantum physics.
This behavior shows that $A$ is akin to the anomalies encountered
in relativistic quantum field theories.

To find an expression for the anomaly $A$, which may be evaluated {\em
without} separately computing the pressure, internal energy, and
densities, we turn to the functional integral representation of the
grand canonical partition function. It proves convenient for this
specific application to use a scaled potential $\tilde \phi(\r) =
\beta \phi(\r) $ so that the interaction terms now involve
\begin{equation}
n_a^0 \, e^{i e_a \beta \phi } = n_a^0 \, e^{i e_a \tilde \phi } \,,
\end{equation}
with no explicit appearance of the inverse temperature $\beta$
(although it does reside in the densities $n_a^0$). Thus the
functional integral takes the form
\begin{equation}
Z[\mu] = {\rm Det}^{1/2} \left[ -\beta^{-1} \nabla^2 \right]
\int [ d \tilde \phi ]\>
\exp\left\{ - {1 \over 2 \beta }  \int (d^\dim\r) \>
   \left(\nabla\tilde\phi(\r)\right)^2
-\Sint[\tilde\phi; \mu ] \right\} \,.
\label{funinttt}
\end{equation}
Although the method that we shall outline is valid for
$\Sint[\tilde\phi; \mu ] $ taken to arbitrary order, to keep the
notation simple, we shall consider only those pieces that contribute
to the three-loop order to which we have calculated,
\begin{eqnarray}
    \Sint[\tilde\phi;\mu]
    &=&
    \int (d^\dim\r) \>
    \biggl[
    - \sum_a \n0a \, e^{i e_a \tilde \phi(\r)}
    +
    \sum_{a,b}
    g^0_{a  b} \,
    \beta^{3} \,
    n_{a}^0 \, e^{i e_{a} \tilde\phi(\r) }
    \,
    n_{b}^0 \, e^{i e_{b} \tilde\phi(\r) }
\nonumber\\
&& \qquad\qquad
- \sum_a \, h^0_a \, \beta^2  \, e_a^2 \,
\left( \nabla \tilde\phi({\bf r}) \right)^2
n_a^0 \, e^{i e_a \tilde\phi({\bf r}) } +
    \cdots \>
    \biggr] \,.
\label {eq:Sint}
\end{eqnarray}

We shall first find an expression for the pressure using its
identification with the response to a change in the volume,
\begin{equation}
\delta \ln Z[\mu] = \beta \, p \>  \delta \vol \,,
\label{pdeff}
\end{equation}
with the change in volume realized by a dilation transformation of
the spatial coordinates within the functional integral. To do this in
a conceptually simple way, we temporarily introduce general
coordinates $x^k$ and a metric tensor $g_{kl}$, so that the physical
distance between neighboring points is given by
\begin{equation}
ds^2 = \sum_{k,l = 1}^\dim g_{kl} \> dx^k dx^l \,.
\label{interval}
\end{equation}
The $(\nabla\tilde\phi)^2$ part of the action in the functional integral now
takes on the generally covariant form
\begin{equation}
 \int (d^\dim x) \> \sqrt{ \det g } \> g^{kl}
\,   \partial_k \tilde\phi(x) \partial_l \tilde\phi(x)
\, \left[ {1 \over 2\beta} - \sum_a \, h^0_a \, \beta^2 \, e_a^2   \,
n_a^0 \, e^{i e_a \tilde\phi(x) }  \right] \,.
\end{equation}
For the terms in $\Sint$ which do not involve derivatives, the
introduction of generalized coordinates is
effected by simply including the factor $ \sqrt{ \det g } $ in the
spatial integration measure.
To effect a dilation or scale change, we take
\begin{equation}
g_{kl} = e^{2 \sigma } \delta_{kl} \,.
\end{equation}
In view of the distance interval (\ref{interval}), this has the effect
of the length alteration $ L \to L e^\sigma $. With this metric, the
determinantal factor and inverse metric are simply
\begin{equation}
\sqrt{ \det g } = e^{\dim \sigma} \,,
\qquad g^{kl} = e^{-2 \sigma} \delta_{kl} \,.
\end{equation}
Finally, taking the constant parameter $\sigma$ to be infinitesimal, $
\sigma \to \delta \sigma$, we have a volume change $ \delta \vol =
\dim \, \delta\sigma \vol$. Thus, the variation of the functional
integral (\ref{funinttt}) brought about by the volume change in the
pressure definition (\ref{pdeff}) gives
\begin{eqnarray}
\dim \beta p  &=& - (\dim {-} 2) \,
 \avg{ \left(\nabla\tilde\phi\right)^2
\left[ {1 \over 2\beta} - \sum_a \,  h^0_a \, \beta^2  \,
e^2_a \, n_a^0 \, e^{i e_a \tilde\phi }  \right]  }
\nonumber\\
&& \qquad +
 \dim \avg { \sum_a \n0a \, e^{i e_a \tilde \phi}
    -
    \sum_{a,b}
    g^0_{a b} \,
    \beta^{3} \,
    n_{a}^0 \, e^{i e_{a} \tilde\phi }
    \,
    n_{b}^0 \, e^{i e_{b} \tilde\phi }
 }  \,.
\label{dilp}
\end{eqnarray}

Using the functional integral representation (\ref{funinttt}) to
evaluate the definitions (\ref{ppdeff}) and (\ref{nndeff})
of the energy and particle number yields
\begin{equation}
- \beta u \vol =  {1 \over 2 \beta }
 \avg{ \left(\nabla\tilde\phi\right)^2 } \vol
- \avg{ \beta { \partial \over \partial \beta } \,
 S_{\rm int}[\tilde\phi; \mu ] } \,.
\end{equation}
and
\begin{equation}
-{\sum}_a \bar n_a \vol =
 {\sum}_a \avg{ { \partial \over \partial \beta \mu_a } \,
 S_{\rm int}[\tilde\phi; \mu ] } \,.
\end{equation}
Since each of the quantities that make up the anomaly (\ref{anomal})
is well defined, we may write it as a $\dim \to 3$ limit in a form that will
prove to be convenient,
\begin{equation}
A = \dim \beta p - \beta u -  { \dim \over 2} \,
{\sum}_a \bar n_a \,.
\end{equation}
The results above express this as
\begin{eqnarray}
A  &=&  \avg{ (3{-}\dim){ 1 \over 2\beta}
\left( \nabla \tilde \phi \right)^2
+ \left[ \dim -2 + \beta { \partial \over \partial \beta }
- { \dim \over 2}
 \,{\sum}_c { \partial \over \partial \beta \mu_c } \right]
\sum_a \,  h^0_a \, \beta^2  \, e^2_a \,
\left( \nabla\tilde\phi \right)^2 \,
n_a^0 \, e^{i e_a \tilde\phi }  }
\nonumber\\
&&   + \avg{
\left[ \dim  + \beta { \partial \over \partial \beta }
- { \dim \over 2}
 \,{\sum}_c { \partial \over \partial \beta \mu_c } \right]
 \left\{ \sum_a \n0a \, e^{i e_a \tilde \phi}
    -
    \sum_{a,b}
    g^0_{a b} \,
    \beta^{3} \,
    n_{a}^0 \, e^{i e_{a} \tilde\phi }
    \,
    n_{b}^0 \, e^{i e_{b} \tilde\phi } \right\}
 } \,.
\end{eqnarray}
The commutation relation
\begin{equation}
 \left[ \dim + \beta { \partial \over \partial \beta }
-  { \dim \over 2}
 \,{\sum}_c { \partial \over \partial \beta \mu_c } \right]
n_a^0 = n_a^0 \left[ \beta { \partial \over \partial \beta }
- { \dim \over 2}
 \,{\sum}_c { \partial \over \partial \beta \mu_c } \right] .
\end{equation}
implies that the classical action part of $S_{\rm int}$,
proportional to $\sum_a \n0a \, e^{i e_a \tilde \phi}$, does not contribute
to the anomaly $A$ (as required).
Moreover,
\begin{equation}
\left[ \beta { \partial \over \partial \beta }
- { \dim \over 2}
 \,{\sum}_c { \partial \over \partial \beta \mu_c } \right]
\beta^3 \, n_b^0 = \beta^3 \, n_b^0
\left[ (3 {-} \dim) +  \beta { \partial \over \partial \beta }
- { \dim \over 2}
 \,{\sum}_c { \partial \over \partial \beta \mu_c } \right] .
\end{equation}
Hence we have, to our three-loop order of accuracy,
\begin{eqnarray}
A  &=&  \avg{ (3 {-}\dim) {1 \over 2\beta}
\left( \nabla \tilde \phi \right)^2 +
\sum_a \, \beta^3 \, { \partial h^0_a \over \partial \beta  } \,
e_a^2 \, n_a^0 \,  \left( \nabla \tilde \phi \right)^2 \,
e^{i e_a \tilde\phi }   }
\nonumber\\
&& -
    \sum_{a b}
  \left[ \left((3 {-} \dim) + \beta { \partial \over \partial \beta}
    \right) g^0_{ab} \right] \beta^{3} \,
\avg{ n_{a}^0 \, e^{i e_{a} \tilde \phi }
    n_{b}^0 \, e^{i e_{b} \tilde\phi } }
    \,.
\end{eqnarray}

To compute the two and three loop contributions to the
anomaly $A$, we first note that since to these orders the
number densities are given by
\begin{equation}
\bar n_a = n_a^0 \avg { e^{i e_a \tilde\phi} }
+  h^0_a \, \beta^2  \, e^2_a \, n_a^0 \,
\avg{ \left(\nabla\tilde\phi\right)^2
e^{i e_a \tilde\phi } }
 - 2 \sum_b \beta^3 \, g^0_{ab}
\, n_a^0 \, n_b^0
\avg { e^{i e_a \tilde\phi}  e^{i e_b \tilde\phi} } \,,
\end{equation}
we may write the pressure (\ref{dilp}) as
\begin{eqnarray}
\dim \beta p  &=& - (\dim {-} 2) {1 \over 2 \beta }
 \avg{ \left(\nabla\tilde\phi\right)^2 }
+  \sum_a  \left[ \dim \bar n_a
+  2 \, e_a^2 \, n^0_a \,h^0_a \, \beta^2 \,
\avg{ \left(\nabla\tilde\phi\right)^2
e^{i e_a \tilde\phi } } \right]
\nonumber\\
&&  \qquad\qquad
+ \dim \sum_{a,b}\beta^3 \,g^0_{ab} \,
n_a^0 \, n_b^0 \avg{ e^{i e_a \tilde\phi}
e^{i e_b \tilde\phi} }  \,.
\end{eqnarray}
The pressure and densities are, of course, well defined in the $\dim
\to 3$ limit.
Hence, in the above expression, the pole in $\dim{-}3$
in the final term,
coming from $g^0_{ab}$, must be canceled by a similar pole,
with opposite residue, in $\avg {(\nabla \tilde \phi)^2}$.
Since the contribution of the coupling $h^0_a$ has a coefficient that
is already of two-loop order, to the order to which we work,
\begin{equation}
 \lim_{\dim\to3} \> (3{-}\dim) \, {1 \over 2 \beta }
\avg{ \left(\nabla\tilde\phi\right)^2 }
=
 3 \sum_{a,b}\beta^3
 \left[\lim_{\dim\to3} \> (3{-}\dim) \, g^0_{ab} \right]
n_a^0 \, n_b^0 \avg{ e^{i e_a \tilde\phi} e^{i e_b \tilde\phi} }  \,.
\end{equation}
Recalling the result (\ref{tworen}) for $g^0_{ab}$, we have
\begin{equation}
\lim_{\dim\to3} \> ( 3 {-} \dim) \, g^0_{ab}
= -
{1 \over 4!} \, { (e_a e_b)^3 \over (4\pi)^2} \,.
\end{equation}
Since $ \lambda_{ab}^2 \sim \beta $,
Eq's.~(\ref{gab}), (\ref {eq:muab}), and (\ref {eq:Gam}) also inform us
that, in the $\dim \to 3$ limit,
\begin{equation}
    \beta  \, {\partial  g^0_{ab} \over \partial \beta }
    =
    - { 1 \over 4!} \, {(e_a e_b)^3 \over (4\pi)^2 }
    \left[ 1 + \half \, \eta_{ab} \, \Gamprime ab \right]
    \,,
\end{equation}
where $\Gamprime ab \equiv d\Gamma_{ab}/d\eta_{ab}$ is given
in Eq.~(\ref {eq:Gamprime}).  Finally, Eq.~(\ref{eq:h0a}) gives
\begin{equation}
\beta^3 {\partial \over \partial \beta} h_a^0 =
- {\lambda_a^2 \over 48 \pi} \,.
\end{equation}

Thus, reverting to the conventionally normalized field $\phi$, and
discarding terms of higher order,
\begin{eqnarray}
A  &=& - {1 \over 4!} \sum_{a,b} \beta^3 \, n_a^0 \, n_b^0 \>
    { (e_a e_b)^3 \over (4\pi)^2 } \,
    \avg{ \exp \left\{ i \beta \left( e_a + e_b \right) \phi(\r) \right\} }
    \Big[ 3 - 2 - \half \, \eta_{ab} \, \Gamprime ab \Big]
\nonumber\\
&& \qquad\qquad
  - \sum_a {\lambda_a^2 \over 48 \pi} \, e_a^2 \, \beta^2 \, n_a^0
\avg{ \left( \nabla\phi \right)^2 } \,.
\end{eqnarray}
In the last line we use the result (\ref{eq:gulp}),
\begin{equation}
\beta \,\avg{ \left( \nabla\phi \right)^2 } =
{\kappa^3 \over 4\pi} \,.
\end{equation}
For the remaining terms, we expand
the exponential involving $\phi$ to second order
to generate the sub-leading (three loop) contribution.
It involves, in the physical $ \dim \to 3$ limit
\begin{equation}
\beta \avg{ \phi^2 } = \lim_{\dim \to 3} G_\dim({\bf 0})
= - {\kappa_0 \over 4\pi} \,.
\end{equation}
Expanding $(e_a{+}e_b)^2$,
the terms involving $e_a^2$ and $e_b^2$ just provide the one-loop
corrections that alter the bare densities $n_a^0$ and $n_a^0$ to the
physical densities $\bar n_a$ and $\bar n_b$. Hence, only the cross
term provides a non-trivial correction, and we have
\begin{eqnarray}
A  &=& - { 1 \over 4! } { 1 \over (4\pi)^2 }
\sum_{a,b} \bar n_a \, \bar n_b \,
(\beta \, e_a e_b)^3 \,
\Big[ 1 - \half \, \eta_{ab} \, \Gamprime ab \Big]
\left( 1 + { \beta \, e_a e_b \, \bar\kappa \over 4 \pi } \right)
\nonumber\\
&& \qquad\qquad
- \sum_a {\lambda_a^2 \kappa^2 \over 48 \pi} \, {\beta  e_a^2
        \kappa \over 4 \pi } \, \bar n_a \,.
\end{eqnarray}
This agrees with Eq.~(\ref{eq:3lanom}).

\ifelsevier
\newpage
\fi
\section {Long Distance Correlations}
\label {sec:longdist}

\begin {figure}[th]
    \begin {center}
    \leavevmode
    \def\epsfsize #1#2{0.4#1}
    \epsfbox {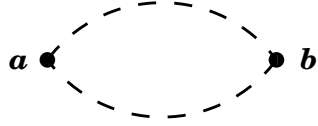}
    \end {center}
\caption
    {%
    Correction to the irreducible density-density correlator $C_{ab}$
    generated by the induced interaction $S_{\rm non-static}$
    involving the non-zero frequency modes of $\phi(\r,\tau)$.
    The dashed lines represent the long-range, unscreened Coulomb
    Green's functions of the non-zero frequency modes.
    }
\label {fig:tail}
\end {figure}

As noted in section III, the interaction (\ref {eq:Shhb}) which
couples the static and non-zero frequency modes of the electrostatic potential
only affects thermodynamic quantities at six-loop order.
However, this term does generate some qualitatively new effects
in correlation functions.
In particular, it destroys the exponential screening of the
quasi-classical theory \cite {alg-decay1,alg-decay2,alg-decay3}.
This effect is easy to calculate using the effective theory
as given in Eq's.~(\ref {eq:neweff}) and (\ref {eq:Shhb}).

We will first examine the workings of this effect on the
single-particle irreducible part $C_{ab}({\bf r} {-} {\bf r}')$ of the
number density correlation function.  The graph of
Fig. \ref{fig:tail} is produced if each of the variational
derivatives in the definition (\ref {eq:Keff}) of the number density
correlator act on $S_{\rm non-static}$ in the functional integral
(\ref{eq:neweff}).  The non-zero-frequency potentials that this
brings down from the exponential become tied together into the
product of two unscreened, long-ranged Coulomb Green's functions.
Since the result is single-particle irreducible, it defines an
$O\left[(\beta e^2 \kappa)(\kappa\lambda)^4\right]$
correction to $C_{ab}({\bf r} {-} {\bf r}')$.%
\footnote
    {%
    More precisely, a correction of relative size
    $O\!\left[(\beta e^2 \kappa)(\kappa\lambda)^4\right]$
    for wave numbers of $O(\kappa)$.
    }
Explicitly, the calculation that we
have just described gives the long-ranged contribution
\begin {equation}
    \Delta C_{ab}(\r {-} \r')
    =
    {1 \over (4\pi)^2} \>
    n_a \, n_b \, e_a^2 \, e_b^2 \, \lambda_a^2 \, \lambda_b^2 \>
    \,  \beta^2 \sum_{m\ne 0} {2 \over (2\pi m)^4} \,
    \Bigl[
	\nabla_k \nabla_l V_C(\r{-}\r')
    \Bigr]^2 \,.
\end{equation}
Here
\begin{equation}
 \sum_{m\ne 0} {2 \over (2\pi m)^4 }
= { \zeta(4) \over 4 \pi^4 } = {1 \over 360 } \,,
\end{equation}
and
\begin{equation}
 \Bigl[ \nabla_k \nabla_l V_C(\r{-}\r') \Bigr]^2 = { 6 \over (4\pi)^2 \, | \r {-} \r' |^6 } \,,
\end{equation}
and so
\begin{equation}
 \Delta C_{ab}(\r {-} \r') =
    {1 \over 60} \,
    { (\beta \, e_a^2 \, n_a \, \lambda_a^2 ) \,
      ( \beta \, e_b^2 \, n_b \,  \lambda_b^2 ) \over
    (4\pi)^4 \, |\r{-}\r'|^6 } \,.
\label{terrible}
\end {equation}
Consequently, density-density correlations do not, in fact,
decay exponentially but rather have long-distance $1/r^6$ tails.

We may use the relation given in  Eq.~(\ref{gginver})
connecting the electrostatic Green's function
${\cal G}({\bf r} {-} {\bf r}')  = \avg {\phi(\r) \phi(\r')}$ and
$C_{ab}$,
\begin {eqnarray}
    \beta \, \tilde {\cal G}(\k)
    =
    \biggl[ k^2 + \beta \sum_{a,b} e_a e_b \,
\tilde C_{ab}(\k) \biggr]^{-1} \,,
\label{remind}
\end {eqnarray}
to find the long-distance tail in ${\cal G}({\bf r} {-} {\bf r})$.
Treating $\Delta C_{ab}(\r {-} \r')$ as a perturbation and
noting that the long-distance limit of the unperturbed Green's
function is given by the Debye screened function, we have%
\footnote
    {%
    Treating $\Delta C_{ab}(\r{-}\r')$ as a perturbation is legitimate,
    even though it determines the leading long distance behavior.
    One may show this rigorously by
    noting that $\Delta \tilde C_{ab}(\k) \sim |k|^3$ for small $k$,
    and that this controls the discontinuity of $\tilde {\cal G}(\k)$
    when $k^2$ is small and negative.
    }
\begin{equation}
\beta \Delta {\cal G}({\bf r} {-} {\bf r}') = - \int (d^3 \r_1) (d^3 \r_2) \>
{ e^{-\kappa |\r {-} \r_1| } \over 4 \pi |\r {-} \r_1| } \>
\beta \sum_{a,b} e_a e_b \Delta C_{ab}(\r_1 {-} \r_2) \,
{ e^{-\kappa |\r_2 {-} \r'| } \over 4 \pi |\r_2 {-} \r'| } \,.
\end{equation}
Since the flanking Debye Green's functions that appear here are of
short range, to obtain the long-distance behavior of
$ \beta \Delta  {\cal G}({\bf r} {-} {\bf r}') $ we may replace
$ \Delta C_{ab}(\r_1 {-} \r_2) $ by $ \Delta C_{ab}(\r {-} \r') $ and use
\begin{equation}
\tilde G({\bf 0}) = \int (d^3 \r) \> { e^{- \kappa r} \over 4 \pi r} =
{ 1 \over \kappa^2 }
\end{equation}
to find that the potential correlator
also acquires a $1/r^6$ tail,
\begin {equation}
  \Delta  {\cal G}(\r {-}\r')
    \sim
    -
    {1 \over 60} \,
    \sum_{a,b} \>
    { (\beta \, e_a^3 \, n_a \, \lambda_a^2 ) \,
      ( \beta \, e_b^3 \, n_b \,  \lambda_b^2 ) \over
    (4\pi \kappa)^4 \, |\r{-}\r'|^6 } \,.
\label {eq:DelG}
\end{equation}
Comparing the magnitude of this $1/r^6$ tail to the original
$e^{-\kappa r} / 4\pi r$ Debye potential,
one finds that the cross-over from exponential to power-law decay
occurs at the parametric scale
\begin {equation}
    \kappa r \sim
    -\ln \!\left[
	(\lambda \kappa)^4 (\beta e^2 \kappa)
    \right].
\label {eq:crossover}
\end {equation}
This characterizes the number of $e$-foldings over which exponential
Debye screening could, in principle,
be observed before the power-law tail takes over.

The function $C_{ab}(\r {-} \r')$ describes the `single-particle'
irreducible part of the density-density correlation function.
The (Fourier transform) of the complete correlator is
given, according to Eq.~(\ref{kstruct}), by
\begin{equation}
\tilde K_{ab}(\k) = \tilde C_{ab}(\k)
- \left[ \beta {\sum}_c e_c \tilde C_{ca}(\k) \right]
 \,\tilde {\cal G}(\k) \,
\left[ \beta {\sum}_c e_c \tilde C_{cb}(\k) \right].
\label{reminded}
\end{equation}
Recalling [Eq.~(\ref{lowestc})] that
$ C_{ab}(\k) = \delta_{ab} \, n_a $
in lowest order, and employing the reasoning just used to find the
long-distance behavior of the potential correlator, we see that
the leading long-distance behavior of the complete density-density correlation
function is given by
\begin{eqnarray}
K_{ab}(\r {-} \r') &\sim&  \Delta C_{ab}(\r {-} \r')
- \beta e_a n_a \, \Delta \tilde {\cal G}(\r {-} \r') \,
	\beta e_b n_b
\nonumber\\
&&
    -  {\sum}_c
\left[
    \Delta C_{ac}(\r {-} \r') \, e_c \, {1 \over \kappa^2 } \, \beta e_b n_b
+
    \beta e_a n_a
    { 1 \over \kappa^2}\, e_c \, \Delta C_{cb}(\r {-} \r')
\right] ,
\end{eqnarray}
or
\begin{eqnarray}
K_{ab}(\r {-} \r') \sim { 1 \over 60} \,
{\beta^2 \over (4\pi)^4 \, | \r - \r' |^6 } \>
\sum_{cd}
\left[ \delta_{ac} - { \beta e_a \, n_a \, e_c \over \kappa^2 } \right]
\left[ \delta_{bd} - { \beta e_b \, n_b \, e_d \over \kappa^2 } \right]
(e^2_c \, n_c \, \lambda^2_c) \, (e^2_d \, n_d \, \lambda^2_d) \,
.
\nonumber\\
\label {eq:asymKab}
\end{eqnarray}

The square brackets appearing in (\ref {eq:asymKab}) function
as projection operators into the subspace orthogonal to the
charge vector $\{ e_a \}$.
Hence, the number-density---charge-density correlator constructed from this
result by multiplying by $e_b$ and summing over $b$ (or
multiplying by $e_a$ and summing over $a$) vanishes.  Consequently, the
number-density---charge-density correlation function must vanish more rapidly
than $1/r^6$ at large distances.  The leading behavior is obtained by
returning to the general forms (\ref{reminded}) and (\ref{remind}) to
write
\begin{equation}
    {\sum}_b \, \tilde K_{ab}(\k) \, e_b  =
k^2 \,  \left[ \beta \, {\sum}_c \, e_c \, \tilde C_{ca}(\k) \right]
 \,\tilde {\cal G}(\k) \,.
\label{return}
\end{equation}
Again using the same reasoning to secure the asymptotic form, with the
factor of $k^2$ in the Fourier transform becoming $ - \nabla^2$ in the
spatial form, gives
\begin{equation}
    {\sum}_b \, e_b \,K_{ab}(\r {-} \r')
\sim -{ 1 \over 2}
    {\beta^2 \over (4\pi)^4 \, \kappa^2 \, | \r - \r' |^8 }  \>
\sum_{cd}
\left[ \delta_{ac} - { \beta e_a \, n_a \, e_c \over \kappa^2 } \right]
(e^2_c \, n_c \, \lambda^2_c) \, (e^3_d \, \lambda^2_d \, n_d) \,.
\label {eq:asymKa}
\end{equation}
The charge-density---charge-density correlation function
$K(\r {-} \r')$ formed
from this result (by multiplying by $e_a$ and
summing over $a$) again vanishes. The charge-density---charge-density
correlator again vanishes more rapidly at infinity. This final correlation
function may be obtained by multiplying Eq.~(\ref{return}) by $e_a$
and summing over $a$. The result is equivalent to the previous relation
Eq.~(\ref {eq:KQM2}), and we find that the charge-density---charge-density
correlator $K(\r {-} \r')$ acquires a $1/r^{10}$ tail,
\begin {equation}
   K(\r {-} \r')
    \sim
     {28 \over |\r-\r'|^{10} } \,
     \left[
    {\sum}_a \>
    { \beta \, e_a^3 \, n_a \, \lambda_a^2 \over (4\pi)^2 \, \kappa^2 }
    \right]^2
\label{kasymp}
\end {equation}
When specialized to the case of a one-component plasma in the presence
of a constant neutralizing background, Eq.~(\ref{kasymp}) becomes
\begin{equation}
    K(\r)
    \sim
     28 \,
    {e^2  \over (4\pi)^4 } \,
    {\lambda^4 \over r^{10} }
    = { 7 e^2 \over (4 \pi)^2 } \left( { \beta \hbar^2 \over m}
  \right)^2 { 1 \over r^{10} } \,.
\end {equation}
The asymptotic results (\ref{eq:asymKab}), (\ref {eq:asymKa}),
and (\ref {kasymp})
for the correlation functions agree with the results of the
calculations of Cornu and Martin~\cite {alg-decay3} and
Cornu~\cite{cornu}.
Note that, once the appropriate effective
field theory is constructed, the single key result (\ref{terrible})
is obtained in only a few lines.

The fact that the charge-density--particle-density correlators
fall off as $1/r^8$, or two powers of $r$
faster than the $1/r^6$ tail of the
potential correlator ${\cal G}$,
and that the charge-density---charge-density correlator falls off
yet faster by another two powers,
is a direct consequence of the Poisson equation
$
    i\nabla^2 \phi(\r) = \rho(\r)
$
relating the charge density and electrostatic potential.%
\footnote
    {
    For the charge-density---particle-density correlator,
    this argument relies on the relation
    connecting the particle-density---potential
    correlator $\avg {n_a(\r) \phi(\r')}$
    with $\cal G$ and $C_{ab}$, namely
    $
	\avg {n_a(\r) \phi(\r')}
	=
	\int (d^3\r'') \>
	\sum_b i\beta \, e_b \, C_{ab}(\r{-}\r'')\, {\cal G}(\r''{-}\r')
    $.
    This relation may be easily derived graphically,
    and also follows from Eq.~(\ref {gamgamb}) and the immediately following
    discussion in Appendix \ref {funmeth}.
    Since $\cal G$ and $C_{ab}$ both have $1/r^6$ tails,
    the same analysis used above, applied to this relation,
    shows that a $1/r^6$ tail is also present in the
    particle-density---potential correlator.
    And hence the Laplacian of this result, which yields
    the particle-density---charge-density correlator,
    will have a $1/r^8$ tail.
    }
The positive overall sign of the asymptotic charge-charge correlator
(\ref {kasymp}) reflects the fact that a charge fluctuation of a given sign
will attract oppositely charged particles and repel
similarly charged particles.
Therefore, the long-distance tail of the charge-charge correlator
shows a positive correlation.
Essentially the same argument may also be used to confirm the
overall sign of the charge-density---number-density correlator
(\ref {eq:asymKa}).
If the charge $e_a$ of species $a$ is arbitrarily small
(so that it is, in effect, a test charge),
then Eq.~(\ref {eq:asymKa}) becomes
\begin{equation}
    {\sum}_b \, e_b \,K_{ab}(\r {-} \r')
    \sim { 1 \over 2}
    {\beta \, n_a \, e_a \over | \r - \r' |^8 }  \>
     \left[
    {\sum}_c \>
    { \beta \, e_c^3 \, n_c \, \lambda_c^2 \over (4\pi)^2 \, \kappa^2 }
    \right]^2
    +
    O(e_a^2)
,
\end{equation}
showing that, in linear response, a test charge of species $a$
creates an induced charge density whose long-distance tail
is positively correlated with the sign of the test charge.%
\footnote
    {
    A variant of this same argument appears in \cite {cornu}.
    }

It should be emphasized
that the above results for long distance tails
were derived for zero-frequency correlator functions.
One may wish to consider equal-time correlators instead.
The difference between the two is determined by the behavior of
non-zero frequency correlations.
If $K_{ab}(\r,\omega_n)$ denotes the density-density
correlator at spatial separation $\r$ and (Matsubara) frequency $\omega_n$,
then the equal-time density-density correlator is
\begin {equation}
    K_{ab}(\r,t {=}0)
    = {1 \over \beta } \,
    \sum_n K_{ab}(\r,\omega_n) \,,
\end {equation}
which differs from the
zero-frequency correlator $K_{ab}(\r) = \beta^{-1} \,
 K_{ab}(\r,\omega{=}0)$
by the sum over all non-zero frequency components,
\begin {equation}
    K_{ab}(\r,t {=}0) - K_{ab}(\r)
    =
    \sum_{n\ne 0} \beta^{-1} \, K_{ab}(\r,\omega_n) \,.
\end {equation}
However, these non-zero frequency number density correlation functions
decrease with increasing
spatial separation {\em faster} than the zero frequency correlation
(\ref{eq:asymKab}), and consequently the equal-time and zero-frequency
correlators have the same leading long distance behavior.
To understand the long-distance behavior of the non-zero frequency correlators,
it is convenient to begin with the definition of the
correlation function at (Matsubara) frequency $\omega_m$ as a
second variational derivative with respect to the $m$'th
Fourier component of the (time-dependent) chemical potentials,
\begin {eqnarray}
   \beta^{-1} \,  K_{ab}(\r{-}\r',\omega_m)
    &\equiv&
    {\delta^2 \ln Z \over \delta \beta \mu_a^m(\r) \>
			    \delta \beta \mu_b^{-m}(\r') }
\nonumber \\&=&
    \avg {
	{\delta S_{\rm non-static} \over \delta \beta \mu_a^m(\r)}
	{\delta S_{\rm non-static} \over \delta \beta \mu_b^{-m}(\r')}
	}
    - \avg {
    {\delta^2 S_{\rm non-static} \over \delta \beta \mu_a^m(\r) \>
					\delta \beta \mu_b^{-m}(\r') }
    } \,.
\label {eq:Keff2}
\end {eqnarray}
\begin {figure}
    \begin {center}
    \vspace*{1cm}
    \leavevmode
    \def\epsfsize #1#2{0.45#1}
    \epsfbox {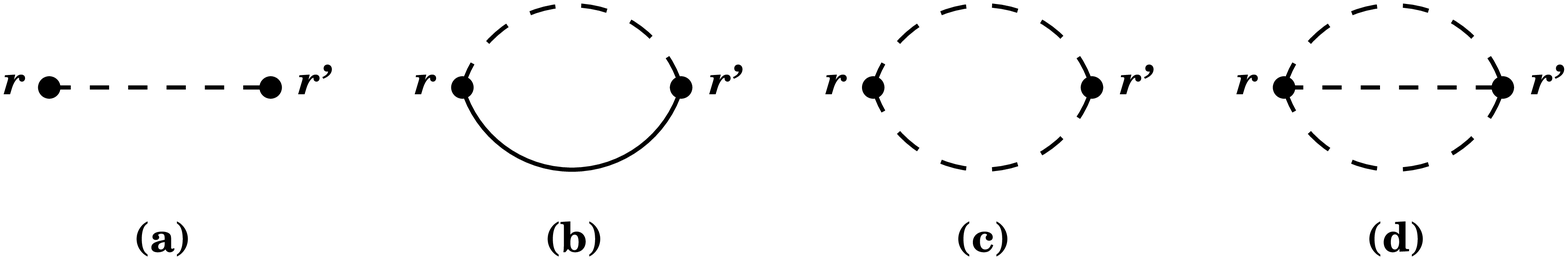}
    \vspace*{1cm}
    \end {center}
\caption
    {
    Contributions to the non-zero frequency components of the
    density-density correlator $K_{ab}(\r{-}\r',\omega_m)$.
    Dashed lines represent long-range, unscreened propagator ${\cal G}_m$
    of non-zero frequency modes,
    while solid lines represent the zero-frequency propagator ${\cal G}$.
    Some non-zero Matsubara frequency $\omega_m$ flows in at the vertex
    labeled $\r$ and out at $\r'$.
    The dashed lines in diagrams (a) and (b) must carry frequency $\omega_m$;
    in diagrams (c) and (d) the non-zero frequencies carried by each of
    the internal lines must sum to $\omega_m$.
    \label {fig:non-static}
    }
\end {figure}%
The leading terms of $S_{\rm non-static}$ are shown
in Eq.~(\ref {eq:non-static}) and further corrections are presented in
Eq.~(\ref{eq:S2}).  The resulting
tree-level, one-loop, and multi-loop contributions
are illustrated schematically in
Fig. \ref {fig:non-static}.
Determining the long-distance behavior of each contribution is
straightforward by
regarding $\r$ and $\r'$ as arbitrarily far apart and
evaluating each diagram in coordinate/frequency space.%
\footnote
    {
    In other words,
    each line of a diagram represents a propagator ${\cal G}_m(\r,\r')$
    which is Fourier transformed in imaginary time, but not in space.
    Each vertex is labeled by a spatial position, and the sum of all
    discrete frequencies (including the external frequency)
    coming into each vertex must vanish.
    }
The leading long distance behavior is determined by the minimal
number of propagators which run from the vicinity of $\r$
to near $\r'$, together with the minimal number of spatial
derivatives which act on these propagators.
As noted in Eq.~(\ref {eq:Gm})] of section~\ref {sec:quasi},
each non-zero frequency propagator is given by
\begin{equation}
{\cal G}_m(\r,\r') = \langle \phi^m(\r) \phi^{-m}(\r')\rangle
= { 1 \over 4 \pi \beta \, |{\bf r} - {\bf r}'| } \,.
\label{propp}
\end{equation}
However, the structure of $S_{\rm non-static}$ (or ultimately,
the gauge invariance of the underlying quantum theory) requires
that only the gradient $\nabla \phi^m$ of the non-zero frequency
components of $\phi$ appear in the effective theory.
Consequently, each non-zero frequency propagator ${\cal G}_m(\r,\r')$
will be acted upon, on either end, by one or more spatial derivatives,
and so each non-zero frequency propagator which runs from near $\r$ to
near $\r'$ will contribute a factor which decreases at least as fast as
$|\r{-}\r'|^{-3}$ to the result.
Taking account of the earlier discussion of Eq.~(\ref {eq:DelG}),
the zero frequency propagator ${\cal G}(\r{-}\r')$
shows exponential Debye screening out to the cross-over distance
(\ref {eq:crossover}), beyond which
it falls off like $|{\bf r}{-}{\bf r}'|^{-6}$.
Furthermore, the contribution of every diagram will also contain
at least one overall gradient with respect to $\r$,
and one with respect to $\r'$,
coming from the fact that the variational derivatives in
Eq~(\ref {eq:Keff2})
can only act on gradients of the (non-zero frequency components of the)
chemical potential appearing in
Eq.~(\ref {eq:non-static}) [augmented by Eq.~(\ref{eq:S2})].
Therefore, diagrams in which $k$ non-zero frequency propagators,
and $l$ zero frequency propagators run from (a neighborhood of) $\r$
to (a neighborhood of) $\r'$ will have
long distance behavior which falls at least as fast as
$|\r{-}\r'|^{-3k-6l-2}$.

If $k$ is even, the fall-off must actually be at least two powers faster.
This is because such contributions can only come from terms in
$S_{\rm non-static}$ which have an odd number of factors of
$\nabla (\mu_a^m + i e_a \phi^m)$.
Such terms must actually involve at least one more derivative
acting on these fields,
since rotation invariance implies that only
even numbers of gradients can appear in any term in the action.
This is shown explicitly in the final term shown in Eq.~(\ref {eq:S2}).
Consequently, in these diagrams the long range propagators are
acted upon by at least one more derivative on either end of the diagram.

Since the total frequency $\omega_m$ flowing through each diagram is,
by assumption, non-zero, every diagram must have at least one
non-zero frequency propagator crossing from $\r$ to $\r'$.
But in any single-particle reducible contribution
where only a single non-zero frequency propagator crosses from $\r$ to $\r'$,
such as diagram (a) of Fig.~\ref {fig:non-static},
the gradients appearing in the non-static action (\ref {eq:non-static})
necessarily generate Laplacians acting on either end of the propagator
so the net contribution is only a local contact term proportional to
$\nabla^2 \delta(\r{-}\r')$.
Consequently, single particle reducible diagrams such as (a)
do not contribute at all to the long-distance behavior.
Diagram (b) containing one zero-frequency, and one non-zero
frequency propagator running from $\r$ to $\r'$ has $|\r{-}\r'|^{-11}$
long-distance behavior, while diagram (c),
with two long-distance non-zero frequency propagators,
has $|\r{-}\r'|^{-10}$ fall-off, since there must be a total of eight
spatial derivatives acting on the two non-zero frequency propagators.
Multi-loop diagrams, such as (d), with three or more long-distance propagators,
necessarily decrease like $|\r{-}\r'|^{-11}$ or faster
(up to possible logarithmic factors).

Therefore, the leading long-distance behavior of the
irreducible part $ C_{ab}(\r{-}\r',\omega_m)$ of the non-zero
frequency number-density correlation function
is generated by diagram (c), and is order $|\r{-}\r'|^{-10}$.
A simple computation using the vertex (\ref{eq:vertex}) presented in
Appendix \ref{app:det}, together with the non-zero frequency
propagator (\ref{propp}), shows that diagram (c) represents the contribution
\begin{eqnarray}
	\Delta C_{ab}(\r , \omega_m) &=&
	\beta^3 \,  n_a \, n_b
	 \left[{ \lambda_a^4 e_a^2 \over (2\pi)^3}\right]
	 \left[{ \lambda_b^4 e_b^2 \over (2\pi)^3}\right]
	H(\omega_m)
	\left( \nabla_k\nabla_l\nabla_{k'}\nabla_{l'} \, {1 \over r} \right) \!
	\left( \nabla_k\nabla_l\nabla_{k'}\nabla_{l'} \, {1 \over r} \right) ,
\end {eqnarray}
where%
\footnote
    {%
    A convenient generating function for evaluating this sum is
    $
	\sum_{-\infty}^\infty \, {1 \over (n - a) (n - b)}
	= \pi \, {\cot \pi b - \cot \pi a \over a-b} \,
    $.
    }
\begin {eqnarray}
    H(\omega_m)
    &\equiv&
	{1 \over (2\pi)^8}
	\sum_{\textstyle {n = -\infty \atop n \ne 0,m}}^\infty \,
	{1 \over 2 m^4 \, n^{2} \, (m{-}n)^{2}}
    =
	{1 \over (2\pi)^8}
	\left[ {\pi^2 \over 3 \, m^6} - {3 \over m^8} \right]
\nonumber\\
    &=&
	{1 \over 12} \, (\beta \omega_m)^{-6} - 3 \, (\beta \omega_m)^{-8}
    \,.
\end{eqnarray}
With the (continuing) neglect of delta function terms which
do not affect the long-range tail,
\begin{eqnarray}
\left(\nabla_k \nabla_l \nabla_{k'} \nabla_{l'} \, {1 \over r} \right)
\left(\nabla_k \nabla_l \nabla_{k'} \nabla_{l'} \, {1 \over r} \right)
	    &\to&
    { 3 \cdot 5 \cdot 7 \over r^9 } \,\,
    r_{l'} \, r_{k'} \, r_l \, r_k \,
    \nabla_k \nabla_l \nabla_{k'} \nabla_{l'} \, {1 \over r}
\nonumber\\
&=&
     { 3 \cdot 5 \cdot 7 \over r^{10} } \> 4! \,,
\end{eqnarray}
where in the last equality repeated use was made of the dimension-counting
properties of the Euler operator, 
namely $ \r \cdot \nabla \, r^{-n} = -n \, r^{-n} $.  Thus
\begin {equation}
    \Delta C_{ab}(\r,\omega_m)
    = \beta^3 \,
	{n_a \, n_b \over r^{10} }
	\left[{ \lambda_a^4 e_a^2 \over (2 \pi)^3}\right]
	\left[{ \lambda_b^4 e_b^2 \over (2 \pi)^3}\right]
	3 \cdot 5 \cdot 7 \cdot 4! \> H(\omega_m) \,.
\label {eq:Cmtail}
\end {equation}
This result
for the long-range tail of the irreducible, non-zero frequency density-density
correlator neglects additional corrections involving higher powers of $1/r$,
and relative corrections to the coefficient of the $r^{-10}$ tail
which are suppressed by further powers of the loop expansion parameter
$\beta e^2 \kappa$ (or equivalently $\kappa \lambda$).

The Fourier transform%
\footnote
    {
    The representation
    $
	r^{-10} = {1 \over 4!} \, \int_0^\infty ds \, s^4 \, e^{-s \, r^2 }
    $
    makes the evaluation of this Fourier transform easy.
    }
of the result (\ref{eq:Cmtail}) gives
\begin {equation}
   \Delta  \tilde C_{ab}(\k,\omega_m)
    =
	|\k|^7 \,
	{n_a \, n_b } \, { \beta^3 \over 2}
	\left[{ \lambda_a^4 e_a^2 \over (4 \pi)^2} \right]
	\left[{ \lambda_b^4 e_b^2 \over (4 \pi)^2} \right]
	H(\omega_m)
	\,.
    \qquad \hbox {($m \ne 0$)}
\label {eq:Cmtail2}
\end {equation}
Note that this result involves the non-analytic term $|{\bf k}|^7 =
(k^2)^3 \, \sqrt{k^2} $. The action (\ref{eq:non-static}) for the
non-zero frequency modes gives the lowest order, analytic contribution
to the irreducible correlation function,
\begin {equation}
    \tilde C_{ab}^{(0)}(\k,\omega_m)
    =
    \delta_{ab} \, \beta \,
   {n_a \lambda_a^2 \over 2\pi} \, {\k^2 \over (2\pi m)^2}
    \,.
    \kern 1.65in \hbox {($m \ne 0$)}
\label{lowest}
\end{equation}

Using the exact relation (\ref {eq:exactk})  of
Appendix \ref {funmeth},
and the same type of analysis employed above,
the result (\ref {eq:Cmtail2}) for the irreducible correlator may be
converted into corresponding results for the long distance behavior,
at non-zero frequency,
of the full number-density---number-density correlator
$\tilde K_{ab}(\k,\omega_m)$.  It is easy to check that the leading
non-analytic part of the
single-particle reducible contribution to
$\tilde K_{ab}(\k,\omega_m)$, the second set of terms
in Eq.~(\ref{eq:exactk}), is also of order $|\k|^7$, and thus gives an
additional contribution to the $1 / r^{10}$ tail. However, these
contributions always involve at least one extra factor of
$ \tilde C_{ac}^{(0)}(\k,\omega_m) \, (e_c e_d / \k^2) $ which is a
correction of relative order
$\beta e^2 \, n \, \lambda^2 \sim (\kappa \lambda)^2 $,
or two powers of the loop expansion parameter.
Hence this correction has the same size as terms that
have already been omitted in the calculation of the irreducible part
of the correlator, and thus to our leading order,
the complete correlator at non-zero frequency has the
same long distance behavior as its irreducible part,
\begin{equation}
    K_{ab}(\r,\omega_m)
    \sim \Delta C_{ab}(\r,\omega_m)
    \sim
	{315 \over 8} \,
	{\beta^3 \over r^{10}}
	\left[{ n_a \lambda_a^4 e_a^2 \over \pi^3}\right]
	\left[{ n_b \lambda_b^4 e_b^2 \over \pi^3}\right]
	H(\omega_m) \,.
\label{eq:Kmtail}
\end{equation}

Since the non-zero frequency
number-density---number-density correlator (\ref {eq:Kmtail})
falls faster than the zero-frequency component (\ref {eq:asymKab}),
this shows that the equal-time and zero-frequency number density
correlators have the same long distance behavior.
This is equally true for the charge-density---number-density
correlator (\ref {eq:asymKa}).
But, from Eq.~(\ref {eq:Kmtail}), the charge-density---charge-density
correlator $K(\r,\omega_m) = \sum_{a,b} e_a e_b \, K_{ab}(\r,\omega_m)$
has a $1/r^{10}$ tail at non-zero frequency,
\begin {equation}
    K(\r,\omega_m)
    \sim
    {315 \over 8} \, {\beta^3 \over r^{10} }
    \left[
	{\sum}_a { n_a \, \lambda_a^4 \, e_a^3 \over \pi^3}
    \right]^2
    H(\omega_m) \,,
\label {eq:Kmtail2}
\end {equation}
just like its zero-frequency counterpart (\ref {kasymp}).
However, the size of the non-zero frequency tail (\ref {eq:Kmtail2})
is a factor of $(\lambda \kappa)^4$
smaller than the zero-frequency tail (\ref {kasymp}).
Consequently, the equal time and zero-frequency
charge-density---charge-density correlators have the same
long distance behavior up to relative corrections suppressed by
four powers of the loop expansion parameter.

When specialized to the case of a one-component plasma,
the expression (\ref {eq:Cmtail2}) for the leading non-analytic
piece of the irreducible correlator at non-zero frequency
agrees to leading order with the corresponding result
of Cornu and Martin \cite {alg-decay3}.
[We have made no effort to retain further corrections to
the $|\k^7|$ coefficient, some of which are included in the result
(3.25) of \cite {alg-decay3}, which are higher order in our
expansion parameter $\kappa\lambda$.]

\section* {Acknowledgments}

The interest of one of the authors (L.S.B) in classical plasma physics
was piqued by R.~F. Sawyer. His work on this paper began while
visiting the Los Alamos National Laboratory and continued at the Aspen
Center for Physics and was largely completed during another visit to the
Los Alamos National Laboratory.
We would like to thank H.~De~Witt
for several clarifying discussions
and particularly for making us aware of various related prior results
which were helpful in resolving interim discrepancies.
Communications with W.-D.~Kraeft and M.~Schlanges were also helpful
in this regard.
We would also like to thank T.~Kahlbaum for informative communications
and useful comments.
This work was supported, in
part, by the U.S. Department of Energy under Grant No.
DE-FG03-96ER40956.

\newpage

\appendix

\section{Functional Methods}
\label{funmeth}

In this Appendix, we define, in the context of our plasma theory, the
generating function of connected correlation functions and its
Legendre transform, the effective action.%
\footnote
    {
    Our discussion of
    the effective action for a plasma parallels that given for quantum
    field theory in Sections 4 and 5 of Brown \cite{brown}, Chapter 6,
    which contains many more details.
    }
 We review relevant properties of these functionals that
are well known in quantum field theory, and then describe how number
densities and density--density correlation functions are related to
them. In particular, we show how the density-density correlator may be
expressed in terms of a ``single-particle irreducible'' function in a
way that explicitly exhibits its structure, particularly its small
wave-number behavior.  We also show how the mean-square fluctuations
in energy, and particle numbers, may be expressed in terms of the same
single-particle irreducible functions.  This formalism is applied to
compute the number densities, density--density correlators, and
equation of state to two loop order in an particularly efficient
manner in a final appendix so as to illustrate methods complimentary
to those employed in the text.

The partition function of our theory,
in either its original quantum form (\ref {eq:Zquant}),
or re-expressed as an effective theory (\ref {eq:neweff}),
has a functional integral representation
\begin {equation}
    Z[\mu] = N \int [d\phi] \> \exp \left\{ -S[\phi;\mu] \right\} ,
\end {equation}
where $[d\phi]$ denotes functional integration over a space and time
dependent potential $\phi(\r,\tau)$ which is periodic,
$\phi(\r,\beta) = \phi(\r,0)$.
The action $S[\phi;\mu]$ has the form
\begin {equation}
    S[\phi;\mu] =
    {1\over2} \int_0^\beta d\tau \int (d^\dim\r) \>
    \bigl[\nabla \phi(\r,\tau) \bigr]^2 + \Sint[\phi;\mu] \,.
\label{actform}
\end{equation}
In the original quantum theory the interaction part of the action
$\Sint$ is (minus the logarithm of) the
functional integral over all charged fields,
\begin {equation}
    \exp \left\{ -\Sint [\phi;\mu] \right\}
    =
    \prod_a \int [d\psi_a^* \, d\psi_a] \>
    \exp \biggl\{ -\int_0^\beta d\tau \int(d^\dim\r) \> {\cal L}_a \biggr\} \,,
\end {equation}
with ${\cal L}_a$ the charged field Lagrangian defined in (\ref
{eq:La}), but with the chemical potentials now extended to be
functions of imaginary time as well as space, $\mu_a(\r) \to mu_a(x) =
\mu_a(\r,\tau)$.  In the effective theory, $\Sint$ is the sum of the
classical interaction and the various induced interactions,
\begin {equation}
    \Sint[\phi;\mu]
    =
    -\beta \int (d^\dim\r)\> \sum_a \n0a \, e^{i \beta e_a \phi^0}
    + \Sind[\phi^0;\mu] + S_{\rm non-static}[\phi^0,\phi^m;\mu] \,,
\end {equation}
with $\Sind$ and $S_{\rm non-static}$ given in
Eq's.~(\ref {Sind}) and (\ref {eq:non-static}), respectively,
and $\{ \phi^m(\r) \}$ denoting the Fourier components of $\phi(\r,\tau)$,
as defined in Eq.~(\ref {eq:fourier-phi}).

In the following formal discussion,
we will allow the generalized chemical potentials $\mu(\r,\tau)$
to vary both in space and (imaginary) time.
The only feature of the
interaction terms which will be relevant is the fact that
$i\phi(\r,\tau)$ couples to the total charge density
via the dependence of $ S_{\rm int}[\phi;\mu] $ on the generalized
chemical potentials $\mu_a(\r,\tau)$,
or
\begin{equation}
    {\delta \over \delta \phi(\r,\tau) } \, S_{\rm int}[\phi;\mu] =
    i \, \sum_a e_a \, {\delta \over \delta \mu_a(\r,\tau) } \,
    S_{\rm int}[\phi;\mu] \,.
\label{chmud}
\end{equation}
This is a reflection of the invariance (\ref {eq:shift-sym})
of the theory under the combined shifts
$\phi\to\phi -i c$ and $\mu_a \to \mu_a -e_a c$.

In the following discussion, for notational convenience,
we will use single symbols $x$, $y$, {\em etc}.,
to denote a (Euclidean) space-time coordinate
so that, for example, $\phi(x) \equiv \phi(\r,\tau)$.
And we will write $\int_x$ as shorthand for
$\int_0^\beta d\tau \int (d^\dim\r)$.

\subsection {Connected Generating Functional}

The addition of an external charge density or source $\sigma(x)$
coupled to the field $\phi(x)$ defines a functional
$W[\sigma;\mu]$ which is the generating functional for connected
$\phi$ field correlation functions ---
correlators whose graphical representations
have no disconnected parts. The definition is
\begin{equation}
    \exp { W[\sigma;\mu] }
    =
    \int [d\phi] \>
    \exp\left\{ - S[\phi;\mu] + \int_x \> \phi(x) \, \sigma(x) \right\} .
\label{wdef}
\end{equation}
In the presence of the source, a normalized thermal expectation value
is defined by
\begin{equation}
    \langle F[\phi] \, \rangle^\sigma_\beta =
    e^{- W[\sigma;\mu] } \,
    \int [d\phi] \> F[\phi] \,
    \exp\left\{ - S[\phi;\mu]
    + \int_x \> \phi(x) \, \sigma(x) \right\} ,
\end{equation}
and in terms of this expectation value
\begin{equation}
{ \delta W[\sigma;\mu] \over \delta \sigma(x) } =
\langle \phi(x)  \rangle^\sigma_\beta \,.
\label{varw}
\end{equation}

The insertion in the functional integrand of the functional derivative
of $-S_{\rm int}[\phi;\mu] $ with respect to a generalized chemical
$\mu_a(x)$ produces the average particle number density,
up to an overall factor of $e^{W[\sigma;\mu]}$.
Thus the properly normalized particle number density of species
$a$ is given by
\begin{equation}
\langle n_a(x) \rangle^\sigma_\beta =
{ \delta W[\sigma;\mu] \over \delta \mu_a(x) } \,.
\label{nden}
\end{equation}
This is the number density in the presence of both
spatially (or temporally) varying chemical potentials $\mu_b(x)$ and the
external charge density $\sigma(x)$. With the chemical potentials
$\{ \mu_a \}$ taken to be constants
and $\sigma$ taken to vanish, Eq.~(\ref{nden}) reduces to the
constant number density $\bar n_a = \langle n_a \rangle_\beta  $ of
particles of species $a$. We shall denote this limit by a vertical bar
with a subscript 0. Thus,
\begin{equation}
\bar n_a = \left.
{ \delta W[\sigma;\mu] \over \delta \mu_a(x) } \right|_0 \,.
\end{equation}
The total charge density in the presence of all
the sources is given by
\begin{equation}
\langle \rho(x)  \rangle^\sigma_\beta =
{\sum}_a \, e_a \, \langle n_a(x)  \rangle^\sigma_\beta =
{\sum}_a \, e_a \, { \delta W[\sigma;\mu] \over \delta \mu_a(x) } \,.
\end{equation}
The partial derivative of $W[\sigma;\mu]$
with respect to the inverse temperature%
\footnote
    {%
    When varying $\beta$, the Fourier components
    $
	\mu_a(\r,\omega_m) \equiv
	\int_0^\beta \mu_a(\r,\tau) \,
	e^{i \omega_m \tau} \> d\tau
    $
    and
    $
	\sigma(\r,\omega_m) \equiv
	\int_0^\beta \sigma(\r,\tau) \,
	e^{i \omega_m \tau} \> d\tau
    $
    (with $\omega_m \equiv 2\pi m / \beta$)
    are to be held fixed.
    }
defines the average energy in the presence of the source $\sigma$,
\begin{equation}
\langle E \rangle_\beta^\sigma =
-  { \partial W[\sigma;\mu] \over \partial \beta } \,.
\label{aveE}
\end{equation}
In the limit of vanishing source and constant chemical potentials,
this reduces to the thermodynamic internal energy,
\begin{equation}
U = \langle E \rangle_\beta =
- \left. { \partial W[\sigma;\mu] \over \partial \beta } \right|_0 \,.
\end{equation}
Second variations with respect to the chemical potentials produce
the number density correlation function,
\begin{equation}
K_{ab}(x {-} x') = \left. {\delta^2 W[\sigma;\mu] \over
    \delta \mu_a(x) \, \delta \mu_b(x') }
\right|_0 \,.
\label{kkabb}
\end{equation}
The static correlator discussed in the text is just the time average
of this space-time dependent correlator,
\begin {equation}
    K_{ab}(\r{-}\r')
    =
    \beta^{-1} \int_0^\beta d\tau \> K_{ab}(\r{-}\r',\tau{-}\tau') \,.
\end {equation}
We shall also have occasion to use the $\phi$ field
correlation function defined by
\begin{equation}
{\cal G}(x {-} x') = \left. {\delta^2 W[\sigma;\mu] \over
    \delta \sigma(x) \, \delta \sigma(x') }
\right|_0 \,.
\end{equation}

Since the functional integral of a total functional derivative
vanishes,
\begin{equation}
0 = \int [d\phi] \, {\delta \over \delta \phi(x) }
\exp\left\{ - S[\phi;\mu] + \int_y \phi(y) \, \sigma(y) \right\} ,
\end{equation}
the functional integral with an extra factor of
\begin{equation}
{\delta S[\phi;\mu] \over \delta \phi(x) } - \sigma(x)
\end{equation}
included in the integrand vanishes.
Hence, in view of the form (\ref{actform}) of the action and
the result (\ref{chmud}),
the expectation value of the field equation is an exact identity:
\begin{eqnarray}
- \nabla^2 \langle \phi(x)  \rangle^\sigma_\beta &=&
i \sum_a e_a {\delta W[\sigma;\mu] \over \delta \mu_a(x) } + \sigma(x)
\nonumber\\
&=&
i \langle \rho(x) \rangle^\sigma_\beta + \sigma(x) \,.
\label{fieldeq}
\end{eqnarray}

\subsection {Effective Action}

The effective action functional
$\Gamma[\bar\phi;\mu]$
is defined by a Legendre transform of the generating functional
$W[\sigma;\mu]$. It generalizes the mean field theory described in
Sec. \ref{mean} to include the effects of thermal and quantum
fluctuations.
The effective action functional has two important properties: Not
only does it contain only connected graphs (as does $W$), it contains
no single-particle reducible graphs --- graphs which can be cut into
two disjoint pieces by cutting a single line.  This is shown
explicitly to two-loop order in Appendix \ref{twoloopyy} below.
This property simplifies calculations.  For example,
when $\Gamma$ is used to compute the free
energy, one can simply delete all ``tadpole'' graphs.
Moreover, as we shall see,
the use of the effective action together with the functional relations
that we are developing reveals the basic structure of the theory
in a very useful form.
The effective action functional is obtained by setting
\begin{equation}
    \langle \phi(x) \rangle^\sigma_\beta = \bar\phi(x) \,,
\label{bpdef}
\end{equation}
that is, by considering the field expectation value rather than the
source to be the independent variable. The effective action
is then defined by the Legendre transformation
\begin{equation}
    \Gamma[\bar\phi;\mu] \equiv
    \int_x \sigma(x) \, \bar\phi(x) - W[\sigma;\mu] \,.
\label{gdef}
\end{equation}
Because of Eq's.~(\ref{varw}), (\ref{nden}), and (\ref{aveE}),
\begin{equation}
    \delta \Gamma[\bar\phi;\mu] = \int_x \left\{ \sigma(x)
    \, \delta \bar\phi(x) - \langle n_a(x)  \rangle^\sigma_\beta
    \, \delta \mu_a(x) \right\} + \langle E \rangle_\beta^\sigma \> d \beta \,.
\label{vargam}
\end{equation}
Thus we may consider $\Gamma$ to be a functional of the independent
variables $\bar\phi(x)$, $\mu_a(x)$, and $\beta$, with the (partial)
functional derivatives
\begin{equation}
    { \delta \Gamma[\bar\phi;\mu] \over \delta \bar\phi(x) }
    = \sigma(x) \,,
\label{dgdp}
\end{equation}
\begin{equation}
- {\delta \Gamma[\bar\phi;\mu] \over \delta \mu_a(x) } =
\langle n_a(x) \rangle^\sigma_\beta \,,
\label{dgdm}
\end{equation}
and the ordinary partial derivative
\begin{equation}
 {\partial \Gamma[\bar\phi;\mu] \over \partial \beta } =
\langle E \rangle^\sigma_\beta \,.
\label{dgdb}
\end{equation}
In view of Eq.~(\ref {dgdp}), evaluating the effective action
at a stationary point, a point where $\delta\Gamma/\delta\bar\phi = 0$,
is the same as setting the source $\sigma$ to zero.
With constant chemical potentials and a vanishing source, the last
equalities reduce to the ordinary number density and internal energy.
As remarked in the text, the grand canonical partition function is
related to the grand potential by $ Z = \exp\{ - \beta \Omega \} $,
and so the grand potential
is given by the effective action evaluated
at its stationary point in the limit of constant chemical potentials,
\begin{equation}
    \beta \Omega = \left. \Gamma[\bar\phi;\mu] \right|_0 \,.
\end{equation}

We return
momentarily to consider $\sigma(x)$ and $\mu_a(x)$ as
independent variables so as to compute the source functional
derivative of Eq.~(\ref{dgdp}):
\begin{equation}
    \int_y
    { \delta^2 \Gamma[\bar\phi;\mu] \over
    \delta \bar\phi(x) \, \delta \bar\phi(y) }
    { \delta \bar\phi(y) \over \delta \sigma(x') } =
    \delta ( x {-} x') \,,
\end{equation}
or, in view of Eq.~(\ref{varw}),
\begin{equation}
    \int_y { \delta^2 \Gamma[\bar\phi;\mu] \over
    \delta \bar\phi(x) \, \delta \bar\phi(y) } \,
    {\delta^2 W[\sigma;\mu] \over
    \delta \sigma(y) \, \delta \sigma(x') }
    =
    \delta ( x {-} x') \,.
\end{equation}
In the limit of constant chemical potentials and a vanishing source,
this becomes
\begin{equation}
    \int_y \left.
    { \delta^2 \Gamma[\bar\phi;\mu] \over \delta \bar\phi(x) \,
    \delta \bar\phi(y) } \right|_0
    {\cal G}(y {-} x' )
    = \delta ( x {-} x') \,.
\label {eq:GGinv}
\end{equation}
Thus
\begin{equation}
\left.
{ \delta^2 \Gamma[\bar\phi;\mu] \over \delta  \bar\phi(x) \,
\delta \bar\phi(x') } \right|_0 = {\cal G}^{-1}( x {-} x' )
\end{equation}
is the operator inverse to the potential correlation function
${\cal G}(x {-} x') $.
After a Fourier transform [in space and (periodic) time],
\begin {equation}
    \tilde{\cal G}(\k,\omega_m)
    \equiv
    \int_0^\beta d\tau \int (d^\dim\r) \>
    e^{-i\omega_m \tau + i \k \cdot \r} \,
    {\cal G}(\r,\tau) \,,
\end {equation}
the linear integral equation (\ref {eq:GGinv})
reduces to the algebraic relation:
\begin{equation}
\tilde {\cal G}^{-1}(\k,\omega_m) \, \tilde {\cal G}(\k,\omega_m) = 1 \,.
\end{equation}

To uncover the structure of the potential correlation
function, we first write the field equation (\ref{fieldeq}) for the
expectation value in terms of the effective action functional. This is
done by using Eq's.~(\ref{bpdef}), (\ref{dgdp}), and (\ref{dgdm}), to
obtain
\begin{equation}
    - \nabla^2 \bar\phi(x) = -i \sum_a \, e_a \,
    {\delta \Gamma[\bar\phi;\mu] \over \delta \mu_a(x) }
    +
    { \delta \Gamma[\bar\phi;\mu] \over \delta \bar\phi(x) } \,.
\label{gamfield}
\end{equation}
Taking the functional derivative of this relation with respect to
$\bar\phi(x')$ and then setting the chemical potentials
constant and the source to zero produces
\begin{equation}
    - \nabla^2 \delta( x {-} x') = -i \sum_a e_a
    \gamma_a( x {-} x') + {\cal G}^{-1}( x {-} x' ) \,,
\end{equation}
where we have defined a two-point vertex or coupling by
\begin{equation}
    \gamma_a(x {-} x') \equiv \left.
{\delta^2 \Gamma \over \delta \mu_a(x) \, \delta \bar\phi(x') }\right|_0  \,.
\label{ddgdmdp}
\end{equation}
Thus in wave number/frequency space
\begin{equation}
    \tilde {\cal G}^{-1}(\k,\omega_m)
    =
    \k^2 + i \, {\sum}_a \, e_a \, \tilde \gamma_a(\k,\omega_m) \,.
\label{ginv}
\end{equation}

The structure of the potential correlation function is intimately
connected to that of the number density correlation function. Hence
it is useful to examine the relationship between the effective action
and the number density correlator. Recalling the expression
(\ref{kkabb}) for this function in terms of $W[\sigma;\mu]$ and then
the fact [Eq.~(\ref{nden})] that one functional derivative defines the
number density, we see that we may write
\begin{equation}
    K_{ab}(x {-} x') =
    \left. {\delta \over \delta \mu_b(x') } \,
   \langle n_a(x)  \rangle^\sigma_\beta \right|_0 \,.
\end{equation}
In this equation, $\sigma$ and $\mu_a$ are taken to
be the independent variables, with $\bar\phi$ a function of
these independent variables. Thus, using Eq.~(\ref{dgdm}) to express
the number density in terms of the effective action, we obtain
\begin{equation}
    K_{ab}(x {-} x') = - \left.
    {\delta^2  \Gamma \over \delta \mu_a(x) \, \delta \mu_b(x') }
    \right|_0
    - \left.  \int_y
    {\delta^2 \Gamma \over \delta \mu_a(x) \, \delta  \bar\phi(y) } \,
    {\delta \bar\phi(y) \over \delta \mu_b(x') }
    \right|_0 \,.
\label{kab}
\end{equation}
We define
\begin{equation}
    C_{ab}(x {-} x')
    \equiv - \left.
    {\delta^2  \Gamma \over \delta \mu_a(x) \, \delta \mu_b(x') }
    \right|_0 \,,
\label{defcab}
\end{equation}
and recall the definition (\ref{ddgdmdp}) to write Eq.~(\ref{kab}) as
\begin{equation}
    K_{ab}(x {-} x') = C_{ab}(x {-} x')
    - \left. \int_y \gamma_a(x {-} y ) \,
    {\delta \bar\phi(y) \over \delta \mu_b(x') }
\right|_0 \,.
\label{kkab}
\end{equation}
To deal with the final variational derivative which appears here, we
note that with $\sigma $ and the $\mu_a$ taken to be
the independent variables,
\begin{equation}
    { \delta \sigma(z) \over \delta \mu_b(x') } = 0 \,,
\end{equation}
and so Eq.~(\ref{dgdp}) implies that
\begin{equation}
    {\delta^2 \Gamma \over \delta \mu_b(x') \, \delta \bar\phi(z)  } +
    \int_y {\delta \bar\phi(y) \over \delta \mu_b(x') } \,
    {\delta^2  \Gamma \over \delta \bar\phi(y) \, \delta \bar\phi(z) }
    = 0 \,.
\label{toofan}
\end{equation}
In the limit of constant chemical potentials and vanishing source, the
first term here is just $\gamma_b(x' {-} z )$ and the
second factor in the integrand is ${\cal G}^{-1}(y {-} z)$.
Hence
\begin{equation}
    \left. {\delta \bar\phi(y) \over \delta \mu_b(x') } \right|_0
    = - \int_z {\cal G}(y {-} z) \, \gamma_b(x' {-} z) \,,
\label{gamgamb}
\end{equation}
and Eq.~(\ref{kkab}) becomes
\begin{equation}
    K_{ab}(x {-} x')
    =
    C_{ab}(x {-} x')
    + \int_{y,z}
    \gamma_a(x {-} y) \, {\cal G}(y {-} z) \, \gamma_b(x' {-} z) \,,
\end{equation}
or, in wave number/frequency space,
\begin{equation}
    \tilde K_{ab}(\k,\omega_m)
    = \tilde C_{ab}(\k,\omega_m)
    + \tilde\gamma_a(\k,\omega_m) \,\tilde {\cal G}(\k,\omega_m) \,
      \tilde\gamma_b(\k,\omega_m) \,.
\label{ftkabr}
\end{equation}
Since the function $ \tilde C_{ab}(\k,\omega_m) $ is a double variational
derivative of the effective action functional
$\Gamma[\bar\phi;\mu_a]$, it is single-particle irreducible. On the
other hand, $ \tilde {\cal G}(\k,\omega_m) $, the potential correlation
function, is not single-particle irreducible.

We have yet to express the number density and potential correlation
functions in the simplest terms. To do so, we return to the
expectation of the field equation (\ref{fieldeq}). With
$\mu_a$ and $\sigma$ taken to be independent variables, the
functional derivative of this equation with respect to a generalized
chemical potential, with the chemical potentials then set constant and
the source to zero, gives
\begin{equation}
    \left. - \nabla^2 {\delta \bar\phi(x) \over \delta \mu_b(x') } \right|_0
    =
    i \sum_a e_a \, K_{ab}( x {-} x') \,.
\end{equation}
With the use of Eq's.~(\ref{gamgamb}) and (\ref{ftkabr}), the Fourier
transform of this constraint may be put in the form
\begin{equation}
    - \left[
    \k^2
    + i \, {\sum}_a \, e_a \, \tilde\gamma_a(\k,\omega_m)
    \right]
    \tilde {\cal G}(\k,\omega_m) \, \tilde\gamma_b(\k,\omega_m)
    = i \, {\sum}_a \, e_a \, \tilde C_{ab}(\k,\omega_m) \,.
\end{equation}
The factor in square brackets on the left-hand side of this result is,
according to Eq.~(\ref{ginv}), just ${\cal G}^{-1}(\k,\omega_m)$.
Hence,
\begin{equation}
    \tilde\gamma_b(\k,\omega_m)
    = - i \, \sum_a \, e_a \tilde C_{ab}(\k,\omega_m) \,.
\label{tildeg}
\end{equation}
Accordingly,
\begin{equation}
    {\cal G}^{-1}(\k,\omega_m)
    = \k^2 + \sum_{a,b} e_a e_b \, \tilde C_{ab}(\k,\omega_m) \,,
\label{eq:ginver}
\end{equation}
and
\begin{equation}
    \tilde K_{ab}(\k,\omega_m)
    = \tilde C_{ab}(\k,\omega_m)
    - \left[ {\sum}_c e_c \, \tilde C_{ca}(\k,\omega_m) \right]
     \,\tilde {\cal G}(\k,\omega_m) \,
    \left[ {\sum}_c e_c \, \tilde C_{cb}(\k,\omega_m) \right] .
\label{eq:exactk}
\end{equation}
We have found that both the potential and number density correlation
functions are determined by the single-particle irreducible function $
\tilde C_{ab}(\k,\omega_m) $. We should note that the definition
(\ref{defcab}) of this function, plus rotation and time reversal
invariance, implies the symmetry%
\footnote
    {
    See, for example, Brown \cite{brown}, Chapter II, Problem 5.
    }
\begin{equation}
    \tilde C_{ab}(\k,\omega_m) = \tilde C_{ba}(\k,\omega_m)
\end{equation}
which thus carries over to the number density correlation function
$\tilde K_{ab}(\k,\omega_m) $.

The above results may also be used to reveal the structure
of correlation functions involving the charge density.
The correlation function of the charge density with the
number density of species $a$ is given by
\begin{eqnarray}
    \sum_b \tilde K_{ab}(\k,\omega_m) \, e_b
    &=&
    \sum_b \tilde C_{ab}(\k,\omega_m) \, e_b
    \left\{ 1 -  \tilde {\cal G}(\k,\omega_m) \,
    \left[ {\sum}_{c,d} \, e_c e_d \, \tilde C_{cd}(\k,\omega_m) \right]
    \right\}
\nonumber\\
    &=& \sum_b \tilde C_{ab}(\k,\omega_m) \,
    e_b \, \k^2 \, {\cal G}(\k,\omega_m) \,.
\end{eqnarray}
The $\k \to 0$ limit gives the correlator of the number density with
the total charge.
This vanishes, as it must for the neutral plasma.
Finally, the charge density -- charge density correlation function
is given by
\begin{eqnarray}
    \tilde K(\k,\omega_m)
    &=& \sum_{a,b} e_a \, \tilde K_{ab}(\k,\omega_m) \, e_b
\nonumber\\
    &=&
    { \k^2 \left[
    \sum_{a,b} \, e_a \,\tilde C_{ab}(\k,\omega_m) \, e_b \right]
    \over
    \k^2 + \sum_{a,b} \, e_a \,\tilde C_{ab}(\k,\omega_m) \, e_b } \,.
\label {eq:Kx}
\end{eqnarray}
This form exhibits explicitly the small wave number behavior
\begin{equation}
    \tilde K(\k,\omega_m) \sim \k^2 \,,
    \qquad \hbox{as $\k \to 0$} \,.
\label {eq:Klim1}
\end{equation}

Static correlators, which are the focus of attention in the main text,
are related to the zero frequency component of the corresponding
time dependent correlator functions by a factor of $\beta^{-1}$:
\begin {equation}
    \tilde K_{ab}(\k) = \beta^{-1} \, \tilde K_{ab}(\k,0) \,,
\end {equation}
and similarly for $\tilde C_{ab}(\k)$, $\tilde {\cal G}(\k)$,
{\em etc}.
Consequently, the static versions of Eq's.~(\ref {eq:ginver}),
(\ref {eq:exactk}),
(\ref {eq:Kx}), and
(\ref {eq:Klim1})
are
\begin {equation}
    {\cal G}^{-1}(\k)
    = \beta \, \k^2 + \beta^2 \sum_{a,b} e_a e_b \, \tilde C_{ab}(\k) \,,
\label{ginver}
\end {equation}
\begin {equation}
    \tilde K_{ab}(\k)
    = \tilde C_{ab}(\k)
    - \left[ \beta \, {\sum}_c e_c \, \tilde C_{ca}(\k) \right]
     \,\tilde {\cal G}(\k) \,
    \left[ \beta \, {\sum}_c e_c \, \tilde C_{cb}(\k) \right] ,
\label{exactk}
\end{equation}
\begin {equation}
    \tilde K(\k)
    =
    { \k^2 \left[
    \sum_{a,b} \, e_a \,\tilde C_{ab}(\k) \, e_b \right]
    \over
    \k^2 + \beta \sum_{a,b} \, e_a \,\tilde C_{ab}(\k) \, e_b } \,,
\end {equation}
and
\begin{equation}
    \tilde K(\k) \sim \beta^{-1} \, \k^2 \,,
    \qquad \hbox{as $\k \to 0$} \,.
\end{equation}

\subsection{Effective Potential, Thermodynamic Quantities}

In quantum field theory, the effective potential (times the space-time
volume) is defined to be the restriction of the effective action to
spatially (and temporally) uniform fields.
We have already remarked that the further restriction to
the stationary point
yields the grand potential (times $\beta$).
With constant chemical potentials, the stationarity condition
$\delta\Gamma/\delta\bar\phi=0$ is just the condition that
charge neutrality hold for a given value of $\bar\phi$.
For convenience, we will assume that physical chemical potentials are
chosen such that this stationary point lies at $\bar\phi=0$, so that
\begin{equation}
    \beta \Omega = \left. \Gamma[\bar\phi;\mu] \right|_{\bar\phi=0} \,.
\end{equation}
This is the function that we have computed to three loops. However,
the charge neutrality constraint is never used in our computations,
and so, in fact, the function $ \Gamma[\bar\phi {=} 0 ;\mu] $ has been
calculated for arbitrary (constant) chemical potentials $\mu_a$. This
extension of the grand potential is needed for the computation of
thermodynamic average numbers and energy and for the correlators of
these quantities. Just as in our previous work, to derive general
relationships it is convenient temporarily to work with
$ \Gamma[\bar\phi;\mu] $ for arbitrary constant $\bar\phi$ and $\mu_a$.
The results of these derivations, however, will depend only upon the
$\bar\phi = 0$ functions that have been computed.

With uniform fields, Eq.~(\ref{dgdp}) reduces to
\begin{equation}
{ \partial \Gamma(\bar\phi;\mu) \over \partial \bar\phi } =
\sigma \beta \vol \,.
\label{dgdpc}
\end{equation}
As we have remarked before, the restriction to a vanishing source,
$\sigma = 0$, determines
\begin{equation}
\bar\phi = \bar\phi(\beta,\{\beta\mu_a\}) \,,
\label{barpd}
\end{equation}
and inserting this value of $\bar\phi$ in $\Gamma$ yields the physical
grand potential $\beta\Omega$. With arbitrary chemical potentials,
$\bar\phi$ is non-vanishing so as to keep a zero charge density in the
plasma.
The previous expressions (\ref{dgdm}) and (\ref{dgdb}),
evaluated with $\bar\phi$ at the stationary point (\ref {barpd}),
gives the physical particle numbers and energy,
\begin{equation}
\bar N_a = \langle n_a \vol \rangle_\beta =
- {\partial \Gamma(\bar\phi;\mu) \over \partial \beta \mu_a } \,,
\label{dgdmc}
\end{equation}
and
\begin{equation}
U = \bar E = \langle E \rangle_\beta =
 {\partial \Gamma(\bar\phi;\mu) \over \partial \beta } \,.
\label{dgdbc}
\end{equation}
To obtain relations for the fluctuations of these quantities, we first
need two results. The derivative of Eq.~(\ref{dgdpc}) with respect to
the inverse temperature keeping $\sigma = 0$ so that $\bar\phi$ is
determined by Eq.~(\ref{barpd}) gives, just as in the previous
analogous calculation of the chemical potential functional derivative
(\ref{toofan}),
\begin{equation}
{ \partial^2 \Gamma(\bar\phi;\mu) \over \partial \beta \,
\partial \bar\phi } +
 \vol \, \tilde {\cal G}^{-1}({\bf 0}) \, { \partial \bar\phi \over \partial
  \beta } = 0 \,.
\label{dgdbdp}
\end{equation}
Note that, from (\ref {ginver}),
\begin{equation}
\tilde {\cal G}^{-1}({\bf 0}) =
\beta^2 \sum_{a,b} e_a \tilde C_{ab}({\bf 0}) e_b \,,
\label{gnau}
\end{equation}
and
\begin{equation}
\tilde C_{ab}({\bf 0}) = { 1 \over \vol} \,
{ \partial^2 \Gamma(\bar\phi;\mu) \over \partial \beta\mu_a \,
\partial \beta\mu_b }
\end{equation}
may be computed directly at $\bar\phi = 0$
with chemical potentials set to values which
satisfy charge neutrality (for $\bar\phi = 0$)
after the derivatives have been performed.
Thus $\tilde C_{ab}({\bf 0})$ can be obtained from the
computation of the grand potential $\beta\Omega$. We may simply write
\begin{equation}
\tilde C_{ab}({\bf 0}) = - { \partial \langle n_a \rangle_\beta
     \over \partial \beta \mu_b }
= - { \partial \langle n_b \rangle_\beta
     \over \partial \beta \mu_a } \,,
\label{Cab0}
\end{equation}
where the partial derivatives are taken at constant temperature or
fixed $\beta$.
In a similar fashion, the derivative with respect to the inverse
temperature of the charge neutrality condition
\begin{equation}
{\sum}_a e_a {\partial \Gamma(\bar\phi;\mu) \over \partial \beta
  \mu_a} = 0
\end{equation}
produces
\begin{equation}
{\sum}_a e_a \left\{ {\partial^2 \Gamma(\bar\phi;\mu) \over \partial \beta
\, \partial \beta\mu_a} + {\partial^2 \Gamma(\bar\phi;\mu) \over
\partial \bar\phi \,\partial \beta\mu_a} \,
{\partial \bar\phi \over \partial\beta}\right\} = 0 \,.
\end{equation}
We use Eq's.~(\ref{ddgdmdp}), (\ref{tildeg}), and (\ref{gnau}) to
write this as
\begin{equation}
{\partial \bar\phi \over \partial\beta} =
- { i \beta \over \vol } \,  \tilde{\cal G}({\bf 0}) \,
{\sum}_a e_a  {\partial^2 \Gamma(\bar\phi;\mu) \over \partial \beta
 \, \partial \beta\mu_a}  \,.
\label{dbpdb}
\end{equation}
After the derivatives in the relations above have been taken, we may again
assume that the chemical potentials are chosen to give charge
neutrality at $\bar\phi = 0$.

With these results in hand, we can examine the fluctuations of energy
and particle numbers. The energy fluctuations in the grand canonical ensemble
are given by
\begin{eqnarray}
\left\langle \left( E - \bar E \right)^2 \right\rangle_\beta
&=& - {\partial \over \partial \beta}
 {\partial \Gamma(\bar\phi;\mu) \over \partial \beta }
\nonumber\\
&=& - {\partial^2 \Gamma(\bar\phi;\mu) \over \partial \beta^2 }
-  {\partial^2 \Gamma(\bar\phi;\mu) \over \partial \beta
\, \partial \bar\phi } \, {\partial \bar\phi \over \partial \beta} \,.
\end{eqnarray}
We can make use of Eq.~(\ref{dgdbdp}) to write this as
\begin{equation}
\left\langle \left( E - \bar E \right)^2 \right\rangle_\beta
= - {\partial^2 \Gamma(\bar\phi;\mu) \over \partial \beta^2 }
+ {\partial^2 \Gamma(\bar\phi;\mu) \over \partial \beta \,
\partial \bar\phi }
{ 1 \over \vol } \,  \tilde{\cal G}({\bf 0})
{\partial^2 \Gamma(\bar\phi;\mu) \over \partial \beta \,
\partial \bar\phi }  \,,
\end{equation}
or alternatively use Eq's.~(\ref{dgdbdp}) and (\ref{dbpdb}) to write
\begin{equation}
\left\langle \left( E - \bar E \right)^2 \right\rangle_\beta
= - {\partial^2 \Gamma(\bar\phi;\mu) \over \partial \beta^2 }
- \left[{\sum}_a e_a  {\partial^2 \Gamma(\bar\phi;\mu) \over \partial \beta
 \, \partial \beta\mu_a} \right]
{ \beta^2 \over \vol } \, \tilde{\cal G}({\bf 0}) \left[
{\sum}_a e_a  {\partial^2 \Gamma(\bar\phi;\mu) \over \partial \beta
 \, \partial \beta\mu_a}  \right] .
\end{equation}
This latter form may be evaluated at $\bar\phi = 0$ with the chemical
potentials chosen to give charge neutrality after their derivatives
have been taken. Thus, this latter form is determined by the
quantities calculated for the grand potential $\beta\Omega$.
The energy --- particle number correlation is given by
\begin{eqnarray}
\left\langle \left( E - \bar E \right) \left( N_a - \bar N_a \right)
 \right\rangle_\beta
= - {\partial \bar N_a \over \partial \beta }
&=& { \partial \over \partial \beta}
{\partial \Gamma(\bar\phi;\mu) \over \partial \beta\mu_a}
\nonumber\\
&=& {\partial^2 \Gamma(\bar\phi;\mu) \over \partial \beta \,
\partial \beta\mu_a}  +
{\partial^2 \Gamma(\bar\phi;\mu) \over \partial\bar\phi \,
\partial \beta\mu_a} \, { \partial\bar\phi \over \partial\beta } \,.
\end{eqnarray}
With the use of Eq's.~(\ref{ddgdmdp}), (\ref{tildeg}), and
(\ref{dbpdb}), this becomes
\begin{equation}
\left\langle \left( E - \bar E \right) \left( N_a - \bar N_a \right)
 \right\rangle_\beta
= {\partial^2 \Gamma(\bar\phi;\mu) \over \partial \beta \,
\partial \beta\mu_a}  - {\sum}_b e_b {\partial^2 \Gamma(\bar\phi;\mu)
\over \partial\beta \, \partial\beta\mu_b} \, \tilde{\cal G}({\bf 0}) \,
\beta^2 \, {\sum}_c \tilde C_{ac}({\bf 0}) e_c \,.
\end{equation}
Again, this result depends only upon quantities involved in the
construction of $\beta\Omega$. Note that, in view of Eq.~(\ref{gnau}),
the charge neutrality condition is obeyed,
\begin{equation}
{\sum}_a \left\langle \left( E - \bar E \right) \left( N_a - \bar N_a \right)
 \right\rangle_\beta e_a = 0 \,.
\label{noqqq}
\end{equation}
Finally, we note that the Fourier transform (\ref{exactk}) evaluated
at zero wave number yields the particle number --- particle number correlators,
\begin{equation}
\left\langle \left( N_a - \bar N_a \right)  \left( N_b - \bar N_b \right)
\right\rangle_\beta = \vol \left\{
\tilde C_{ab}({\bf 0})
- \left[ \beta {\sum}_c e_c \tilde C_{ca}({\bf 0}) \right]
 \tilde {\cal G}({\bf 0})
\left[ \beta {\sum}_c e_c \tilde C_{cb}({\bf 0}) \right]
\right\} .
\end{equation}

The results that we have obtained may be used to compute the specific
heat at constant volume. This is simply related to the derivative of
the average energy with respect to the inverse temperature at constant
particle numbers,
\begin{equation}
C_{\rm V} = - \beta^2 \left. {\partial \bar E \over \partial \beta }
\right|_{\{\bar N_a\}} .
\end{equation}
Thus the chemical potentials must change as the temperature
is varied in order to maintain constant numbers. That is, we have
\begin{eqnarray}
d \bar E &=& { \partial \bar E \over \partial \beta } \, d \beta
+ {\sum}_a
 {\partial \bar E \over \partial \beta \mu_a } \, d(\beta
\mu_a)
\nonumber\\
&=& \left\{ {\partial^2 \Gamma(\bar\phi;\mu) \over \partial \beta^2 }
+ \left[{\sum}_a e_a  {\partial^2 \Gamma(\bar\phi;\mu) \over \partial \beta
 \, \partial \beta\mu_a} \right]
{ \beta^2 \over \vol } \, \tilde{\cal G}({\bf 0}) \left[
{\sum}_a e_a  {\partial^2 \Gamma(\bar\phi;\mu) \over \partial \beta
 \, \partial \beta\mu_a}  \right] \right\} d \beta
\nonumber\\
&& {} + {\sum}_a
\left\{ {\partial^2 \Gamma(\bar\phi;\mu) \over \partial \beta \,
\partial \beta\mu_a}  - \left[{\sum}_b e_b {\partial^2 \Gamma(\bar\phi;\mu)
\over \beta \, \partial\beta\mu_b} \right] \tilde{\cal G}({\bf 0})
\beta^2 \left[ {\sum}_c \tilde C_{ac}({\bf 0})
e_c \right] \right\} d(\beta \mu_a)
 \,,
\label{dbare}
\end{eqnarray}
with the chemical potential changes constrained by
\begin{eqnarray}
0 &=& d \bar N_a = { \partial \bar N_a \over \partial \beta } \, d \beta
+ {\sum}_b {\partial \bar N_a \over \partial \beta \mu_b} \,
d (\beta\mu_b)
\nonumber\\
&=& - \left\{ {\partial^2 \Gamma(\bar\phi;\mu) \over \partial \beta \,
\partial \beta\mu_a}  - \left[ {\sum}_b e_b {\partial^2 \Gamma(\bar\phi;\mu)
\over \partial\beta \, \partial\beta\mu_b} \right] \tilde{\cal G}({\bf
0})
\beta^2 \left[ {\sum}_c \tilde C_{ac}({\bf 0}) e_c \right] \right\} d \beta
\nonumber\\
&& {} + {\sum}_b
\vol \left\{
\tilde C_{ab}({\bf 0})
- \left[ \beta {\sum}_c e_c \tilde C_{ca}({\bf 0}) \right]
 \,\tilde {\cal G}({\bf 0})
\left[ \beta {\sum}_c e_c \tilde C_{cb}({\bf 0}) \right]
\right\} d(\beta\mu_b)
\,.
\label{dbarn}
\end{eqnarray}
Introducing the inverse matrix $\tilde C^{-1}({\bf 0})$,
\begin{equation}
{\sum}_b \tilde C^{-1}({\bf 0})_{ab}\, \tilde C_{bc}({\bf 0}) = \delta_{ac} \,,
\end{equation}
which is a symmetric matrix since
$||\tilde C_{ab}({\bf 0})||$
is symmetric, we may rewrite Eq.~(\ref{dbarn})
as
\begin{eqnarray}
d(\beta\mu_a) &=& { 1 \over \vol} {\sum}_b \tilde C^{-1}({\bf 0})_{ab}
 {\partial^2 \Gamma(\bar\phi;\mu) \over \partial \beta \,
\partial \beta\mu_b} \, d\beta
\nonumber\\
&& {} - e_a \left\{ { 1 \over \vol}
   \left[ {\sum}_b e_b {\partial^2 \Gamma(\bar\phi;\mu)
\over \partial\beta \, \partial\beta\mu_b} \right]
\tilde{\cal G}({\bf 0})
\beta^2 \,  d \beta
 - \tilde {\cal G}({\bf 0}) \,
 \beta^2 \sum_{bc} e_b \, \tilde C_{bc}({\bf 0}) \, d(\beta\mu_c)
\right\} .
\end{eqnarray}
Because of the charge neutrality condition (\ref{noqqq}), a change
$d(\beta\mu_a)$ proportional to $e_a$ does not alter Eq.~(\ref{dbare}).
Hence
\begin{equation}
\left. { \partial \bar E \over \partial \beta } \right|_{\{N_a\}} =
{\partial^2 \Gamma(\bar\phi;\mu) \over \partial \beta^2 }
+ { 1 \over \vol} \sum_{a,b}
{\partial^2 \Gamma(\bar\phi;\mu) \over \partial \beta \,
\partial \beta\mu_a} \, \tilde C^{-1}({\bf 0})_{ab}
{\partial^2 \Gamma(\bar\phi;\mu) \over \partial \beta \,
\partial \beta\mu_b}  \,.
\end{equation}

\subsection{Time-Dependent Correlations}
\label{timeapp}

We noted in section \ref {sec:freq} that
although the static two-point potential correlation function, the zero
frequency part of the general correlator, describes a Debye screened
potential (except for the very long-distance tail elucidated in
section \ref{sec:longdist}), the non-zero frequency parts of this
correlation function are not Debye screened. Recalling the general
result (\ref{eq:ginver}):
\begin{equation}
    {\cal G}^{-1}(\k,\omega_m)
    = \k^2 + \sum_{a,b} e_a e_b \, \tilde C_{ab}(\k,\omega_m) \,,
\label{ginvv}
\end{equation}
this lack of Debye screening for $\omega_m \ne 0$ is the statement
that, for this case,
\begin{equation}
\k^2 \to 0 \,: \qquad  \tilde C_{ab}(\k,\omega_m) \sim k^2 \,,
\label{ksquar}
\end{equation}
which implies that $ {\cal G}^{-1}({\bf r} - {\bf r}' , \omega_m) $
behaves as $|{\bf r} - {\bf r}' |^{-1}$ for large $|{\bf r} - {\bf r}'
|$. In this section we shall show how this follows from the
conservation of the number currents or, equivalently, from the gauge
invariance of the coupling of the basic theory to a set of
[($\dim {+} 1$)-dimensional] vector potentials.

Number-current correlation functions are generated by coupling a
vector potential $A_\mu^a(x) \equiv (A_4^a(x), {\bf A}^a(x) ) $ for
each particle species $a$. This is done by augmenting the Lagrangian
(\ref{eq:La}) for each basic charged field to read
\begin{eqnarray}
    {\cal L}_a
    =
    \psi_a^*(\r, \tau)
\left\{ { \partial \over \partial \tau}  - A_4^a(\r,\tau) - {1 \over 2 m_a}
\left[ \nabla -i {\bf A}^a(\r,\tau) \right]^2  -
    \mu_a - i e_a \phi(\r , \tau) \right\} \psi_a(\r, \tau) \,.
\label{extl}
\end{eqnarray}
Connected correlation functions of $n$ space-time currents $J^a_\mu(x)
\equiv ( n_a(\r,\tau) , {\bf J}_a(\r,\tau) )$ are produced by $n$ functional
derivatives $ \delta / \delta A_\mu^a(x)$ acting on the generating
functional $W$. In particular, the connected number-density
correlation function (\ref{kkabb}) is now extended to the space-time
correlation function
\begin{equation}
K_{ab}^{\mu\nu}(x {-} x') = \left. {\delta^2 W[\sigma;A] \over
    \delta A_\mu^a(x) \, \delta A_\nu^b(x') }
\right|_0 \,.
\end{equation}
The corresponding connected, single-particle irreducible function is
given by the same functional derivatives of the Legendre transform of
$W$, the effective action $\Gamma$. This extension of
Eq.~(\ref{defcab}) reads
\begin{equation}
    C_{ab}^{\mu\nu}(x {-} x')
    = - \left.
    {\delta^2  \Gamma \over \delta A_\mu^a(x) \, \delta A_\nu^b(x') }
    \right|_0 \,.
\end{equation}

The actions formed from the extended Lagrangians (\ref{extl}) are
invariant under local phase rotations of the charged fields
\begin{equation}
\psi_a^*(x) \to \psi_a^*(x) \exp\{ -i \lambda^a(x) \} \,, \qquad
\psi_a(x) \to \exp\{ i \lambda^a(x) \} \psi_a(x) \,,
\label{phrot}
\end{equation}
coupled with the gauge transformations of the external potentials
\begin{equation}
A_4^a(x) \to A_4^a(x) + i { \partial \lambda^a \over \partial \tau} \,,
\qquad
A_k^a(x) \to A_k^a(x) +  { \partial \lambda^a \over \partial x^k} \,.
\label{gtrans}
\end{equation}
The integration measures of the charged field functional integrals are
unchanged by the phase rotation (\ref{phrot}). Hence the connected
generating functional $W[A]$ is invariant under the gauge
transformation (\ref{gtrans}). This invariance carries over to the
effective action $\Gamma[A]$ since the Legendre transformation which
relates it to $W[A]$ involves only neutral fields that are not altered
by the phase rotation or gauge transformation.  In the limit of an
infinitesimal transformation, the gauge invariance gives functional
differential statements
\begin{equation}
\partial_\mu { \delta \over \delta A_\mu^a(x) } W[A] = 0 =
\partial_\mu { \delta \over \delta A_\mu^a(x) } \Gamma[A] \,,
\label{diver}
\end{equation}
where we have adopted the shorthand notation
\begin{equation}
\partial_\mu = \left(
    i {\partial \over \partial  \tau}
    \,,
    { \partial \over \partial x^k }
\right) \,.
\end{equation}
Taking additional functional derivatives of Eq's.~(\ref{diver}) shows
that any number current correlation function has a transverse form. In
particular, one additional functional derivative yields
\begin{equation}
\partial_\mu K_{ab}^{\mu\nu}(x {-} x') = 0 =
\partial_\nu K_{ab}^{\mu\nu}(x {-} x') \,,
\end{equation}
and
\begin{equation}
   \partial_\mu C_{ab}^{\mu\nu}(x {-} x') = 0 =
  \partial_\nu C_{ab}^{\mu\nu}(x {-} x') \,.
\end{equation}
In terms of Fourier components,
\begin{equation}
-i \omega_m \, \tilde C_{ab}^{4 \nu}(\k,\omega_m)
+ k^l \, \tilde C_{ab}^{l \nu}(\k,\omega_m) = 0 \,.
\label{fdiv}
\end{equation}

We are now in a position to demonstrate that the potential correlation
function at non-zero frequency has no Debye screening.  Three
paragraphs ago, we remarked that this correlator is determined by $
\tilde C_{ab}(\k,\omega_m) = \tilde C_{ab}^{4 4}(\k,\omega_m) $.
Because of rotational invariance, $\tilde C_{ab}^{l4}(\k,\omega_m) =
k^l \, f_{ab}(k^2 , \omega_m) $ and the $\nu = 4$ component of the
Fourier form (\ref{fdiv}) of the divergence condition becomes
\begin{equation}
-i \omega_m \, \tilde C_{ab}(\k,\omega_m)
+ k^2 \, f_{ab}(k^2 , \omega_m) = 0 \,.
\end{equation}
This demonstrates the assertion (\ref{ksquar}) that $ \tilde C_{ab}
(\k,\omega_m) \sim k^2 $ as $ k^2 \to 0 $ when $ \omega_m \ne 0$ and
thus that there is no Debye screening in the $ \omega_m \ne 0$ potential
correlation function $ {\cal G}(\k , \omega_m) $.

The fact that, for small $k^2$, $ \tilde C_{ab}(\k,\omega_m) = O(k^2)
$ when $\omega_m \ne 0$ but $ \tilde C_{ab}(\k, 0) = O(1) $ might
appear a bit odd since $ \tilde C_{ab}(\k,\omega_m) $ is equal to an
analytic function of $\omega$, $F_{ab}(k^2 , \omega) $, evaluated at
discrete points on the imaginary axis,%
\footnote
    {
    See, for example, the
    discussion in Problem 4 of Chapter II of Brown \cite{brown}.
    }
$\omega = i \omega_m = i 2 \pi m / \beta $.
Thus one might expect a uniform
behavior in $\omega$ which would require that $ \tilde C_{ab}(\k,0) =
O(k^2) $ for small $k^2$ and no Debye screening. In fact, the behavior
of the analytic function $F_{ab}(k^2 , \omega) $ is not uniform in $
k^2 $ when $\omega$ is small.  This non-uniform behavior is
illustrated by the simple one-loop contribution of the charged fields
to $ \tilde C_{ab}(\k,\omega_m) $.  To further simplify the result,
we also take the $\hbar \to$ classical limit but with the frequency $
\omega_m / \hbar$ kept fixed to obtain
\begin{equation}
 \tilde C_{ab}(\k,\omega_m) =  \tilde C_{ab}^{44}(\k,\omega_m)  =
\delta_{ab} \int { (d^\dim p) \over (2\pi)^\dim } \> { 1 \over
i ( \omega_m / \hbar ) - ( {\bf p} \cdot \k / m_a ) } \>
\k \cdot { \partial \over \partial {\bf p} } n^0_a({\bf p}) \,,
\label{cloop}
\end{equation}
where
\begin{equation}
n^0_a({\bf p}) = g_a \, \exp\left\{
- \beta \left[ { {\bf p}^2 \over 2 m_a } - \mu_a \right] \right\}
\end{equation}
is the Maxwell-Boltzmann density of particles in momentum space.  This
result is obtained by taking the indicated limits of Eq.~(\ref{PI})
in the following Appendix.  Taking $\omega_m = 0$ gives
\begin{eqnarray}
 \tilde C_{ab}(\k, O ) &=&
\delta_{ab} \int { (d^\dim p) \over (2\pi)^\dim } \> { 1 \over
  (- {\bf p} \cdot \k / m_a ) } \,
( - \beta {\bf p} \cdot \k / m_a )  n^0_a({\bf p})
\nonumber\\
&=& \delta_{ab} \, \beta n^0_a \,,
\end{eqnarray}
which produces the leading order contribution to the
Debye wave number,
\begin{equation}
{\sum}_{ab} e_a e_b  \tilde C_{ab}(\k, O ) =
{\sum}_a \beta e^2_a n^0_a = \kappa_0^2 \,,
\end{equation}
yielding for small $k^2$ [{\it c.f.} Eq.~(\ref{ginvv})]
\begin{equation}
{\cal G}^{-1}(\k ,0) \approx k^2 + \kappa^2_0 \,,
\end{equation}
On the other hand, for $\omega_m \ne 0$, the linear term in $\k$ in
Eq.~(\ref{cloop}) vanishes, and expanding the denominator to first
order in $\k$ together with a partial integration of $\k \cdot {
\partial \over \partial {\bf p} }$ gives the small $k^2$ limit
\begin{equation}
 \tilde C_{ab}(\k, \omega_m ) =
\delta_{ab} \left( { \hbar \over \omega_m } \right)^2 \,
{ k^2 \over m_a} \, n_a^0 \,.
\end{equation}
The corresponding small $k^2$ contribution to the potential correlator
produces
\begin{equation}
{\cal G}^{-1}(\k , \omega_m) \approx k^2
\left[ 1 + { \hbar^2 \omega_P^2 \over \omega_m^2 } \right] \,,
\end{equation}
where
\begin{equation}
\omega_P^2 = {\sum}_a { e_a^2 n^0_a \over m_a }
\end{equation}
is the lowest order contribution to the plasma frequency.
This is the result (\ref{nodebye}) given in the text.

To see how the non-uniform behavior of the one-loop correlator
(\ref{cloop}) is in accord with the conservation (\ref{fdiv}) of the
number current correlators, we note that the classical limit of a
one-loop calculation also gives
\begin{equation}
 \tilde C_{ab}^{l4}(\k,\omega_m)  =
\delta_{ab} \int { (d^\dim p) \over (2\pi)^\dim } \> { p^l \over m_a} \,
{ 1 \over i ( \omega_m / \hbar ) - ( {\bf p} \cdot \k / m_a ) } \>
\k \cdot { \partial \over \partial {\bf p} } \, n^0_a({\bf p}) \,.
\end{equation}
We see that these contributions to $ -i \omega_m \, \tilde
C_{ab}^{44}(\k,\omega_m) + k^l \, \tilde C_{ab}^{l 4}(\k,\omega_m) $
combine to form the integral of a total derivative which vanishes, and
so the current conservation is confirmed.  We also note the
non-uniform limits
\begin{equation}
 \tilde C_{ab}^{l4}(\k, 0)  = 0 \,,
\end{equation}
while for $\omega_m \ne 0$,
\begin{equation}
 k^2 \to 0 \, : \qquad \tilde C_{ab}^{l4}(\k,\omega_m)  =
\delta_{ab} \,  { i \hbar \over \omega_m } \,
{ n^0_a \over m_a } \, k^l \,.
\end{equation}

For the sake of completeness, we note that the calculations leading to
Eq.~(\ref{eq:exactk}) are easily generalized to relate the number
current correlation functions to their single-particle irreducible
counterparts.  The result is
\begin{equation}
    \tilde K_{ab}^{\mu\nu}(\k,\omega_m)
    = \tilde C_{ab}^{\mu\nu}(\k,\omega_m)
    - \left[ {\sum}_c e_c \, \tilde C_{ca}^{4\mu}(\k,\omega_m) \right]
     \,\tilde {\cal G}(\k,\omega_m) \,
    \left[ {\sum}_c e_c \, \tilde C_{cb}^{4\nu}(\k,\omega_m) \right] .
\end{equation}
We also note that time-reversal and spatial-rotation invariance
together with the current conservation imply the symmetries
\begin{equation}
\tilde C_{ab}^{\mu\nu}(\k,\omega_m) = \tilde C_{ab}^{\nu\mu}(\k,\omega_m)
= \tilde C_{ba}^{\mu\nu}(\k,\omega_m) \,,
\end{equation}
and
\begin{equation}
\tilde C_{ab}^{\mu\nu}(\k,\omega_m)^* =
\tilde C_{ab}^{\mu\nu}(-\k,-\omega_m) \,.
\end{equation}

\newpage

\section{Green's Functions and Determinants}
\label {app:det}

The result (\ref{eq:Zquant}) in the text involves a product of path
integrals of the form%
\footnote
  {We use the notation $Z[V]$ because,
  when $V$ is independent of imaginary time $\tau$,
  this functional integral is a representation
  the grand canonical partition function
  for a gas of particles with no mutual interactions but moving
  in the external potential $V$.}
\begin{equation}
Z[V] = \int [d\psi^* d\psi] \exp\left\{ - \int_0^\beta d\tau \int (d^\dim
 \r) \> \psi^*(\r,\tau) \left[ { \partial \over \partial \tau } - { \nabla^2
     \over 2m } - \mu + V(\r,\tau) \right] \psi(\r,\tau) \right\} ,
\end{equation}
with $m$ one of the masses $\{ m_a \}$ and $ V({\bf r}, \tau) = - i e
\phi({\bf r},\tau) $, with $e$ the corresponding charge $e_a$.
When the chemical potential is generalized to contain spatial
or temporal variation, so as to generate number density correlation functions,
its spatially or temporally varying part will be implicitly included in
the potential $V$.
The field $\psi(\r,\tau)$ is either periodic (for Bosons) or
antiperiodic (for Fermions) in $\tau$ with period $\beta$.  The
external potential $V(\r , \tau )$ is initially defined in the
interval $ 0 < \tau < \beta $, but may be extended to all real $\tau$
by regarding it as a periodic function with period $\beta$. The
functional integral produces an inverse determinant in the Bose case
and a determinant in the Fermi case,
\begin{equation}
Z[V] = {\rm Det}^{\mp 1} \left[ { \partial \over \partial \tau } - { \nabla^2
     \over 2m } - \mu + V(\r,\tau) \right] .
\end{equation}
In this appendix, we shall show how the determinant $Z[V]$ is related
to a sum of ordinary, single-particle quantum-mechanical amplitudes.
We shall then make use of this result to derive approximate
evaluations of $Z[V]$ that become valid in the limit in which the
dynamics may be treated classically, approximations that are used in
the calculations of the text. These needed results could perhaps be
obtained more quickly with other methods, but the development given
here hopefully illuminates the character of the theory and the
intermediate results that are obtained may be useful in other
contexts.

The determinant can be constructed by integrating its variation.
The familiar form for the variation of the determinant gives
\begin{equation}
\delta \ln Z[V] = \mp \int_0^\beta \int (d^\dim\r) \> G_\beta( \r,
\tau ; \r, \tau {+} 0 ) \> \delta V(\r,\tau) \,,
\label{ddet}
\end{equation}
in which the thermal Green's function $G_\beta$ is defined by
\begin{equation}
\left[ { \partial \over \partial \tau } - { \nabla^2
     \over 2m } - \mu + V(\r,\tau) \right] G_\beta( \r,
\tau ; \r' \tau' )  = \delta( \tau {-} \tau') \,
\delta ( \r {-} \r') \,,
\label{geq}
\end{equation}
together with the boundary conditions that it be periodic for Bosons
and antiperiodic for Fermions with a period of $\beta$,
\begin{equation}
G_\beta( \r, \tau{+}\beta ; \r' \tau' ) =
G_\beta( \r, \tau ; \r', \tau'{+}\beta ) =
\pm G_\beta( \r, \tau ; {\bf  r}' , \tau' )  \,.
\label{period}
\end{equation}
The coincident time limit used in the variation (\ref{ddet}),
in which $\tau' \to \tau$ from above,
is needed to give the proper operator ordering $ \psi^\dagger \psi $
that represents the density operator.

To construct the thermal Green's function, it is convenient to
introduce the quantum-mechanical
transformation function in imaginary time $\langle \r ,\tau |
\r' , \tau' \rangle $ whose dynamics is governed by the external
potential. It is defined by
\begin{equation}
\left[ { \partial \over \partial \tau } - { \nabla^2
     \over 2m } + V(\r,\tau) \right]
\langle \r ,\tau | \r' , \tau' \rangle = 0 \,,
\label{scheq}
\end{equation}
together with the boundary condition
\begin{equation}
\langle \r, \tau | \r' , \tau \rangle =
\delta ( \r {-} \r' ) \,.
\label{trbc}
\end{equation}
We now assert that the thermal Green's function in the interval $ -\beta \le
\tau , \tau' \le \beta$ has the construction (akin to an image
construction in electrostatics)
\begin {equation}
    G_\beta (\r,\tau;\r',\tau')
    =
    \theta(\tau{-}\tau') \, \langle \r ,\tau | \r', \tau' \rangle
    +
    \sum_{n=1}^\infty (\pm 1)^n e^{\mu ( \tau {-} \tau' + n \beta ) }
    \langle \r, \tau {+} n \beta | \r', \tau' \rangle \,,
\end {equation}
where $\theta(\tau)$ is the unit step function.
The proof is as follows.
Since
\begin{equation}
G_\beta( \r, \tau' {+} 0 ; \r', \tau' ) -
G_\beta( \r, \tau' {-} 0 ; \r' ,\tau' ) =
\langle \r, \tau' | \r', \tau' \rangle = \delta ( {\bf
  r} {-} \r' ) \,,
\end{equation}
Eq.~(\ref{scheq}) implies that the inhomogeneous Green's function
equation (\ref{geq}) is obeyed. And the construction is easily seen to
satisfy the periodicity condition (\ref{period}).

The coincident time limit of the Green's function which enters into
the variation (\ref{ddet}) thus has the representation
\begin{equation}
G_\beta( \r ,\tau ; \r, \tau {+} 0 )  = \sum_{n=1}^\infty (\pm
1)^n e^{n \beta \mu }
\langle \r, \tau {+} n \beta | \r ,\tau \rangle \,.
\end{equation}
Thus
\begin{equation}
\delta \ln Z[V] = - \sum_{n=1}^\infty (\pm 1)^{n{+}1}
 e^{n \beta \mu } \int_0^\beta d \tau \int (d^\dim\r) \>
\langle \r, \tau {+} n \beta | \r ,\tau \rangle \> \delta V(\r,\tau) \,.
\end{equation}
Since the potential is periodic,
\begin{equation}
V(\r, \tau + k \beta ) = V(\r , \tau) \,,
\end{equation}
so is the transformation function in the presence of this potential,
\begin{equation}
\langle \r, \tau {+} k \beta {+} n \beta | \r , \tau {+} k \beta \rangle
= \langle \r, \tau {+} n \beta | \r, \tau \rangle \,.
\end{equation}
Hence, since we may add $n$ equal copies if we divide by $n$, we may
write
\begin{equation}
\delta \ln Z[V] = - \sum_{n=1}^\infty  (\pm 1)^{n{+}1}
{ e^{n \beta \mu } \over n} \int_0^{n\beta} d \tau \int (d^\dim\r) \>
\langle \r, \tau {+} n \beta | \r, \tau \rangle \> \delta V(\r,\tau) \,.
\label{var}
\end{equation}
To integrate this variational statement, we introduce a complete set
of intermediate states and write
\begin{equation}
\langle \r, \tau {+} n\beta | \r , \tau \rangle = \int (d^\dim{\bf
  \bar r}) \> \langle \r, \tau {+} n\beta | {\bf \bar r}, n \beta \rangle \,
\langle {\bf \bar r}, n\beta | \r , \tau \rangle \,,
\end{equation}
and again use the periodicity of the external potential to write
\begin{equation}
\langle \r, \tau {+} n\beta | {\bf \bar r}, n \beta \rangle =
\langle \r , \tau | {\bf \bar r}, 0 \rangle \,.
\end{equation}
Hence, the variational statement may be expressed as
\begin{eqnarray}
\delta \ln Z[V] &=& - \sum_{n=1}^\infty  (\pm 1)^{n{+}1}
{e^{n\beta \mu } \over n} \int_0^{n\beta} d \tau \int (d^\dim\r)
(d^\dim{\bf \bar r}) \> \langle {\bf \bar r}, n\beta | {\bf  r} ,\tau \rangle \>
\delta V(\r,\tau) \>
\langle \r ,\tau | {\bf \bar r} ,0 \rangle
\nonumber\\
&=& \sum_{n=1}^\infty  (\pm 1)^{n{+}1}  {e^{n \beta \mu } \over n}
 \int (d^\dim{\bf \bar r}) \> \delta \langle {\bf \bar r}
, n\beta | {\bf \bar r} ,0 \rangle \,,
\end{eqnarray}
where the second equality recognizes that this is just the variation
of the transformation function when the potential is varied. Hence,
\begin{equation}
\ln Z[V] =  \sum_{n=1}^\infty  (\pm 1)^{n{+}1}
{ e^{n \beta \mu } \over n} \int (d^\dim{\bf r}) \>  \langle {\bf r}
,n\beta | {\bf r}, 0 \rangle \,,
\label{lnz}
\end{equation}
which expresses the determinant in terms of an expansion in powers $n$
of the fugacity $ z = e^{\beta\mu} $ whose coefficients are traces of
single-particle transformation functions over the imaginary time
interval $(0,n\beta)$.  To confirm that the correct integration
constant has been secured, we note that when the external potential
$V(\r, \tau)$ vanishes, this form immediately gives the free-particle
partition function since in this case
\begin{equation}
\langle \r , n\beta | \r ,0 \rangle^0 = \int { (d^\dim{\bf p}) \over
  (2\pi)^\dim } \>
    e^{ - n\beta p^2 / 2m} \,,
\label{freeatlast}
\end{equation}
and so
\begin{eqnarray}
\ln Z[V] &=&  \sum_{n=1}^\infty { (\pm 1)^{n{+}1} \over n} \,
e^{n \beta \mu }  \int (d^\dim{\bf r}) \>
\int { (d^\dim{\bf p}) \over
  (2\pi)^\dim } \>
    e^{ - n\beta p^2 / 2m}
\nonumber\\
&=& \mp  \int { (d^\dim{\bf r}) (d^\dim{\bf p}) \over
  (2\pi)^\dim } \> \ln \left[ 1 \mp  e^{ \beta \mu }
    e^{ - \beta p^2 / 2m}  \right] \,,
\end{eqnarray}
which is the well-known result for the quantum-statistical
free-particle partition function.

The single-particle transformation functions that appear here have a
convenient path integral representation
\begin{equation}
\int (d^\dim{\bf r}) \> \langle {\bf r} ,n\beta | {\bf r}, 0 \rangle
 = \int [d{\bf r}] \> \exp\left\{ - \int_0^{n\beta} d\tau \left[ {
      m\over 2 }  \,
{ d {\bf r} \over d\tau } \cdot { d {\bf r} \over d\tau } + V({\bf r}(\tau)
, \tau) \right] \right\} \,.
\label{funfunint}
\end{equation}
Here the functional integral is over all paths that begin and end at
position ${\bf r}$, ${\bf r}(0) = {\bf r} = {\bf r}(n\beta)$, with
${\bf r}$ then integrated over the large spatial volume $\vol\/$.
In other words, the integral is over all paths which are
periodic with period $n\beta\hbar$.
In
the limit in which the quantum-mechanical aspects of the particle's
dynamics is not important, the classical limit for the dynamics which
is equivalent to the large mass $m$ limit, the dominant path is just
the constant path ${\bf r}(\tau) = {\bf r}$ so that, in this limit,
\begin{equation}
\int (d^\dim{\bf r}) \>
\langle {\bf r} , n\beta | {\bf r},  0 \rangle
= \langle {\bf r} , n\beta | {\bf r},  0 \rangle^0
 \int (d^\dim {\bf r}) \> \exp\left\{ - n \int_0^\beta
   d \tau \> V({\bf r},\tau) \right\} \,,
\label{heavier}
\end{equation}
where the overall constant is determined by the
free-particle limit (\ref{freeatlast}), and the periodicity of the
potential has been used to write the integral from $0$ to $n\beta$ as
$n$ times the integral from $0$ to $\beta$. Placing this approximation
in the general result (\ref{lnz}) gives
\begin{equation}
\ln Z[V] =  \mp  \int { (d^\dim{\bf r}) (d^\dim{\bf p}) \over
  (2\pi)^\dim } \> \ln \left[ 1 \mp \exp\left\{\beta \mu
  - \beta { p^2\over 2m} -  \int_0^\beta d \tau \> V({\bf r},\tau)
    \right\} \right] \,.
\end{equation}
In this expression, the quantum Bose-Einstein or Fermi-Dirac
statistics are treated exactly, but the dynamics is treated entirely
classically.  In the limit of classical statistics, $- \beta \mu \gg
1$, and only the first term in the expansion of the logarithm is
significant. Replacing $V$ by $-ie\phi$ and remembering the definition
of the bare particle density puts this classical limit in the form
\begin{equation}
\ln Z[V] = \int (d^\dim \r)\> n^0 \, \exp\left\{ i e \int_0^\beta d\tau \>
  \phi(\r,\tau) \right\} \,.
\end{equation}
This is the formula used in the text and derived there so as to
obtain the correct Coulomb classical partition function. Here we have
obtained it as the classical limit of the many-particle, quantum
mechanical system.

To find sub-leading corrections to the large mass limit,
it is convenient first to derive an exact series representation.
The representation is obtained by placing the Fourier
transform representation of the potential
\begin{equation}
V({\bf r},\tau) = \int { (d^\dim {\bf k}) \over (2 \pi)^\dim} \>
\tilde V({\bf k},\tau) \, e^{i {\bf k} \cdot {\bf r} }
\end{equation}
in the exponent of the functional integral (\ref{funfunint}) and
expanding the exponential in powers of the potential. Interchanging
the orders of integration then yields
\begin{eqnarray}
&&
\int (d^\dim{\bf r}) \> \langle {\bf r} ,n\beta | {\bf r}, 0 \rangle
\nonumber\\
&& \quad =
\sum_{l=0}^\infty { ( -1)^l \over l! }
\int_0^{n\beta} d \tau_1 \int { (d^\dim {\bf k}_1) \over (2 \pi)^\dim} \>
\tilde V({\bf k}_1,\tau_1) \> \cdots  \>
\int_0^{n\beta} d \tau_l \int { (d^\dim {\bf k}_l) \over (2 \pi)^\dim} \>
\tilde V({\bf k}_l,\tau_l)
\; z_n[{\bf F}] \,,
\label {funyfun}
\end{eqnarray}
in which
\begin{equation}
 z_n[{\bf F}] =
 \int [d{\bf r}] \> \exp\left\{ - \int_0^{n\beta} d\tau \left[ {
      m\over 2 } \,
{ d {\bf r} \over d\tau } \cdot { d {\bf r} \over d\tau }
+ {\bf F}(\tau) \cdot {\bf r}(\tau) \right] \right\} ,
\label {eq:pathint}
\end{equation}
with
\begin{equation}
{\bf F}(\tau) = -i \sum_{a=1}^l \> {\bf k}_a \, \delta( \tau {-} \tau_a) \,.
\end{equation}
The remaining path integral (\ref {eq:pathint}) describes free-particle
motion (in imaginary time) between ``kicks'' introduced by
the impulsive force ${\bf F}(\tau)$.
To evaluate this path integral explicitly, we write the path $\r(\tau)$
as a constant mean position $\r$ plus a deviation
whose integral over the interval $(0,n\beta)$ vanishes.
The integration measure factors into an ordinary integral over
the mean position $(d^\dim{\bf r})$
and a constrained measure $[d{\bf r}]'$ which denotes integration
over the space of periodic functions with vanishing mean.
The integration over the mean position produces a delta-function,
\begin{equation}
\int (d^\dim {\bf r}) \> \exp\left\{  i \sum_{a=1}^l {\bf k}_a \cdot
{\bf r} \right\} = (2\pi)^\dim \,
    \delta
    \!\left( \sum_{a=1}^l {\bf k}_a \right) ,
\end{equation}
reflecting the spatial translational invariance of the theory.
Hence the time integral of the impulsive force must vanish,
\begin{equation}
\int_0^{n\beta} d \tau \> {\bf F}(\tau) = 0 \,.
\label{intF}
\end{equation}

The remaining functional integral
can be evaluated by `completing the square'. This is done with the aid
of a Green's function $f_n(\tau {-}\tau')$ defined in the space of
periodic functions with vanishing mean.
We take this function to be dimensionless so that it obeys
\begin{equation}
   n \beta {d^2 \over d \tau^2} \,  f_n(\tau ) =
    \delta( \tau ) - (n\beta)^{-1} \,,
\label{fgreen}
\end{equation}
together with the periodicity condition
\begin{equation}
    f_n(\tau + n\beta) = f_n(\tau) \,.
\label{fperiod}
\end{equation}
The solution, when $-n\beta \le \tau \le n\beta$, is
\begin{equation}
f_n(\tau) = {|\tau| \over 2n\beta} \left( 1 - {|\tau| \over n\beta} \right) ,
\label {eq:f(tau)}
\end{equation}
up to an additive constant.
For the formulas below, it is convenient to choose the particular
solution (\ref {eq:f(tau)}) which vanishes at $\tau = 0$.
The square is completed by shifting the functional integration
variable to
$
    \Delta {\bf r}(\tau) \equiv {\bf r}(\tau) - {\bf \bar r}(\tau)
$,
where
\begin{equation}
{\bf \bar r}(\tau) \equiv { n\beta \over m}
 \int_0^{n\beta} d \tau' \, f_n(\tau {-} \tau') \, {\bf F}(\tau')  \,.
\end{equation}
Since the Green's function $f_n(\tau {-} \tau')$ is periodic,
${\bf \bar r}(\tau)$ is periodic, and since ${\bf r}(\tau)$ is
periodic, so is $\Delta{\bf r}(\tau)$.
Moreover, since $f_n(\tau {-} \tau')$ obeys the Green's function equation
(\ref{fgreen}) and ${\bf F}(\tau')$ has a vanishing mean [Eq.~(\ref{intF})],
\begin{equation}
m { d^2 \over d \tau^2 } \, {\bf \bar r}(\tau) = {\bf F}(\tau) \,.
\end{equation}
Hence we may make the shift and freely integrate by
parts with no boundary contributions to evaluate the remaining
functional integral and obtain
\begin{equation}
    z_n[{\bf F}]
    =
    (2\pi)^\dim \, \delta\biggl(i\int_0^{n\beta} d\tau \> {\bf F}(\tau)\biggr)
    \exp\left\{ -  { n\beta \over 2 m}
     \int_0^{n\beta} d\tau \, d\tau' \, f_n(\tau {-} \tau') \,
    {\bf F}(\tau) \cdot {\bf F}(\tau') \right\}
z_n[{\bf 0}] \,.
\label{whiz}
\end{equation}
The final factor $z_n[{\bf 0}]$ is a free particle path integral in the
absence of any external force.
This is just a constant whose precise value
is of no concern since
the overall normalization will be trivially determined {\em a posteriori}
by requiring that our result exhibit the correct free particle limit
when the potential $V$ vanishes.
With these results in hand, we now see that the series (\ref{funyfun})
may be written as
\begin{eqnarray}
&&
\int (d^\dim{\bf r}) \> \langle {\bf r} ,n\beta | {\bf r}, 0 \rangle
\nonumber\\
&& \quad =
  \int (d^\dim {\bf r}) \> \langle {\bf r} ,n\beta | {\bf r}, 0 \rangle^0 \>
\sum_{l=0}^\infty { ( -1)^l \over l! }
\int_0^{n\beta} d \tau_1 \int { (d^\dim {\bf k}_1) \over (2 \pi)^\dim} \>
e^{i {\bf k}_1 \cdot {\bf r} } \, \tilde V({\bf k}_1,\tau_1) \cdots
\nonumber\\
&& \qquad\qquad {}
\int_0^{n\beta} d \tau_l \int { (d^\dim {\bf k}_l) \over (2 \pi)^\dim} \>
e^{ i {\bf k}_n \cdot {\bf r} } \, \tilde V({\bf k}_l,\tau_l) \>
\exp\Biggl\{ { n \beta \over m }  \sum_{b>a=1}^l
{\bf k}_a \cdot {\bf k}_b \> f_n( \tau_a {-} \tau_b )  \Biggr\} \,.
\label{greatest}
\end{eqnarray}

To illustrate the working of our results and to make contact with more
familiar forms, we examine the two-point, charge density -- charge
density correlation function. This function is given by the double
functional derivative of Eq.~(\ref{greatest}) with respect to $\tilde
V$ with $\tilde V$ then taken to vanish, the result summed over $n$ as
in Eq.~(\ref{lnz}), and multiplied by the square of the charge of the
particle which we denote simply as $e^2$.  We also take the Fourier
transform in the imaginary time as well as space. In view of the
time-translation invariance of the result, this Fourier transform is
given by one imaginary time integral over the interval $0,n\beta$ with
a factor $\exp\{ i \omega \tau\}$ while the other imaginary time
integral just provides a factor of $n\beta$, with the factor of
$\beta$ removed by the Fourier transform conventional normalization.
Thus the correlation function is given by
\begin{eqnarray}
\Pi({\bf k}, \omega) &=& e^2 \sum_{n=1}^\infty (\pm 1)^{n+1} { e^{n\beta\mu}
  \over n} \langle \r , n\beta | \r , 0 \rangle^0 \,
n  \int_0^{n\beta} d\tau \> e^{i\omega\tau} \exp\left\{ - n \beta \, { k^2
    \over m} \, f_n(\tau) \right\} \,.
\end{eqnarray}
In order to perform the sum and the Fourier transform, we recall
Eq's.~(\ref{freeatlast}) and  (\ref{eq:f(tau)}) to  write
\begin{eqnarray}
&&\langle \r , n\beta | \r , 0 \rangle^0 \>
\exp\left\{ - n \beta { k^2  \over m} f_n(\tau) \right\}
= \int{ (d^\dim {\bf p}) \over (2\pi)^\dim } \exp\left\{ - n \beta {
    p^2 \over 2m} \right\} \exp\left\{ - { k^2 \over 2m } \, \tau \left(
    1 - {\tau \over n\beta } \right) \right\}
\nonumber\\ && \kern 0.7in {}
= \int{ (d^\dim {\bf p}) \over (2\pi)^\dim } \exp\left\{ - n \beta {
    p^2 \over 2m} \right\} \exp\left\{ - { \tau \over 2m }
 \left(  k^2 - 2 {\bf k} \cdot {\bf p} \right) \right\} \,,
\end{eqnarray}
where the second equality%
\footnote
    {%
    This later form is
    the result obtained by using operator methods to evaluate
    $$
    {\rm Tr} e^{ -n \beta p^2 /2m} e^{i \k \cdot \r (\tau) }
    e^{- i \k \cdot \r (0) } \,,
    $$
    where $ \r (\tau) = \r (0) -i {\bf p} \tau/m$ is the operator
    free-particle motion in imaginary time.
    }
follows by making the translation
${\bf p}
\to {\bf p} - \tau {\bf k} / n \beta $. Since the frequency $\omega$
is a positive or negative integer multiple of $ 2\pi/\beta$, we find
that
\begin{eqnarray}
\Pi({\bf k}, \omega) &=&  e^2 \sum_{n=1}^\infty (\pm 1)^{n+1}
\int{ (d^\dim {\bf p}) \over (2\pi)^\dim } \> e^{n \beta\mu}\,
\exp\left\{ - n \beta {
    p^2 \over 2m} \right\}
{ 1 - \exp\left\{ - { n \beta \over 2m } \left( k^2 - 2 \,{\bf p} \cdot
      {\bf k} \right) \right\} \over ( 1 /2m) \left( k^2 - 2 \, {\bf p}
    \cdot {\bf k} \right) - i \omega }
\nonumber\\
&=&
e^2 \int{ (d^\dim {\bf p}) \over (2\pi)^\dim } \> {
F_\pm ( {\bf p} - \k /2 ) - F_\pm ({\bf p} + \k /2 ) \over
{\bf p} \cdot {\bf k} / m - i \omega } \,,
\label{PI}
\end{eqnarray}
where
\begin{equation}
F_\pm( {\bf p} ) = \left[ \exp\left\{ \beta { p^2 \over 2m} -
      \beta \mu  \right\} \mp 1 \right]^{-1}
\end{equation}
are the free-particle Bose or Fermi distributions, and we have made
a  further translation $ {\bf p} \to - {\bf p} + \k /2 $. This is the
familiar form for the density-density correlator in the `random phase'
or single-ring approximation.%
\footnote
    {
    See, for example, Eq.~(30.9) and
    the discussion about it, in Fetter and Walecka \cite{fetter}.
    }

Let us now restrict the discussion to the limit of classical,
Maxwell-Boltzmann statistics where
\begin{equation}
\Pi({\bf k}, \omega) =
 e^2 \int { (d^\dim {\bf p}) \over (2\pi)^\dim } \>
 e^{\beta\mu} \, \exp\left\{ - { \beta
    \over 2m } \left( p^2 + k^2 /4 \right) \right\}
{2 \sinh \left( { \beta {\bf p} \cdot {\bf k} \over 2m } \right)  \over
{\bf p} \cdot {\bf k} / m - i \omega } \,.
\label{classst}
\end{equation}
Taking the frequency to vanish and expanding in powers of the wave
number gives
\begin{eqnarray}
\Pi({\bf k}, 0) &\simeq&  e^2 \beta n^0 \left[ 1 - { \beta k^2 \over
    12m} \right]
=
\kappa_e^2 \left[ 1 - { \lambda^2 k^2 \over 24 \pi } \right] \,.
\end{eqnarray}
Here in the second form we have written
$ \kappa_e^2 = e^2 \beta n^0$,
which is the contribution to the squared Debye wave number of a
particle of generic charge $e$ and density $n^0$, and $\lambda^2 = 2 \pi
\hbar^2 \beta / m $ for the corresponding thermal wave length. We have
explicitly included the factor of $\hbar^2$ here to emphasize that
this is a quantum correction. On the other hand,
expanding in the wave number
with the frequency non-zero gives
\begin{equation}
\Pi({\bf k}, \omega) \simeq  {e^2  n^0 \over m } { k^2 \over
  \omega^2 } =  \omega_e^2 \, { k^2 \over \omega^2} = \kappa_e^2 \,
 {  \lambda^2  k^2
   \over 2\pi  (\beta \hbar \omega )^2 } \,,
\end{equation}
in which we have identified the generic contribution to the squared
plasma frequency $\omega_e^2 = e^2 n^0 / m$. The plasma frequency is,
of course, purely a classical quantity. However, the discrete
frequencies that enter here are the quantum Matsubara frequencies that are
integer multiples of $ 2 \pi / \hbar \beta $
(with $\beta$ taken to have the units of inverse energy).

The original form of the classical statistics limit is
\begin{equation}
\Pi({\bf k}, \omega) = e^2 n^0
  \int_0^{\beta} e^{i\omega\tau} \exp\left\{ - { \lambda^2 k^2
    \over 2 \pi } \, f(\tau) \right\} \,,
\label{original}
\end{equation}
where we now write
\begin{equation}
f(\tau) = f_1(\tau)
= {|\tau| \over 2\beta} \left( 1 - {|\tau| \over \beta} \right) \,.
\end{equation}
Since this is periodic in $\tau$ with period $\beta$, it has the
Fourier series representation
\begin {equation}
    f(\tau) = \sum_{m=-\infty}^\infty f_m \, e^{-i \omega_m \tau} \,,
\end{equation}
with $\omega_m = 2 \pi m / \beta $. Expanding Eq.~(\ref{original}) to
order $k^2$ and comparing with the results above, we conclude that
\begin{equation}
    f_m = \cases {
			-1/(2 \pi m )^2 \,,& $m \ne 0$; \cr
			\phantom-1/12 \,,& $m = 0$. }
\label {eq:f_m}
\end {equation}
These coefficients are, of course, the same as those obtained directly
from the Fourier transformation of $f(\tau)$.

We now return to the heavy mass limit, or equivalently the classical
limit $\hbar \to 0$,
in which the thermal wave length becomes small,
$ \lambda^2 =2 \pi \hbar^2 \beta/ m\to 0$.
In this limit,
the final exponential in Eq.~(\ref{greatest}) is set to one, and the
resulting series may be trivially summed to reproduce the previous
result (\ref{heavier}).
To obtain systematic corrections to this limit,
it is worth noting that the sum which forms the
integrand in Eq.~(\ref{greatest}) has the same
structure as that of a classical grand canonical partition function
for a system with pairwise interactions given by
$
    V_{ab} \equiv
    -( n / m )
    {\bf k}_a \cdot {\bf k}_b \> f_n( \tau_a {-} \tau_b )
$.
The imaginary-time integrals in Eq.~(\ref{greatest}) take the place of
the spatial integrations that appear in a classical partition
function, while the
remaining factors in (\ref{greatest}) may be
interpreted as defining the single-particle measure.
Consequently the usual ``linked cluster'' theorem of classical
statistical mechanics shows that the sum appearing
in the integrand of Eq.~(\ref{greatest}) equals the
{\it exponential} of the sum of ``connected clusters''.%
\footnote
    {%
    See, for example Ref.~\cite{huang}, Section 10.1.
    The connected nature of the expansion implies that, if the potentials
    $\tilde V({\bf k},\tau)$ were independent of $\tau$, then each term
    in the cluster sum would behave as $n\beta$ in the large $\beta$ limit.
    This is analogous to the appearance of a single volume factor in the
    usual statistical mechanical cluster expansion.
    }
In other words, if
\begin {equation}
    d\mu_a \equiv
    d\tau_a \>
    {(d^\dim \k_a) \over (2\pi)^\dim} \>
    e^{i \k_a \cdot \r} \, \tilde V(\k_a,\tau_a)
\end {equation}
denotes the ``single-particle'' measure, and
\begin {equation}
    g_{ab} \equiv
    \exp\Biggl\{ { n \beta \over m } \,
    {\bf k}_a \cdot {\bf k}_b \> f_n( \tau_a {-} \tau_b )  \Biggr\}
    -1 \,,
\end {equation}
then the sum appearing in Eq.~(\ref {greatest}) may be expressed as
the cluster expansion
\begin {eqnarray}
    &&
    \sum_{l=0}^\infty
	{(-1)^l \over l!}
	\int
	\left[
	    \prod_{a=1\vphantom{a>b}}^l
	    d\mu_a
	\right]
	\left[ \prod_{b>a=1}^l (1+g_{ab}) \right]
\nonumber\\ && {} =
    \exp \left\{
	\int d\mu_1
	+ {1 \over 2!} \int d\mu_1 \, d\mu_2 \> g_{12}
	+ {1 \over 3!} \int d\mu_1 \, d\mu_2 \, d\mu_3 \,
	    \left[ 3 \, g_{12} \, g_{13} + g_{12} \, g_{13} \, g_{23} \right]
	+ \cdots
    \right\}.
\label{cluster}
\end {eqnarray}
An expansion in powers of wave numbers is equivalent to an expansion
in spatial gradients of the potential.
Since $g_{ab}$ starts out proportional to ${\bf k}_a \cdot {\bf k}_b$,
the $j$-th term in the cluster expansion is of order wave-number to the
$2(j{-}1)$-th power. Thus the cluster representation (\ref{cluster}) is a
convenient vehicle for developing a systematic gradient expansion,
in which each gradient will be accompanied by a factor of
$\lambda = \sqrt{2 \pi \hbar^2 \beta/ m }$.
Although the expansion is easily done for an arbitrary term
$n$ in the fugacity expansion, we shall need only the $n{=}1$
result corresponding to the classical limit of Maxwell-Boltzmann
statistics. Hence we now restrict the discussion to $n{=}1$ and write
$f_1(\tau) = f(\tau)$ as before.  A little calculation, taking account
of the remarks that we have just made, shows that to order $\lambda^6$,
\begin {eqnarray}
&&
\int (d^\dim{\bf r}) \> e^{\beta \mu}
\langle {\bf r} , \beta | {\bf r}, 0 \rangle
     =  \int (d^\dim {\bf r})  \> n^0 \,
    \exp \,\Biggl\{ -\int_0^\beta d \tau \> V({\bf r},\tau)
\nonumber\\
&&  \qquad\qquad {}
    - {\lambda^2 \over 4\pi}
    \int_0^\beta d \tau_1 \> d\tau_2 \> f(\tau_1{-}\tau_2) \>
     \nabla_k V(\r,\tau_1) \, \nabla_k V(\r,\tau_2)
\nonumber\\
&&  \qquad\qquad {}
    + \left( {\lambda^2 \over 4\pi} \right)^2
    \int_0^\beta d \tau_1 \> d\tau_2 \> f(\tau_1{-}\tau_2)^2 \>
     \nabla_k \nabla_l V(\r,\tau_1) \, \nabla_k \nabla_l V(\r,\tau_2)
\nonumber\\
&& \qquad\qquad {}
    - 2  \left( {\lambda^2 \over 4\pi} \right)^2
    \int_0^\beta d \tau_1 \> d\tau_2 \> d\tau_3 \>
	f(\tau_1{-}\tau_2) \> f(\tau_1{-}\tau_3) \>
    \nabla_k \nabla_l V(\r,\tau_1) \nabla_k V(\r,\tau_2)
	\nabla_l V(\r,\tau_3)
\nonumber\\
&& \qquad\qquad {}
    + O(\lambda^6)
    \Bigg\} \,.
\label{series-1}
\end {eqnarray}

To apply this result, so as to obtain the corresponding effective
action interaction terms, we must replace, for each ion species $a$,
$
    -V(\r,\tau) \to  \mu_a(\r,\tau) + i e_a \phi(\r,\tau)
$.
We are now allowing arbitrary space-time dependence in the chemical potential
so that the resulting effective action will serve as a generating functional
for frequency and wave-vector dependent number density correlation functions.
After summing over the various particle species,
and inserting the Fourier series decompositions
\begin{equation}
    \phi(\r,\tau) = \sum_{m=-\infty}^\infty \phi^m(\r)\, e^{-i\omega_m \tau}
    \,,
    \qquad
    \mu_a(\r,\tau) = \sum_{m=-\infty}^\infty \mu_a^m(\r)\, e^{-i\omega_m \tau}
    \,,
\end{equation}
one finds the action contribution
\begin{eqnarray}
    \Wone &=&
    -
    \int (d^\dim \r) \>
    \sum_a
    \n0a(\r)
    \> \exp \>
    \Biggl\{
    i \beta e_a \phi^0
\nonumber\\ && {}
    -
    { \beta^2 \lambda_a^2 \over 4 \pi} \,
    \sum_m \, f_m \,
    \nabla_k \, [ \mu_a^m {+} i e_a\phi^m ]
    \nabla_k \, [ \mu_a^{-m} {+} i e_a\phi^{-m} ]
\nonumber\\ && {}
    +
    { \beta^2 \lambda_a^4 \over 16 \pi^2} \,
    \sum_m \, {\it ff\!}_m \,
    \nabla_k \nabla_l \, [ \mu_a^m {+} i e_a\phi^m ] \,
    \nabla_k \nabla_l \, [ \mu_a^{-m} {+} i e_a\phi^{-m} ]
\nonumber\\ && {}
    +
    { \beta^3 \lambda_a^4 \over 8 \pi^2} \,
    \sum_{m,n} \, f_m f_n \,
    \nabla_k \nabla_l \, [ \mu_a^{m{+}n} {+} i e_a\phi^{m{+}n} ] \,
    \nabla_k \, [ \mu_a^{-m} {+} i e_a\phi^{-m} ] \,
    \nabla_l \, [ \mu_a^{-n} {+} i e_a\phi^{-n} ]
\nonumber\\ && {}
    +
    O(\lambda^6)
    \Biggr\} \,.
\label{eq:S2}
\end{eqnarray}
Here $\{ f_m \}$ are the Fourier series coefficients
(\ref {eq:f_m}) for the function $f(\tau)$,
and $\{ {\it ff\!}_m \}$ denote the Fourier series
coefficients for $f(\tau)^2$, namely
\begin {equation}
    {\it ff\!}_m = \cases {
			-6/(2 \pi m )^4 \,,& $m \ne 0$; \cr
			\phantom-1/120 \,,& $m = 0$.
		     }
\label{ffm}
\end {equation}
In Eq.~(\ref {eq:S2}),
the bare density $\n0a(\r)$ is to be understood as containing
the zero-frequency part of the chemical potential, so that
$
    \n0a(\r) = g_a \, \lambda_a^{-\dim} \, \exp {\beta \mu_a^0(\r)}
$.

In the final portion of Section \ref {sec:longdist},
we need the contribution to the time-dependent
number density operator generated by the action (\ref {eq:S2}).
This is the variational derivative of $-S_{\rm int}$ with respect
to $\beta \mu_a (\r,\tau)$.
However, for the specific application of that section that
involves the contribution to the long-distance tail of non-zero frequency
correlators, specifically diagram~(c) of Fig.~\ref {fig:non-static},
we may omit all terms involving $\nabla^2\phi^n$, as such terms lead to
spatial delta function contributions which cannot affect the leading
long distance behavior. The required correction to the number density
operator arises only from the fourth line of Eq.~(\ref {eq:S2}) and,
when Fourier transformed in time, becomes
\begin {equation}
   \Delta n_a(\r,\omega_m)
   =
   - \n0a \, { \beta^2 \, \lambda_a^4 \, e_a^2 \over 8 \pi^2} \,
    \sum_n
    \left( f_n f_{m-n} {-} f_n f_{-m} {-} f_{m-n} f_{-m} \right)
    \nabla_k \nabla_l \, \phi^{-n} \, \nabla_k \nabla_l \, \phi^{n-m} .
\end {equation}
Only terms involving exclusively non-zero frequency
components of $\phi$ will be needed
[as only these components have long-range $1/r$ correlations],
which allows one to exclude the $n = 0$ and $n = m$ terms from
the above sum and express the result as
\begin {equation}
   \Delta n_a(\r,\omega_m \ne 0)
   =
   - \n0a \, { \beta^2 \, \lambda_a^4 \, e_a^2 \over (2 \pi)^6} \,
    \sum_{{n = -\infty \atop n \ne 0,m}}^\infty \,
    [ n (m{-}n) m^2 ]^{-1} \,
    \nabla_k \nabla_l \, \phi^{-n} \, \nabla_k \nabla_l \, \phi^{n-m} \,.
\label {eq:vertex}
\end {equation}

\section {Required Integrals}
\label{required}

\subsection {Coulomb Integrals}
\label {coulomb ints}

The pure Coulomb potential for unit charges in $\dim$ dimensions may
be expressed as the Fourier transform
\begin{equation}
    V_\dim(\r) = \int { (d^\dim \k) \over (2\pi)^\dim } \>
    {e^{i \k \cdot \r } \over k^2 } \,.
\label {eq:V2}
\end{equation}
To evaluate the potential explicitly, it is convenient to use the
representation
\begin{equation}
{ 1 \over k^2 } = \int_0^\infty ds \> e^{ - s k^2 } \,,
\end{equation}
interchange the $s$ and $\k$ integrations, and perform
the resulting Gaussian $\k$ integral.
Writing $s = 1/t$
converts the result to the standard form of a $\Gamma$ function,
and yields
\begin{equation}
    V_\dim(\r) = { \Gamma\left( { \dim \over 2} {-} 1 \right)
    \over 4\pi^{\dim/2} }
    \left( {1 \over r^2 } \right)^{ {\dim \over 2} - 1} \,.
\label {eq:V3}
\end{equation}

\subsubsection {Powers of $V$}

The same procedure may be used to evaluate
Fourier transforms of powers of the Coulomb potential,
\begin {equation}
    C^{(n)}_\dim(\k) \equiv
    \int (d^\dim\r) \> e^{-i \k\cdot\r} \> V_\dim(\r)^n \,.
\end {equation}
We insert the form (\ref {eq:V3}) for the Coulomb potential,
use
\begin{equation}
    r^{-a}
    =
    \Gamma( \coeff a2 )^{-1}
    \int_0^\infty ds \> s^{{a\over2} - 1} e^{ - s r^2 }
    \,,
\label {eq:rsub}
\end{equation}
to represent the resulting power of $r$,
interchange integrals and evaluate the Gaussian $\r$ integral.
The variable change $s = 1/t$ once again produces the
standard representation of the $\Gamma$ function, yielding
\begin{eqnarray}
    C^{(n)}_\dim(\k)
    &=&
    { \Gamma\Bigl( \coeff\dim2 {-} 1 \Bigr)^n \over
      \Gamma\Bigl( n ( \coeff \dim2 {-} 1 ) \Bigr) } \>
    { \Gamma\Bigl(n{-}\coeff\dim2 (n{-}1)\Bigr) \over (4\pi)^n}
    \left({k^2 \over 4\pi}\right)^{{\dim\over2}(n-1)-n}.
\label {eq:Cn}
\end{eqnarray}

To obtain the $\dim \to 3$ limit of this result for various powers of
the potential $n$, we make use of
\begin{equation}
\Gamma(z) \, \Gamma(1-z) = { \pi \over \sin \pi z} \,,
\end{equation}
(from which follows $\Gamma(1/2) = \sqrt \pi$), use
\begin{equation}
\psi(z) \equiv {d \over dz} \ln \Gamma(z) \,,
\end{equation}
with $\psi(1) = - \gamma$, where $\gamma = 0.57721 \cdots $ is Euler's
constant, and Legendre's duplication formula
\begin{equation}
\Gamma(2z) = 2^{2z - 1} \, \pi^{-1/2} \, \Gamma(z) \, \Gamma(z + 1/2) \,,
\end{equation}
(which shows that $\psi(1/2) = -\gamma - \ln 4$, a result that
will also be needed). Using these ingredients,
we find that $C^{(2m)}_\dim(\k)
$ has a smooth limit as $\dim\to 3$,
\begin {equation}
    C^{(2m)}_3(\k)
    = { (-1)^{m+1} \over 4}
    \left({1 \over 4 \pi}\right)^m
    {  \sqrt \pi \over \Gamma(2m-1) }
    \left({k^2 \over 4\pi}\right)^{m-{3\over2}} \,.
\end {equation}
In particular, we will need
\begin{equation}
C^{(4)}_3(\k) = {-1 \over (16 \pi)^2 } \,\sqrt{ k^2 } \,.
\label{cfour}
\end{equation}
For odd powers (greater than 1) there is a simple
pole in $3-\dim$ arising from the last gamma function in
(\ref {eq:Cn}), and one finds that
\begin {eqnarray}
    C^{(2m+3)}_\dim(\k)
    &=& (-1)^m
    \left({1 \over 4\pi}\right)^{m+2}
    \left({k^2 \over 4\pi}\right)^{m+(\dim-3)(m+1)}
    {1 \over \Gamma(2m+3)}
\nonumber
\\ && {}
    \times \left\{
	{1 \over 3{-}\dim}
	+ (\coeff 32{+}m)\left[ \gamma+\ln4+\psi(\coeff 32{+}m)\right]
	+ (1{+}m) \, \psi(1{+}m)
	+ O(\dim{-}3)
    \right\} .
\nonumber\\
&&
\end {eqnarray}
In particular,
\begin{eqnarray}
    C^{(3)}_\dim(\k)
    &=&
    {1 \over 2\,(4\pi)^2}
    \left( {k^2 \over 4 \pi} \right)^{\dim-3}
    \left\{ { 1 \over 3-\dim} + 3 - \gamma + O(\dim{-}3) \right\} ,
\label{ftvcube}
\\
    C^{(5)}_\dim(\k)
    &=&
    {k^2 \over 4!\,(4\pi)^4}
    \left( {k^2 \over 4 \pi} \right)^{2(\dim-3)}
    \left\{ -{ 1 \over 3-\dim} - {26\over3} + 2\gamma + O(\dim{-}3) \right\} .
\label{ftvfive}
\end{eqnarray}

\subsection {Debye Integrals}
\label {debye ints}

The Debye potential for a point charge in $\dim$ spatial dimensions
has the Fourier transform representation
\begin{equation}
    G_\dim(\r)
    =
    \int { (d^\dim{\k}) \over (2\pi)^\dim} \>
    { e^{i\k\cdot\r} \over k^2 + \kappa^2 } \,.
\end{equation}
Writing the denominator as
\begin{equation}
    {1 \over k^2 + \kappa^2} = \int_0^\infty ds \> e^{-(k^2 + \kappa^2) s} ,
\end{equation}
interchanging integrals, performing the resulting Gaussian integral in
$\k$, and scaling the resulting parameter integration variable by $ s
= t (r /2\kappa) $ expresses $G_\dim(\r)$ in terms of a standard
representation for a modified Bessel function,
\begin{eqnarray}
    G_\dim(\r) &=&
    {1 \over (2\pi)^{\dim/2}}
    \left( { \kappa \over r } \right)^{ { \dim \over 2} - 1}
    {1 \over 2} \int_0^\infty dt \> t^{-\dim/2}
    \exp\left\{ - { \kappa r \over 2}
	\left( t + {1 \over t} \right) \right\}
    \nonumber
    \\ &=&
    { 1 \over (2\pi)^{\dim/2} }
    \left( { \kappa \over r } \right)^{ { \dim \over 2} -1}
    K_{ { \dim \over 2} - 1} (\kappa r) \,.
\label {eq:Gbessel}
\end{eqnarray}
The power series development of the modified Bessel function yields
\begin{eqnarray}
    G_\dim(\r)
    &=&
    { 1 \over 2 \, (2\pi)^{\dim/2} } \sum_{m=0}^\infty
    {(-1)^m \over m!} \, \left({\kappa r\over 2}\right)^{2m} \>
    \Biggl[ \left( { 2 \over r^2} \right)^{{\dim\over2} - 1}
    \Gamma(-m+\coeff\dim2-1)
\nonumber
\\ && \kern 2in {}
    + \left({\kappa^2 \over 2}\right)^{{\dim\over2} - 1}
    \Gamma(-m-\coeff\dim2+1)
    \Biggr] \,,
\label {eq:Kser}
\end {eqnarray}
which displays the singular and regular terms for small $r$.

\subsubsection {Powers of $G$}
\label {sec:debye ints}

\noindent
Let $D^{(n)}_\dim(\k) $ denote the Fourier transform of the $n$-th
power of the Debye potential,
\begin {equation}
    D^{(n)}_\dim(\k) \equiv \int (d^\dim\r) \>
    e^{-i \k\cdot\r} \> G_\dim(\r)^n \,.
\label {eq:Dn}
\end {equation}
The density-density correlation function at $l$-loop order
requires $D_\dim^{(n)}(\k)$ for $n$ up to $l{+}1$,
and the $\k=0$ limits, $D^{(n)}_\dim({\bf 0})$ for $n \le l{+}1$,
are needed for the $l$-loop free energy.

$D^{(1)}_\dim(\k)$ is just the Fourier transformed Debye potential,
\begin {equation}
    D^{(1)}_\dim(\k) = \tilde G(\k) = {1 \over \k^2 + \kappa^2} \,,
\end {equation}
while $D^{(2)}_\dim(\k)$ may be evaluated directly in three dimensions
(and was already computed in section 2),
\begin {equation}
    D^{(2)}_3(\k)
    = {1 \over 4\pi} \int_{2\kappa}^\infty {d\mu \over k^2 + \mu^2}
    = {1 \over 4\pi k} \, \arctan {k \over 2\kappa} \,.
\label {eq:D2}
\end {equation}
It has the vanishing wave number limit
\begin {equation}
    D^{(2)}_3({\bf 0})
    = {1 \over 8\pi\kappa} \,.
\label {d20}
\end {equation}

For $ D_\dim^{(n)}(\k)$ with $n \ge 3$, one must work in $\dim < 3$
dimensions and separate out the terms which diverge as $\dim \to 3$,
terms which arise from the small $r$ region of the Fourier
transform (\ref{eq:Dn}). Since the Coulomb potential in $\dim$
dimensions, $V_\dim(\r)$, is the $\kappa \to 0$ limit of $G_\dim(\r)$,
the short-distance limit of the expansion (\ref{eq:Kser}) may be
written as
\begin{equation}
G_\dim(\r) = V_\dim(\r) \left[ 1 + O \left( (\kappa r)^2 \right)
\right] + { \kappa^{\dim -2} \over (4 \pi )^{\dim/2} } \,
\Gamma ( 1 - \coeff\dim2 ) \left[ 1 + O \left( (\kappa r)^2 \right)
\right] .
\label{smrg}
\end{equation}
To compute $D_\dim^{(3)}(\k)$, we note that as $ r \to 0$, $
V_\dim(\r) \sim (1 / r)^{\dim -2} $, and so $ [ G_\dim(\r)^3 -
V_\dim(\r)^3 ] $ is less singular than $ 1 / r^3$ when $\dim \to
3$. Hence the Fourier transform of this difference may be evaluated
directly in $\dim = 3$ dimensions, and we may write
\begin{equation}
D_\dim^{(3)}(\k) = \int (d^3\r) e^{-i \k \cdot \r} [ G_3(\r)^3 -
V_3(\r)^3 ] + C_\dim^{(3)}(\k) + O(\dim{-}3) \,,
\end{equation}
where $C_\dim^{(3)}(\k)$ is the Fourier transform of the cube of the
Coulomb potential previously evaluated in Eq.~(\ref{ftvcube}). To
compute the integral of the difference of the cube of the Debye and
Coulomb potentials, we represent
\begin{equation}
G_3(\r)^3 = { e^{-3 \kappa r} \over ( 4 \pi r)^3 } =
{1 \over (4 \pi)^2 } \int_{3\kappa}^\infty d \mu \> ( \mu - 3 \kappa ) \,
{ e^{- \mu r} \over 4 \pi r} \,,
\end{equation}
and use its $\kappa \to 0$ limit to represent $V_3(\r)^3$. Placing an
upper bound $\mu = M$ on the these parametric integrals, with the
limit $ M \to \infty$ reserved until the end of the computation,
allows separate Fourier transforms to be taken, with the result,
using Eq.~(\ref{ftvcube}), that
\begin{eqnarray}
D_\dim^{(3)}(\k) &=& { 1 \over (4 \pi)^2 } \lim_{M \to \infty} \left\{
\int_{3\kappa}^M d\mu \> {\mu - 3\kappa \over k^2 + \mu^2 }  -
\int_{0}^M d\mu \> {\mu \over k^2 + \mu^2 }  \right\}
\nonumber\\
&& {} + {1 \over (4\pi)^2}
    \left( {k^2 \over 4 \pi} \right)^{\dim-3} { 1 \over 2}
    \left\{ { 1 \over 3-\dim} + 3 - \gamma + O(\dim{-}3) \right\} .
\end{eqnarray}
To keep the result in a dispersion relation or spectral form,
we write
\begin{equation}
    {\mu \over k^2 + \mu^2}
    =
    {1\over \mu}
    -
    {1\over \mu} \, {k^2 \over k^2+\mu^2}
\end{equation}
in the first (Debye) integral and add the part
\begin{equation}
    \int_{3\kappa}^M {d\mu \over \mu}
    =
    \ln \left({M \over 3\kappa}\right)
\end{equation}
to the $-\ln (M / k)$ produced by the second (Coulomb) integral.
The limit $M \to \infty$ can then be taken, and
these two pieces reduce to $ \half \ln ( k^2 / 9 \kappa^2 ) $.
Since in the $\dim \to 3$ limit
\begin{equation}
    \left( {k^2 \over 4 \pi} \right)^{\dim-3} { 1 \over 3 - \dim } =
    \left( {\kappa^2 \over 4 \pi} \right)^{\dim-3} { 1 \over 3 - \dim }
    + \ln \left( {\kappa^2 \over k^2} \right) ,
\end{equation}
the $ \ln ( k^2 / \kappa^2 ) $ terms cancel, as they must, and there
remains
\begin{eqnarray}
D_\dim^{(3)}(\k) = { 1 \over (4\pi)^2 } \left( { 9 \kappa^2 \over 4
    \pi} \right)^{\dim -3} { 1 \over 2}
\Bigg\{ { 1 \over 3 - \dim} + 3 - \gamma -
2 \int_{3\kappa}^\infty d \mu \left( { k^2 \over \mu} + 3 \kappa
\right) { 1 \over k^2 + \mu^2 } + O(\dim{-}3)\bigg\} \,.
\nonumber\\
\label{d3kd}
\end{eqnarray}
We have written an overall factor of $(\kappa^2)^{\dim -3}$ so as
to keep the dimensions correct when $ \dim - 3 \neq 0$ although this
factor may be replaced by unity when it multiplies regular terms.
It is a simple matter to evaluate the final integral and
obtain the explicit result
\begin{eqnarray}
    D^{(3)}_\dim(\k) =
    { 1 \over (4\pi)^2}
    \left( {9 \kappa^2 \over 4 \pi } \right)^{\dim-3}
    {1 \over 2}
    \left\{
    {1 \over 3-\dim} + 3 - \gamma
    -{6\kappa \over k} \arctan{k \over 3\kappa}
    - \ln \biggl[1 + {k^2 \over 9 \kappa^2}\biggr]
    + O(\dim{-}3)
    \right\} ,
\nonumber\\
\label {eq:D3}
\end{eqnarray}
whose $\k \to 0$ limit is equal to
\begin{equation}
    D^{(3)}_\dim({\bf 0}) =
    { 1 \over (4\pi)^2}
    \left( {9 \kappa^2 \over 4 \pi } \right)^{\dim-3}
    {1 \over 2}
    \left\{
    {1 \over 3-\dim} + 1 - \gamma + O(\dim{-}3)
    \right\} .
\label{d30}
\end{equation}

The computation of $D_\dim^{(4)}(\k)$ may be performed in a similar
fashion. Again referring to Eq. (\ref{smrg}), it is easy to check that
\begin{equation}
    G_\dim(\r)^4 - V_\dim(\r)^4 - 4 V_\dim(\r)^3 {\kappa^{\dim -2}
    \over (4 \pi)^{\dim/2} } \, \Gamma( 1 {-} \coeff \dim 2)
\end{equation}
is less singular than $ 1 / r^3 $ when $ \dim \to 3$. Hence,
\begin{eqnarray}
    D_\dim^{(4)}(\k) &=&
    C_\dim^{(4)}(\k) +
    4 \, C_\dim^{(3)}(\k) \,
    {\kappa^{\dim -2} \over (4 \pi)^{\dim/2} } \, \Gamma( 1 {-} \coeff\dim 2 )
\nonumber\\
&& {} + \,
    \int (d^3\r) \> e^{-i \k \cdot \r} { 1 \over (4 \pi)^4 }
    \left[
	{ e^{-4 \kappa r} \over r^4 } - { 1 \over r^4} + {4 \kappa \over r^3 }
    \right]
    + O(\dim{-}3)
  \,.
\end{eqnarray}
As before, we write the terms in the square brackets in the Fourier
transform integral as parametric integrals over $ e^{- \mu r} / r$ and
interchange integrals to obtain
\begin{eqnarray}
D_\dim^{(4)}(\k) &=& { 1 \over (4 \pi)^3 } \lim_{M \to \infty} \left\{
  {1 \over 2} \int_{4 \kappa}^M d \mu \> { ( \mu - 4 \kappa )^2
    \over k^2 + \mu^2 } - { 1 \over 2} \int_0^M d \mu \> { \mu ( \mu - 8
    \kappa) \over k^2 + \mu^2 } \right\}
\nonumber\\
&& {} + \,
C_\dim^{(4)}(\k) + 4 C_\dim^{(3)}(\k) \, {\kappa^{\dim -2}
    \over (4 \pi)^{\dim/2} } \, \Gamma( 1 {-} \coeff\dim 2)
  \,.
\end{eqnarray}
With the aid of the results (\ref{cfour}) and (\ref{ftvcube}) for
$C_\dim^{(4)}(\k)$ and $C_\dim^{(3)}(\k)$, it is a straightforward
matter to compute $D_\dim^{(4)}(\k)$. Since we need only
$D_\dim^{(4)}({\bf 0})$, we shall simply state that
\begin{equation}
D_\dim^{(4)}({\bf 0}) = - { 2 \kappa \over (4 \pi)^3 } \left( {
    \kappa^2 \over 4 \pi } \right)^{3 (\dim -3)/2} \left\{ { 1 \over 3
    - \dim } + 4 - \coeff 32 \gamma - 5 \ln 2 \right\} .
\label{d40}
\end{equation}

\subsubsection {Convolution integrals}

\noindent
The Fourier transforms
\begin{equation}
    D^{(lmn)}_\dim(\k)
    = \int (d^\dim\r)(d^\dim\r_1) \>
	e^{-i \k\cdot\r}\>
	G_\dim(\r {-} \r_1)^l \, G_\dim(\r_1)^m \, G_\dim(\r)^n \,,
\end{equation}
and
\begin{equation}
    D^{(klmn)}_\dim(\k) =
	\int (d^\dim\r)(d^\dim\r_1)(d^\dim\r_2) \>
	e^{-i \k\cdot\r}\>
	G_\dim(\r{-}\r_1)^k \, G_\dim(\r_1{-}\r_2)^l \,
	G_\dim(\r_2)^m \, G_\dim(\r)^n \,,
\end{equation}
were defined in the text in Eq's.~(\ref{Dnlm}) and (\ref{Dnlmq}). The
two-loop correlators require the evaluation of $D^{(111)}(\k)$,
$D^{(211)}(\k)$, and $D^{(1211)}(\k)$, while the three-loop free
energy involves $D^{(211)}({\bf 0})$, $D^{(221)}({\bf 0})$, and
$D^{(2121)}({\bf 0})$. All of these quantities are well defined and
may be evaluated directly in $\dim = 3$ dimensions.

The Fourier transform representation of $D^{(111)}(\k)$ reads
\begin{equation}
    D^{(111)}_3(\k)
    =
    \int {(d^3 \q) \over (2\pi)^3} \>
    [\q^2+\kappa^2]^{-2} \, [(\k{-}\q)^2 + \kappa^2]^{-1} \,.
\label{d111k}
\end{equation}
This is just the derivative with
respect to the (squared) Debye wave number of the Fourier transform of
the square of the Debye Green's function, $D^{(2)}_3(\k)$,
\begin{equation}
    D^{(111)}_3(\k)
    =
    - {1 \over 2} \, {d D^{(2)}_3(\k)\over d\kappa^2}
    =
    {1 \over 8\pi k} \,
    {1 \over k^2 + 4 \kappa^2} \,.
\end{equation}

The other needed integrals are most easily evaluated
using the spectral representation for the square of a Debye propagator
in three dimensions,
\begin {equation}
    G_3(\r)^2
    =
    \int_{2 \kappa}^\infty {d\mu \over 4\pi} \>
    { e^{- \mu r} \over 4 \pi r}
    =
    \int {(d^3\k)\over(2\pi)^3}
    \int_{2 \kappa}^\infty {d\mu \over 4\pi} \>
    {e^{i\k\cdot\r} \over k^2 + \mu^2}
    \,.
\end {equation}
Inserting this form into the definitions of $D^{(211)}_3(\k)$
and $D^{(1211)}_3(\k)$, Fourier transforming,
and interchanging orders of integration produces
\begin{equation}
    D^{(211)}_3(\k) = \Delta(\k;\kappa,\kappa) \,,
\end{equation}
and
\begin{equation}
    D^{(1211)}_3(\k) = -\left.\!{\partial \over \partial m^2} \>
    \Delta(\k;\kappa,m) \right|_{m=\kappa}\,,
\end {equation}
in which
\begin{equation}
    \Delta(\k;\kappa,m) \equiv
    \int_{2\kappa}^\infty {d\mu \over 4\pi}
    \int {(d^3\q)\over (2\pi)^3} \>
    {1 \over (\q^2+ \mu^2)} \, {1 \over (\q^2+m^2)} \,
    { 1 \over (\k{-}\q)^2+\kappa^2 } \,.
\end{equation}
The partial fraction decomposition
\begin{equation}
 {1 \over \q^2+ \mu^2} \> {1 \over \q^2+m^2} =
{1 \over \mu^2 - m^2 } \left[ {1 \over \q^2+m^2} -  {1 \over \q^2+ \mu^2}
    \right]
\end{equation}
yields two convolution integrals
\begin{equation}
\int { (d^3\q) \over (2\pi)^3}
\left[ {1 \over \q^2+m^2} -  {1 \over \q^2+ \mu^2} \right]
 { 1 \over (\k{-}\q)^2+\kappa^2 }
\end{equation}
which just represent the Fourier transform of the difference of two
products in coordinate space,
\begin{equation}
\left[ { e^{-mr} \over 4\pi r} - { e^{-\mu r} \over 4 \pi r} \right] {
  e^{- \kappa r} \over 4 \pi r} = \int_{\kappa + m}^{\kappa + \mu} {d
  u_1 \over 4 \pi } { e^{-\mu_1 r} \over 4\pi r} \,.
\end{equation}
Hence
\begin{equation}
    \Delta(\k;\kappa,m) =
    {1\over(4\pi)^2}
    \int_{2\kappa}^\infty {d\mu\over \mu^2-m^2}
    \int_{\kappa+m}^{\kappa+\mu} {d\mu_1 \over \mu_1^2 + k^2} \,.
\end{equation}
Using
\begin{equation}
{  1 \over \mu^2 - m^2 } = - { 1 \over 2m} {d\over d\mu} \ln \left[ {
    \mu + m \over \mu - m} \right] ,
\end{equation}
and integrating by parts gives
\begin{eqnarray}
    \Delta(\k;\kappa,m) &=& { 1 \over (4\pi)^2 } {1 \over 2m}
    \left\{
	\ln\left[ { 2\kappa + m \over 2\kappa - m} \right]
	\int_{\kappa + m}^{3\kappa}  { d \mu \over k^2 + \mu^2 }
	+
	\int_{3\kappa}^\infty
	{ d\mu \over k^2 + \mu^2 } \>
	\ln \left[ \mu + m - \kappa \over \mu - m - \kappa \right]
    \right\} .
\end{eqnarray}

This result yields
\begin{eqnarray}
    D_3^{(211)}(\k)  &=& { 1 \over (4\pi)^2 } {1 \over 2\kappa}
    \left\{
    \ln 3
    \int_{2\kappa }^{3\kappa}  { d \mu \over k^2 + \mu^2 }
    +
    \int_{3\kappa}^\infty
    { d\mu \over k^2 + \mu^2 } \>
    \ln \left[ \mu \over \mu  - 2 \kappa \right]
    \right\} ,
\label {eq:d211}
\end{eqnarray}
and
\begin{eqnarray}
D_3^{(1211)}(\k) &=&
{ 1 \over (4 \pi)^2 } {1 \over 4 \kappa^2 } \left\{ { \ln 3 \over
  k^2 + 4 \kappa^2 }
- { 4 \over 3 \kappa} \int_{2\kappa}^{3\kappa} { d \mu \over
  k^2 + \mu^2 }
  - \int_{3\kappa}^\infty
  { d\mu \over k^2 + \mu^2}
  \left[ { 1 \over \mu} + { 1 \over \mu - 2 \kappa} \right]
  \right\}
\nonumber\\
&& {}
+
{ 1 \over 2 \kappa^2 } \, D_3^{(211)}(\k)
    \,.
\label {eq:d1211}
\end{eqnarray}
Simple integrations give the $\k=0$ limits
\begin{equation}
D_3^{(121)}({\bf 0}) =
D_3^{(211)}({\bf 0}) = { 1 \over (4\pi)^2 } \, { 1 \over 6 \kappa^2} \,,
\label{d2110}
\end{equation}
and
\begin{equation}
D_3^{(1211)}({\bf 0}) = { 1 \over (4\pi)^2 } \, { 1 \over 18 \kappa^4} \,,
\label{d12110}
\end{equation}
where in Eq.~(\ref{d2110}) we have noted that at zero wave number
$D_3^{(121)}({\bf 0}) = D_3^{(211)}({\bf 0}) $.
Again, we have placed the results (\ref {eq:d211}) and (\ref {eq:d1211})
in dispersion relation form. They
may also be expressed in terms of elementary functions and Euler's dilogarithm
\begin {equation}
    \Li_2(-z) \equiv -\int_0^z {dt\over t} \> \ln(1+t) \,.
\end {equation}
The dilogarithm contributions are exhibited by changing the dispersion
relation integration variable to $s = 1/\mu$, and then making partial fraction
decompositions and further linear transformations on the $s$
integration variable. The results are:
\begin {eqnarray}
    D^{(211)}_3(\k)
    &=&
    {1 \over (4\pi)^2 }{1  \over 4k \kappa}
    \left\{
	 i \, \Li_2\biggl(-2+{ik\over\kappa}\biggr)
	-i \, \Li_2\biggl(-2-{ik\over\kappa}\biggr)
    \right.
\nonumber\\&& \qquad\qquad \left. {}
	+i \, \Li_2\biggl(-{ik \over 3\kappa}\biggr)
	-i \, \Li_2\biggl({ik \over 3\kappa}\biggr)
	+ 2\ln 3 \, \arctan {k \over 2\kappa}
    \right\},
\label {eq:D211}
\end{eqnarray}
and
\begin{eqnarray}
    D^{(1211)}_3(\k)
    &=&
    {1 \over (4\pi)^2 } { 1 \over 8 k \kappa^3}
    \left\{
	- { 8 \over 3} \, \arctan {k \over 2\kappa}
	+ \left({8\over3} + {4 \kappa^2 \over k^2 + 4\kappa^2}\right)
	    \arctan {k \over 3\kappa}
    \right.\nonumber\\&& \quad\qquad\qquad \left. {}
	- {2 \kappa \over k} \,
	    \biggl({k^2 + 2 \kappa^2 \over k^2 + 4\kappa^2}\biggr)
		\ln\biggl[1+{k^2\over9\kappa^2}\biggr]
    \right\}
    +
    {1 \over 2\kappa^2} D^{(211)}_3(\k) \,.
\label{eq:D1211}
\end {eqnarray}

The same techniques may be used to compute
$D^{(221)}({\bf 0})$ and $D^{(2121)}({\bf 0})$.
It is easy to see that
\begin {equation}
    D^{(221)}_3({\bf 0}) = T(\kappa,\kappa) \,,
\end{equation}
and
\begin{equation}
    D^{(2121)}_3({\bf 0}) = -\left. \!{\partial \over \partial m^2} \>
    T(\kappa,m) \right|_{m=\kappa} \,,
\end {equation}
where
\begin{equation}
    T(\kappa,m)
     \equiv
    \int_{2\kappa}^\infty {d\mu_1 \over 4\pi}
    \int_{2\kappa}^\infty {d\mu_2 \over 4\pi}
    \int {(d^3\q)\over (2\pi)^3} \>
    {1 \over (\q^2 + \mu_1^2) (\q^2 + \mu_2^2) (\q^2 + m^2)} \,.
\end{equation}
The three-dimensional $\q$ integral can be readily evaluated using
spherical coordinates. Since the radial integral is even in $q$, it
may be extended to run over $-\infty < q < + \infty$ if it is
multiplied by $1/2$. The resulting integral over an infinite range is
trivially evaluated by contour integration.
A little algebra puts the result in the form
\begin{equation}
    T(\kappa,m)
     = { 1 \over (4 \pi)^3 }
    \int_{2\kappa}^\infty d\mu_1 \> d\mu_2 \>
    { 1 \over (\mu_1+m) (\mu_2+m) (\mu_1+\mu_2) } \,.
\end{equation}
The change of variables
\begin{equation}
\mu_1 = (m + 2 \kappa) (y^{-1} -1 )x + 2 \kappa \,, \qquad
\mu_2 = (m + 2 \kappa) (y^{-1} -1 )(1-x) + 2 \kappa \,,
\end{equation}
converts the integration region to $ 0 < x,y < 1$. The $x$ integration
is easily performed with the result that
\begin{equation}
    T(\kappa,m)
     = { 2 \over (4 \pi)^3 }
    \int_0^1 dy \, { 1 \over (2 \kappa + m) + y ( 2\kappa - m) } \,
{ 1  \over 1 + y } \, \ln { 1 \over y} \,.
\end{equation}
A partial fraction decomposition, integration by parts, and
a simple scale change for the integration variable in one of the
terms gives the final form
\begin{equation}
    T(\kappa,m) =
{ 1 \over (4\pi)^3 } { 1  \over m} \left[
  \Li_2\biggl(-{2\kappa-m\over 2\kappa+m}\biggr) + \Li_2(-1)
    \right] .
\end {equation}
Hence, using $ \Li_2(-1) = \pi^2 / 12$, we have
\begin{equation}
    D^{(221)}_3({\bf 0}) =
    {1 \over (4\pi)^3 } { 1 \over \kappa}
    \left[
	\Li_2\Bigl(-\coeff13\Bigr)
	+ {\textstyle{\pi^2 \over 12}}
    \right] ,
\label{d2210}
\end{equation}
and
\begin{equation}
    D^{(2121)}_3({\bf 0})
    =
    {1 \over (4\pi)^3 } {1 \over 2\kappa^3}
    \left[
	\Li_2\Bigl(-\coeff13\Bigr)
	+ {\textstyle{\pi^2 \over 12}}
	+\coeff 43 \ln \coeff 34
    \right] .
\label{d21210}
\end {equation}

\subsubsection {Even worse integrals}

The final integral needed for the three loop free energy is
the ``Mercedes'' integral
\begin {eqnarray}
    D_M &=&
	\int (d^3\r)(d^3\r')(d^3\r'') \>
	G_3(\r)\,
	G_3(\r')\,
	G_3(\r'')\,
	G_3(\r{-}\r')\,
	G_3(\r'{-}\r'')\,
	G_3(\r''{-}\r) \,,
    \\ &\equiv&
    {C_M \over(4\pi\kappa)^3} \,.
\end {eqnarray}%
The pure number $C_M$ may be shown
to be given by%
\footnote
    {%
    The first integral form in (\ref {eq:CM-1})
    was derived by Rajantie \cite {rajantie}.
    The second integral form and the relation to Clausen integrals
    was ``experimentally'' deduced (and verified to 1000 place precision)
    by Broadhurst \cite {broadhurst}.
    This reduction to Clausen integrals has been proven analytically by
    Almkvist \cite {almkvist}.
    }
\begin {eqnarray}
    C_M &=&
    {1 \over \sqrt 2}
    \int_0^1
    {dx \over \sqrt {3-x^2}}
    \left[
	\ln {3\over4}
	+ \ln {3+x \over 2+x}
	-{x^2 \over 4-x^2} \,\ln {4 \over 2+x}
	+{x \over 2+x} \,\ln {3+x \over 3}
    \right]
\label {eq:CM-1}
\\ &=&
    -{1 \over \sqrt 2} \int_{2\alpha}^{4\alpha}
    d \theta \> \ln \left( \textstyle 2 \sin {\theta \over 2} \right)
\nonumber
\\ &=&
    {1 \over \sqrt 2} \left[ {\rm Cl}_2(4\alpha) - {\rm Cl}_2(2\alpha) \right]
\nonumber
\\ &=&
    0.0217376 \cdots \,.
\label {eq:C_M}
\end {eqnarray}
Here
$
    {\rm Cl}_2(\theta)
    \equiv
	{\rm Im} \, \Li_2(e^{i\theta})
    \equiv \sum_{n=1}^\infty \sin(n\theta)/n^2
$
is the Clausen integral, and $\alpha \equiv \arcsin {1 \over 3}$.

The final integral needed for the two-loop self energy is
\begin {eqnarray}
    D_J(\k) &=&
	\int (d^3\r)(d^3\r_1)(d^3\r_2) \>
	e^{-i \k \cdot \r} \,
	G_3(\r{-}\r_1) \,G_3(\r{-}\r_2) \,G_3(\r_1{-}\r_2) \,
	G_3(\r_1) \,G_3(\r_2)
\\ &\equiv&
    {J(k/\kappa) \over (4\pi)^2 \, \kappa^4} \,.
\end {eqnarray}%
This integral is related to the discontinuity of the Mercedes integral
if the screening length in one of the Debye potentials is analytically
continued.
It is not (so far as we know) expressible in terms of standard functions.
However, Rajantie \cite{rajantie} has shown that it may be reduced
to the one-dimensional form
\begin {eqnarray}
    J(z)
    =
    {z^{-2}\over \sqrt{z^2+3}}
    \int_0^1 \! {dx \over \sqrt{z^2 + 4 - x^2}}
    \left\{
	{2z\over 2+x}
	\left[
	    \arctan {z \over 2+x}-\arctan {z\over 2}
	\right]
	+
	\ln \Biggl[{z^2 + (2+x)^2\over (2+x)^2}\Biggr]
    \right\}.
\nonumber\\
\label {eq:J}
\end {eqnarray}
For small $z$,
\begin {equation}
    J(z) = {1 \over 36} - {2 \, z^2 \over 243} + O(z^4) \,.
\label{dj0}
\end {equation}

\section{Quantum Coulomb $su(1,1)$ Symmetry Exploited}
\label {app:SU(1,1)}

As discussed in the text, the ultraviolet divergences of classical
two-loop order quantities are tamed by quantum fluctuations.
The value of the first induced coupling which must be added to the
classical theory can be inferred from the
computation of the quantum-mechanical, two-particle,
finite-temperature correlation function.
With the center-of-mass
motion factored out as done in the text
[Eq.~(\ref{eq:qmcorr})], the Fourier transform of the
direct contribution to the
relative motion correlation for particle species $a,b$ reads
\begin{equation}
F_+(\k) = \int (d^3 \r) \> e^{- i \k \cdot \r } \, \langle \r |
e^{- \beta H } | \r \rangle \,.
\label{corrfn}
\end{equation}
while the exchange contribution is
\begin{equation}
F_-(\k) = \int (d^3 \r) \> e^{- i \k \cdot \r } \langle \r |
e^{- \beta H } | {-} \r \rangle \,.
\label{correx}
\end{equation}
Here
\begin{equation}
H = { {\bf p}^2 \over 2 m_{ab} } + { e_a e_b \over 4\pi r } \,,
\end{equation}
is the Hamiltonian for the relative motion, with
\begin{equation}
{ 1 \over m_{ab} } = { 1 \over m_a } + { 1 \over m_b}
\end{equation}
the reduced mass of the two particles. To temporarily simplify the
notation, we shall write $ m_{ab} = m$ and $e_a e_b / 4 \pi = e^2$
so that the Hamiltonian reads
\begin{equation}
H = { {\bf p}^2 \over 2m} + { e^2 \over r} \,.
\end{equation}
Placing the factor of $e^{- i \k \cdot \r }$ inside the
matrix element in Eq.~(\ref{corrfn}) and treating the coordinate
$\r$ as an operator allows one to express the correlation function as a
quantum-mechanical trace,
\begin{equation}
F_+(\k) = {\rm Tr} \> e^{- \beta H} e^{- i \k \cdot \r } \,.
\label {F=tr}
\end{equation}
For the exchange contribution (\ref{correx}),
we may write $ | {-} \r \rangle = {\cal P} \, | \r \rangle$, where
${\cal P}$ is the parity operator, so that
\begin{equation}
F_-(\k) = {\rm Tr} \> {\cal P} \, e^{- \beta H} e^{- i \k \cdot \r } \,.
\label {F-=tr}
\end{equation}
The evaluation of $F_-(\k)$ closely parallels that of $F_+(\k)$.
To keep the presentation as simple as possible we will focus on $F_+(\k)$,
and then summarize the analogous results for the exchange contribution $F_-(\k)$
at the end of this appendix.

It proves convenient to write the correlation function $F_+(\k)$
as a contour integral involving the Green's function
\begin{equation}
G(\k,E) = {\rm Tr} \> { 1 \over H - E } \, e^{- i \k \cdot \r } \,;
\label {G=tr}
\end{equation}
namely%
\footnote
    {%
    \label {fn:sing}%
    This is slightly cavalier.
    Although the trace defining $F_+(\k)$ in (\ref {F=tr}) is well-defined,
    the corresponding trace in (\ref {G=tr})
    has a high-energy divergence in two or more dimensions.
    This divergence, which merely reflects the growth of the
    density of states at high energy, is independent of the charge $e^2$.
    Therefore, we should really subtract the $e^2 \to 0$ limit inside the
    trace defining $G(\k,E)$ and
    write $F_+(\k) = F_+^0(\k) + \Delta F_+(\k)$,
    where $F_+^0(\k) = \lambda_{ab}^{-3} \> (2\pi)^3 \, \delta(\k)$
    is the $e^2 = 0$ limit,
    so that the contour integral (\ref {contint}) becomes a
    representation just for the difference $\Delta F_+(\k)$.
    But to keep the notation as simple as possible,
    we will not indicate this subtraction explicitly.
    }
\begin{equation}
F_+(\k) = \int_{\rm C} { dE \over 2\pi i } \, e^{- \beta E } \, G(\k,E) \,,
\label{contint}
\end{equation}
\begin {figure}[t]
   \begin {center}
      \leavevmode
      \def\epsfsize #1#2{0.3#1}
      \epsfbox {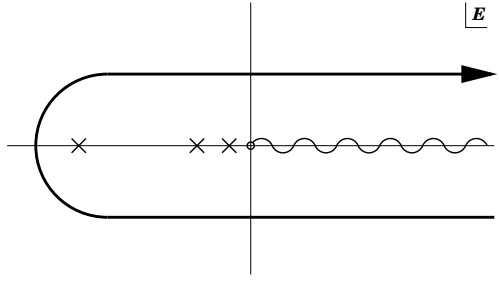}
   \end {center}
   \caption
       {%
       Integration contour for $F_+(\k)$.
       The Green's function $G(\k,E)$ has a cut along the positive
       real axis and, in the case of an attractive potential,
       bound state poles at $E_n = -m e^4 / (2 n^2)$ for $n = 1$, 2, ... .
       }
    \label {fig:contour1}
\end {figure}
where the contour C, shown in Fig.~\ref {fig:contour1}, wraps
clockwise about the cut along the positive real $E$ axis
and also encircles all the bound-state poles which occur when $e^2 < 0 $,
corresponding to an attractive Coulomb potential.
We shall first compute $G(\k,E)$ when the energy $E$ is real and
sufficiently negative so that $E$ lies to the left of all singularities,
and only later analytically continue to energies lying on the contour C.
Thus at first we write
\begin{equation}
    E = - { \gamma^2 \over 2m} \,,
\end{equation}
with $\gamma$ real and further restricted by $ \gamma > |e^2| \, m $ when
the potential is attractive. In view of the spherical symmetry of the
problem, we may average over the orientations of $\k$ and use
\begin{equation}
    G(\k,E) = {\rm Tr} \, { 1 \over H - E } \, {\sin kr \over kr} \,.
\end{equation}
In view of the cyclic symmetry of the trace, this may be expressed as
\begin{equation}
    G(\k,E) =
    { 1 \over k} \> {\rm Tr} \, { 1 \over \sqrt{r}( H - E) \sqrt{r} } \,\sin kr \,.
\label{smart}
\end{equation}

\subsection {Coulomb $su(1,1)$ Symmetry}

This latter form permits a remarkably simple evaluation by group theory.%
\footnote
    {%
    This $su(1,1)$ symmetry is a subgroup of a larger $so(4,2)$
    ``dynamical'' symmetry of the hydrogen which was noted
    many years ago by Barut, Fronsdal, Nambu, and others
    \cite {ancient}.
    The explicit construction used here of the generators in terms
    of canonical variables was, to our knowledge,
    first done by one of the authors (LSB) and G.~J.~Maclay
    and appears in the latter's Ph.D. dissertation \cite {thesis}.
    Although we know of no other references,
    this construction may well appear elsewhere in the literature.
    }
To do this, we first define the Hermitian operator
\begin{equation}
{\scr J}_0 = { 1 \over 2 \gamma} \, \sqrt{r} \, {\bf p}^2 \sqrt{r} + {\gamma
  \over 2 } \, r \,,
\end{equation}
so that
\begin{equation}
\sqrt{r} \, ( H - E) \sqrt{r} = { \gamma \over m} \, {\scr J}_0 + e^2 \,.
\label{denom}
\end{equation}
The $\sqrt{r}$ transformation converts the energy eigenvalue problem to
a coupling eigenvalue problem. To see this, we consider the Coulomb
bound states $| nlm\rangle $ which have the fixed energy $ - \gamma^2 / 2m $
that corresponds to (mutually attractive) charges $\pm e_n$ obeying the
Bohr formula
\begin{equation}
{ \gamma^2 \over 2m } = { e_n^4 \, m \over 2 n^2 } \,,
\end{equation}
or
\begin{equation}
e_n^2 = n \, { \gamma \over m} \,.
\end{equation}
Hence,
\begin{eqnarray}
{\scr J}_0 \, { 1 \over \sqrt{r} } \, | nlm \rangle &=& {m \over \gamma}
\sqrt{r} \left[ { {\bf p}^2 \over 2m } + { \gamma^2 \over 2m} \right]
| nlm \rangle
\nonumber\\
&=& {m \over \gamma} \sqrt{r} \, { e_n^2 \over r} \, | nlm \rangle
\nonumber\\
&=& n \, { 1 \over \sqrt{r} } \, | nlm \rangle \,,
\label{eigenket}
\end{eqnarray}
and so the eigenvalues $j_0$ of ${\scr J}_0$ are the positive integers,
\begin{equation}
    j_0 = n \,, \qquad\qquad n = 1 ,\, 2, \, \ldots \,.
\end{equation}
For a fixed principal quantum number $n$, $l$ ranges over $ 0 \ge l \ge
n-1$ and $m$ in turn varies through $ - l \ge m \ge + l $. Thus the
degeneracy of the $n$'th eigenvalue is
\begin{equation}
\sum_{l=0}^{n-1} \> (2l + 1) = n^2 \,.
\label{degenerate}
\end{equation}

To exploit the latent group properties, we introduce the Hermitian
dilation operator which is conveniently labeled as
\begin{equation}
{\scr J}_2 = { 1 \over 2 } \left( \r \cdot {\bf p} +
     {\bf p}  \cdot \r \right) ,
\end{equation}
and denote the commutator of ${\scr J}_0$ and ${\scr J}_2$
as ${\scr J}_1$ (times $-i$),
\begin {equation}
    [ {\scr J}_0, {\scr J}_2 ] = -i {\scr J}_1 \,.
\end {equation}
Since
\begin{equation}
i \left[ {\bf p} , {\scr J}_2 \right] = \p \,, \qquad
i \left[ \r , {\scr J}_2 \right] = - \r \,,
\end{equation}
${\scr J}_1$ differs from ${\scr J}_0$ merely by a sign change,
\begin{equation}
{\scr J}_1 = { 1 \over 2 \gamma} \, \sqrt{r} \, {\bf p}^2 \sqrt{r} - {\gamma
  \over 2 } \, r \,.
\end{equation}
Moreover, a further commutation with ${\scr J}_2$
restores the original signs,
\begin{equation}
\left[ {\scr J}_1 , {\scr J}_2 \right] = -i {\scr J}_0 \,.
\end{equation}
And a straight forward computation of the final commutator
shows that the algebra closes,
\begin{equation}
\left[ {\scr J}_0 , {\scr J}_1 \right] = i {\scr J}_2 \,.
\end{equation}
The three Pauli spin matrices $\sigma_k$ obey the $su(2)$ Lie algebra
\begin{equation}
\left[ \sigma_k , \sigma_l \right] = 2 i \, \epsilon_{klm} \, \sigma_m \,.
\end{equation}
Thus, as far as the commutation relations go, we have the
correspondences
\begin{equation}
{\scr J}_0 \leftrightarrow {1 \over 2 } \sigma_3 \,, \quad
{\scr J}_1 \leftrightarrow {i \over 2 } \sigma_1 \,, \quad
{\scr J}_2 \leftrightarrow {i \over 2 } \sigma_2 \,,
\end{equation}
which identifies the commutators of the ${\scr J}_a$
with the Lie algebra $su(1,1)$.
This, of course, corresponds to a non-compact group which has
infinite-dimensional irreducible representations.

With these results in hand, we return to our computation. Since
\begin{equation}
{\scr J}_0 - {\scr J}_1 = \gamma \, r \,,
\end{equation}
the sine function in Eq.~(\ref{smart}) may be written in terms of
group generators. Using this and the expression (\ref{denom}) for the
denominator in Eq.~(\ref{smart}), we obtain
\begin{equation}
G(\k,E) = { m \over 2i \gamma  k} \, {\rm Tr} \> { 1 \over {\scr J}_0 +
  (me^2 / \gamma) } \left[ e^{ik ( {\scr J}_0 - {\scr J}_1) / \gamma}
  -  e^{-ik ( {\scr J}_0 - {\scr J}_1) / \gamma}  \right] .
\end{equation}
Representing the denominator in terms of the integral of an exponential
now places the result in terms of the trace of the product of group
elements:
\begin{equation}
G(\k,E) = { m \over 2i \gamma  k} \int_0^\infty ds \> e^{-(me^2 /
  \gamma) s} \> {\rm Tr} \> e^{ - s {\scr J}_0 }
\left[ e^{ik ( {\scr J}_0 - {\scr J}_1) / \gamma}
  -  e^{-ik ( {\scr J}_0 - {\scr J}_1) / \gamma} \right] .
\end{equation}
The products of two group elements may be expressed as a third group
element. Since the trace is invariant under similarity
transformations, this third group element may be ``rotated'' into one
involving only the generator ${\scr J}_0$,
\begin{equation}
{\rm Tr} \, e^{ - s {\scr J}_0 }
  e^{\pm ik ( {\scr J}_0 - {\scr J}_1) / \gamma}
= {\rm Tr} \, e^{- s_\pm {\scr J}_0 } \,.
\label{group}
\end{equation}
The required parameters $s_\pm$ will be determined momentarily.
Evaluating the trace using the known eigenvalues $j_0{=}n$
of ${\scr J}_0$ with multiplicity $n^2$ yields
\begin{eqnarray}
{\rm Tr} \, e^{- s_\pm {\scr J}_0 } &=& \sum_{n=1}^\infty
n^2 \, e^{- s_\pm n}
= \left( { \partial \over \partial s_\pm } \right)^2 { 1 \over
     e^{s_\pm} - 1 }
\nonumber\\
   &=& {1 \over 4} \, { \cosh s_\pm /2 \over \sinh^3 s_\pm /2 } \,.
\end{eqnarray}
Therefore
\begin{equation}
G(\k,E) = { m \over 8 i \gamma k } \int_0^\infty ds \,  e^{ - ( me^2 /
  \gamma) s } \left[ { \cosh s_+/2 \over \sinh^3 s_+/2 }
   -  { \cosh s_-/2 \over \sinh^3 s_-/2 } \right] .
\label{geee}
\end{equation}

In view of the algebraic isomorphism between the group generators and
the Pauli matrices, the parameters $s_\pm$ may be found by replacing the
generators in Eq.~(\ref{group}) by the equivalent $2\times 2$ Pauli matrices.
Hence,
\begin{eqnarray}
2 \cosh s_\pm /2 &=& {\rm tr} \exp\left\{ - s_\pm {1 \over 2} \sigma_3
  \right\}
\nonumber\\
&=& {\rm tr} \exp\left\{ - s {1\over2} \sigma_3 \right\}
\exp\left\{ \pm { ik \over 2 \gamma}
( \sigma_3 -i \sigma_1 ) \right\}
\nonumber\\
&=& {\rm tr} \exp\left\{ - s {1\over2} \sigma_3 \right\}
\left[ 1 \pm { ik \over 2 \gamma }
\left( \sigma_3 - i \sigma_1 \right) \right]
\nonumber\\
&=& e^{ - s /2} \left[ 1 \pm { ik \over 2 \gamma } \right]
   + e^{ + s /2} \left[ 1 \mp { ik \over 2 \gamma } \right] .
\label{spm}
\end{eqnarray}
We write
\begin{equation}
\left[ 1 + { ik \over 2\gamma } \right] = e^{ i \theta /2} \sqrt{ 1 +
  { k^2 \over 4 \gamma^2 } }
\end{equation}
so that Eq.~(\ref{spm}) becomes
\begin{equation}
\cosh s_\pm /2 = \sqrt{ 1 + { k^2 \over 4
    \gamma^2 } } \, \cosh ( s \mp i \theta )/2 \,.
\label{recall}
\end{equation}
A short calculation yields
\begin{equation}
\sinh^2 s_\pm/2 = \left( 1 + { k^2 \over 4 \gamma^2 } \right) \,
\left[ \sinh^2 ( s \mp i \theta ) / 2 + \sin^2
  \theta/2 \right] .
\end{equation}
For later use, note that
\begin {equation}
    \theta
    = 2 \arctan (k/2\gamma)
    = k / \gamma + O(k^3) \,.
\end {equation}
Hence
\begin{eqnarray}
G(\k,E) &=& { m \over 8 i \gamma k } { 1 \over 1 + (k/2\gamma)^2 }
\int_0^\infty ds  \> e^{ - ( me^2 / \gamma) s }
\left\{ { \cosh (s -i \theta) / 2 \over [ \sinh^2(s -i \theta)/2 +
    \sin^2 \theta /2 ]^{3/2}  }
\right.
\nonumber\\ && \kern 2.3in
\left. {}
   -  { \cosh (s +i \theta) / 2 \over [ \sinh^2(s +i \theta)/2 +
    \sin^2 \theta /2 ]^{3/2}  } \right\} .
\label {eq:Gs}
\end{eqnarray}
As noted in footnote \ref {fn:sing}, all along we should have
subtracted the $e^2 = 0$ contribution from $G(\k,E)$.
In the integral representation above, this simply
means replacing $e^{-(m e^2 /\gamma)s}$ by the subtracted exponential
$[e^{-(m e^2 /\gamma)s}-1]$.
As anticipated, this subtraction removes what would otherwise
be a singularity in the integral at $s = 0$.

\begin {figure}
   \begin {center}
      \leavevmode
      \def\epsfsize #1#2{0.4#1}
      \epsfbox {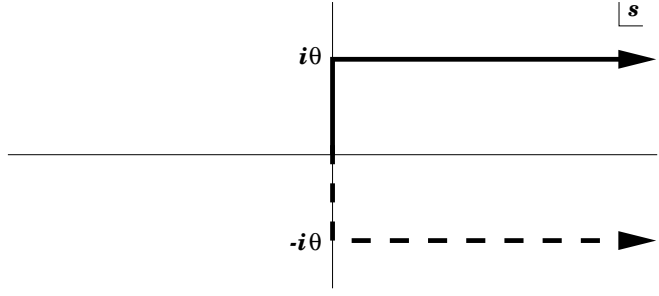}\hspace*{1in}
   \end {center}
   \caption
       {%
       Integration contours for $G(\k,E)$.
       }
    \label {fig:contour2}
\end {figure}

\subsection {Direct Contribution}

To compute the integral (\ref {eq:Gs}) (with the $e^2{=}0$ piece removed),
it is convenient to deform the path of integration into the contours shown in
Fig.~\ref {fig:contour2}. For the first term in
braces in the integrand, the contour is taken to run first over a
portion of the imaginary axis, $s = i \phi, \,\, 0 < \phi < \theta $,
and then to continue along the line parallel to the real axis, $ s \to s
+ i \theta, \,\, 0 < s < \infty$. The integration contour for the
second term in the braces is the complex conjugate of the
first. These contour deformations produce
\begin{equation}
    G(\k,E) =  { m \over 4 \gamma k } \, { 1 \over 1 + (k/2\gamma)^2 }
    \left[
	J\!\left(\theta, \coeff {m e^2}{\gamma}\right) -
	\sin \!\left(\coeff {m e^2 }{\gamma} \, \theta\right) \,
	I\!\left(\theta, \coeff {m e^2}{\gamma}\right)
    \right] ,
\label{wonder}
\end {equation}
where
\begin {equation}
    J(\theta, z) =
    \int_0^\theta d\phi \, \left[\cos  z \phi  -1 \right]
    { \cos (\theta - \phi ) / 2 \over [ \sin^2 \theta /2
     - \sin^2(\theta - \phi )/2]^{3/2}  } \,,
\end{equation}
and
\begin{equation}
    I(\theta, z) =
    \int_0^\infty ds  \> e^{ - z s }
    { \cosh s / 2 \over [ \sinh^2s /2 + \sin^2 \theta /2 ]^{3/2}  } \,.
\end {equation}

Although this general result may be of interest in other contexts,
here we are interested in the small $k^2$ behavior,
since this determines the induced couplings in the effective theory.
In the first integral $J(\theta,z)$,
it is convenient to make the variable change $\phi = \theta ( 1 {-} x) $
and write the integral as
\begin {equation}
    J(\theta, z) =
    \theta  \int_0^1 dx \, \left[\cos  z \theta(1{-}x)  - 1 \right]
    { \cos \theta x / 2 \over [ \sin^2 \theta /2
     - \sin^2\theta x  /2]^{3/2}  } \,.
\label{jtheta}
\end{equation}
Recalling that $\theta \simeq k/\gamma$, we may expand
the trigonometric functions in the integrand in Eq.~(\ref{jtheta})
and keep only the leading terms to obtain
\begin{equation}
    J(\theta,z) =
    - 4 z^2
    \int_0^1 dx \>
     { ( 1 - x )^{1/2} \over (1 + x )^{3/2} } + O(\theta^2) \,.
\end{equation}
Writing $ (1+x)^{-3/2} = - 2 (d /dx) (1+x)^{-1/2} $ and integrating by
parts produces an end-point contribution and an integral made trivial
by the substitution $x = \sin \chi$, and one finds that
\begin{equation}
    J(\theta,z) = -
    z^2 \> (8 - 2\pi) \; [1 + O(\theta^2)] \,.
\end{equation}
If $I(\theta,z)$ is expanded in powers of $z$, the first three terms
are singular as $\theta \to 0$, while all remaining terms have finite
$\theta\to0$ limits.
It is convenient to separate the singular terms by writing
\begin {equation}
    I(\theta,z) =
       I_0(\theta) -z \, I_1 (\theta) + \half z^2 \, I_2(\theta) +
    \bar I(\theta,z) \,,
\end {equation}
where
\begin {equation}
    I_k(\theta) \equiv
    \int_0^\infty ds \>
    {s^k \, \cosh s/2 \over [\sinh^2 s/2 + \sin^2 \theta/2]^{3/2}} \,,
\end{equation}
and
\begin{equation}
    \bar I(\theta,z) \equiv
    \int_0^\infty ds \,
    \left[ e^{-z s} - 1 + z s - \half z^2 s^2 \right]
    {\cosh s/2 \over [\sinh^2 s/2 + \sin^2 \theta/2]^{3/2}} \,.
\end {equation}

Since $d(\sinh s/2) = (ds/2) \cosh s/2$, the change of variable
$\sinh s/2 = \sin (\theta/2) \, \tan \chi $ makes the integral $I_0(\theta)$
elementary,
\begin {equation}
    I_0(\theta)
    =
	{2 \over \sin^2\theta/2} \int_0^{\pi/2} d\chi \>
	\cos \chi
    =
	{2 \over \sin^2 \theta/2}
    =
	{8 \over \theta^2} + {2 \over 3} + O(\theta^2) \,.
\end {equation}
To evaluate $I_1(\theta)$ we write
\begin {eqnarray}
    I_1(\theta)
    &=&
    \int_0^\infty ds \>
    {8s \over [s^2 + \theta^2]^{3/2}}
    +
    \int_0^\infty ds \> s
    \left(
    {\cosh s/2 \over [\sinh^2 s/2 + \sin^2 \theta/2]^{3/2}}
    -
    {8 \over [s^2 + \theta^2]^{3/2}}
    \right) .
\end {eqnarray}
The first integral, which is easy to evaluate,
contains the piece which is singular as $\theta \to 0$,
while the second integral is finite as $\theta \to 0$
and may be evaluated directly at $\theta = 0$.
Therefore,
\begin {eqnarray}
    I_1(\theta)
    &=&
    {8 \over \theta} +
    \int_0^\infty ds \>
    \left(
	s \, {\cosh s/2 \over \sinh^3 s/2} - {8 \over s^2}
    \right)
    + O(\theta)
\nonumber\\
    &=&
    {8 \over \theta} -
    \left.\left(
     {s \over \sinh^2 s/2 } + 2 \, {\cosh s/2 \over \sinh s/2}
    - { 8 \over s} \right)\right|_0^\infty
    + O(\theta)
\nonumber\\
    &=&
    {8 \over \theta} - 2 + O(\theta) \,.
\end {eqnarray}
A similar approach may be used for $I_2(\theta)$ if one
first splits the integral into the contributions
from $s < 1$ and $s > 1$,
\begin {eqnarray}
    I_2(\theta)
    &=&
    \int_0^1 ds \>
    {8s^2 \over [s^2 + \theta^2]^{3/2}}
    +
    \int_0^1 ds \> s^2 \,
    \left(
    {\cosh s/2 \over [\sinh^2 s/2 + \sin^2 \theta/2]^{3/2}}
    -
    {8 \over [s^2 + \theta^2]^{3/2}}
    \right)
\nonumber\\ && \kern 1.2in {}
    +
    \int_1^\infty ds \> s^2 \,
    {\cosh s/2 \over [\sinh^2 s/2 + \sin^2 \theta/2]^{3/2}}
\nonumber\\
    &=&
    -8 (1+\ln \theta/2)
    +
    \int_0^1 ds \>
    \left(s^2 {\cosh s/2 \over \sinh^3 s/2 } - {8 \over s} \right)
    +
    \int_1^\infty ds \> s^2 \,
    {\cosh s/2 \over \sinh^3 s/2 }
    +
    O(\theta)
\nonumber\\ &=&
    -8 (1+\ln \theta/2)
    +
    \lim_{\epsilon \to 0}
    \left[
    8\ln\epsilon
    +
    \left.\left(
    8 \ln\sinh s/2 - {s^2 \over \sinh^2 s/2} - 4 s \, {\cosh s/2 \over
     \sinh s/2}
    \right)\right|_\epsilon^\infty
    \right]
    +
    O(\theta)
\nonumber\\ &=&
    4 -8 \ln \theta/2 + O(\theta) \,.
\end {eqnarray}
The final integral $\bar I(\theta,z)$ is non-singular as $\theta\to0$,
and so we may simply set $\theta$ equal to zero and then integrate-by-parts
twice,
\begin{eqnarray}
    \bar I(0,z)
    &=&
    -\int_0^\infty ds \,
    \left[ e^{-z s} - 1 + z s - \half z^2 s^2 \right]
    {d \over ds} \, {1 \over \sinh^2 s/2}
\nonumber\\ &=&
    -z \int_0^\infty ds \,
    \left[ e^{-z s} - 1 + z s \right]
    {1 \over \sinh^2 s/2}
\nonumber\\ &=&
    4z \int_0^\infty ds \,
    \left[ e^{-z s} - 1 + z s \right]
    {d \over ds} \, {1 \over e^s-1}
\nonumber\\ &=&
    4 z^2 \int_0^\infty ds \>
    { e^{-z s} -1  \over e^s-1} \,.
\label {eq:Ibar-int}
\end{eqnarray}
The denominator may be expanded in a geometric series and
the resulting $s$ integrals performed to give
\begin{eqnarray}
    \bar I(0,z) =  -4 z^3 \sum_{l=1}^\infty { 1 \over l ( l + z ) } \,.
\label{psisum}
\end{eqnarray}
Using
\begin{equation}
\psi(z{+}1) + \gamma = z \sum_{l=1}^\infty { 1 \over l ( l + z ) } \,,
\label{psisumm}
\end{equation}
where $\psi(z)$ is the
logarithmic derivative of the gamma function and $\gamma$ is Euler's
constant,
yields the closed-form result
\begin {equation}
    \bar I(0,z) = -4 z^2 \left[ \psi(z{+}1) + \gamma \right] .
\label {eq:Ibar}
\end {equation}
This form may be used to make contact with the literature on
quantum Coulomb corrections \cite{ebeling,book}.
A power series expansion in $z$ is obtained if the denominator in the
sum (\ref{psisum}) is expanded in powers of $z$ and the
order of the resulting double sum interchanged. This process gives
\begin {equation}
    \bar I(0,z) = 4 z^2 \sum_{n=1}^\infty (-z)^n \zeta(n{+}1) \,,
\label {eq:I-series}
\end {equation}
where
\begin{equation}
\zeta(n) = \sum_{l=1}^\infty { 1 \over l^n }
\label{riemann}
\end{equation}
is the Riemann $\zeta$ function.

Assembling the various pieces contributing
to the Green's function $G(\k,E)$ and inserting
\begin {equation}
    \theta = (k/\gamma) - \coeff 1{12} (k/\gamma)^3 + O(k/\gamma)^5
\end {equation}
produces
\begin {equation}
    G(\k,E)
    =
    m \left\{
    -{2 z \over k^2}
    +{\pi \over 2} {z^2 \over \gamma k}
    +{1 \over \gamma^2}
    \left[
	{z \over 6}
	-{z^2 \over 2}
	- z^3 \ln \biggl({2\gamma \over k}\biggr)
	-{z^3 \over 6}
	-{z \over 4} \, \bar I(0,z)
    \right]
    + O(k)
    \right\} ,
\label {eq:GfromI}
\end {equation}
where $z = m e^2 / \gamma$.
This result is to be inserted into the contour integral
(\ref{contint}) relating $G(\k,E)$ to the thermal correlator $F_+(\k)$
which, with the $e^2 =0$ subtraction made explicit, reads
\begin{equation}
    \Delta F_+(\k) \equiv F_+(\k) -F_+^0(\k)
    = \int_{\rm C} { dE \over 2\pi i } \, e^{- \beta E } \, G(\k,E) \,.
\label{contintt}
\end{equation}
Inserting the power series representation (\ref {eq:I-series})
for $\bar I(0,z)$
and recalling that $\gamma^2 = -2 m E$,
the required contour integrals are easily performed using
Hankel's formula%
\footnote{See, for example, p.~245 of Whittaker and Watson \cite{W&W}.}
\begin{equation}
    { 1 \over \Gamma(\alpha) } =
    \int_C {dt\over 2\pi i} \> (-t)^{-\alpha} \, e^{-t} \,,
\label {eq:Hankel}
\end{equation}
and its derivative with respect to $\alpha$,
\begin{equation}
    { \psi(\alpha) \over \Gamma(\alpha) } =
    \int_C {dt\over 2\pi i} \> \ln (-t) \, (-t)^{-\alpha} \, e^{-t} \,.
\label {eq:lnHankel}
\end{equation}
The result, neglecting $O(k)$ contributions, is
\begin {eqnarray}
    \Delta F_+(\k)
    &=&
    \biggl({m \over 2\pi\beta}\biggr)^{3/2}
    \Bigg\{
    -
    {4\pi \beta e^2 \over k^2}
    +
    {\pi^2 \beta^2 e^4 \over k}
\nonumber\\ && \quad\qquad\qquad {}
    +  {\pi \beta^3 e^6  \over 3} \,
      \Biggl[
	 \ln \biggl({\beta k^2 \over 2m}\biggr)
	+ \gamma - 3
	+ {1 \over \beta m e^4}
	+ f \Biggl(\sqrt{{\beta m e^4 \over 2\pi}}\Biggr)
      \Biggr]
    \Bigg\} \,,
\end {eqnarray}
where
\begin {eqnarray}
    f(y)
    &\equiv&
    -
    {3 \over 4y}
    -
    {3\sqrt \pi \over 2}
    \sum_{n=1}^\infty
    \left(-\sqrt \pi \, y\right)^n
    {\zeta(n{+}1) \over \Gamma((n{+}5)/2)} \,.
\label{labeled}
\end {eqnarray}

Returning to our rationalized units with
$ e^2 \to e_a e_b / 4\pi$, replacing the mass parameter by the reduced
mass, $ m \to m_{ab}$, and writing the result in terms of the thermal
wavelength corresponding to the reduced mass
\begin{equation}
\lambda_{ab} = \hbar \left({ 2\pi \beta \over m_{ab} }\right)^{1/2}
\end{equation}
gives
\begin {eqnarray}
    \Delta F_+(\k)
    &=&
    \lambda_{ab}^{-3} \>
    \Bigg\{
    -
    {\beta e_a e_b \over k^2}
    +
    {(\beta e_a e_b)^2 \over 16k}
\nonumber\\
&& \quad {}
  +
    \biggl({\beta e_a e_b \over 4\pi}\biggr)^3 \, {\pi\over3}
    \left[
    \ln \biggl({\lambda_{ab}^2 k^2 \over 4\pi}\biggr)
      + \gamma - 3
      + {8\pi \lambda_{ab}^2 \over (\beta e_a e_b)^2}
      + f \biggl({\beta e_a e_b \over 4\pi \lambda_{ab}}\biggr)
    \right]
    + O(k)
    \Bigg\} .
\label {suresult}
\end {eqnarray}

Evaluating $f(y)$ using the power series representation (\ref
{labeled}) is appropriate if $\beta e_a e_b/\lambda_{ab}$ is order one
or smaller.  But if $\beta e_a e_b/\lambda_{ab} $ is large, which
corresponds to the formal $m \to \infty$ limit, one needs the
asymptotic form of $f(y)$ for large argument.  The result differs
depending on whether the Coulomb interactions are attractive or
repulsive.  Consider the repulsive case first,
where $z = m e^2 /\sqrt{-2mE}$ is positive on the negative real $E$ axis.
In this case
$\bar I(0,z)$ has no poles on the negative real $E$ axis, which
reflects the absence of bound states for repulsive potentials.
Thus, for repulsive interactions the contour integral (\ref{contint}) only wraps
about the positive $E$ axis, and $\bar I(0,z)$ appears with $ |\arg z| < \pi$.
Hence the large $m$ limit may be obtained by
using the large $z$ asymptotic behavior of the $\psi$ function,
\begin {equation}
    \psi(z+1) \sim \ln z + {1 \over 2z}
    - \sum_{n=1}^\infty {B_{2n} \over 2 n z^{2n}} \,,
    \kern 1.05in \hbox {($|\arg z| < \pi$)}
\end {equation}
where $B_{2n}$ are the Bernoulli numbers,
to write the asymptotic form
of Eq.~(\ref {eq:Ibar}) as
\begin {equation}
    \bar I(0,z) \sim -4 z^2
    \left( \ln z + \gamma + {1 \over 2z}
    - \sum_{n=1}^\infty {B_{2n} \over 2 n z^{2n}} \right).
    \qquad \hbox {($|\arg z| < \pi$)}
\end {equation}
Using this form for $\bar I(0,z)$
and re-evaluating the contour integral (\ref{contintt})
yields the asymptotic expansion for large positive argument,
\begin{equation}
    f(y)
    \sim
      2 \,  \ln (2 \sqrt{\pi} y ) + 3 \gamma - \frac {8}3
    -
    {3 \over 2 \sqrt \pi}
    \sum_{n=1}^\infty
    (-1)^n {B_{2n} \over 2n \pi^n y^{2n}} \, \Gamma(n{-}\coeff 32) \,.
\label {eq:fpos}
\end{equation}
Evaluating the first term in the sum with $B_2 = 1/6$ and
$\Gamma(-1/2) = - 2 \sqrt{\pi}$ yields
\begin{equation}
    f(y)
    =
      2 \,  \ln (2 \sqrt{\pi} y ) + 3 \gamma - \frac {8}3
    -
    {1 \over 4 \pi  y^2}
    + O( 1 / y^4 ) \,.
\label {eq:ffpos}
\end{equation}

To obtain the corresponding limit in the attractive case, note that
Eq.~(\ref{labeled}) gives
\begin {equation}
f(y) - f(-y) = - { 3 \over 2 y} + 3 \pi
\sum_{m=0}^\infty \,
{\zeta(2m+2) \over (m+2)! } \, \pi^m \, y^{2m+1} \,.
\end {equation}
We insert the definition (\ref{riemann}) of the $\zeta$ function and
interchange the order of the summations. The sum over $m$ now produces
an exponential with its first two expansion coefficients removed, and
we obtain
\begin{equation}
    f(y) - f(-y)
    =
    - { 3 \over 2 y}
    +
    {3 \over \pi \, y^3}
    \sum_{n=1}^\infty \, n^2
	\left[\,
	    \exp\left\{ {\pi \,  y^2 \over n^2 } \right\}
	    - 1 - { \pi \, y^2 \over n^2 }
	\right] .
\end{equation}
As $y \to -\infty$, the first term in the sum, which
corresponds to the lowest bound state contribution, dominates,
\begin {equation}
    f(y)
    \sim
    {3 \over \pi \, y^3} \, \exp \left\{ \pi \, y^2 \right\} \,,
\label {eq:fneg}
\end{equation}
with exponentially small corrections.

\subsection {Exchange Contribution}

The same approach may be used to evaluate the exchange contribution
\begin {equation}
    F_-(\k) = {\rm Tr} \> {\cal P} \, e^{-\beta H} e^{-i \k\cdot\r} \,.
\end {equation}
Since the parity operator $\cal P$ commutes
with all the $su(1,1)$ group generators,
all the previous formulas hold for this
exchange term with the trivial change of an insertion of
${\cal P}$ in the trace defining the Green's function.
To evaluate the final trace
$$
    {\rm Tr} \, {\cal P} \, e^{-s_{\pm} {\scr J}_0 } \,,
$$
we note that the $|nlm\rangle$ basis which diagonalizes ${\scr J}_0$
as shown in Eq.~(\ref{eigenket}) has the familiar parity assignment of
the hydrogen atom states,
\begin{equation}
{\cal P} \, | nlm \rangle = | nlm \rangle \, (-1)^l \,.
\end{equation}
Hence, we have essentially the same evaluation of
the trace as before except that the previous degeneracy factor
(\ref{degenerate})
is replaced by
\begin{equation}
\sum_{l=0}^{n-1} \, (-1)^l \, (2l+1) = - (-1)^n \, n \,.
\end{equation}
Thus we now have
\begin{eqnarray}
{\rm Tr} \, {\cal P} \, e^{-s_\pm {\cal J}_0 } &=& -\sum_{n=1}^\infty
(-1)^n \, n \, e^{-s_\pm n}
\nonumber\\
&=& { 1 \over 4 \cosh^2 s_\pm / 2 } \,,
\end{eqnarray}
and Eq.~(\ref{geee}) is replaced by
\begin{equation}
G_-(\k,E) = { m \over 8 i \gamma k }
\int_0^\infty ds \,  e^{ - ( me^2 /
  \gamma) s } \left[ { 1 \over \cosh^2 s_+/2 }
   -  { 1 \over \cosh^2 s_-/2 } \right] \,.
\end{equation}

We are interested in the (finite) $\k=0$ limit.
Recalling Eq.~(\ref{recall}) and that $\theta \sim k/\gamma$,
this is given by
\begin{eqnarray}
    G_-({\bf 0},E) &=&
    { m \over 8  \gamma^2 }
    \int_0^\infty ds \>  e^{ - ( me^2 / \gamma) s } \,
    \lim_{\theta \to 0} \>
    { 1 \over i \theta}
    \left[ { 1 \over \cosh^2 (s - i \theta ) /2 }
    - { 1 \over \cosh^2 (s + i \theta ) /2 } \right] \,,
\nonumber\\ &=&
    - { m \over 4  \gamma^2 }
    \int_0^\infty ds \,  e^{ - ( me^2 / \gamma) s } \,
    { d \over ds} \,  { 1 \over \cosh^2 s  /2 } \,.
\end{eqnarray}
Expressing the hyperbolic cosine in terms of exponentials and
performing two partial integrations produces
\begin {equation}
    G_-({\bf 0},E)
    =
    {m \over \gamma^2} \> h\biggl({m e^2 \over \gamma}\biggr)
\end {equation}
with
\begin{equation}
    h(z)
    =
    {1 \over 4} - {z \over 2}
    + z^2 \int_0^\infty ds \> e^{ - z s } \, { 1 \over e^s + 1 } \,.
\label{hintrep}
\end{equation}
Expanding the denominator and performing the $s$ integration gives
\begin{equation}
    h(z)
    =
    {1 \over 4} - {z \over 2}
    + { z^2 \over 2} \sum_{l=1}^\infty \left\{
      { 1 \over l + ( z {-} 1 ) /2 } - { 1 \over l + z/2 } \right\} \,.
\label{sumdiff}
\end{equation}
Adding and subtracting $1/l$ in the sum, combining denominators,
and referring to the representation (\ref{psisumm}),
identifies the sum as the difference of two $\psi$ functions,
and yields the closed form result
\begin{equation}
h(z) =
    {1 \over 4} - {z \over 2}
    + {z^2 \over 2}
   \left[ \psi\Biggl({z{+}2\over 2}\Biggr) -
   \psi\Biggl({z{+}1\over 2}\Biggr) \right].
\end{equation}
Expanding the denominators in Eq.~(\ref{sumdiff}) in powers of $z$
and using
\begin{equation}
\sum_{l=1}^\infty \, { (-1)^{l+1} \over l^z}
= \left[ 1 - 2^{1-z} \right] \zeta (z)
\label{oddzeta}
\end{equation}
gives the useful power series expansion
\begin{equation}
    h(z) = {1 \over 4} - { z \over 2} +
    z^2 \, \sum_{n=0}^\infty
    \left(-z\right)^n
    (1 - 2^{-n}) \, \zeta (n{+}1) \,.
\end{equation}
Here the $n=0$ member of the sum is to be understood as containing
\begin{equation}
\lim_{z\to1} \left[ 1 - 2^{1-z} \right] \zeta (z)
= \sum_{l=1}^\infty \, { (-1)^{l+1} \over l} = \ln 2 \,.
\end{equation}
Inserting this series into the contour integral
\begin {equation}
    F_-({\bf 0})
    =
    \int_{\rm C} {dE \over 2\pi i} \> e^{-\beta E} \, G_-({\bf 0},E) \,,
\label{f-int}
\end {equation}
and again making use of Hankel's formula (\ref {eq:Hankel}) produces
\begin{equation}
    F_-({\bf 0}) =
    \biggl( {m \over 2\pi\beta} \biggr)^{3/2} \, \beta^3 e^6 \,
    {\pi \over 3} \, \tilde f\Biggl(\sqrt{{\beta m e^4 \over 2 \pi}}\Biggr) \,,
\label{Biggl}
\end{equation}
where
\begin{eqnarray}
    \tilde f(y)
    =
   { 3 \over 8 \pi y^3 } - { 3 \over 2 \pi y^2 } + { 3 \ln 2 \over 2 y }
    -{ 3\sqrt\pi \over 2 } \sum_{n=0}^\infty \,
    \left(-\sqrt\pi \, y\right)^n \left[ 1 - { 1 \over 2^{1+n}} \right] \,
    { \zeta(n{+}2) \over \Gamma((n{+}5)/2)} \,.
\end{eqnarray}

To obtain the behavior for the case of strong repulsion,
that is, the large $y^2 = \beta m e^4 /2$ limit,
we return to the integral expression
(\ref{hintrep}) for $h(z)$. Writing
\begin{equation}
{ 1 \over e^s + 1 } = { 1 \over 2} - {1 \over 2} \tanh {s\over 2} \,,
\end{equation}
performing simple integrals, and rescaling the integration variable
casts this integral representation in the form
\begin{equation}
h(z) = - { 1 \over 2} \int_0^\infty du \, e^{-u}
    \left[ z \tanh \biggl({u\over 2z} \biggr) - { u \over 2} \right] .
\label{goodint}
\end{equation}
This result shows explicitly that $h(z)$ is an even function of $z =
me^2 / \gamma $. Writing $ E = p^2 / 2m $  sets $\gamma =
\sqrt{-2mE} = i p$, with no problem with the sign of $i$ since only
even functions of $\gamma$ appear.
And, again because only even functions of $p$ appear,
we may replace the contour integral (\ref{f-int}) in the
$E$ plane by a contour integral in the $p$ plane having the exactly
the same contour. Thus
\begin {equation}
    F_-({\bf 0})
    =
  -  \int_{\rm C} {dp \over 2\pi i} \,
     { 1 \over p}  \> e^{-\beta p^2 / 2m} \,
     h\!\left( { me^2 \over ip } \right) .
\end {equation}
Introducing the integral representation (\ref{goodint}) into this
contour integral, interchanging the integration order, and rescaling
the contour integration variable by writing $ p = 2me^2 \zeta / u$
yields
\begin{equation}
    F_-({\bf 0})
    =  { 1 \over 4} \int_0^\infty du \, u \, e^{-u}
    \int_{\rm C} {d\zeta \over 2\pi i} \, { 1 \over \zeta}
      \> \exp\left\{ - 2 \beta m e^4 \zeta^2 / u^2 \right\} \,
    \left[ { \tan \zeta \over \zeta } - 1 \right] .
\end {equation}
The integrand of the contour integral has no pole at $\zeta = 0$ since
the quantity in the square brackets vanishes there.
Thus the only singularities of the integrand come from the factor
in square brackets,
which has a series of simple poles at odd integer multiples of $ \pi/2$
with residue $-1$.
Since these poles are encircled in a negative, clockwise sense, we obtain
\begin{equation}
    F_-({\bf 0})
    =  { 1 \over 4} \int_0^\infty du \, u \, e^{-u}
   \sum_{n=0}^\infty \left[ { 2 \over (2n{+}1) \pi } \right]^2
      \> \exp\left\{ - \beta  m e^4 (2n{+}1)^2 \pi^2  / 2 u^2 \right\} .
\end {equation}
The leading asymptotic behavior is obtained by
evaluating the $u$ integral, term-by-term,
using the method of steepest descents.
Only the $n=0$ term of the sum is relevant, since the remaining terms
are exponentially smaller.
Writing the result in terms of the function
$\tilde f(y)$ defined in Eq.~(\ref{Biggl}) with $y^2 = \beta m e^4
/2\pi $ gives
\begin{equation}
   \tilde f(y)
    \sim  { 3 \over  y^3 \, \pi^3 } \int_0^\infty du \, u \, e^{-u}
      \> \exp\left\{ - y^2 \pi^3  /  u^2 \right\} ,
\end {equation}
whose steepest descent evaluation yields
\begin{equation}
   \tilde f(y)
    =  { 2 \sqrt 3  \over y^2 \, \pi }
      \> \exp\left\{ - {3 \pi \over 2} \left( 2 y^2  \right)^{1/3}
      \right\}
      \; \left[ 1 + { 17 \over 18 \pi }
{ 1 \over (2 y^2 )^{1/3} } + \cdots \right] \,.
\label {eq:tilde-f-asym}
\end {equation}

Since the exchange term involves interactions of a single particle
type, the reduced mass $m$ appearing in the above formulae is $m_a /
2$ for species $a$.  Reverting to our rationalized units gives
\begin{equation}
    F_-({\bf 0})
    =
    \lambda_{aa}^{-3} \, { 3 \over \pi } \,
    \biggl({\beta e_a^2 \over 4\pi}\biggr)^3 \,
    \tilde f \biggl({\beta e_a^2 \over 4\pi
     \lambda_{aa}} \biggr)  \,,
\label {eq:Fminus}
\end{equation}
where $\lambda_{aa} = \sqrt 2 \lambda_a$. Note that here the argument of
$\tilde f(y) $ is always positive.

\section{First Quantum Correction to Classical One-Component Plasma}
\label{qfluck}

Here we shall derive the leading, order $\hbar^2$, quantum correction to
the $N$-particle canonical partition function of the classical,
one-component plasma.
This result appears in the literature [Eq.~(24) of \cite {hansen}],
but we will give a self-contained pedagogical treatment.
To do so, it is convenient first to examine
the path integral representation of the single-particle
partition function previously given in Eq.~(\ref{funfunint}), namely
\begin{equation}
\int (d^3{\bf r}) \> \langle {\bf r} ,\beta | {\bf r}, 0 \rangle
 = \int [d{\bf r}] \> \exp\left\{ - { 1 \over \hbar}
 \int_0^{\hbar\beta} d\tau \left[ {
      m\over 2 }  \,
{ d {\bf r} \over d\tau } \cdot { d {\bf r} \over d\tau }
+ V({\bf r}(\tau)) \right] \right\} \,.
\end{equation}
We have explicitly displayed the factors of $\hbar$ which appear when
$\beta$ and $\tau$ have their conventional units of inverse energy
and time, respectively.
We state again that the functional integral is over all paths
which are periodic with period $\beta\hbar$.
It is therefore convenient to use a Fourier series representation for
the path,
$
    \r(\tau) = \r + \bxi(\tau) \,,
$
in which
\begin{equation}
    \bxi(\tau) = \sum_{{n=-\infty \atop n\ne 0}}^{\infty} \, \bxi_n \,
    \exp\left\{ - {2 \pi i \, n \, \tau \over \hbar \beta } \right\}
\end{equation}
contains the non-zero frequency fluctuations of the path about
its mean position $\r$.
As we shall see, the size of the fluctuations $\bxi(\tau)$ are of order $\hbar$.
Since the (imaginary) time average of these fluctuations vanish,
the leading quantum-mechanical correction appears in
\begin{eqnarray}
    \int (d^3{\bf r}) \> \langle {\bf r} ,\beta | {\bf r}, 0 \rangle
     &=& \int (d^3{\bf r})  \> \exp\left\{ - \beta
    V({\bf r}) \right\}
    \int [d \xi] \exp\left\{ - { \beta \, m \over 2}
    \sum_{{n=-\infty \atop n\ne 0}}^{\infty} \,
    \left( { 2 \pi n \over \hbar \beta } \right)^2 \bxi_n \cdot \bxi_{-n} \,
    \right\}
\nonumber\\
&& \kern 1in {} \times
    \left\{ 1 - \beta \sum_{{n=-\infty \atop n\ne 0}}^{\infty} \,
    \xi^k_n \, \xi^l_{-n} \, \nabla_k \, \nabla_l \, V({\bf r})
    + O(\bxi^4)
    \right\} .
\end{eqnarray}
The path integral over the fluctuations defines a correlator which is
just the inverse of the matrix defining the quadratic form in the
exponential,
\begin{equation}
\langle \xi^k_n \, \xi^l_{-n'} \rangle = \delta_{n,n'} \, \delta^{kl}
\,  { 1 \over \beta m } \left( { \hbar \beta \over 2\pi n } \right)^2
\,.
\end{equation}
Using
\begin{equation}
 \sum_{{n=-\infty \atop n\ne 0}}^{\infty} { 1 \over 2 n^2 } = \zeta(2)
   = { \pi^2 \over 6 } \,,
\end{equation}
one immediately finds
\begin{equation}
    \int (d^3{\bf r}) \> \langle {\bf r} ,\beta | {\bf r}, 0 \rangle
    =
    \langle {\bf r} ,\beta | {\bf r}, 0 \rangle^0
    \int (d^3{\bf r})  \> \exp\left\{ - \beta
  V({\bf r}) \right\}
 \left\{ 1 - { \beta^2 \hbar^2 \over 24 \, m } \,
	\nabla^2  \, V({\bf r}) + O(\hbar^4) \right\} \,.
\label{wonderr}
\end{equation}
In other words, to $O(\hbar^2)$, the effect of quantum fluctuations
is equivalent to a shift in the potential energy $V(\r)$
appearing in the classical partition function by
\begin {equation}
    \delta V(\r) =
    { \beta \hbar^2 \over 24 \, m } \, \nabla^2  \, V({\bf r}) \,.
\label{wonderrr}
\end {equation}
As a check on the result (\ref {wonderr}),
we note that a partial integration of one
of the gradients in the $\nabla^2$ term,
together with the identification $\lambda^2 = 2\pi \hbar^2 \beta/ m$
and other minor notational changes, places
this result in precisely the form of the first line of
Eq.~(\ref{eq:S2}).

This result is easily extended to the case of the canonical partition
function for $N$ particles. This case is represented by a path integral
over the variables ${\bf r}_a(\tau)$, where $ a = 1,\, 2 ,\, \cdots ,\, N$,
\begin {equation}
    Z_N
    =
    \int \prod_a [d\r_a] \> \exp \left\{
     - { 1 \over \hbar} \int_0^{\hbar\beta} d\tau \left[ { m\over 2 } \, \sum_a
    { d {\bf r}_a \over d\tau } \cdot { d {\bf r}_a \over d\tau }
    + {1\over2} \sum_{a \neq b} V({\bf r}_a(\tau){-}{\bf r}_b(\tau)) \right] \right\}
    \,.
\end {equation}
The leading quantum, order $\hbar^2$, corrections come from the quadratic
fluctuations in the coordinates ${\bf r}_a(\tau)$ and
${\bf r}_b(\tau)$ in each of the potential terms. Expanding these
terms out from the exponential is performed in a compact fashion if the
$N$-particle number densities are introduced,
\begin{equation}
    n({\bf r}) = \sum_{a=1}^{N} \, \delta( {\bf r} {-} {\bf r}_a ) \,,
\end{equation}
along with the canonical two-particle correlator
\begin{equation}
    K_N({\bf r} - {\bf r}')
    =
    \biggl\langle \sum_{a\ne b} \delta(\r{-}\r_a) \, \delta(\r{-}\r_b)
    \biggr\rangle
    =
    \Bigl\langle
	n({\bf r}) \, n({\bf r}') - \delta( {\bf r} {-} {\bf r}' ) \, n({\bf r})
    \Bigr\rangle
    \,.
\end{equation}
With this notation, the change in the $N$-particle canonical partition
function $Z_N$
for a general variation in the interparticle potential is given by
\begin{equation}
    \delta \, \ln Z_N = - {\beta \over 2} \int (d^3{\bf r}) (d^3{\bf r}') \>
    K_N({\bf r} {-} {\bf r}') \, \delta V({\bf r} {-} {\bf r}') \,.
\end{equation}
Thus, in view of the previous one-particle result (\ref {wonderrr}),
but keeping in mind that both the coordinates in the potential
undergo fluctuations, we see that the leading quantum
correction is given by
\begin{equation}
    \delta \, \ln Z_N = - {\beta \over 2} \int (d^3{\bf r}) (d^3{\bf r}') \>
    K_N({\bf r} {-} {\bf r}') \,
    { \beta \hbar^2 \over 12 m } \, \nabla^2  \,
    V({\bf r} {-} {\bf r}' ) \,.
\end{equation}
Taking account of translational invariance, which gives an overall
factor of the system volume $\vol$, and remembering the definition
$ -\beta F = \ln Z_N$ of the Helmholtz free energy $F$, we have
\begin{equation}
    \delta F =  { \beta \hbar^2 \over 24 \, m} \,
    \vol \int (d^3{\bf r}) \, K_N({\bf r}) \, \nabla^2 V({\bf r}) \,.
\label{fluck}
\end{equation}

This general result may be applied to a one-component plasma
in the presence of a uniform neutralizing background charge density
[since a strictly classical limit exists in this special case where
the charge carriers all have a common sign of their charge].
However, one must be careful to properly handle the effect of the
neutralizing background charge density before taking the thermodynamic limit.
The easiest way to do so is to regard the interaction potential
for the one-component plasma as the $\mu \to 0$ limit of
the regularized potential
\begin{equation}
    V({\bf r})
    =
    e^2 \left( {e^{-\mu r} \over 4\pi r} - {1 \over \vol \, \mu^2} \right) .
\end{equation}
The integral of this potential over the large volume $\vol$ of the
system vanishes, reflecting that a proper subtraction of the
uniform background charge density has been performed.
[Equivalently, this amounts to using a regularized Coulomb potential
$e^{-\mu r} / (4\pi r)$
with a total charge density of $e[n(\r) {-} \bar n]$, where $\bar n = N/\vol$
is the fixed average particle density.]

Now using
\begin{equation}
- \nabla^2 V({\bf r}) = e^2 \left( \delta({\bf r}) - \mu^2 \,
	{ e^{-\mu r} \over 4 \pi r} \right) ,
\end{equation}
Eq.~(\ref{fluck}) becomes
\begin{equation}
    \delta \, F
    =
    { \beta\hbar^2 e^2 \over 24 \, m } \, \vol
    \left[
	-K_N({\bf 0}) +
	\lim_{\mu\to 0}
	\mu^2
	\int (d^3\r) \> K_N(\r) \, {e^{-\mu r} \over 4\pi r}
    \right] .
\label {eq:dF}
\end{equation}
Because of the
singularity of the Coulomb potential when $r \to 0$,
the two-particle correlation $K_N(\r)$ vanishes as $\r\to0$,
\begin{eqnarray}
    K_N({\bf r}) \propto \exp\left\{ - \beta V({\bf r}) \right\}
    \to 0 \; \hbox {as $\r\to0$,}
\end{eqnarray}
and therefore the first term in Eq.~(\ref {eq:dF}) identically vanishes.
On the other hand, at large separations, the number densities are not
correlated, and so
\begin{eqnarray}
    K_N({\bf r}) \to \left( { N \over \vol } \right)^2 = \bar n^2
    \; \hbox {as $\r\to\infty$.}
\end{eqnarray}
For infinitesimal $\mu$,
the integral in the second term of Eq.~(\ref {eq:dF})
is dominated by arbitrarily large distances,
and hence one may simply replace $K_N(\r)$ in the integrand by its
asymptotic value of $\bar n^2$.
When multiplied by $\mu^2$,
the resulting error one is making in the short distance part of the integral
has no affect on the $\mu\to0$ limit.
Consequently,
\begin{eqnarray}
    \delta \, F
    &=&  { \beta \hbar^2 e^2 \over 24 \, m} \, \vol \bar n^2 \>
    \lim_{\mu\to0} \> \mu^2 \int (d^3{\bf r}) \> { e^{-\mu r} \over 4 \pi r}
\nonumber\\
    &=&  { \beta \hbar^2 e^2 \over 24 \, m} \, \vol \bar n^2 \,.
\label{ffluck}
\end{eqnarray}

Finally, the pressure is given by $p = - \partial F / \partial \vol $ with $N$
and $\beta$ fixed. Thus, since $ \vol \bar n^2 = N^2 / \vol $, we
find that the first quantum correction to the pressure of classical
one-component plasma is given by
\begin{equation}
\delta \left( {\beta p \over n} \right) =
	 { \beta \hbar^2 \over 24 \, m} \, \beta e^2 \, \bar n \,.
\end{equation}
Recalling the definitions $\kappa^2 = \beta e^2 n $,
$g = \beta e^2 \kappa / 4 \pi $, $ \lambda_{ee}^2 = 4 \pi \beta \hbar^2 / m $,
and $\eta_{ee} = \beta e^2 / (4\pi \lambda_{ee})$, we may rewrite this result as
\begin{equation}
\delta \left( {\beta p \over n} \right) =
	 g^2 { \beta \hbar^2 \over 24 \, m}
	\left( { 4 \pi \over \beta e^2} \right)^2 =
	{g^2 \over 24} \left( { 1 \over 4 \pi \eta_{ee}^2 } \right) .
\label{pfluck}
\end{equation}
This correction agrees with that which appears in Eq.~(\ref{good}) in
the text, as well as with the discussion of Eq.~(24) in Ref. \cite {hansen}.
Note that since the $O(\hbar^2)$ correction is proportional to $g^2$, it is
entirely contained in the two-loop contribution of the equation of state;
no $O(\hbar^2)$ corrections are contained in any higher-loop contributions.

\section{Some Elements of Quantum Field Theory}
\label{someqft}

Here we shall briefly derive and review some of the methods of quantum
field theory that are used in the text. For simplicity we shall
explicitly treat the case of the functional integral representation of
the classical plasma without the induced interactions that provide the
quantum corrections which make the theory finite. These additional
interactions entail no essential changes in the techniques that we are
about to outline, and their effects are easily described by including
the appropriate additional terms in the functional integrand. In the
same vein, we shall also neglect the quantum-mechanical imaginary time
dependence discussed in Sec. \ref{sec:freq}. Thus we shall
examine the generating functional
\begin{equation}
Z[\mu] = {\rm Det}^{1/2} \left[ -\beta \nabla^2 \right] \int [ d \phi ]
\exp\left\{ -S[\phi; \mu ] \right\} ,
\end{equation}
in which
\begin{equation}
S[\phi; \mu ] = {\beta \over 2} \int (d^\dim\r)
   \phi(\r) \left( - \nabla^2 \right) \phi(\r)
+ S_{\rm int}[\phi; \mu ] \,,
\end{equation}
where
\begin{equation}
S_{\rm int}[\phi; \mu] = - {\sum}_a \int(d^\dim\r)
 \,  \n0a(\r) \,  e^{i\beta e_a \phi(\r) } \,.
\end{equation}
With the generalized chemical potential functions $\mu_a(\r)$
taken to be constants, the generating functional $Z[\mu]$ reduces to
the grand canonical partition function. Functional derivatives with
respect to the generalized chemical potentials, with these potentials
then taken to be constants, yield number density correlation
functions.

\subsection{Perturbation Theory}

Perturbation theory developments of correlation functions can be done
in essentially either of two ways: One can first perform the functional
derivatives with respect to the generalized chemical potentials to
bring down extra factors in the functional integrand that represent
the particle densities and then set the
chemical potentials to be constants and make a perturbative expansion.
Or one can make a formal perturbative expansion of the functional
integral with spatially varying, generalized chemical potentials
$\mu_a(\r)$, and
then expand the result in a spatial varying part of the chemical
potentials to identify the correlation functions.  Either case is
subsumed in a slight generalization of the second way in which we
write the functional integration field as $ \phi(\r) = \bar
\phi(\r) + \phi'(\r)$, where $\bar\phi(\r)$ is some
suitable background field.  We then take out and
explicitly display the pieces of zeroth and second order in the
fluctuation field $\phi'(\r)$. Since the background field $\bar\phi$
appears as a `constant' translation in the (dummy) functional
integration variable, $[d\phi] = [d\phi']$, and so with this separation
and with an operator or infinite matrix notation,
\begin{eqnarray}
Z[\mu] &=& {\rm Det}^{1/2} \left[ -\beta \nabla^2 \right] \,
\exp\left\{ -S[\bar\phi; \mu ] \right\}
\nonumber\\
&&
\int [ d \phi' ] \exp\left\{ - (\beta / 2 ) \phi' \left[ - \nabla^2 +
    {\cal V}(\bar\phi;\mu) \right] \phi' \right\} F[\phi'] \,.
\label{zzz}
\end{eqnarray}
Here
\begin{equation}
 {\cal V}(\bar\phi;\mu;\r) =  {\sum}_a \beta e_a^2
\,   \n0a(\r) \, e^{i\beta e_a \bar\phi(\r) } \,,
\label{calv}
\end{equation}
and
\begin{equation}
F[\phi'] =
\exp\left\{ - \tilde S_{\rm int}[\phi'; \bar\phi, \mu ] \right\}
\cdots \,,
\label{ellip}
\end{equation}
where $ - \tilde S_{\rm int}[\phi'; \bar\phi ,\mu ] $ contains the
linear, cubic, and higher order terms in $\phi'$ in the exponential.
The ellipsis $\cdots$ stands for possible insertions in the integrand
of the factors of the form $ \n0a e^{i\beta e_a \phi(\r) } $ that
result in the first case above when functional derivative are first
taken to construct correlation functions. We shall soon work out
explicit examples that should make this perhaps somewhat abstract
formulation clear.

To obtain the perturbative development, we first note that by completing the
square to obtain a Gaussian functional integral which produces an
infinite, Fredholm determinant, we have, using again an operator
notation,  the evaluation:
\begin{eqnarray}
X[\zeta] &=&
{\rm Det}^{1/2} \left[ \beta (- \nabla^2 ) \right]
\int [d \phi'] \exp\left\{ - (\beta / 2 ) \phi' \left[ - \nabla^2 +
    {\cal V}(\bar\phi;\mu) \right] \phi' + i \phi' \zeta \right\}
\nonumber\\
&=& {\rm Det}^{1/2} \left[ \beta (- \nabla^2 ) \right]
{\rm Det}^{-1/2} \left[ \beta (- \nabla^2 + {\cal V} ) \right]
\exp\left\{ - ( 1 / 2 \beta ) \zeta G \zeta \right\}
\nonumber\\
&=& {\rm Det}^{-1/2} \left[ 1 + {1 \over - \nabla^2}  {\cal V} \right]
 \exp\left\{ - { 1 \over 2 \beta} \int (d^\dim\r) (d^\dim\r' )
\zeta(\r) G(\r ,\r') \zeta(\r') \right\} ,
\label{xxx}
\end{eqnarray}
where in the last line we have noted that the product of determinants
is the determinant of the product of operators and used ordinary
notation with $ G(\r,\r') $ the Green's function defined by
\begin{equation}
\left[- \nabla^2 + {\cal V}(\bar\phi;\mu;\r) \right]
G(\r ,\r') = \delta^{(\dim)}(\r {-} \r') \,.
\end{equation}
We next note that%
\footnote
    {
    The essence of the proof of this relation is obtained by
    replacing the functions by numbers, and by observing that
    \begin{eqnarray}
    \left. \exp\left\{ {d \over dx} g {d \over dx} \right\} \exp\left\{ ipx
      \right\} \right|_{x=0} &=& \exp\left\{ - ipx \right\}
    \exp\left\{ {d \over dx} g {d \over dx} \right\} \exp\left\{ ipx \right\}
    \nonumber\\
    &=& \exp\left\{ e^{-ipx} {d \over dx } e^{ipx} g e^{-ipx}
    {d \over dx} e^{ipx}  \right\}
    \nonumber\\
    &=&  \exp\left\{ \left( {d \over dx} +ip \right)  g
    \left( {d \over dx} + ip \right) \right\} =
     \exp\left\{ - p g p  \right\} .
    \nonumber
    \end{eqnarray}
    }
\begin{eqnarray}
&&
 \exp\left\{ - { 1 \over 2 \beta} \int (d^\dim\r) (d^\dim\r' )
\zeta(\r) G(\r ,\r') \zeta(\r') \right\} =
\nonumber\\
&& \qquad
\left. \exp\left\{ { 1 \over 2 \beta} \int (d^\dim\r)
(d^\dim\r' ) { \delta \over \delta \phi(\r) }
 G(\r,\r') {\delta \over \delta \phi(\r') } \right\}
\exp\left\{ i \int (d^\dim\r) \phi(\r) \zeta(\r) \right\}
\right|_{\phi = 0} \,.
\end{eqnarray}
Hence the functional integral (\ref{xxx}) defining $X[\zeta]$
may instead be replaced by an exponential functional derivative
operation which, in the operator notation, reads
\begin{equation}
X[\zeta] = {\rm Det}^{-1/2} \left[ 1 + {1 \over - \nabla^2}  {\cal V} \right]
\left. \exp\left\{ { 1 \over 2 \beta}  { \delta \over \delta \phi' }
 G {\delta \over \delta \phi' } \right\}
\exp\left\{ i \phi' \zeta \right\} \right|_{\phi' = 0} \,.
\end{equation}
Now the functional $X[\zeta]$ defined by the functional integral (\ref{xxx})
has precisely the same form as the functional integral (\ref{zzz})
defining the thermodynamic generating functional $Z[\mu]$ except
that $ F[\phi'] $ is replaced by the functional Fourier transform
factor
$ \exp\{ i \phi' \zeta \} $.
Since this functional Fourier transform
factor can be used to generate any functional, we conclude that
\begin{eqnarray}
Z[\mu] &=&
{\rm Det}^{-1/2} \left[ 1 + {1 \over - \nabla^2}  {\cal V} \right]
\exp\left\{ -S[\bar\phi; \mu ] \right\}
\nonumber\\
&&
\left. \exp\left\{ { 1 \over 2 \beta} \int (d^\dim\r)
(d^\dim\r' ) { \delta \over \delta \phi'(\r) }
 G(\r,\r') {\delta \over \delta \phi'(\r') } \right\}
F[\phi'] \right|_{\phi' = 0} \,.
\label{funfun}
\end{eqnarray}
This is the functional form than lends itself to a perturbative
development by expanding the exponential of the functional
derivatives.  Performing the functional derivatives produces the
``Wick contractions'' that are familiar in quantum field theory and
lead to the familiar graphical representation. The exact analytic
form with the proper numerical factors associated with a given graph
is easily obtained from the expansion of the exponential and the
operation of the functional derivatives.%
\footnote
    {
    The functional
    derivatives may be viewed as `pacmen' that eat up fields sprouting
    from vertices with each pair of devoured fields connected by a line
    that represents the Green's function $ G(\r,\r')$.
    }

\subsection{Straightforward Expansions}

We consider the ordinary partition function in which all the
chemical potentials are constants. This will illustrate the use of
this functional derivative formulation in a straightforward fashion.
In this case we take the background field to vanish, $\bar\phi = 0 $,
and so the quadratic part of the action involves the lowest-order
Debye (squared) wave number
\begin{equation}
\kappa_0^2 = \beta \sum_{a=1}^A e_a^2 \n0a \,,
\end{equation}
with ${\cal V} \to \kappa_0^2 $, and so the Green's function reduces to
the Debye Green's function $G_\dim(\r {-} \r') $  that was defined
previously in Eq.~(\ref{frep}). With constant chemical
potentials and with $\bar\phi =0$,
\begin{equation}
- S[0;\mu] = {\sum}_a \n0a \int (d^\dim \r) \,,
\end{equation}
while the determinantal prefactor reduces to that evaluated previously
in Eq.~(\ref{dett}) of the text,
\begin{equation}
{\rm Det}^{-1/2} \left[ 1 +  { 1 \over - \nabla^2 } \kappa_0^2
\right] =
 \exp\left\{ - G_\dim( {\bf 0} ) { 1 \over \dim}
\kappa_0^2 \int (d^\dim\r) \right\} .
\end{equation}
Therefore, the first two factors in the general perturbative formula
(\ref{funfun}) yield the partition function valid up to one-loop
order,
\begin{equation}
Z_1 = \exp\left\{ \left[ \sum \n0a  -  G_\dim( {\bf 0} )
{ \kappa_0^2 \over \dim}\right] \int (d^\dim\r) \right\} ,
\end{equation}
the result (\ref{zzone}) in the text, and hence
\begin{equation}
Z[\mu] =  Z_1
\left. \exp\left\{ { 1 \over 2 \beta} \int (d^\dim\r)
(d^\dim\r' ) { \delta \over \delta \phi(\r) }
 G_\dim(\r{-}\r') {\delta \over \delta \phi(\r') } \right\}
\exp\left\{ - \tilde S_{\rm int}[\phi; \mu ] \right\}
\right|_{\phi = 0} \,.
\label{pertresl}
\end{equation}
Here
\begin{equation}
- \tilde S_{\rm int}[\phi; \mu ] = \int (d^\dim \r)
{\sum}_a \, \n0a \left\{ e^{i\beta e_a \phi(\r) } - 1 + {1 \over
    2} \left[ \beta e_a \phi(\r) \right]^2 \right\} .
\label{pertint}
\end{equation}

As a first application of of this method, we derive the result
(\ref{phione}) for $ \avg {\phi}^{(1)} $ given in the text. To all
orders
\begin{eqnarray}
\avg { \phi({\bf 0}) } &=& Z[\mu]^{-1} Z_1
 \exp\left\{ { 1 \over 2 \beta} \int (d^\dim\r_1)
(d^\dim\r_2 ) { \delta \over \delta \phi(\r_1) }
 G_\dim(\r_1-\r_2) {\delta \over \delta \phi(\r_2) } \right\}
\nonumber\\ && \qquad \left. \Bigg\{ \phi({\bf 0}) \exp\left\{ -
\tilde S_{\rm int}[\phi; \mu ] \right\} \Bigg\} \right|_{\phi = 0} \,.
\end{eqnarray}
To the desired one-loop order, with the linear coupling
to $\phi$ counted as itself of one-loop order as explained in the
discussion of Eq.~(\ref{phione}) of the text,
\begin{eqnarray}
\avg {\phi({\bf 0}) }^{(1)} &=&
 \exp\left\{ { 1 \over 2 \beta} \int (d^\dim\r_1)
(d^\dim\r_2 ) { \delta \over \delta \phi(\r_1) }
 G_\dim(\r_1-\r_2) {\delta \over \delta \phi(\r_2) } \right\}
\nonumber\\
&& \qquad\qquad
\left. \Bigg\{ \phi({\bf 0}) \sum_{a=1}^A \n0a \int(d^\dim\r) \left[
	(i\beta e_a \phi(\r))
	+
	\coeff 1{3!}
	(i\beta e_a \phi(\r))^3 \right] \Bigg\}
\right|_{\phi = 0}
\nonumber\\
&=&
\left\{ { 1 \over 2 \beta} \int (d^\dim\r_1)
(d^\dim\r_2 ) { \delta \over \delta \phi(\r_1) }
 G_\dim(\r_1-\r_2) {\delta \over \delta \phi(\r_2) }
\right\}
\nonumber\\
&& \qquad\qquad
\phi({\bf 0}) \sum_{a=1}^A \n0a \int(d^\dim\r)
	(i\beta e_a \phi(\r))
\nonumber\\
&& +
{ 1 \over 2}
\left\{ { 1 \over 2 \beta} \int (d^\dim\r_1)
(d^\dim\r_2 ) { \delta \over \delta \phi(\r_1) }
 G_\dim(\r_1-\r_2) {\delta \over \delta \phi(\r_2) } \right\}^2
\nonumber\\
&& \qquad\qquad
\phi({\bf 0}) \sum_{a=1}^A \n0a \int(d^\dim\r)
	\coeff 1{3!}
	(i\beta e_a \phi(\r))^3
\nonumber\\
&=& i \int (d^\dim\r) G_\dim( {\bf 0} - \r )
 \sum_{a=1}^A e_a \n0a \left[ 1 - {1 \over 2} \beta e_a^2 G_\dim({\bf 0})
     \right] .
\end{eqnarray}
In view of the Fourier representation (\ref{frep}) of the Debye
Green's function,
\begin{equation}
\int (d^\dim\r) G_\dim( {\bf 0} - \r ) = { 1 \over \kappa_0^2
  } \,,
\end{equation}
and so
\begin {equation}
    \avg {\phi}^{(1)} = { i \over \kappa_0^2 }
 \sum_{a=1}^A e_a \n0a \left[ 1 - {1 \over 2} \beta e_a^2 G_\dim({\bf 0})
     \right] ,
\label{phionee}
\end{equation}
which is just the result (\ref{phione}) of the text.

The perturbative expansions of the density and density--density
correlations discussed in the text follow from
\begin{eqnarray}
\avg { n_a(\r) } &=& Z[\mu]^{-1} Z_1
 \exp\left\{ { 1 \over 2 \beta} \int (d^\dim\r_1)
(d^\dim\r_2 ) { \delta \over \delta \phi(\r_1) }
 G_\dim(\r_1-\r_2) {\delta \over \delta \phi(\r_2) } \right\}
\nonumber\\
&& \qquad
\left. \Bigg\{ \n0a \, e^{i\beta e_a \phi(\r) } \,
\exp\left\{ - \tilde S_{\rm int}[\phi; \mu ] \right\} \Bigg\}
\right|_{\phi = 0} \,,
\end{eqnarray}
and
\begin{eqnarray}
\avg { n_a(\r) n_b(\r') } &=& Z[\mu]^{-1} Z_1
 \exp\left\{ { 1 \over 2 \beta} \int (d^\dim\r_1)
(d^\dim\r_2 ) { \delta \over \delta \phi(\r_1) }
 G_\dim(\r_1-\r_2) {\delta \over \delta \phi(\r_2) } \right\}
\nonumber\\
&& \quad
\left. \Bigg\{ \n0a \, e^{i\beta e_a \phi(\r) }
\n0b \,\, e^{i\beta e_b \phi(\r') }  \,
\exp\left\{ - \tilde S_{\rm int}[\phi; \mu ] \right\} \Bigg\}
\right|_{\phi = 0} \,.
\end{eqnarray}

\subsection{Loop Parameter}
\label{loopy}

It was emphasized in the text that the size of loop corrections is
measured, in $\dim$ dimensions, by the dimensionless parameter $ \beta
e^2 G_\dim({\bf 0}) \sim \beta e^2 \kappa_0^{\dim-2}$, which reduces to $
\beta e^2 \kappa_0$ in three dimensions: A perturbative term
corresponding to a graph containing $\ell$ loops is of order $ [ \beta
e^2 \kappa_0^{\dim-2} ]^\ell $ , or in three dimensions, $ [ \beta e^2
\kappa_0]^\ell $. That is, the power of $[\beta e^2 \kappa_0]$ counts
the loop order of the expression. It should be noted that these loop
graphs are connected and single-particle irreducible.  In this
counting, $e^2$ denotes a generic, typical charge of any of the
particle species, or, equivalently, one could write $ e_a = Z_a e $,
and $e$ is the electron charge. Here we shall sketch the proof of this
assertion.

To do this, we examine the expansion of the perturbative
formula (\ref{pertresl}) in powers of the unperturbed densities, which
we write in the schematic form
\begin{eqnarray}
Z_N[\mu] &\sim& { 1 \over N!}
 \exp\left\{ { 1 \over 2 \beta} \int (d^\dim\r)
(d^\dim\r' ) { \delta \over \delta \phi(\r) }
 G_\dim(\r{-}\r') {\delta \over \delta \phi(\r') } \right\}
\nonumber\\
&&  \left.
\int (d^\dim \r_1) \, n^0(\r)  e^{i\beta e \phi(\r_1) }
\int (d^\dim \r_2) \, n^0(\r)  e^{i\beta e \phi(\r_2) }
\cdots
\int (d^\dim \r_N) \, n^0(\r) e^{i\beta e \phi(\r_N) }
\right|_{\phi = 0} \,.
\nonumber\\
&&
\label{schem}
\end{eqnarray}
This corresponds to a graph with $N$ vertices. Functional derivatives
with respect to $\mu(\r)$ may be taken to give number density
correlators. We have omitted the subtraction of the unit and $\phi^2$
terms which appear in the interaction part of the action
(\ref{pertint}), which we may do with the understanding that at least
three $\phi$ functional derivatives are taken at each vertex or that
each vertex emits at least three propagator lines.

Let us first assume that functional derivatives have been taken so
that each vertex is connected by a single propagator line. At this
stage, we have a graph which is a polygon with $N$ lines and $N$
vertices. Note that since our counting applies to connected,
single-particle irreducible graphs, any of these graphs must have such
a perimeter polygon. To exhibit the parameters, we introduce
dimensionless spatial coordinates by writing $ \r = {\bf q} /
\kappa_0$. Then the propagator $G_\dim(\r {-} \r')$ becomes
$\kappa_0^{\dim-2}$ times a dimensionless function of the dimensionless
variable $({\bf q} - {\bf q}')$. The functional derivative operations
in Eq.~(\ref{schem}) produce a propagator line times $\beta^{-1}$ with
a factor of $\beta e$ at each end of the line. Thus for each
propagator line and the vertex factors associated with both ends of
the line we have an overall factor of $\beta e^2 \kappa_0^{\dim-2}$. Each
vertex involves $\int (d^\dim\r) \sim \kappa_0^{-\dim} $ times $n^0$
(a dimensionless product), and so, all together for our skeleton
polygon, we have $N$ factors of $ \beta e^2 n^0 \kappa_0^{-2} $. But $
\beta e^2 n^0 \sim \kappa_0^2 $, and so these are just $N$ factors of
$1$. If we measure the size of this skeleton one-loop graph in term of
the unperturbed grand potential $ \beta \Omega \sim \int (d^\dim\r)
n^0 $, then one factor of $\beta e^2 \kappa_0^{\dim-2}$ remains to
characterize the order of the one-loop graph.

The remainder of the proof is now trivial. Each additional propagator
line added to the skeleton polygon gives a factor $\beta e^2
\kappa_0^{\dim-2}$ and increases the number of loops by one.

\section{Calculations Using Functional Methods}
\label{funcal}

We turn now to apply the functional methods
using the alternative background field method mentioned in the
preceding Appendix.  We choose the background field $\bar\phi(\r)$
used there to be the solution
$\phi_{\rm cl}(\r)$ of the classical field equation of the total
action which now contains a source:
\begin{eqnarray}
S_{\rm tot}[\phi;\sigma;\mu] &=& S[\phi;\mu] - \beta \int (d^\dim\r)
\phi(\r) \sigma(\r)
\nonumber\\
&=& \int (d^\dim\r) \left\{ {\beta \over 2}
\phi \left( - \nabla^2 \right) \phi - {\sum}_a n_a(\r)
\exp\{ i \beta e_a \phi \} - \beta \phi \sigma \right\} .
\label{totdef}
\end{eqnarray}
Thus $\phi_{\rm cl}(\r)$ is defined by
\begin{equation}
- \nabla^2 \phi_{\rm cl}(\r) = i \, {\sum}_a e_a
   n_a^0(\r) \exp\{ i \beta e_a \phi(\r) \} +
     \sigma(\r) \,.
\label{pcldef}
\end{equation}
This choice is made because, since the action is stationary for
variations about the solution of the classical field equation, with
$\phi = \phi_{\rm cl} + \phi'$, there are no linear terms in the
fluctuation field $\phi'$ and the result (\ref{funfun}) of the
previous Appendix takes the form
\begin{eqnarray}
\exp\{ W[\sigma;\mu] \} &=&
{\rm Det}^{-1/2} \left[ 1 + {1 \over - \nabla^2}
{\cal V}[\phi_{\rm cl};\mu] \right]
\exp\left\{ -S_{\rm tot}[\phi_{\rm cl} ; \sigma ; \mu ] \right\}
\nonumber\\
&&
\left. \exp\left\{ { 1 \over 2 \beta} \int (d^\dim\r)
(d^\dim\r' ) { \delta \over \delta \phi'(\r) }
 G(\r,\r') {\delta \over \delta \phi'(\r') } \right\}
\exp\{ - \tilde S_{\rm int}[\phi'; \phi_{\rm cl},\mu] \}
 \right|_{\phi' = 0} \,.
\nonumber\\
&&
\label{funform}
\end{eqnarray}
Here
\begin{equation}
 {\cal V}(\phi_{\rm cl};\mu;\r) =  {\sum}_a \beta e_a^2
\, \n0a(\r ) \, e^{i\beta e_a \phi_{\rm cl}(\r) } \,,
\label{calvdef}
\end{equation}
with
\begin{equation}
\left\{ - \nabla^2 +  {\cal V}(\phi_{\rm cl};\mu;\r) \right\}
 G(\r,\r') = \delta ( \r {-} \r') \,,
\label{ggdef}
\end{equation}
and
\begin{eqnarray}
 - \tilde S_{\rm int}[\phi'; \phi_{\rm cl},\mu] &=&
 \int (d^\dim\r) {\sum}_a \, \n0a(\r) \,
e^{i\beta e_a \phi_{\rm cl}(\r) }
\left\{ e^{i\beta e_a \phi'(\r) } - 1 - i\beta e_a \phi'(\r)
+ { 1 \over 2} \left[ \beta e_a \phi'(\r) \right]^2 \right\} \,.
\nonumber\\
&&
\label{sintt}
\end{eqnarray}

\subsection{Results Through One Loop}
\label{ffunone}

We shall make use of this general result in the next section where
the two-loop correction will be evaluated. Here we note that action of
the exponential of functional derivatives on the exponential of
$\tilde S_{\rm int} $ produces only two and higher order loops since
$\tilde S_{\rm int} $ contains no linear terms in $\phi'$. Hence, to
the one-loop order with which we are concerned here, we have
\begin{equation}
 W[\sigma;\mu] =  W^{(1)}[\sigma;\mu] =
 -S_{\rm tot}[\phi_{\rm cl} ; \sigma ; \mu ] - {1 \over 2} \ln {\rm Det}
 \left[ 1 + {1 \over - \nabla^2}  {\cal V}[\phi_{\rm
      cl};\mu] \right] .
\label{onea}
\end{equation}
To obtain the effective action described in Appendix \ref{funmeth},
we first need the relation between
\begin{equation}
\langle \phi(\r) \rangle_\beta^\sigma = { \delta W[\sigma;\mu]
  \over \delta \beta \sigma(\r) }
\end{equation}
and $\phi_{\rm cl}(\r)$. Since the action
$S_{\rm tot}[\phi_{\rm cl} ; \sigma ; \mu ] $ is stationary for
field variations about $\phi_{\rm cl}$, the induced variation in $\phi_{\rm
  cl}$ when the source $\sigma$ is varied does not contribute to
$\delta S_{\rm tot}$ and only the explicit source variation contributes,
giving
\begin{equation}
- { \delta S_{\rm tot}[\phi_{\rm cl} ; \sigma ; \mu ] \over \delta
  \beta\sigma(\r) } = \phi_{\rm cl}(\r) \,.
\end{equation}
This is the classical or tree contribution. Using the formula $\delta
\ln {\rm Det} X = {\rm Tr} X^{-1} \delta X $, the one loop contribution
is contained in
\begin{eqnarray}
\delta \ln {\rm Det} \left[ 1 + {1 \over - \nabla^2}  {\cal V}[\phi_{\rm
      cl};\mu] \right] &=& {\rm Tr}
\left[ 1 + {1 \over - \nabla^2} {\cal V}[\phi_{\rm cl};\mu]
\right]^{-1} {1 \over - \nabla^2}  \delta {\cal V}[\phi_{\rm cl};\mu]
\nonumber\\
 &=& {\rm Tr} \left[ - \nabla^2 +  {\cal V}[\phi_{\rm cl};\mu]
\right]^{-1}  \delta {\cal V}[\phi_{\rm cl};\mu]
\nonumber\\
&=& \int (d^\dim\r) G(\r,\r) i \beta g_3(\r) \,
\delta \phi_{\rm cl}(\r) \,,
\label{dlndet}
\end{eqnarray}
where
\begin{equation}
 g_3(\r) =
{\sum}_a \beta e_a^3
\, \n0a(\r) \,
e^{i\beta e_a \phi_{\rm cl}(\r) } \,.
\end{equation}
The variation of the equation (\ref{pcldef}) defining the classical
solution yields
\begin{equation}
\left\{ - \nabla^2 +  {\cal V}(\phi_{\rm cl};\mu;\r) \right\}
 \delta \phi_{\rm cl}(\r)  = \delta \sigma(\r) \,,
\end{equation}
and so
\begin{equation}
{\delta \phi_{\rm cl}(\r) \over \delta \sigma(\r') }
= G(\r, \r') \,.
\end{equation}
Hence, to one-loop order,
\begin{equation}
\langle \phi(\r) \rangle^{(1)} =
{ \delta W^{(1)}[\sigma;\mu]
  \over \delta \beta \sigma(\r) }
= \phi_{\rm cl}(\r) + \Delta\phi(\r) \,,
\label{funphi}
\end{equation}
where
\begin{equation}
\Delta\phi(\r) = - { i \over 2}
\int (d^\dim\r') G(\r, \r') g_3(\r')
 G(\r',\r') \,.
\label{dphidef}
\end{equation}
The correction $ \Delta\phi(\r) $ is a one-loop contribution
corresponding to a ``tadpole'' graph. This graph is just the same as
the second graph in Fig.~\ref{fig:tadpole} except that now the vertex
is given by $g_3(\r')$ and the lines represent the Green's
function $ G(\r,\r')$. With the source vanishing, and with $ \sum e_a
n_a^0 $ taken to be of one-loop order, $ \phi_{\rm cl} =
(i / \kappa_0^2) \sum e_a n_a^0 $ to this order. Thus, to one-loop
order, $\phi_{\rm cl}$ can be taken to vanish in the explicitly
one-loop term $\Delta \phi$, and the Green's functions there can be
replaced by the Debye functions. With these remarks in mind, it is
easy to check that general one-loop result (\ref{funphi}) reduces to
the previous result (\ref{phionee}).

The effective potential is taken to be a functional of the field
expectation value which we relabel as $\bar\phi(\r)$.  The
one-loop action (\ref{onea}) is a functional of the classical field
$\phi_{\rm cl}(\r) $ which differs from the expectation value by
the one-loop correction $ \Delta\phi(\r) $. Since the classical
action is stationary for variations about $\phi_{\rm cl}(\r)$,
replacing $\phi_{\rm cl}(\r)$ in it by $\bar\phi(\r)$
entails a correction involving $ \Delta\phi(\r)^2 $, which is of
two-loop order. Since the determinantal contribution is already of
first order, replacing $\phi_{\rm cl}(\r)$ in it by
$\bar\phi(\r)$ also gives a two-loop correction. Thus, to
one-loop order, we may replace $\phi_{\rm cl}(\r)$ by
$\bar\phi(\r)$ in the action functional (\ref{onea}). [The
explicit form for $\Delta\phi$ given in the previous
paragraph is, of course, not needed to reach this conclusion. We made
this explicit calculation because the result will be used in the next
section on the two-loop effective action.] The effective action for
the time-independent field $\bar\phi$ is defined by simply restricting
the Legendre transformation (\ref{gdef}) to involve time-independent
quantities so that the imaginary time integral is replaced by a factor
of $\beta$. The source--field product in the Legendre transformation
cancels the source term in the relation (\ref{totdef}) between $S_{\rm
  tot}[\phi;\sigma\mu]$ and $S[\phi;\mu]$, and we have to
one-loop order
\begin{equation}
\Gamma^{(1)}[\bar\phi;\mu] =  S[\bar\phi; \mu ]
+ {1 \over 2} \ln {\rm Det} \left[ 1 + {1 \over - \nabla^2}
{\cal V}[\bar\phi;\mu] \right] .
\end{equation}

It proves convenient to rewrite the determinant in the form:
\begin{eqnarray}
{\rm Det} \left[ 1 + {1 \over - \nabla^2}  {\cal V} \right]
&=&
{\rm Det} \left[ {1 \over - \nabla^2}\left\{ - \nabla^2
+  \kappa_0^2 + \left( {\cal V} - \kappa_0^2 \right) \right\} \right]
\nonumber\\
&=&
{\rm Det} \left[ 1 + {1 \over - \nabla^2} \kappa_0^2 \right]
{\rm Det} \left[ 1 + G_\dim \left({\cal V} - \kappa_0^2 \right) \right]
\,.
\label{ddett}
\end{eqnarray}
Here we have added and subtracted the (squared) Debye wave number
$\kappa_0^2$ for the (lowest-order) densities when the generalized
chemical potentials reduce to the their standard, spatially uniform
form, $\mu_a(\r) \to \mu_a $, and $G_\dim$ is the Debye Green's
function for this wave number. The first factor in Eq.~(\ref{ddett})
is the one-loop correction to the standard partition function; in the
limit in which the generalized chemical potentials become constant
and $\bar\phi = 0$, ${\cal V} = 0 $, and we see that since
\begin{equation}
\left. S[\bar\phi;\mu] \right|_0 = - \int (d^\dim\r) \n0a \,,
\end{equation}
the grand partition function to one-loop order is given by
\begin{eqnarray}
\ln Z_1 &=& - \left. \Gamma^{(1)}[\bar\phi;\mu] \right|_0
\nonumber\\
&=&  \n0a \vol - {1 \over 2} \ln
{\rm Det} \left[ 1 + {1 \over - \nabla^2} \kappa_0^2 \right] ,
\end{eqnarray}
in agreement with Eq.~(\ref{eq:Z1}).

To compute the number densities and number-density correlation
functions to one-loop order, we first note that
\begin{eqnarray}
\ln {\rm Det} \left[ 1 + G_\dim \left({\cal V} - \kappa_0^2 \right) \right]
&=& {\rm Tr} \ln \left[ 1 + G_\dim \left({\cal V} - \kappa_0^2 \right) \right]
\nonumber\\
&=& {\rm Tr} G_\dim \, \left({\cal V} - \kappa_0^2 \right) - { 1 \over 2}
{\rm Tr} \left[ G_\dim \left({\cal V} - \kappa_0^2 \right) \right]^2 +
\cdots
\nonumber\\
&=& G_\dim({\bf 0}) \int (d^\dim\r_1) \left( {\cal V}(\bar\phi;\mu;\r_1)
       - \kappa_0^2 \right)
\nonumber\\
 - { 1 \over 2} \int (d^\dim\r_1) (d^\dim\r_2)
\!\!\!&\!\!&\!\!\! \left[ G_\dim( \r_1 - \r_2 ) \right]^2
\left( {\cal V}(\bar\phi;\mu;\r_1 ) -\kappa_0^2 \right)
\left( {\cal V}(\bar\phi;\mu;\r_2) -\kappa_0^2 \right)
\nonumber\\
&& \qquad + \cdots \,.
\label{trexpan}
\end{eqnarray}
Since $ \left( {\cal V}(\bar\phi;\mu;\r_2) -\kappa_0^2 \right)
\to 0 $ when the generalized chemical potentials take on their constant
values and the field $\bar\phi$ vanishes, only the first term on the
right-hand side of the last equality contributes to the number density
which involves a single functional derivative before this limit is
taken, only the first two terms contribute to the number density
correlation function, and so forth for the higher correlators.

To compute the density -- chemical potential relation to one-loop
order, we note that
since $ {\cal V}(\bar\phi;\mu;\r_2) $ is related to
  $\n0a(\r)$ by Eq.~(\ref{calvdef}) and
\begin{equation}
{\delta \n0b(\r') \over \delta \beta \mu_a(\r) }
= \delta_{ab} \delta ( \r {-} \r' ) \n0a(\r) \,,
\end{equation}
it is easy to compute
\begin{equation}
\langle n_a \rangle_\beta^{(1)} =
\left. { \delta \Gamma[\bar\phi;\mu] \over \delta \beta
    \mu_a(\r) } \right|_0 = \n0a \left[ 1 - { 1 \over 2}
\beta e_a^2 G_\dim({\bf 0}) \right] ,
\end{equation}
which is the result (\ref{denone}) in the text.

As discussed in Appendix \ref{funmeth}, the density-density
correlation function is determined by
\begin{equation}
C_{ab}(\r {-} \r') = - \left.
{ \delta \over \delta \beta \mu_a(\r) }
{ \delta \over \delta \beta \mu_b(\r') }
\Gamma[\bar\phi;\mu] \right|_0 \,.
\end{equation}
In the leading, tree approximation,
\begin{equation}
C_{ab}^{\rm tree}(\r {-} \r') = - \left.
{ \delta \over \delta \beta \mu_a(\r) }
{ \delta \over \delta \beta \mu_b(\r') }
S[\bar\phi;\mu] \right|_0 = \delta_{ab} \, \n0a \, \delta ( \r - {\bf
r}' ) \,.
\end{equation}
The correction arising from the first trace term in
Eq.~(\ref{trexpan}) is just to replace the lowest-order, chemical
potential -- density relation $\n0a$ here with the corrected
functional form $n_a^{(1)}$. Of course, to whatever order we work,
at the end we replace the chemical potential -- density relation by the
actual densities $\bar n_a$. The contribution from the second trace in
Eq.~(\ref{trexpan}) is obtained with the same ingredients used in the
number density evaluation, and we find that, through one-loop order,
\begin{equation}
C_{ab}^{(1)}(\r {-} \r')  = \delta_{ab} \, \bar n_a \,
\delta ( \r {-} \r' ) + { 1 \over 2 } \, \beta e^2_a \, \bar
n_a \, G_\dim(\r {-} \r' )^2 \, \beta e^2_b \, \bar n_b \,.
\end{equation}
This is precisely Eq.~(\ref{cone}) of the text.

\subsection{Two-Loop Effective Action}
\label{twoloopyy}

We turn now to compute the effective action to two-loop order. Before
obtaining the terms that contribute to two-loop order, it is
instructive to examine some two-loop order terms that cancel among
themselves. As was discussed in Appendix \ref{funmeth},
the effective action is single
particle irreducible. We can now see explicitly how this works out to
the two-loop order. In the preceding section, we noted that the
replacement of $\phi_{\rm cl}$ by
$
\bar\phi = \phi_{\rm cl} +  \Delta \phi
$
entailed two-loop corrections. First we note that, since the
action is stationary for variations about the classical
solution, we have, to order $\Delta\phi^2$,
\begin{eqnarray}
S_{\rm tot}[\bar\phi;\sigma;\mu] &\simeq&
S_{\rm tot}[\phi_{\rm cl};\sigma;\mu]
+ {1 \over2}\int (d^\dim\r) (d^\dim\r')
\left. { \delta^2 S_{\rm tot}[\phi;\sigma;\mu] \over \delta
  \phi(\r) \delta\phi(\r') } \right|_{\phi_{\rm cl}}
\Delta\phi(\r) \Delta\phi(\r') \,.
\end{eqnarray}
Since $\Delta \phi$ is already of one-loop order, we can replace
$\phi_{\rm cl}$ by $\bar\phi$ in the second term here.
Since the second variation of the classical action brings in $ {\cal
  V}(\bar\phi;\mu;\r) $ and produces the inverse Green's function,
to two-loop order,
\begin{equation}
S_{\rm tot}[\phi_{\rm cl};\sigma;\mu] =
S_{\rm tot}[\bar\phi_{\rm cl};\sigma;\mu]
- {\beta \over 2} \int (d^\dim\r) \Delta\phi(\r) \left[
- \nabla^2 + {\cal V}(\bar\phi;\mu;\r) \right] \Delta\phi(\r) \,.
\end{equation}
Again to two-loop order, the correction to the determinant
contribution is given, in view of Eq.~(\ref{dlndet}),  by
\begin{eqnarray}
{1\over2}\ln {\rm Det} \left[ 1 + {1 \over - \nabla^2}
{\cal V}[\phi_{\rm cl};\mu] \right] &=&
{1\over2}\ln {\rm Det} \left[ 1 + {1 \over - \nabla^2}
{\cal V}[\bar\phi;\mu] \right]
\nonumber\\
&& - {1\over2} \int (d^\dim\r) G(\r,\r) i \beta g_3(\r) \,
\Delta \phi(\r) \,,
\end{eqnarray}
We define $\Delta\Gamma[\bar\phi;\mu]$ by the sum of the two-loop
corrections which appear above. Using the definitions (\ref{dphidef})
of $\Delta\phi$ and (\ref{ggdef}) of $G(\r,\r')$, and a little
algebra, we find that
\begin{equation}
\Delta\Gamma[\bar\phi;\mu] = - { \beta \over 8}
\int (d^\dim\r) (d^\dim\r') G(\r,\r) g_3(\r)
G(\r,\r') G(\r',\r') g_3(\r') \,.
\label{dumbbell}
\end{equation}%
This corresponds to the ``dumbbell'' graph,
\begin {equation}
\begin {picture}(40,25)(-40,10)
    \thicklines
    \put(-15,20){\circle{30}}
    \put(1,20){\circle*{5}}
    \put(0,20){\line(1,0){40}}
    \put(39,20){\circle*{5}}
    \put(55,20){\circle{30}}
\end {picture}
\end {equation}%
This graph is obviously single-particle reducible. Hence it must cancel the
single-particle reducible piece of the remaining part
$\Delta\Gamma^{(2+)}[\bar\phi;\mu]$ of the effective action which we now
turn to compute.

The variational derivative expression (\ref{funform}) for
$\exp\{ W[\sigma;\mu]\} $ is a convenient tool to use to calculate this
remaining part of the effective action. For the two-loop terms of
interest, the exponential of the interaction terms (\ref{sintt})
can be approximated by
\begin{eqnarray}
\exp\{ - \tilde S_{\rm int}[\phi'; \phi_{\rm cl},\mu] \}
&\simeq&  - { \beta^4 \over 2} \left( { 1 \over 3! } \right)^2
\int (d^\dim\r_1)  g_3(\r_1) \phi'(\r_1)^3
  \int (d^\dim\r_2) g_3(\r_2) \phi'(\r_2)^3
\nonumber\\
&& \qquad\qquad
+ { \beta^3 \over 4! } \int (d^\dim\r_1) g_4(\r_1)
   \phi'(\r_1)^4 \,,
\end{eqnarray}
where we recall that (to our order)
\begin{equation}
 g_3(\r) =
{\sum}_a \beta e_a^3
\, \n0a(\r) \,
e^{i\beta e_a \bar\phi(\r) } \,.
\end{equation}
and define
\begin{equation}
 g_4(\r) =
{\sum}_a \beta e_a^4
\, \n0a(\r) \,
e^{i\beta e_a \bar\phi(\r) } \,.
\end{equation}
To our order of interest, the Legendre transform relation between $W$
and $\Gamma$ reduces to simply $ W = - \Gamma$, the classical action
and determinantal terms do not contribute, and Eq.~(\ref{funform})
gives
\begin{eqnarray}
- \Delta\Gamma^{(2+)}[\bar\phi;\mu] &=&
 \exp\left\{ { 1 \over 2 \beta} \int (d^\dim\r)
(d^\dim\r' ) { \delta \over \delta \phi'(\r) }
 G(\r,\r') {\delta \over \delta \phi'(\r') } \right\}
\nonumber\\
&& \qquad \Bigg\{ - { \beta^4 \over 2} \left( { 1 \over 3! } \right)^2
\int (d^\dim\r_1)  g_3(\r_1) \phi'(\r_1)^3
  \int (d^\dim\r_2) g_3(\r_2) \phi'(\r_2)^3
\nonumber\\
&& \qquad \qquad
+ { \beta^3 \over 4! } \int (d^\dim\r_1) g_4(\r_1)
   \phi'(\r_1)^4 \Bigg\} \Bigg|_{\phi' = 0} \,.
\end{eqnarray}
It is a straightforward matter to carry out the functional derivatives
and verify that one set of terms precisely cancels the previous
``dumbbell'' piece (\ref{dumbbell}). Thus, we prove explicitly to
two-loop order that the effective action functional has no
single-particle reducible terms.

The remaining terms give
\begin{eqnarray}
 \Delta\Gamma^{(2)}[\bar\phi;\mu] &=&
 { \beta \over 2}  { 1 \over 3! }
\int (d^\dim\r_1)  \int (d^\dim\r_2)  g_3(\r_1)
   G(\r_1,\r_2)^3 g_3(\r_2)
\nonumber\\
&& \qquad
- { \beta \over 4 } \int (d^\dim\r_1) g_4(\r_1)
    G(\r_1,\r_1)^2 + S^{(2)}_{\rm ind}[\bar\phi;\mu] \,,
\end{eqnarray}
where the last term stands for the two-loop contribution of the
induced interaction (\ref{Sind}) that we have belatedly added. To our
two-loop order, this additional term is given by the $p=2$ piece of
Eq.~(\ref{Sind}) evaluated in the tree approximation which replaces the
potential by its expectation value $\bar\phi$. Using the result
(\ref{eq:Zab}) for the coupling constant gives
\begin{eqnarray}
S^{(2)}_{\rm ind}[\bar\phi;\mu] &=&
{\sum}_{ab}  \, \left[ g_{ab}(\mu)
- { \pi \over 6} { 1 \over 3-\dim }
\,  \left( {\beta e_a e_b \over 4\pi } \right)^3 \right]
 \, \mu^{2(\dim - 3)}
\nonumber\\
&& \qquad\qquad
\int(d^\dim\r)
\n0a(\r) \,
\exp\left\{ i \beta e_a \bar\phi(\r) \right\}
\n0b(\r) \,
\exp\left\{ i \beta e_b \bar\phi(\r) \right\}
\nonumber\\
\nonumber\\
&& \qquad {} +
    \sum_{a} h^0_{a} \, \beta^{4}
     \int (d^\dim\r) \>
    \,\Bigl[ \nabla
	\Bigl( \mu_{a}(\r) {+}i e_{a} \bar\phi(\r) \Bigr)
    \Bigr]^2
    n_{a}^0(\r) \, e^{i\beta e_{a} \bar\phi(\r) } \,.
\end{eqnarray}
To make use of the first part of this interaction to present an
explicitly finite result in the $\dim \to 3$ limit,
we write the single particle irreducible two-loop effective
action as a sum of two parts,
\begin{equation}
 \Delta\Gamma^{(2)}[\bar\phi;\mu] =  \Delta\Gamma^{(2)}_1[\bar\phi;\mu]
 +  \Delta\Gamma^{(2)}_2[\bar\phi;\mu] \,.
\end{equation}
The first part, defined by
\begin{eqnarray}
 \Delta\Gamma^{(2)}_1[\bar\phi;\mu] &=&
 { \beta \over 2}  { 1 \over 3! }
\int (d^\dim\r_1)  \int (d^\dim\r_2)  g_3(\r_1)
  \left[ G(\r_1,\r_2)^3 -  G_3(\r_1 {-} \r_2)^3
\right] g_3(\r_2)
\nonumber\\
&& \qquad
- { \beta \over 4 } \int (d^\dim\r_1) g_4(\r_1)
    G(\r_1,\r_1)^2 \,,
\label{gamone}
\end{eqnarray}
may be evaluated
directly at $\dim=3$ since the subtraction of the cube of the
three-dimensional Debye Green's function $G_3(\r_1 {-} \r_2)$
in the first double integral renders it finite while (with dimensional
continuation) the remaining single integral is well-behaved in the
$\dim \to 3$ limit.
Making a convenient rearrangement of the remaining part gives
\begin{eqnarray}
 \Delta\Gamma^{(2)}_2[\bar\phi;\mu] &=&
 { \beta \over 2}  { 1 \over 3! }
 \int (d^\dim\r_1)  \int (d^\dim\r_2)  g_3(\r_1)
 G_\dim(\r_1 {-} \r_2)^3 \left[ g_3(\r_2) -  g_3(\r_1)
    \right]
\nonumber\\
&&
+  { \beta \over 2}  { 1 \over 3! }
\int (d^\dim\r_1)  \left[ g_3(\r_1) \right]^2
\int (d^\dim\r_2)   G_\dim(\r_1 {-} \r_2)^3  +
S^{(2)}_{\rm ind}[\bar\phi;\mu] \,.
\label{gamtwo}
\end{eqnarray}
The $\dim \to 3 $ limit may be taken in the first line on the right-hand
side of this equation since a subtraction has been make that gives an
integrable singularity when $ \r_1 = \r_2 $. The result
(\ref{d30}) gives
\begin{equation}
\int (d^\dim\r_2)   G_\dim(\r_1 {-} \r_2)^3
 = { 1 \over (4\pi)^2} {1 \over 2} \left\{ \left( { \kappa_0^2 \over
4 \pi } \right)^{\dim-3} { 1 \over 3 - \dim} + 1 - \gamma - 2 \ln 3 \right\} .
\end{equation}
Thus the pole terms on the second line on the right-hand side of
Eq.~(\ref{gamtwo}) combine the give the well-defined $ \dim \to 3$ limit
\begin{equation}
{ 1 \over 3 -\dim } \left( {\kappa_0^2 \over 4\pi} \right)^{\dim-3}
\left[ 1 - \left( {\kappa_0^2 \over 4\pi\mu^2 } \right)^{3-\dim}
\right] \to - \ln \left( {\kappa_0^2 \over 4\pi\mu^2} \right)
\,.
\end{equation}
Thus, taking the $ \dim \to 3$ limit and writing the first line in
Eq.~(\ref{gamtwo}) in a symmetrical manner, we arrive at
\begin{eqnarray}
 \Delta\Gamma^{(2)}_2[\bar\phi;\mu] &=&
-  { \beta \over 4}  { 1 \over 3! }
\int (d^\dim\r_1)  \int (d^\dim\r_2)
  G_\dim(\r_1 - \r_2)^3 \left[ g_3(\r_2) -  g_3(\r_1)
    \right]^2
\nonumber\\
&&
{\sum}_{ab}  \, \left\{ g_{ab}(\mu) - { \pi \over 6}
\,  \left( {\beta e_a e_b \over 4\pi } \right)^3
\left[ \ln \left( {\kappa_0^2 \over 4\pi \mu^2} \right) - 1
+ \gamma + 2 \ln 3 \right] \right\}
\nonumber\\
&& \qquad
 \int(d^\dim\r)
\n0a(\r) \,
\exp\left\{ i \beta e_a \bar\phi(\r) \right\}
\n0b(\r) \,
\exp\left\{ i \beta e_b \bar\phi(\r) \right\}
\nonumber\\
&&  {} +
    \sum_{a} h^0_{a} \, \beta^{4}
     \int (d^\dim\r) \>
    \,\Bigl[ \nabla
	\Bigl( \mu_{a}(\r) {+}i e_{a} \bar\phi(\r) \Bigr)
    \Bigr]^2
    n_{a}^0(\r) \, e^{i\beta e_{a} \bar\phi(\r) } \,.
\end{eqnarray}

The sum
\begin{equation}
 \Gamma[\bar\phi;\mu] \approx
  \Gamma^{(1)}[\bar\phi;\mu]  +
 \Delta\Gamma^{(2)}_1[\bar\phi;\mu]
+  \Delta\Gamma^{(2)}_2[\bar\phi;\mu]
\end{equation}
is the generating functional for all connected, single-particle
irreducible contributions through two-loop order. For example, the
double functional derivative of this result with respect to the
chemical potentials, with the chemical potentials then taken constant
and $\bar\phi = 0 $,
produces the irreducible number density correlation function $C_{ab}$
through two-loop order, the result summarized in Eq.~(\ref{genform})
of the text.  The grand potential is given by the effective action
with the generalized chemical potentials taking on constant values and
with $\bar\phi = 0$,
The two-loop contribution
to the grand potential plus the previous lower-order terms  give
\begin{equation}
\beta \Omega^{(2)}(\beta,\mu) =
\left. \Gamma^{(1)}[\bar\phi;\mu] \right|_0 +
\left. \Delta\Gamma^{(2)}_1[\bar\phi;\mu] \right|_0
+ \left. \Delta\Gamma^{(2)}_2[\bar\phi;\mu] \right|_0
\end{equation}
In this limit, the first piece of
$\Delta\Gamma^{(2)}_1[\bar\phi;\mu]$ in Eq.~(\ref{gamone}) vanishes
while the second piece involves $G_3({\bf 0})$ which has the value $ -
\kappa_0 / 4 \pi $ according to Eq.~(\ref{perf}). Moreover, in this limit,
the first line on the right-hand side of the equation above for
$\Delta\Gamma^{(2)}_2[\bar\phi;\mu] $
also vanishes. With these remarks in mind, it is a simple matter to
verify that our effective action results agree with the result
(\ref{ztwolps}) of the text.

\newpage

\ifelsevier

\else
\end {references}
\fi

\end{document}